\documentclass{cernyrep}

\usepackage{graphicx,amssymb,wasysym}
\usepackage{amsmath,amsthm,amsfonts,amssymb,multirow}
\usepackage{epsfig,cite}

\include{epsf}

\def\dbar{{/\mkern-13mu{D}}}

\def\beq{\begin{equation}}
\def\eeq{\end{equation}}
\def\bea{\begin{eqnarray}}
\def\beaa{\begin{eqnarray*}}
\def\eea{\end{eqnarray}}
\def\eeaa{\end{eqnarray*}}
\def\bq{\begin{quote}}
\def\eq{\end{quote}}
\def\gappeq{\mathrel{\rlap {\raise.5ex\hbox{$>$}}
{\lower.5ex\hbox{$\sim$}}}}
\def\lappeq{\mathrel{\rlap{\raise.5ex\hbox{$<$}}
{\lower.5ex\hbox{$\sim$}}}}
\newcommand{\beqn}{\begin{eqnarray}}
\newcommand{\eeqn}{\end{eqnarray}}
\newcommand{\bsg}{BR($b \to s \gamma$)}
\newcommand{\bmm}{BR($B_s \to \mu^+\mu^-$)}

\newcommand{\gev}{\,\, \mathrm{GeV}}
\newcommand{\Mh}{M_h}
\newcommand{\tb}{\tan\beta}

{\begin{list}{}%
         {\setlength{\leftmargin}{#1}}%
         \item[]%
}
{\end{list}}

\def\NP{ Nucl.\ Phys. }
\def\PL{ Phys.\ Lett. }
\def\PR{ Phys.\ Rev. }
\def\PRL{ Phys.\ Rev.\ Lett. }

\begin{document}

\pagestyle{plain}

\title{Beyond the Standard Model for Monta\~neros\footnote{Based on lectures by John~Ellis at the
2009 CERN--CLAF School of High-Energy Physics, Medell\'{\i}n, Colombia.}}

\author{M.~Bustamante$^1$, L.~Cieri$^2$ and J.~Ellis$^3$}

\institute{$^{1}$ Pontificia Universidad Cat\'olica del Per\'u, Lima, Peru\\
$^{2}$ Universidad de Buenos Aires, Buenos Aires, Argentina\\
$^{3}$ CERN, Geneva, Switzerland}

\maketitle

\begin{abstract}
These notes cover
(i) {\it electroweak symmetry breaking} in the Standard Model (SM) and the Higgs boson,
(ii) {\it alternatives to the SM Higgs boson} including an introduction to composite Higgs
models and Higgsless models that invoke extra dimensions, (iii) the theory and
phenomenology of {\it supersymmetry}, and
(iv) various {\it further beyond topics}, including Grand Unification, proton decay and 
neutrino masses, supergravity, superstrings and extra dimensions.
\end{abstract}

\maketitle


\section{The Standard Model, electroweak symmetry breaking and the Higgs boson}

In this first Lecture, we review the electroweak sector of the
Standard Model (SM) (for more detailed accounts, see, 
e.g.,~\cite{Hung:1980rh,Weinberg:2008zzb,Quigg:2009vq}), with
particular emphasis on the nature of electroweak symmetry breaking. The theory grew out of 
experimental information on charged-current weak interactions, and of the realisation 
that the four-point Fermi description ceases to be valid above 
$\sqrt{s} = 600$~GeV~\cite{Quigg:2009vq}. Electroweak theory was able to predict the 
existence of neutral-current interactions, as discovered by the Gargamelle
Collaboration in 1973~\cite{GGM}. One of its greatest subsequent successes was the detection in 1983 
of the $W^\pm$ and $Z^0$ bosons~\cite{Arnison:1983rp,Banner:1983jy,Bagnaia:1983zx,Rubbia:1985pv}, whose existences it had predicted. Over time, thanks to the accumulating 
experimental evidence, the $SU(2)_L \otimes U(1)_Y$ electroweak theory and $SU(3)_C$ 
quantum electrodynamics, collectively known as the Standard Model, 
have come to be regarded as 
the correct description of electromagnetic, weak and strong interactions up to the energies that 
have been probed so far. However, although the SM has many successes, it also has
some shortcomings, as we also indicate. In subsequent Lectures 
we discuss ideas for rectifying (at least some of) these defects: see
also~\cite{Ellis:1998eh,JE-SUSY,Welzel:2005cb}.

The particle content of the SM is summarized in Table \ref{TabSMPartCont}.
Within the SM, the electromagnetic and weak interactions are described by a Lagrangian that is symmetric under local weak isospin and hypercharge gauge transformations, described using the $SU(2)_L \otimes U(1)_Y$ group (the $L$ subindex refers to the fact that the weak $SU(2)$ group 
acts only the left-handed projections of fermion states; $Y$ is the hypercharge). We can write the 
$SU(2)_L \otimes U(1)_Y$ part of the SM Lagrangian as
\begin{eqnarray}\label{EqSMLag}
 \nonumber
 \mathcal{L} &=& -\frac{1}{4} \mathbf{F}_{\mu\nu}^a \mathbf{F}^{a\mu\nu} \\ \nonumber
             &+& i \overline{\psi} \dbar \psi + h.c. \\ \nonumber
             &+& \psi_i y_{ij} \psi_j \phi + h.c. \\
             &+& \lvert D_\mu \phi \rvert^2 - V\left(\phi\right) ~.
\end{eqnarray}
This is short enough to write on a T-shirt!

The first line is the kinetic term for the gauge sector of the electroweak theory, 
with $a$ running over the total number of gauge fields: 
three associated with $SU(2)_L$, which we shall call $B_\mu^1$, $B_\mu^2$, $B_\mu^3$, 
and one with $U(1)_Y$, which we shall call $\mathcal{A}_\mu$. Their field-strength tensors are
\begin{eqnarray}
 F^a_{\mu\nu} &=& \partial_\nu B_\mu^a - \partial_\mu B_\nu^a + g \varepsilon_{bca} B_\mu^b B_\nu^c  \; \; {\rm for~a = 1,2, 3}
 \label{EqFSTSU2} \\
 f_{\mu\nu} &=& \partial_\nu \mathcal{A}_\mu - \partial_\mu \mathcal{A}_\nu \label{EqFSTU1} ~.
\end{eqnarray}
In Eq.~(\ref{EqFSTSU2}), $g$ is the coupling constant of the weak-isospin group $SU(2)_L$, 
and the $\varepsilon_{bca}$ are its structure constants. The last term in this equation stems 
from the non-Abelian nature of $SU(2)$. At this point, all of the gauge 
fields are massless, but we will see later that specific linear combinations of the four
electroweak gauge fields acquire masses 
through the Higgs mechanism.  

The second line in Eq.~(\ref{EqSMLag}) describes the interactions between the matter fields 
$\psi$, described by Dirac equations, and the gauge fields. 

The third line is the Yukawa sector and incorporates the interactions between the matter fields 
and the Higgs field, $\phi$, which are responsible for giving fermions their masses when 
electroweak symmetry breaking occurs. 

The fourth and final line describes the scalar or Higgs sector. The first piece is the
kinetic term with the covariant derivative defined here to be
\begin{equation}
 D_\mu = \partial_\mu + \frac{ig^\prime}{2} \mathcal{A}_\mu Y + \frac{ig}{2} 
 \mathbf{\tau} \cdot \mathbf{B}_\mu ~,
\end{equation}
where $g^\prime$ is the $U(1)$ coupling constant, and $Y$ and 
$\mathbf{\tau} \equiv \left(\tau_1, \tau_2, \tau_3\right)$ (the Pauli matrices) are, respectively, the generators of $U(1)$ and $SU(2)$. The second piece of the final line of (\ref{EqSMLag}) is
the Higgs potential $V\left(\phi\right)$. 

{\it Whereas the first two lines of (\ref{EqSMLag}) have been confirmed in many different
experiments, there is {\bf no} experimental evidence for the last two lines and one
of the main objectives of the LHC is to discover whether it is right, needs modification,
or is simply wrong.}

\begin{table}
 \caption{Particle content of the Standard Model with a minimal Higgs sector.}
 \label{TabSMPartCont}
 \centering
 \begin{tabular}{cc}
  \hline
  \hline
  \textbf{Bosons} & \textbf{Scalars} \\ 
  \hline
  $\gamma$, $W^+$, $W^-$, $Z^0$, $g_{1 \ldots 8}$ & $\phi$ (Higgs) \\
  \hline
  \multicolumn{2}{c}{\textbf{Fermions}} \\
  \hline
  Quarks (each with 3 colour charges) & Leptons \\
  $\begin{array}{r}
   2/3: \\
   -1/3:
  \end{array} ~
  \left(\begin{array}{c}
   u \\ d       
  \end{array}\right) ~,
  \left(\begin{array}{c}
   c \\ s       
  \end{array}\right) ~,
  \left(\begin{array}{c}
   t \\ b       
  \end{array}\right)$ 
  &
  $\begin{array}{r}
   \text{neutral}: \\
   -1:
  \end{array} ~
  \left(\begin{array}{c}
   \nu_e \\ e^-       
  \end{array}\right) ~,
  \left(\begin{array}{c}
   \nu_\mu \\ \mu^-       
  \end{array}\right) ~,
  \left(\begin{array}{c}
   \nu_\tau \\ \tau^-       
  \end{array}\right)$ \\
  \hline
  \hline
 \end{tabular}
\end{table}

\subsection{The Higgs mechanism in $U(1)$}

To explain the Higgs mechanism of mass generation, we first apply it to the gauge group $U(1)$, and then extend it to the full electroweak group $SU(2)_L \otimes U(1)_Y$.
Thus, we first consider the following Lagrangian for a single complex scalar field:
\begin{equation}\label{EqU1Lag}
 \mathcal{L} = \left(\partial_\mu \phi\right)^\ast \left(\partial^\mu \phi\right) - V\left(\phi^\ast \phi\right) ~,
\end{equation}
with the potential defined as
\begin{equation}\label{EqHiggsPotU1}
 V\left(\phi^\ast \phi\right) = \mu^2 \left(\phi^\ast \phi\right) + \lambda \left(\phi^\ast \phi\right)^2 ~,
\end{equation}
where $\mu^2$ and $\lambda > 0$ are real constants. 
This Lagrangian is clearly invariant under global $U(1)$ phase transformations
\begin{equation}
 \phi \rightarrow e^{i\alpha} \phi ~,
\end{equation}
for $\alpha$ some rotation angle. Equivalently, it is invariant under a $SO(2)$ rotational symmetry, which is made evident by writing $\mathcal{L}$ in terms of the decomposition of the complex
scalar field into two real fields $\phi_1$ and $\phi_2$:
$\phi \equiv \phi_1 + i\phi_2$.

If we choose $\mu^2 > 0$
in (\ref{EqVU1}), the sole vacuum state has $\langle \phi \rangle = 0$. Perturbing 
around this vacuum reveals that, in this case, the scalar-sector Lagrangian simply 
factors into two Klein--Gordon 
Lagrangians, one for $\phi_1$ and the other for $\phi_2$, with a common mass. The symmetry 
of the original Lagrangian is preserved in this case.

However, when $\mu^2 <0$, the Lagrangian (\ref{EqU1Lag}) exhibits
spontaneous breaking of the $U(1)$ global symmetry,
which introduces a massless scalar particle known as a Goldstone boson,
as we now show.
In order to make manifest this breaking of the $U(1)$ symmetry present in Eq.~(\ref{EqU1Lag}), 
we first minimize the potential (\ref{EqHiggsPotU1}) so as to identify the 
vacuum expectation value, or v.e.v., of the scalar field. 
To do this, we first write the Higgs potential as
\begin{equation}\label{EqVU1}
 V\left(\phi^\ast \phi\right) = \mu^2 \left(\phi_1^2 + \phi_2^2\right) + \lambda \left(\phi_1^2 + \phi_2^2\right)^2 ~,
\end{equation}
and note that minimization with respect to $\phi^\ast \phi$ yields the value
\begin{equation}
 \phi_1^2 + \phi_2^2 = -\mu^2/\left(2\lambda\right) ~,
\end{equation}
i.e., there is a set of equivalent
minima lying around a circle of radius $\sqrt{-\mu^2/\left(2\lambda\right)}$,
when $\mu^2 < 0$ as assumed. The quanta of the Higgs field arise when a particular ground state is chosen and perturbed. Reflecting the appearance of spontaneous symmetry 
breaking we may, without loss of generality, choose for instance
\begin{equation}
 \phi_{1,\text{vac}} = \sqrt{-\mu^2/\left(2\lambda\right)} \equiv v/\sqrt{2} ~~~, ~~~~ \phi_{2,\text{vac}} = 0 ~.
\end{equation}
Perturbations around this vacuum may be parametrized by
\begin{equation}\label{EqPerturb}
 \eta/\sqrt{2} \equiv \phi_1 - v/\sqrt{2} ~~~, ~~~~ \xi/\sqrt{2} \equiv \phi_2 ~,
\end{equation}
so that the perturbed complex scalar is $\phi = \left(v + \eta + i \xi\right)/\sqrt{2}$,
where $\eta$ and $\xi$ are real fields. In terms of these, the Lagrangian becomes
\begin{eqnarray}\label{EqLagGoldstone}
 \nonumber
 \mathcal{L} &=& \left[ \frac{1}{2} \left(\partial^\mu \eta\right) \left(\partial_\mu \eta\right) - \frac{\mu^2}{2} \eta^2 \right] 
                 + \frac{1}{2} \left(\partial^\mu \xi\right) \left(\partial_\mu \xi\right) \\
             &-& \frac{\lambda}{2} \left[ \left(v+\eta\right)^2 + \xi^2 \right]^2 - \mu^2 v \eta - \frac{\mu^2}{2} \xi^2 - \frac{1}{2} \mu^2 v^2 ~.
\end{eqnarray}
The first and second terms describe two scalar particles: the first, $\eta$, is massive 
with $m^2_\eta = -\mu^2 > 0$ (we recall that $\mu^2 < 0$),
and the second, $\xi$, is massless, 
the Goldstone boson.

We now discuss how this spontaneous symmetry breaking manifests
itself in the presence of a $U(1)$ gauge field. For this purpose, we make the 
Lagrangian (\ref{EqU1Lag}) invariant under local $U(1)$ phase transformations, i.e.,
\begin{equation}
 \phi \rightarrow e^{i\alpha\left(x\right)} \phi ~.
\end{equation}
This requires the introduction of a gauge field $\mathcal{A}_\mu$
that transforms as follows
under $U(1)$:
\begin{equation}
\mathcal{A}_\mu^\prime \rightarrow \mathcal{A}_\mu + \left(1/q\right) \partial_\mu \alpha\left(x\right) ~,
\label{gaugeU1}
\end{equation}
and replacing the space-time derivatives by covariant derivatives
\begin{equation}
 D_\mu = \partial_\mu + i q \mathcal{A}_\mu ~,
\end{equation}
where $q$ is the conserved charge. Replacing the derivatives in Eq.~(\ref{EqU1Lag}) 
and adding a kinetic term for the $\mathcal{A}_\mu$ field, the Lagrangian becomes
\begin{equation}\label{EqU1Lag2}
 \mathcal{L} = \left[\left(\partial_\mu - iq\mathcal{A}_\mu\right)\phi^\ast\right] \left[\left(\partial^\mu + iq\mathcal{A}^\mu\right)\phi\right]
             - V\left(\phi^\ast \phi\right) - \frac{1}{4} F^{\mu\nu}F_{\mu\nu} ~.
\end{equation}
The last term in this equation, $\left(1/4\right) F^{\mu\nu}F_{\mu\nu}$, with 
$F_{\mu\nu} \equiv \partial_\nu \mathcal{A}_\mu - \partial_\mu \mathcal{A}_\nu$, 
is the kinetic term, which is separately invariant under the transformation (\ref{gaugeU1})
of the gauge field. 

We now repeat the minimization of the potential $V\left(\phi\right)$ and write the Lagrangian in terms of the perturbations around the ground state, Eqs.~(\ref{EqPerturb}):
\begin{eqnarray}
 \nonumber
 \mathcal{L} 
  &=& \left\{
      \frac{1}{2} \left[ \left(\partial^\mu \eta\right) \left(\partial_\mu \eta\right) - \mu^2 \eta^2 \right]
      + \frac{1}{2} \left(\partial^\mu \xi\right) \left(\partial_\mu \xi\right) 
      -\frac{1}{4}F^{\mu\nu}F_{\mu\nu} + \frac{1}{2} q^2 v^2 \mathcal{A}^\mu \mathcal{A}_\mu \right\} \\ \nonumber
  &+& v q^2 {A}^\mu \mathcal{A}_\mu \eta  + \frac{q^2}{2} \mathcal{A}^\mu \mathcal{A}_\mu \eta^2 + q \left(\partial^\mu \xi\right) \mathcal{A}_\mu \left(v+\eta\right) 
      - q \left(\partial^\mu \eta\right) \mathcal{A}_\mu \xi \\ 
  &-& \mu^2 v \eta - \frac{\mu^2}{2} \xi^2 - \frac{\lambda}{2} \left[ \left(v+\eta\right) + \xi^2 \right]^2 - \frac{\mu^2 v}{2} ~.
\end{eqnarray}
The first three terms again describe a (real) scalar particle, $\eta$, of 
mass $\sqrt{-\mu^2}$ and a massless Goldstone boson, $\xi$. The fourth term describes the free gauge field. However, whereas previously the Lagrangian described a massless boson field [see Eq.~(\ref{EqLagGoldstone})], now it contains a term proportional to $\mathcal{A}_\mu \mathcal{A}^\mu$, which gives the gauge field a mass of
\begin{equation}
 m_\mathcal{A} = q v ~,
\end{equation}
from which we see that the boson field has acquired a mass that is 
proportional to the vacuum expectation value of the Higgs field.
Indeed, the last two terms in the first line of Eq.~(\ref{EqLagGoldstone}) 
are identical with the Proca Lagrangian for a $U(1)$ gauge boson of mass $m$.

The rest of the terms in Eq.~(\ref{EqLagGoldstone}) 
define couplings between the fields $A^\mu, \eta$ and $\xi$,
among which is a bilinear interaction coupling $A^\mu$ and $\partial_\mu \xi$.
In order to give the correct propagating particle interpretation of
(\ref{EqLagGoldstone}), we must diagonalize the bilinear terms and
remove this term.
This is easily done by exploiting the gauge freedom of the $\mathcal{A}_\mu$ field to replace
\begin{equation}
 \mathcal{A}_\mu \rightarrow \mathcal{A}_\mu^\prime = \mathcal{A}_\mu + \frac{1}{qv} \partial_\mu \xi ~,
\end{equation}
which is accompanied by the local phase transformation
\begin{equation}
 \phi \rightarrow \phi^\prime = e^{-i\xi\left(x\right)/v} \phi = \left( v + \eta\right) / \sqrt{2} ~.
\end{equation}
After making this transformation, the field $\xi$ no longer appears, and the Lagrangian
(\ref{EqLagGoldstone}) takes the simplified form
\begin{equation}
\label{eaten}
 \mathcal{L} = \frac{1}{2} \left[ \left(\partial^\mu\right) \left(\partial_\mu\right) - \mu^2 \eta^2 \right] 
               - \frac{1}{4} F^{\mu\nu} F_{\mu\nu} + \frac{q^2 v^2}{2} \mathcal{A}^{\mu ~ \prime} \mathcal{A}_\mu^\prime + \ldots ~.
\end{equation}
where the $\ldots$ represent trilinear and quadrilinear interactions.

The interpretation of (\ref{eaten}) is that the Goldstone boson $\xi$
that appeared when the global $U(1)$ symmetry was broken by the choice of an
asymmetric ground state when $\mu^2 < 0$ has been absorbed (or `eaten') by the gauge field 
$\mathcal{A}_\mu$, with the effect of generating a mass. Another way to understand this is to 
recall that, whereas a massless gauge boson has only two degrees of freedom, or polarization
states (which are transverse), a massive gauge boson must have a third (longitudinal) polarization
state. In the Higgs mechanism, this is supplied by the Goldstone boson of the spontaneously-broken
$U(1)$ global symmetry.

At first sight, the Higgs mechanism may seem somewhat artificial.
From one point of view, it is merely a description 
of the breaking of electroweak symmetry, rather than an explanation
of how a massless gauge boson may become massive. As Quigg 
says~\cite{Quigg:1997}, the electroweak symmetry is broken because $\mu^2 < 0$, and 
we must choose $\mu^2 < 0$,  because otherwise electroweak symmetry is not broken. 
From another point of view, the {\it only} consistent formulation of an
interacting massive gauge boson is {\it via} the Higgs mechanism,
and the spontaneous breaking of symmetry is a mathematical ruse
for describing this phenomenon.

\subsection{The Higgs mechanism in $SU(2)_L \otimes U(1)_Y$}

Following closely in both spirit and notation
the book by Quigg~\cite{Quigg:1997}, we now consider the weak-isospin doublet
\begin{equation}
 \text{L} = 
 \left(\begin{array}{c}
  \nu \\
  e
 \end{array}\right)_L ~,
\end{equation}
with the left-handed neutrino and electron states defined by
\begin{equation}
 \nu_L = \frac{1}{2} \left(1-\gamma_5\right) \nu ~~~, ~~~~ e_L = \frac{1}{2} \left(1-\gamma_5\right) e ~.
\end{equation}
The operator $\left(1-\gamma_5\right)/2$ is of course the left-handed helicity projector, and $\nu$, $e$ are solutions of the free-field Dirac equation. Within the SM, we consider the neutrino to be massless, and
it does not have a corresponding right-handed component, i.e.,
\begin{equation}
 \nu_R = \frac{1}{2} \left(1+\gamma_5\right) \nu = 0 ~. 
\end{equation}
Hence, the only right-handed lepton, $e_R$, constitutes a weak-isospin singlet, i.e.,
\begin{equation}
 \text{R} = e_R = \frac{1}{2} \left(1+\gamma_5\right) e ~.
\end{equation}
We write initially the Lagrangian as
\begin{eqnarray}
 \mathcal{L} &=& \mathcal{L}_{\text{gauge}} + \mathcal{L}_{\text{leptons}}  \\
 \mathcal{L}_{\text{gauge}} &=& -\frac{1}{4} F_{\mu\nu}^a F^{a\mu\nu} - \frac{1}{4} f_{\mu\nu} f^{\mu\nu}  \\
 \mathcal{L}_{\text{leptons}} &=& \overline{\text{R}} \left( \partial_\mu + i\frac{g^\prime}{2} \mathcal{A}_\mu Y \right) \text{R}
                                  + \overline{\text{L}} i\gamma^\mu \left( \partial_\mu + i \frac{g^\prime}{2} \mathcal{A}_\mu Y 
                                    + i \frac{g}{2} \mathbf{\tau} \cdot \mathbf{B_\mu} \right) \text{L} ~,
\end{eqnarray}
where the field-strength tensors, $F_{\mu\nu}$ and $f_{\mu\nu}$, were defined in Eqs.~(\ref{EqFSTSU2}) and (\ref{EqFSTU1}), respectively. Here, $g^\prime/2$ is the coupling constant associated to the hypercharge group $U(1)_Y$, and $g/2$ is the coupling to the weak-isospin group $SU(2)_L$. So far, we are presented with four massless bosons ($\mathcal{A}_\mu$, $B_\mu^1$, $B_\mu^2$, $B_\mu^3$); the Higgs mechanism will select linear combinations of these to produce three massive bosons 
($W^\pm$, $Z^0$) and a massless one ($\gamma$). 

The Higgs field is now a complex $SU(2)$ doublet
\begin{equation}
 \phi = 
 \left(\begin{array}{c}
  \phi^+ \\
  \phi^0 
 \end{array}\right) ~,
\end{equation}
with $\phi^+$ and $\phi^0$ scalar fields. We need to add the Lagrangian
\begin{equation}\label{EqLagHiggs}
 \mathcal{L}_{\text{Higgs}} = \left(D_\mu \phi\right)^\dagger \left(D^\mu \phi\right) - V\left(\phi^\dagger \phi\right) ~,
\end{equation}
with the Higgs potential given by analogy to Eq.~(\ref{EqHiggsPotU1}) as
\begin{equation}\label{EqScalarPot}
 V\left(\phi^\dagger \phi\right) = \mu^2 \left( \phi^\dagger \phi \right) + \lambda \left(\phi^\dagger \phi\right)^2 ~,
\end{equation}
with $\lambda > 0$. We should also include the interaction Lagrangian between this scalar field and the fermionic matter fields, which occurs through Yukawa couplings,
\begin{equation}\label{EqYukawa}
 \mathcal{L}_{\text{Yukawa}} = 
 - G_e \left[ \overline{\text{R}} \phi^\dagger \text{L} + \overline{\text{L}} \phi \text{R} \right] ~.
\end{equation}
As we see later, these terms give rise to masses for the matter fermions.

\begin{figure}[h!]
  \begin{center}
    \scalebox{0.6}{\includegraphics{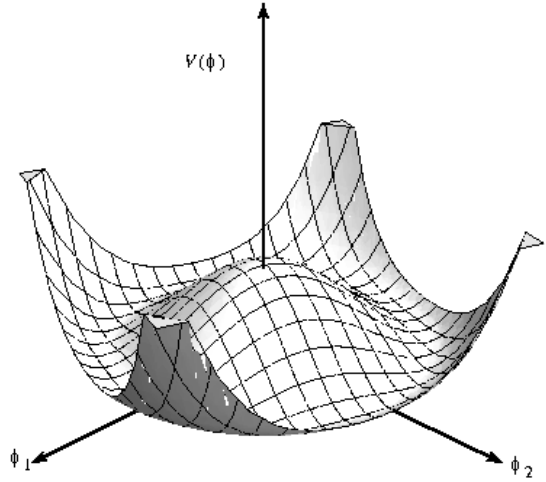}} 
    \label{FigMexHat}
  \end{center}
\caption{Scalar potential $V\left(\phi^\dagger \phi\right)$ with $\lambda > 0$ and $\mu^2 < 0$}
\end{figure}

A plot of the Higgs potential is presented in Fig.~\ref{FigMexHat}, where we see that
$\langle \phi \rangle = 0$ is an unstable local minimum of the effective potential if
$\mu^2 < 0$, and that the minimum is at some $\langle \phi \rangle \not= 0$ with
an arbitrary phase, leading to spontaneous symmetry breaking. 
Minimizing the Higgs potential, we obtain
\begin{equation}
 \frac{\partial}{\partial\left(\phi^\dagger \phi\right)} V\left(\phi^\dagger \phi\right)
 = \mu^2 + 2 \lambda \langle\phi\rangle_0 = \mu^2 + 2 \lambda \left[ \left(\phi_{\text{vac}}^+\right)^2 + \left(\phi_{\text{vac}}^0\right)^2 \right] = 0 ~.
\end{equation}
Choosing $\phi_{\text{vac}}^+ = 0$ and $\phi_{\text{vac}}^0 = \sqrt{-\mu^2/\left(2\lambda\right)}$, the v.e.v. of the scalar field becomes
\begin{equation}
 \langle \phi \rangle_0 = 
 \left(\begin{array}{c}
  0 \\
  v/\sqrt{2}      
 \end{array}\right) ~,
\end{equation}
with $v \equiv \sqrt{-\mu^2/\lambda}$. Selecting a particular v.e.v. breaks,
of course, both $SU(2)_L$ and $U(1)_Y$ symmetries.
Nevertheless, an invariance under the $U(1)_{\text{EM}}$ symmetry is preserved, with the charge operator as the generator. In the preceding section, we saw one example of the general
theorem that, for every broken generator (i.e., every generator that does not leave the vacuum 
invariant), there would (in the absence of the Higgs mechanism) be a Goldstone boson. 

In general, a generator $\mathcal{G}$ leaves the vacuum invariant if
\begin{equation}
 e^{i\alpha\mathcal{G}} \langle\phi\rangle_0 \simeq \left(1+i\alpha\mathcal{G}\right) \langle\phi\rangle_0 = \langle\phi\rangle_0 ~,
\end{equation}
which is satisfied when $\mathcal{G} \langle\phi\rangle_0 = 0$. Let's test whether the generators of $SU(2)_L \otimes U(1)_Y$ satisfy this condition:
\begin{eqnarray}
 \tau_1 \langle\phi\rangle_0 
 &=& 
 \left(\begin{array}{cc}
  0 & 1 \\
  1 & 0
 \end{array}\right)
 \left(\begin{array}{c}
  0 \\
  v/\sqrt{2}
 \end{array}\right) 
 =
 \left(\begin{array}{c}
  v/\sqrt{2} \\
  0
 \end{array}\right) \label{EqGenBroken1} \\
 \tau_2 \langle\phi\rangle_0 
 &=& 
 \left(\begin{array}{cc}
  0 & -i \\
  i & 0
 \end{array}\right)
 \left(\begin{array}{c}
  0 \\
  v/\sqrt{2}
 \end{array}\right) 
 =
 \left(\begin{array}{c}
  -i v/\sqrt{2} \\
  0
 \end{array}\right) \\
 \tau_3 \langle\phi\rangle_0 
 &=& 
 \left(\begin{array}{cc}
  1 & 0 \\
  0 & -1
 \end{array}\right)
 \left(\begin{array}{c}
  0 \\
  v/\sqrt{2}
 \end{array}\right) 
 =
 \left(\begin{array}{c}
  0 \\
  -v/\sqrt{2}
 \end{array}\right) \\
 Y \langle\phi\rangle_0 &=&  \langle\phi\rangle_0 \label{EqGenBroken4} ~.
\end{eqnarray}
Thus, none of the generators leave the vacuum invariant. However, we note that
\begin{equation}
 Q \langle\phi\rangle_0 = \frac{1}{2} \left(\tau_3 + Y\right) \langle\phi\rangle_0 = 0 ~,
\end{equation}
which is what we expected: the linear combination of generators corresponding to electric charge remains unbroken. Correspondingly, as we shall now see,
whilst the photon remains massless, the other three gauge bosons acquire mass.

To see this,
we now consider perturbations around the choice of vacuum. 
The full perturbed scalar field is
\begin{equation}
 \phi = \exp\left(\frac{i \mathbf{\xi} \cdot \mathbf{\tau}}{2v}\right) 
        \left(\begin{array}{c}
         0 \\
         \left( v + \eta \right) / \sqrt{2}
        \end{array}\right) ~.
\end{equation}
However, in analogy to what we did for the $U(1)$ Higgs in the previous section to rotate the Goldstone boson $\xi$ away, we are also able here to gauge-transform the scalar $\phi$ and the gauge and matter fields, i.e.,
\begin{eqnarray}
 \phi &\rightarrow& \phi^\prime 
 = \exp\left(\frac{-i \mathbf{\xi} \cdot \mathbf{\tau}}{2v}\right) \phi 
 = \left(\begin{array}{c}
    0 \\
    \left( v + \eta \right) / \sqrt{2}
   \end{array}\right) ~. \\
 \mathbf{\tau} \cdot \mathbf{B}_\mu &\rightarrow& \mathbf{\tau} \cdot \mathbf{B}_\mu^\prime \\
 \text{L} &\rightarrow& \text{L}^\prime = \exp\left(\frac{-i \mathbf{\xi} \cdot \mathbf{\tau}}{2v}\right) \text{L} ~,
\end{eqnarray} 
while the $\mathcal{A}_\mu$ and R remain invariant. It is possible to show that $\mathbf{\tau} \cdot \mathbf{B}_\mu^\prime = \mathbf{\tau} \cdot \mathbf{B}_\mu - \mathbf{\xi} \times \mathbf{B}_\mu \cdot \mathbf{\tau} - \left(1/g\right) \partial_\mu \left( \mathbf{\xi} \cdot \mathbf{\tau} \right)$.

In the unitary gauge, we can write the perturbed state as
\begin{equation}
 \langle \phi \rangle_0 \rightarrow \phi = 
 \left(\begin{array}{c}
  0 \\
  \left(v + \eta\right)/\sqrt{2}
 \end{array}\right) ~,
\end{equation}
and the Lagrangian in the Yukawa sector, Eq.~(\ref{EqYukawa}), becomes
\begin{equation}\label{EqLagYukawa}
 \mathcal{L}_{\text{Yukawa}}
 = - G_e \left[ \overline{e}_R \phi^\dagger 
   \left(\begin{array}{c}
    \nu_L \\
    e_L
   \end{array}\right)
   +
   \left( \overline{\nu}_L ~ \overline{e}_L \right) \phi e_R \right]
 = - G_e \frac{v+\eta}{\sqrt{2}} \left( \overline{e}_R e_L + \overline{e}_L e_R \right) ~.
\end{equation}
Defining $\overline{e} \equiv \left( \overline{e}_R,  \overline{e}_L \right)$ 
and $e \equiv \left( e_L,  e_R \right)^T$ yields
\begin{equation}
 \mathcal{L}_{\text{Yukawa}} = -\frac{G_e v}{\sqrt{2}} \overline{e}e - \frac{G_e \eta}{\sqrt{2}} \overline{e}e ~,
\end{equation}
so that the electron has acquired a mass
\begin{equation}
 m_e = G_e v / \sqrt{2} ~.
\end{equation}
Clearly, this mechanism may be applied to all the SM fermions, with the
general feature that their masses are proportional to their Yukawa couplings
to the Higgs field~\footnote{The Higgs couplings to quarks also induce
their Cabibbo--Kobayashi--Maskawa mixing --- see Eq.~(\ref{fmix}) below.}. 
This implies that the preferred decays of a Higgs boson into
generic fermions $f$ are into heavier species, as long as the Higgs mass $> 2 m_f$.

To see the effect of spontaneous symmetry breaking on the scalar-sector Lagrangian, 
$\mathcal{L}_{\text{Higgs}}$ in Eq.~(\ref{EqLagHiggs}), it is useful to calculate first
\begin{equation}
 \phi^\dagger \phi = \left( \frac{v+\eta}{\sqrt{2}} \right)^2 ~,
\end{equation}
so that
\begin{equation}
 V\left( \phi^\dagger \phi \right)
 = \mu^2 \left( \frac{v+\eta}{\sqrt{2}} \right)^2 + \lambda \left( \frac{v+\eta}{\sqrt{2}} \right)^4 ~,
\end{equation}
and we also need
\begin{equation}
 D_\mu \phi = \partial_\mu \phi + \frac{ig^\prime}{2} \mathcal{A}_\mu Y \phi + \frac{ig}{2} \mathbf{\tau} \cdot \mathbf{B}_\mu \phi ~,
\end{equation}
whose first term is simply
\begin{equation}
 \partial_\mu \phi
 = \left(\begin{array}{c}
  0 \\
  \partial_\mu \eta / \sqrt{2}
 \end{array}\right) ~.
\end{equation}
Using Eqs.~(\ref{EqGenBroken1})--(\ref{EqGenBroken4}), we calculate the second and third terms, i.e.,
\begin{eqnarray}
 \frac{ig^\prime}{2} \mathcal{A}_\mu Y \phi &=& \frac{ig^\prime}{2} \mathcal{A}_\mu \phi = \frac{ig^\prime}{2} \mathcal{A}_\mu 
 \left(\begin{array}{c}
  0 \\
  \left(v + \eta\right) / \sqrt{2}
 \end{array}\right) ,\\
 \left( \mathbf{\tau} \cdot \mathbf{B}_\mu \right) \phi &=&
 B_\mu^1 
 \left(\begin{array}{c}
  \left(v + \eta\right) / \sqrt{2} \\
  0
 \end{array}\right)
 +
 B_\mu^2
 \left(\begin{array}{c}
  -i \left(v + \eta\right) / \sqrt{2} \\
  0
 \end{array}\right)
 +
 B_\mu^3 
 \left(\begin{array}{c}
  0 \\
  -\left(v + \eta\right) / \sqrt{2}
 \end{array}\right) ~.
\end{eqnarray}
Hence,
\begin{equation}
 D_\mu \phi =
 \left(\begin{array}{c}
  \frac{ig}{2} \left(\frac{v+\eta}{\sqrt{2}}\right) \left(B_\mu^1 - iB_\mu^2\right) \\
  \frac{1}{\sqrt{2}} \partial_\mu \eta + \left(\frac{v+\eta}{\sqrt{2}}\right) \frac{i}{2}\left( ig^\prime \mathcal{A}_\mu - ig B_\mu^3 \right)
 \end{array}\right) 
\end{equation}
and
\begin{equation}
 \left(D^\mu\phi\right)^\dagger \left(D_\mu \phi\right) =
 \frac{g^2}{8} \left(v+\eta\right)^2 \lvert B_\mu^1 - i B_\mu^2 \rvert ^2 
 + \frac{1}{2} \left(\partial_\mu \eta\right) \left(\partial^\mu \eta\right)
 + \frac{1}{8} \left(v+\eta\right)^2 \left( g^\prime \mathcal{A}_\mu - g B_\mu^3 \right)^2 ~.
\end{equation}
With this, the scalar-sector Lagrangian becomes
\begin{eqnarray}\label{EqLagHiggsFinal}
 \mathcal{L}_{\text{Higgs}} &=&
 \left\{
 \frac{1}{2} \left(\partial_\mu \eta\right) \left(\partial^\mu \eta\right) - \frac{\mu^2}{2} \eta^2
 + \frac{v^2}{8} \left[ g^2 \lvert B_\mu^1 - i B_\mu^2 \rvert^2 + \left( g^\prime \mathcal{A}_\mu - g B_\mu^3 \right)^2 \right]
 \right\} \nonumber \\
 &+&
 \left\{
 \frac{1}{8} \left(\eta^2+2v\eta\right) \left[ g^2 \lvert B_\mu^1 - i B_\mu^2 \rvert^2  
                                               + \left( g^\prime \mathcal{A}_\mu - g B_\mu^3 \right)^2 \right] \right. \nonumber \\
 &-& \left. \frac{1}{4} \eta^4 - \lambda v \eta^3 - \frac{3}{2} \lambda v^2 \eta^2 - \left( \lambda v^3 + \mu^2 v \right) \eta
 - \left( \frac{\lambda v^4}{4} + \frac{\mu^2 v^2}{2} \right)
 \right\} ~.
\end{eqnarray}
From the second term inside the first curly brackets, we see that the $\eta$ field has acquired a mass; indeed, it is the Higgs boson, with non-zero mass. The terms inside the second curly brackets either describe interactions between the gauge and Higgs fields, or are constants that do not affect the physics.

It is convenient to define the charged gauge fields $W_\mu^{\pm}$ as linear combinations of the massless fields $B_\mu^1$ and $B_\mu^2$, i.e.,
\begin{equation}
 W_\mu^{\pm} \equiv \frac{B_\mu^1 \mp i B_\mu^2}{\sqrt{2}} ~,
\end{equation}
and, analogously,
\begin{eqnarray} 
 Z_\mu &\equiv& \frac{-g^\prime \mathcal{A}_\mu + g B_\mu^3}{\sqrt{g^2+g^{\prime~2}}} \label{EqDefA} ~, \\ 
 A_\mu &\equiv& \frac{g \mathcal{A}_\mu + g^\prime B_\mu^3}{\sqrt{g^2+g^{\prime~2}}} \label{EqDefZ} ~.
\end{eqnarray}
Writing the original fields $\mathcal{A}_\mu$, $B_\mu^i$ in terms of the new fields, we have
\begin{eqnarray}
 B_\mu^1 &=& \frac{\sqrt{2}}{2} \left(W_\mu^- + W_\mu^+\right) ~~, ~~~
 B_\mu^2 = \frac{\sqrt{2}}{2} \left(W_\mu^- - W_\mu^+\right) ~, \\
 B_\mu^3 &=& \frac{g^\prime}{\sqrt{g^2+g^{\prime ~ 2}}} \left( A_\mu + \frac{g}{g^\prime} Z_\mu \right) ~~, ~~~
 \mathcal{A}_\mu = \frac{g}{\sqrt{g^2+g^{\prime ~ 2}}} \left( A_\mu - \frac{g^\prime}{g} Z_\mu \right) ~.
\end{eqnarray}
Making these replacements in the broken scalar-sector Lagrangian, Eq.~(\ref{EqLagHiggsFinal}), leads to
\begin{eqnarray}
 \mathcal{L}_{\text{Higgs}}
 = \left[\frac{1}{2} \left( \partial^\mu \eta \right) \left( \partial_\mu \eta \right) - \frac{\mu^2}{2} \eta^2\right] 
   &+& \frac{v^2 g^2}{8} W^{+ ~ \mu}W^+_\mu + \frac{v^2 g^2}{8} W^{- ~ \mu}W^-_\mu 
   + \frac{\left( g^2 + g^{\prime ~ 2} \right) v^2}{8} Z^\mu Z_\mu \nonumber \\
   &+& ... ~,
\end{eqnarray}
and it is evident now that while the photon field $\mathcal{A}_\mu$ is massless due to the unbroken 
$U(1)_{\text{EM}}$ symmetry (i.e., the symmetry under $e^{iQ\alpha\left(x\right)}$ rotations), the 
vector bosons $W^\pm$ and $Z^0$ have masses 
\begin{equation}
 m_W = gv/2 ~~~, ~~~~ m_Z = \left(v / 2\right) \sqrt{g^2 + g^{\prime ~ 2}} ~.
\end{equation}
We see again that the Higgs couplings to other particles, in this case the
$W^\pm$ and $Z^0$, are related to their masses.

We also see that the masses of the neutral and charged weak-interaction bosons are 
related through
\begin{equation}
 m_Z = m_W \sqrt{1 + g^{\prime ~ 2} / g^2} ~.
 \label{MWMZ}
\end{equation}
Experimentally, the weak gauge boson masses are known to high accuracy to be \cite{Amsler:2008zzb}
\begin{equation}
 m_W = 80.399 \pm 0.023 ~\text{GeV} ~~~~, ~~~~~
 m_Z = 91.1875 \pm 0.0021 ~\text{GeV} ~,
\end{equation}
which can be compared in detail with (\ref{MWMZ}) only after the inclusions of
radiative corrections.
Meanwhile, the current experimental upper limit on the photon mass, based on plasma physics, is very stringent: $m_\gamma < 10^{-18}$ eV \cite{Ryutov:2007zz}. For the Higgs mass, we see
from (\ref{EqLagHiggsFinal}) that
\begin{equation}
 m_H = - 2 \mu^2 ~.
\end{equation}
{\it A priori}, however, there is no theoretical prediction within the Standard Model, since $\mu$
is not determined by any of the known parameters of the Standard Model. Later we will see
various ways in which experiments constrain the Higgs mass.

We can introduce a weak mixing angle $\theta_W$ to parametrize the mixing of the neutral gauge bosons, defined by
\begin{equation}
 \tan\left(\theta_W\right) = g^\prime/g ~,
\end{equation}
so that
\begin{equation}
 \cos\left(\theta_W\right) = \frac{g}{\sqrt{g^2+g^{\prime ~ 2}}} ~~~, ~~~~
 \sin\left(\theta_W\right) = \frac{g^\prime}{\sqrt{g^2+g^{\prime ~ 2}}} ~.
\end{equation}
With this, we can write, from Eqs.~(\ref{EqDefA}) and (\ref{EqDefZ}), 
\begin{eqnarray}
 Z_\mu &=& -\sin\left(\theta_W\right) \mathcal{A}_\mu + \cos\left(\theta_W\right) B_\mu^3 ~,\\
 A_\mu &=& \cos\left(\theta_W\right) \mathcal{A}_\mu + \sin\left(\theta_W\right) B_\mu^3 ~.
\end{eqnarray}
The relation (\ref{MWMZ}) between the masses of $W^\pm$ and $Z^0$ becomes
\begin{equation}
 m_W = m_Z \cos\left(\theta_W\right) ~,
\end{equation}
and it is common practice to define the ratio
\begin{equation}\label{EqRhoDef}
 \rho = \frac{m_W^2}{m_Z^2 \cos^2\left(\theta_W\right)} ~.
\end{equation}
According to the Standard Model, this is equal to unity at the tree level, a prediction that has been well tested by experiment, including radiative corrections. The value of $\sin^2\left(\theta_W\right)$ is obtained from measurements of the $Z$ pole and neutral-current processes, and depends on the renormalization prescription. The 2008 Particle Data Group review \cite{Amsler:2008zzb} states 
values of $\sin^2\left(\theta_W\right) = 0.2319(14)$ and $\rho = 1.0004_{-0.0004}^{+0.0008}$.

Therefore, after the spontaneous breaking of the electroweak $SU(2)_L \otimes U(1)_Y$ symmetry, we have ended up with what we desired: three massive gauge bosons ($W^\pm$, $Z^0$) that mediate weak interactions, one massless gauge boson ($A$) corresponding to the photon, and an extra, massive, Higgs boson ($H$).

\subsection{QCD}

The QCD Lagrangian has a structure similar to that of the electroweak 
Lagrangian~\cite{Amsler:2008zzb}, being also a gauge theory, but based on the group $SU(3)$
and without spontaneous symmetry breaking:
\begin{eqnarray}
 \mathcal{L}_{\text{QCD}} 
 &=& - \frac{1}{4} F_{\mu\nu}^a F^{a ~ \mu\nu} + i \sum_q \overline{\psi}_q^i \gamma^\mu \left( D_\mu \right)_{ij} \psi_q^j 
   - \sum_q m_q \psi_q^i \psi_{qi} ~, \\
 F_{\mu\nu}^a &=& \partial_\mu A_\nu^a - \partial_\nu A_\mu^a - g_s f_{abc} A_\mu^b A_\nu^c ~, \\
 \left( D_\mu \right)_{ij} &=& \delta_{ij} \partial_\mu + i g_s \sum_a \frac{\lambda_{i,j}^a}{2} A_\mu^a ~,
\label{LQCD}
\end{eqnarray}
with $g_s$ the strong coupling constant, $f_{abc}$ the $SU(3)$ structure constants, and $\lambda_i$ ($i=1,\ldots,8$) the generators of $SU(3)$ (which can be taken to be the eight traceless Gell-Mann matrices). Note also that $\psi_q^i$ is the free-field Dirac spinor representing a quark of colour $i$ 
and flavour $q$ and the $A_\mu^a$ ($a=1,\ldots,8$) are the eight gluon fields. As is well known,
QCD and non-Abelian gauge theories possess the property of asymptotic freedom:
$\alpha_s \equiv g_s^2/4\pi$ obeys the renormalization-group equation (RGE) that determines
its evolution as a function of the effective scale $Q$:
\begin{equation}
Q\frac{d \alpha_s}{d Q} \; = \; 2 \beta_0 \alpha_s + ... \ ,
\end{equation}
where
\begin{equation}
\beta_0 \; = \; 11 - \frac{2}{3} n_q 
\end{equation}
and $n_q$ is the number of quark flavours with masses $\ll Q$. In addition to (\ref{LQCD}),
which specifies QCD at the perturbative level, its full specification of its vacuum
at the non-perturbative level requires an additional angle parameter, $\theta_{QCD}$,
that violates both parity P and CP~\cite{Theta}~\footnote{The upper limit on the electric dipole moment
of the neutron tells us that $|\theta_{QCD}| < {\cal O}(10^{-9})$~\cite{Amsler:2008zzb}.}.

\subsection{Parameters of the Standard Model}

The transformation from being one of the possible explanations of electromagnetic, weak and strong phenomena into a description in outstanding agreement with experiments is reflected in the dozens of electroweak precision measurements available 
today~\cite{EWWG09,Flacher:2008zq,Amsler:2008zzb}. These are sensitive to quantum 
corrections at and beyond the one-loop level, which are essential for obtaining agreement with
the data. The calculations of these corrections rely upon the renormalizability
(calculability) of the SM~\footnote{A crucial aspect of this is cancellation of anomalous
triangle diagrams between quarks and leptons, which may be a hint of an underlying
Grand Unified Theory --- see Lecture 4.}, and depend on the masses of heavy virtual particles, 
such as the top
quark and the Higgs boson and possibly other particles beyond the SM. The consistency with
the data may be used to constrain the masses of these particles.

\begin{figure}[htbp]
\begin{center}
  \mbox{\includegraphics[width=0.45\linewidth]{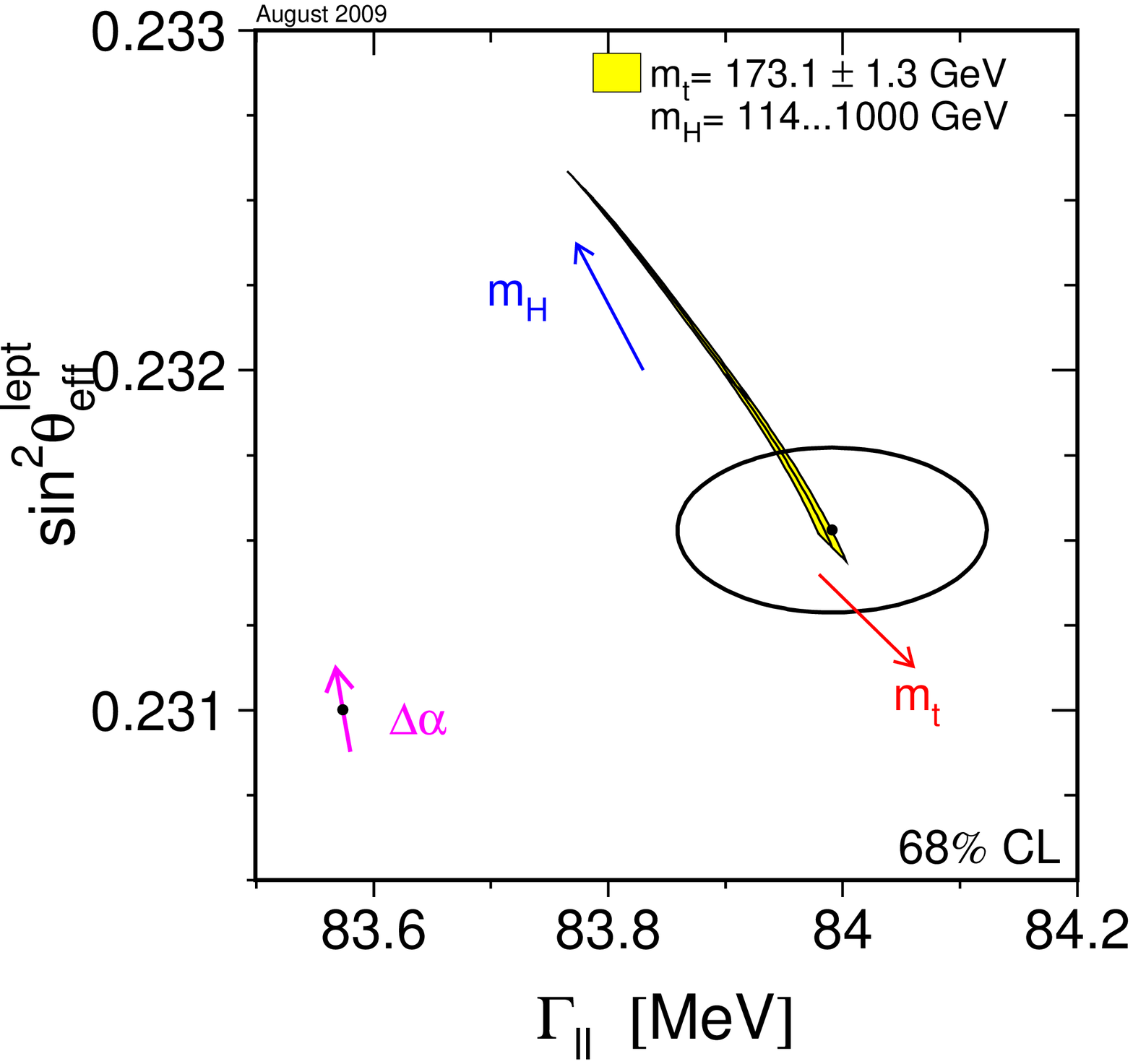}}
  \mbox{\includegraphics[width=0.45\linewidth]{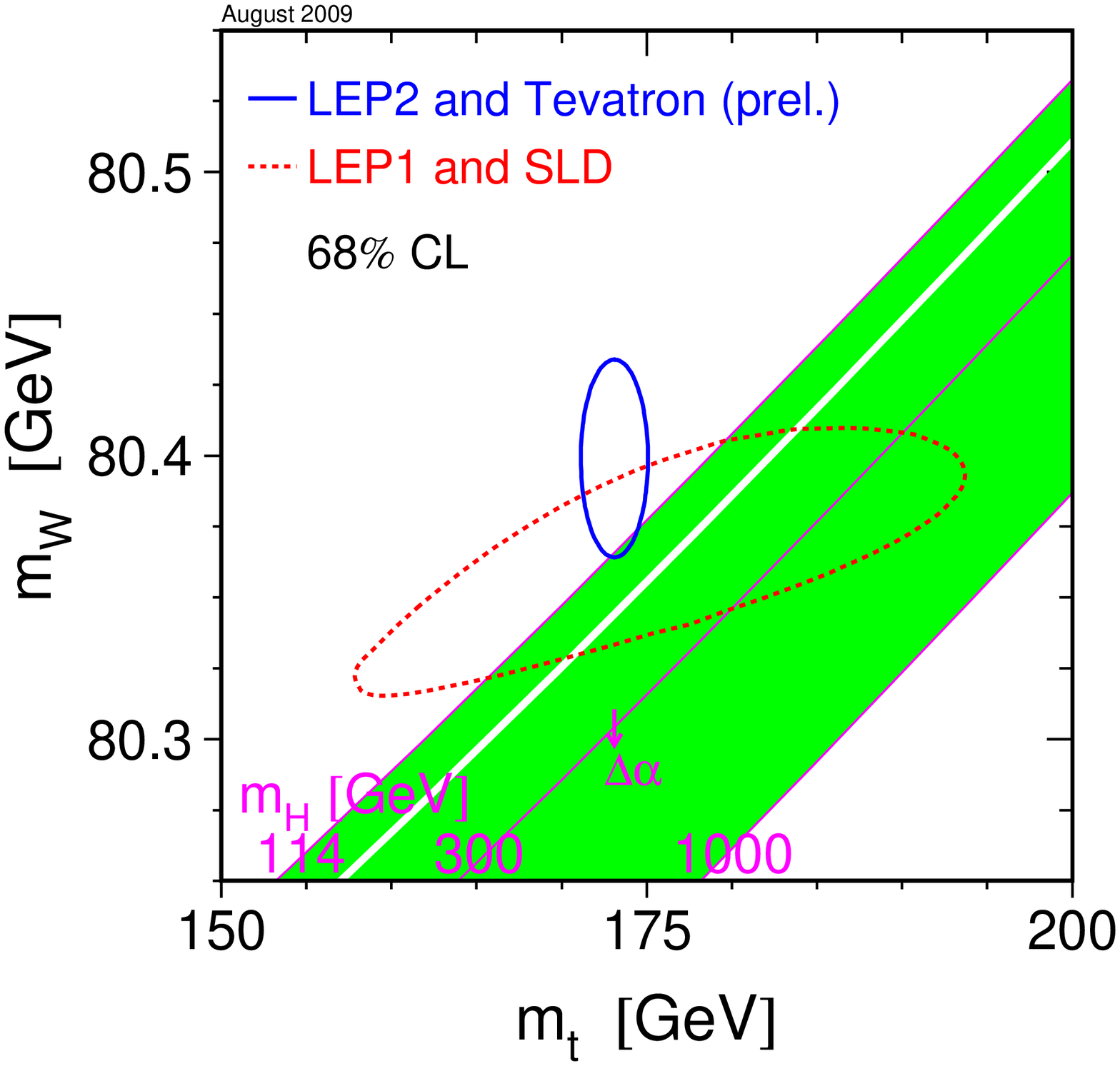}}
\end{center}
\vskip -1cm
\caption{Left: LEP and SLD measurements of $\sin^2 \theta_W$ and the leptonic
decay width of the $Z^0$, $\Gamma_{ll}$, compared with the SM prediction for
different values of $m_t$ and $m_H$. Right: The predictions for $m_t$ and $m_W$
made in the SM using LEP1 and SLD data (dotted mango-shaped contour) for different
values of $m_H$, compared with the LEP2 and Tevatron measurements (ellipse). 
The arrows show the additional effects of the uncertainty in
the value of $\alpha_{em}$ at the $Z^0$ peak~\protect\cite{EWWG09}.}
\label{fig:EWWG}
\end{figure}

Many of these observables have quadratic sensitivity to the mass of the top quark, e.g.,
\begin{equation}
s_W^2 \; \equiv \; 1 - m_W^2/m_Z^2 \; \ni \; - \frac{2\alpha}{16\pi\sin^2\left(\theta_W\right)} \frac{m_t^2}{m_Z^2} ~.
\label{topquark}
\end{equation}
This effect was used before the discovery of the top quark to predict successfully its mass~\cite{EF},
and the consistency of the prediction with experiment can be used to constrain possible
new physics beyond the SM, particularly mass-squared differences between isospin partner
particles, that would contribute analogously to (\ref{topquark}). Many electroweak observables
are also logarithmically sensitive to the mass of the Higgs boson, e.g.,
\begin{equation}
s_W^2 \; \ni \; \frac{5\alpha}{24\pi} \ln\left(\frac{m_H^2}{m_W^2}\right) 
\label{Higgsboson}
\end{equation}
when $m_H \gg m_W$. If there were no Higgs boson, or nothing to do its 
job~\footnote{See Lecture~2 for a discussion of possible alternatives.}, 
radiative corrections such as (\ref{Higgsboson}) would
diverge, and the SM calculations would become meaningless. Two examples of precision
electroweak observables, namely the coupling of the $Z^0$ boson to leptons and the
mass of the $W$ boson, are shown in Fig.~\ref{fig:EWWG}.

\begin{figure}[h!]
  \begin{center}
    \scalebox{0.6}{\includegraphics{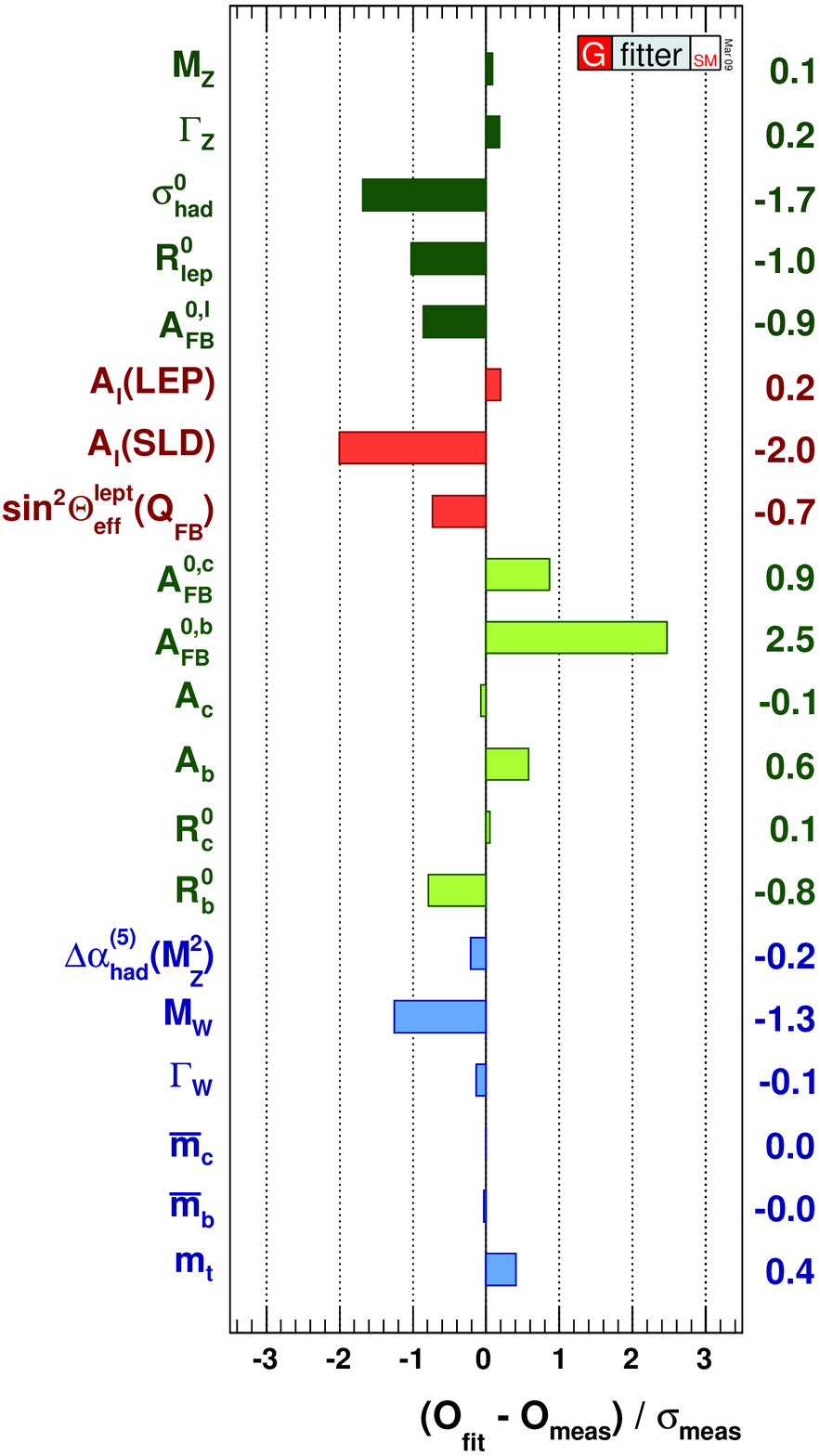}} 
    \label{FigFitExpComp}
  \end{center}
    \caption{Comparison between direct measurements and the results of
    a fit using the \texttt{Gfitter} 
    package~\cite{Flacher:2008zq}}
\end{figure}

Table~\ref{TabFitExpComp} and 
Fig.~\ref{FigFitExpComp}~\cite{Flacher:2008zq} compare the predicted (fitted) and experimentally measured values for several parameters of the Standard Model; the agreement is usually better than $1\sigma$. This is a remarkable success
for a theory that, as we have seen, can be written down in only a few lines. 

\begin{table}
 \caption{Fit and experimental values of some SM quantities, 
 as obtained using the~\texttt{Gfitter} package~\cite{Flacher:2008zq}. 
 For all the observables listed, except $A_l ~\text{(LEP)}$ and $A_l ~\text{(SLD)}$, 
 the fit values shown are the results of `complete fits', i.e., the results of using all 
 the inputs, including the input value of the parameter that is being fit, to calculate the result. 
 For the two exceptions, the fit values shown were calculated using all inputs except their own. 
 Consult~\cite{Flacher:2008zq} for a description of each observable.}
 \label{TabFitExpComp}
 \centering
 \begin{tabular}{ccc}
  \hline
  \hline
  \textbf{Parameter}                                              & \textbf{Input value}   & \textbf{Fit value}              \\
  \hline
  $M_Z$ [GeV]                                                     & $91.1875 \pm 0.0021$   & $91.1876 \pm 0.0021$            \\
  $\Gamma_Z$ [GeV]                                                & $2.4952 \pm 0.0023$    & $2.4956 \pm 0.0015$             \\
  $\sigma_{\text{had}}^0$                                         & $41.540 \pm 0.037$     & $41.478 \pm 0.014$              \\
  $R_l^0$                                                         & $20.767 \pm 0.025$     & $20.741 \pm 0.018$              \\
  $A_{\text{FB}}^{0,l}$                                           & $0.0171 \pm 0.0010$    & $0.01624 \pm 0.0002$            \\
  \hline
  $A_l ~\text{(LEP)}$                                             & $0.1465 \pm 0.0033$    & $0.1473 \pm 0.0009$             \\
  $A_l ~\text{(SLD)}$                                             & $0.1513 \pm 0.0021$    & $0.1465_{-0.0010}^{+0.0007}$    \\
  $\sin^2 \phi_{\text{eff}}^l \left(Q_{\text{FB}}\right)$         & $0.2324 \pm 0.0012$    & $0.23151_{-0.00012}^{+0.00010}$ \\
  \hline
  $A_{\text{FB}}^{0,c}$                                           & $0.0707 \pm 0.0035$    & $0.0737 \pm 0.0005$             \\
  $A_{\text{FB}}^{0,b}$                                           & $0.0992 \pm 0.0016$    & $0.1032_{-0.0006}^{+0.0007}$    \\
  $A_c$                                                           & $0.670 \pm 0.027$      & $0.6679_{-0.00036}^{+0.00042}$  \\
  $A_b$                                                           & $0.923 \pm 0.020$      & $0.93463_{-0.00008}^{+0.00007}$ \\
  $R_c^0$                                                         & $0.1721 \pm 0.0030$    & $0.17225 \pm 0.00006$           \\
  $R_b^0$                                                         & $0.21629 \pm 0.00066$  & $0.21577 \pm 0.00005$           \\
  \hline
  $\Delta \alpha_{\text{had}}^{\left(5\right)}\left(M_Z^2\right)$ & $2768 \pm 22$          & $2764_{-21}^{+22}$              \\
  $M_W$ [GeV]                                                     & $80.399 \pm 0.023$     & $80.371_{-0.011}^{+0.008}$      \\
  $\Gamma_W$ [GeV]                                                & $2.098 \pm 0.048$      & $2.092 \pm 0.001$               \\
  $\overline{m}_c$ [GeV]                                          & $1.25 \pm 0.09$        & $1.25 \pm 0.09$                 \\
  $\overline{m}_b$ [GeV]                                          & $4.20 \pm 0.07$        & $4.20 \pm 0.07$                 \\
  $m_t$ [GeV]                                                     & $173.1 \pm 1.3$        & $173.6 \pm 1.2$                 \\
  \hline
  \hline
 \end{tabular}
\end{table}

\begin{figure}[htbp]
\begin{center}
  \mbox{\includegraphics[width=0.6\linewidth]{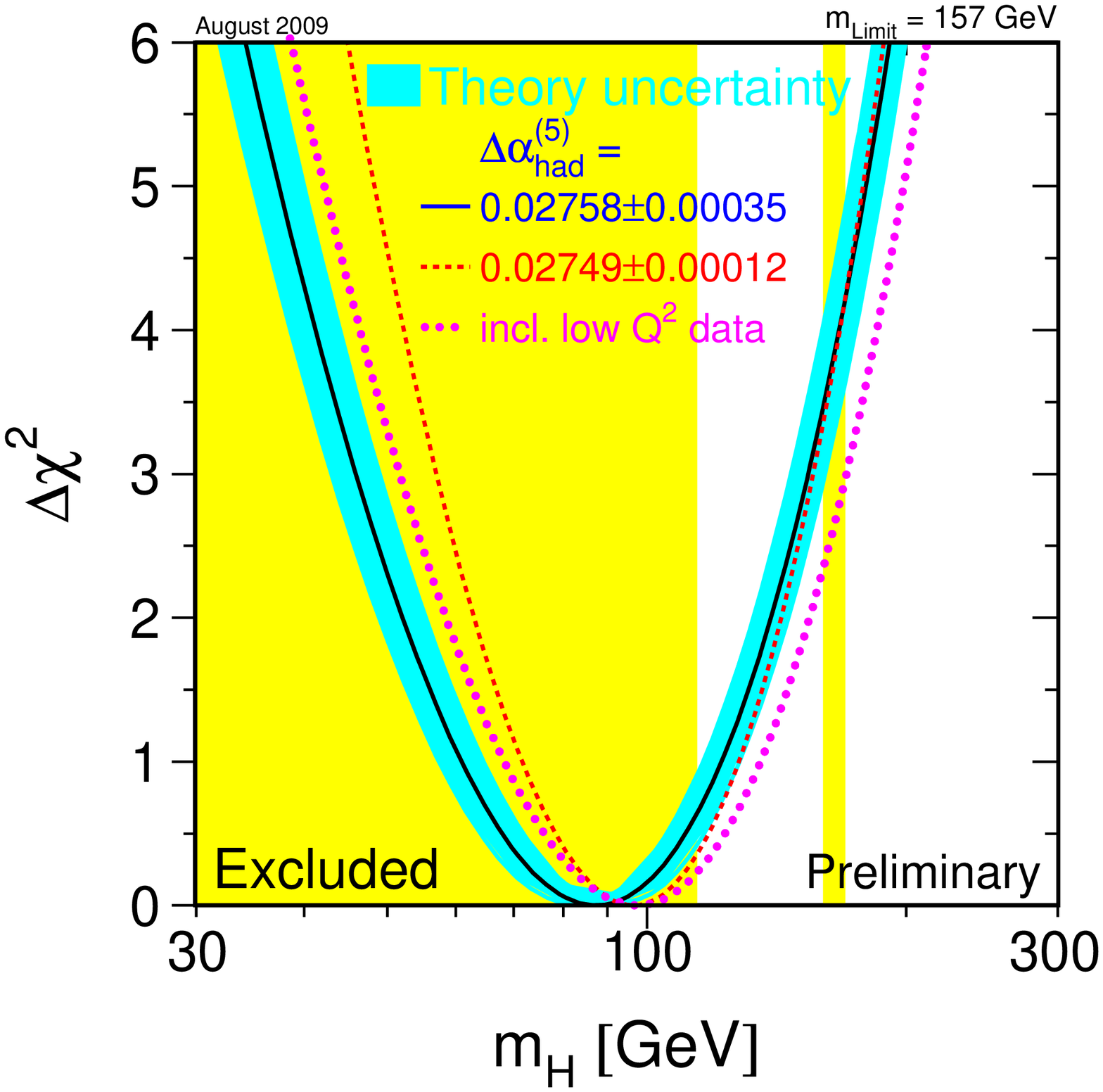}}
\end{center}
\vskip -1cm
\caption{The $\chi^2$ likelihood function for $m_H$ in a global electroweak fit. The blue band
around the (almost) parabolic solid curve represents the theoretical uncertainty: the other curves
indicate the effects of different calculations of the renormalization of $\alpha_{em}$ and of
including low-energy data. The shaded regions are those excluded by LEP and by the 
Tevatron~\protect\cite{EWWG09}.}
\label{fig:EWWGH}
\end{figure}

The agreement of the precision electroweak observables with the SM can be used to
predict $m_H$, just as it was used previously to predict $m_t$. Since the early 1990s~\cite{EFL},
this method has been used to tighten the vise on the Higgs, providing ever-stronger
indications that it is probably relatively light, as hinted in Fig.~\ref{fig:EWWGH}. 
The latest estimate of the Higgs mass is~\cite{EWWG09}
\begin{equation}
m_H \; = \; 89^{+35}_{-26}~~{\rm GeV} .
\label{Higgsmass}
\end{equation}
Although the central value is somewhat below the lower limit of 114.4~GeV set by
direct searches at LEP~\cite{LEPsearch}, there is consistency at the 1-$\sigma$ level, 
and no significant
discrepancy. {\it A priori},
the relatively light mass range (\ref{Higgsmass}) suggests that the Higgs boson
interacts relatively weakly, with a small quartic coupling $\lambda$, though there is
no theoretical consensus on this: see the discussion in the next Lecture.

%


This success is very impressive. However, our rejoicing is muted by the fact that to specify the SM we need at least 19 input parameters in order to calculate physical processes, namely:
\begin{itemize}
 \item three coupling parameters, which we can choose to be the strong coupling constant, $\alpha_s$, the fine structure constant, $\alpha_{\text{em}}$, and the weak mixing angle, $\sin^2\left(\theta_W\right)$;
 \item two parameters that specify the shape of the Higgs potential, $\mu^2$ and $\lambda$ (or, equivalently, $m_H$ and $m_W$ or $m_Z$);
 \item six quark masses (or the six Yukawa couplings for the quarks);
 \item four parameters (three mixing angles and one weak CP-violating angle) for the Cabibbo-Kobayashi-Maskawa matrix [see Eq.~(\ref{fmix}) below];
 \item three charged-lepton masses (or the corresponding Yukawa couplings);
 \item one parameter to allow for non-perturbative CP violation in QCD, $\theta_{\text{QCD}}$.
\end{itemize}
Moreover, because we now know that neutrinos have mass and that they mix (see, e.g., \cite{Kayser:2005cd,Mohapatra:2005wg}), the Standard Model must be extended to incorporate this fact.
Therefore, we also need to specify three neutrino masses and three mixing angles plus a CP-violating phase for the neutrino mixing matrix, bringing the grand total to 26 parameters. 
Additionally, if neutrinos turn out to be Majorana particles, so that they are their own antiparticles, 
two more CP-violating phases need to be specified. Notice that at least 20 of the parameters relate to flavour physics.

Many of the ideas for physics beyond the SM that are discussed later have been
motivated by attempts to reduce the number of its parameters, or understand their
origins, or at least to make them seem less unnatural, as discussed in subsequent Lectures.

\subsection{Bounds on the Standard Model Higgs boson mass}

\subsubsection{Upper bounds from unitarity}

As already emphasized, if there were no Higgs boson, and nothing analogous to replace it, the
Standard Model would not be a calculable, renormalizable theory.
This incompleteness is reflected in the behaviours of physical quantities
as the Higgs mass increases. The most basic example of this is
$W^+ W^-$ scattering~\cite{LQT}, whose high-energy $s$-wave amplitude grows with $m_H$:
\begin{equation}
T \; \sim \; - \frac{4 G_F}{\sqrt{2}} m^2_H .
\label{LQT1}
\end{equation}
Imposing the unitarity bound $|T| < 1$, one finds the upper limit
$M_H^2 < 4 \pi \sqrt{2}/G_F$, which is strengthened to
\begin{equation}
M_H^2 \; < \; \frac{8 \pi \sqrt{2}}{3 G_F} \sim 1~{\rm TeV}^2
\label{LQT}
\end{equation}
when one makes a coupled analysis including the $Z^0 Z^0$ channel.

\begin{figure}[t]
\begin{center}
  \mbox{\includegraphics[width=0.65\linewidth]{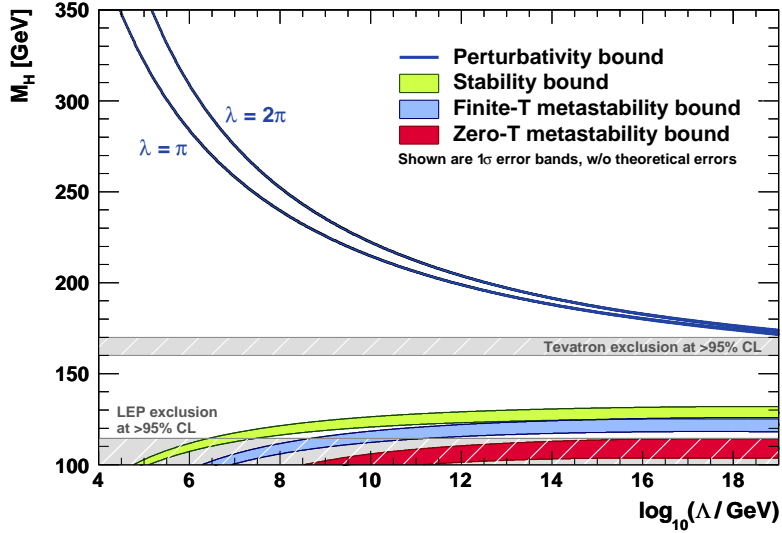}}
\end{center}
  \vspace{-0.5cm}
\caption{The scale 
         $\Lambda$ at which the two-loop RGEs drive the quartic SM Higgs 
         coupling non-perturbative (upper curves), and the scale $\Lambda$ at which the RGEs 
         create an instability in the electroweak vacuum (lower curves). 
         The widths of the bands reflect the uncertainties in $m_t$ and $\alpha_s(m_Z)$ (added quadratically).         
         The perturbativity upper bound (sometimes referred to as 
         `triviality' bound) is given for $\lambda = \pi$ 
         (lower bold line [blue]) and $\lambda =2\pi$ 
         (upper bold line [blue]). Their difference indicates the 
         theoretical uncertainty in this bound. The absolute 
         vacuum stability bound is displayed by the light shaded [green] band,
         while the less restrictive finite-temperature and zero-temperature 
         metastability bounds are medium [blue] and dark shaded [red], respectively.
         The grey hatched areas indicate the LEP~\protect\cite{LEPsearch} and 
         Tevatron~\protect\cite{TeVsearch} exclusion domains. 
         Figure taken from~\protect\cite{EEGHR}.}
\label{fig:Andreas}
\end{figure}

A related effect is seen in the behaviour of the quartic self-coupling $\lambda$
of the Higgs field. Like any of the Standard Model parameters, $\lambda$ 
is subject to renormalization {\it via} loop corrections. Loops of fermions,
most importantly the top quark,
tend to {\it decrease} $\lambda$ as the renormalization scale $\Lambda$ increases,
whereas loops of bosons tend to {\it increase} $\lambda$. In particular, if the Higgs
mass $\gtrsim m_t$, the positive renormalization due to the Higgs self-coupling
itself is dominant, and $\lambda$ increases uncontrollably with $\Lambda$. The larger
the value of $m_H$, the larger the low-energy value of $\lambda$, and the
smaller the value of $\Lambda$ at which $\lambda$ blows up. 
In general, one should
regard the limiting value of $\Lambda$, also for smaller $m_H$, as a scale where
novel non-perturbative dynamics must set in. This behaviour is seen
in the upper part of Fig.~\ref{fig:Andreas}, where we see, for example, that if
$m_H = 170$~GeV, then $\Lambda \sim 10^{19}$~GeV, whereas 
if $m_H = 300$~GeV, the coupling $\lambda$ blows up at a scale $\Lambda \sim 10^6$~GeV.
One may ask: under what circumstances does $m_H \sim \Lambda$ itself? The
answer is when $m_H \sim 700$~GeV: if the Higgs boson were heavier than this mass,
the Higgs self-coupling would blow up at a scale smaller than its mass. In this case,
Higgs physics would necessarily be described by some new strongly-interacting
theory, cf., the technicolour theories described in Lecture~2.

\subsubsection{Lower bounds from vacuum stability}

Looking at lower values of $m_H$ in Fig.~\ref{fig:Andreas}, we see an uneventful
range of $m_H$ extending down to $m_H \sim 130$~GeV, where (as far as we
know) the SM could continue to be valid all the way to the Planck scale. At lower
$m_H$, there is a band below which the present electroweak vacuum becomes
unstable at some scale $\Lambda < 10^{19}$~GeV. For example, if the Higgs is
slightly above the present experimental lower limit from LEP, $m_H \sim 115$~GeV,
the present electroweak vacuum is unstable against decay into a vacuum with
$\langle |\phi| \rangle \sim 10^7$~GeV. This instability is due to the negative
renormalization of $\lambda$ by the top quark, which overcomes the positive
renormalization due to $\lambda$ itself, and drives $\lambda < 0$~\footnote{The
widths of the boundary bands indicate the uncertainties in these calculations.}.

If $m_H$ is only slightly below the top band, and above the middle band, it is
true that the present electroweak vacuum is in principle unstable against
decay into a state with $\langle |\phi| \rangle > \Lambda$, but it would
not have decayed during the conventional thermal expansion of the Universe
at finite temperatures. Below the middle band but above the lowest band, the
vacuum would have decayed to a correspondingly large value of 
$\langle |\phi| \rangle$ at some finite temperature, but its present-day
(low-temperature) lifetime is longer than the age of the Universe. Below the
lowest band, the lifetime for decay to a vacuum with $\langle |\phi| \rangle > \Lambda$
would be less than the present age of the Universe
at low temperatures, and we should really watch out!

In fact, as we see shortly, such low values of $m_H$ are almost excluded
by LEP searches for the SM Higgs boson, as also seen in Fig.~\ref{fig:Andreas}.

One could in principle avoid this vacuum instability by introducing some
new physics at an energy scale $< \Lambda$: what type of physics~\cite{ERoss}?
One needs to overcome the negative effects of renormalization of $\lambda$
by loops with the top quark circulating. The sign of renormalization could be 
reversed by loops with some boson circulating, potentially restoring the
stability of the electroweak vacuum. However, then one should consider
the renormalization of the quartic coupling between the Higgs and the new 
boson. It turns out that the renormalization of this coupling is in turn very unstable,
and that the best way to stabilize this coupling would be to introduce a new
fermion. 

These new scalars and fermions look very much like the partners of the top
quark and Higgs bosons, respectively, that are predicted by supersymmetry~\cite{ERoss}.
In Lecture 3 we will study in more detail the renormalization of mass and vacuum
parameters in a supersymmetric theory.

\subsubsection{Results from searches at LEP and the Tevatron}

\begin{figure}[t]
\begin{center}
  \epsfig{file=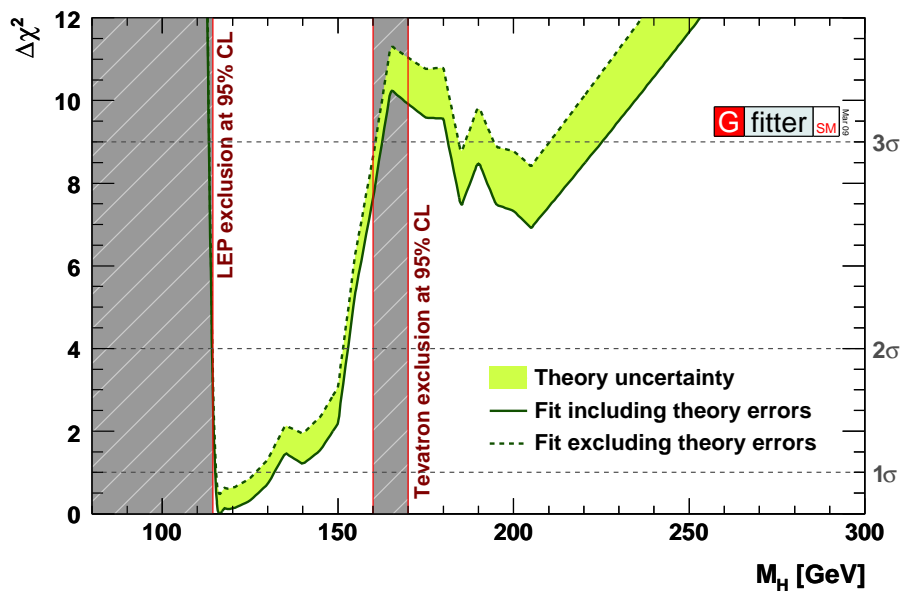, scale=0.6}
\end{center}
  \vspace{-0.5cm}
\caption{Dependence on $M_H$ of the $\Delta \chi^2$ function obtained 
         from the global fit of the SM parameters to precision 
         electroweak data~\protect\cite{EEGHR}, excluding (left) or including (right) the 
         results from direct searches at LEP and the Tevatron
\label{fig:basis}}
\end{figure}

As seen in Fig.~\ref{fig:EWWG},
searches for the reaction $e^+ e^- \to Z^0 + H$ at LEP established a lower
limit on the possible mass of a SM Higgs boson~\cite{LEPsearch}:
\begin{equation}
m_H \; > \; 114.4~{\rm GeV}
\label{LEPmH}
\end{equation}
at the 95\% confidence level. The lower limit (\ref{LEPmH}) is
somewhat higher than the central value of the SM Higgs mass preferred by the 
global precision electroweak fit (\ref{Higgsmass}), 
but there is no significant tension between
these two pieces of information. Figure~\ref{fig:basis} shows the
$\chi^2$ likelihood function obtained by combining the LEP search and the global
electroweak fit. At the 95\% confidence level, one finds~\cite{LEPsearch}
\begin{equation}
m_H \; < 157~{\rm GeV} , \; 186~{\rm GeV} ,
\label{uppermH}
\end{equation}
depending whether one uses precision electroweak data alone, or includes
also the lower limit (\ref{LEPmH}) from the direct search at LEP.
The $\chi^2$ function obtained by combining the LEP limit (\ref{LEPmH})
with the precision electroweak fit is shown in Fig.~\ref{fig:basis}.
Notice the little blip at $m_H \sim 115$~GeV, reflecting the hint
of a signal found in the last run at the highest LEP energies: this was only 
at the 1.7-$\sigma$ level, insufficient to claim any evidence. 

\begin{figure}[t]
\begin{center}
  \epsfig{file=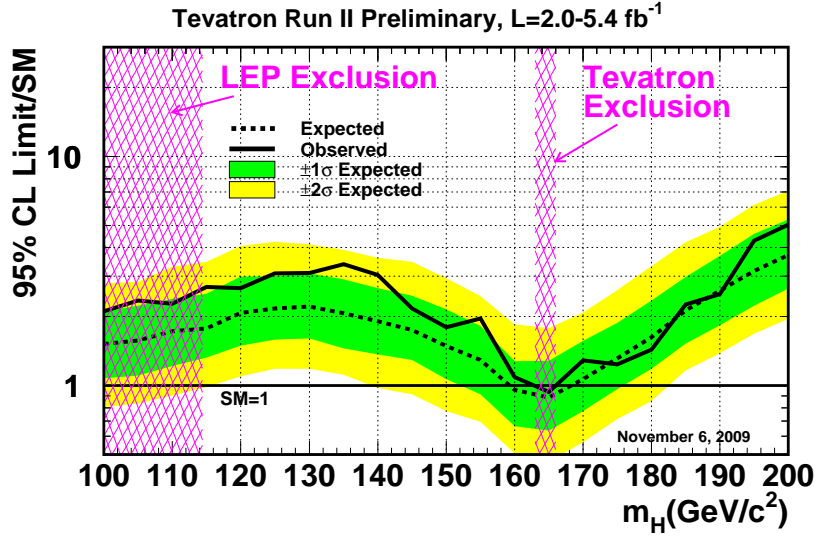, scale=0.6}
\end{center}
  \vspace{-0.5cm}
\caption{Combined 95\% confidence level upper limit from searches by 
CDF and D0 for the Higgs boson
at the Tevatron collider~\protect\cite{TeVsearch}, compared with the SM expectation
\label{fig:TeVH}}
\end{figure}

Searches at the Fermilab Tevatron collider have recently started to exclude
a region of mass for the SM Higgs boson, as also seen in Figs.~\ref{fig:EWWG},
\ref{fig:Andreas} and \ref{fig:basis}.
At the time of writing, these searches exclude~\cite{TeVsearch}
\begin{equation}
163~{\rm GeV} \; < \; m_H \; < \; 166~{\rm GeV}
\label{TeVmH}
\end{equation}
at the 95\% confidence level, as seen in Fig.~\ref{fig:TeVH}. At smaller masses,
the Tevatron 95\% confidence level upper limit on Higgs production and decay
is only a few times bigger than the SM expectations, and the integrated luminosity is
expected to double over the next couple of years.

Figure~\ref{fig:basis} also includes the effect on the $\chi^2$ likelihood function of combining the 
Tevatron search with the global electroweak fit and the LEP search. 
We see from this that the `blow-up' region $m_H > 180$~GeV is strongly
disfavoured: above the 99\% confidence level if the Tevatron data are included,
compared with 96\% if they are dropped~\cite{EEGHR}. The combination of all the data
yields a 68\% confidence level range~\cite{Flacher:2008zq}
\begin{equation}
m_H \; = \; 116^{+ 16}_{-1.3}~{\rm GeV} .
\label{mHrange}
\end{equation}
The Tevatron is expected to continue running until
late 2011, accumulating ${\cal O}(10)$/fb of integrated luminosity. 
That could be sufficient to exclude a SM Higgs boson over all the mass range between 
(\ref{LEPmH}) and (\ref{TeVmH}), which would exclude all the preferred range
(\ref{uppermH}) --- a very intriguing possibility! Alternatively, perhaps the Tevatron
will find some evidence for a Higgs boson with a mass within this range?

\subsubsection{LHC prospects}

The search for the Higgs boson is one of the main raisons d'\^etre of the LHC. 
Many mechanisms may make important contributions to SM
Higgs production at the LHC. If the Higgs boson is relatively light, as suggested above,
the dominant production mechanisms are expected to be $gg \to H$ and
$W^+ W^- \to H$, where the $W^\pm$ are radiated off incoming quarks: $q \to W q'$.

\begin{figure}[t]
\begin{center}
  \epsfig{file=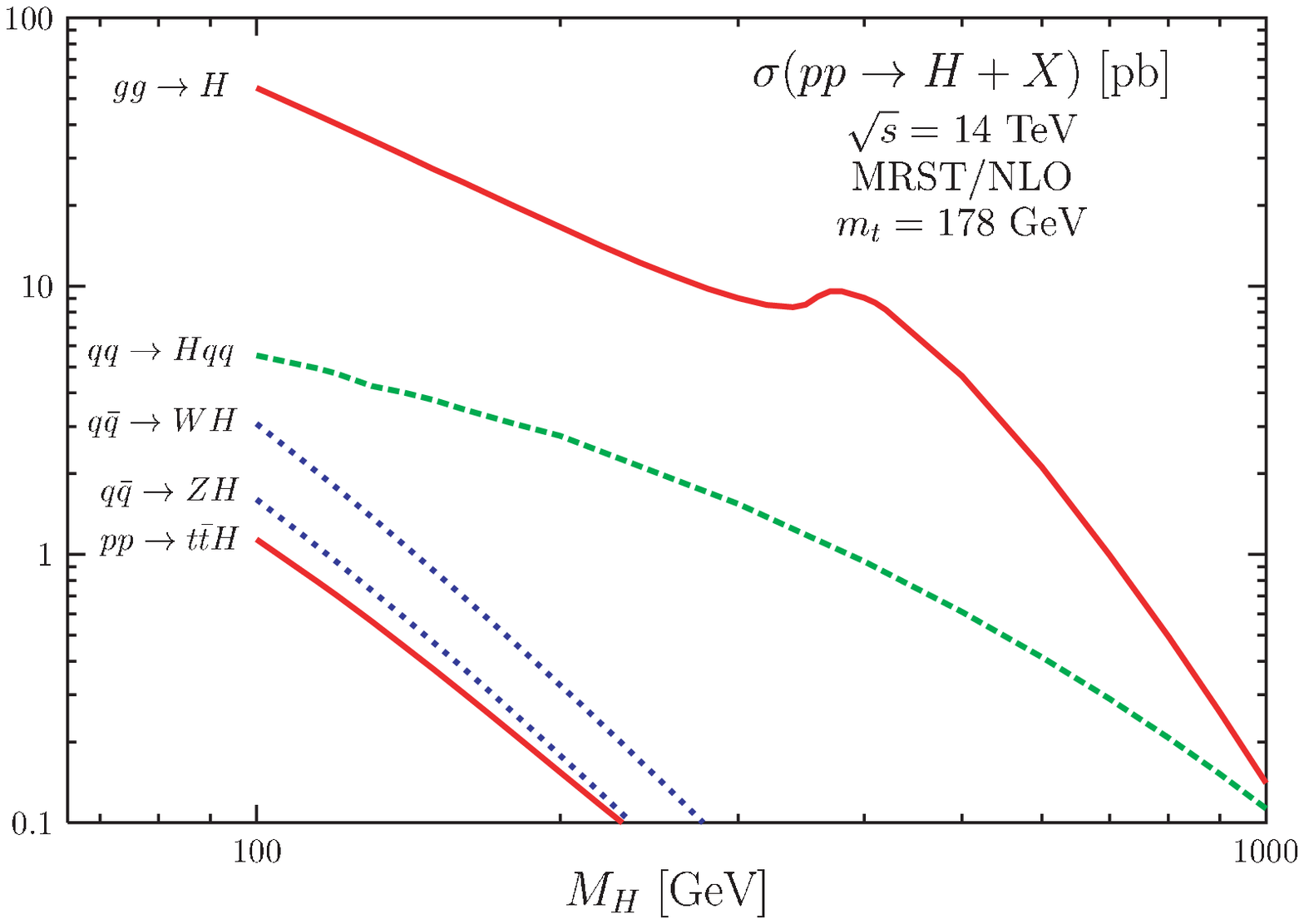, scale=0.47}
  \epsfig{file=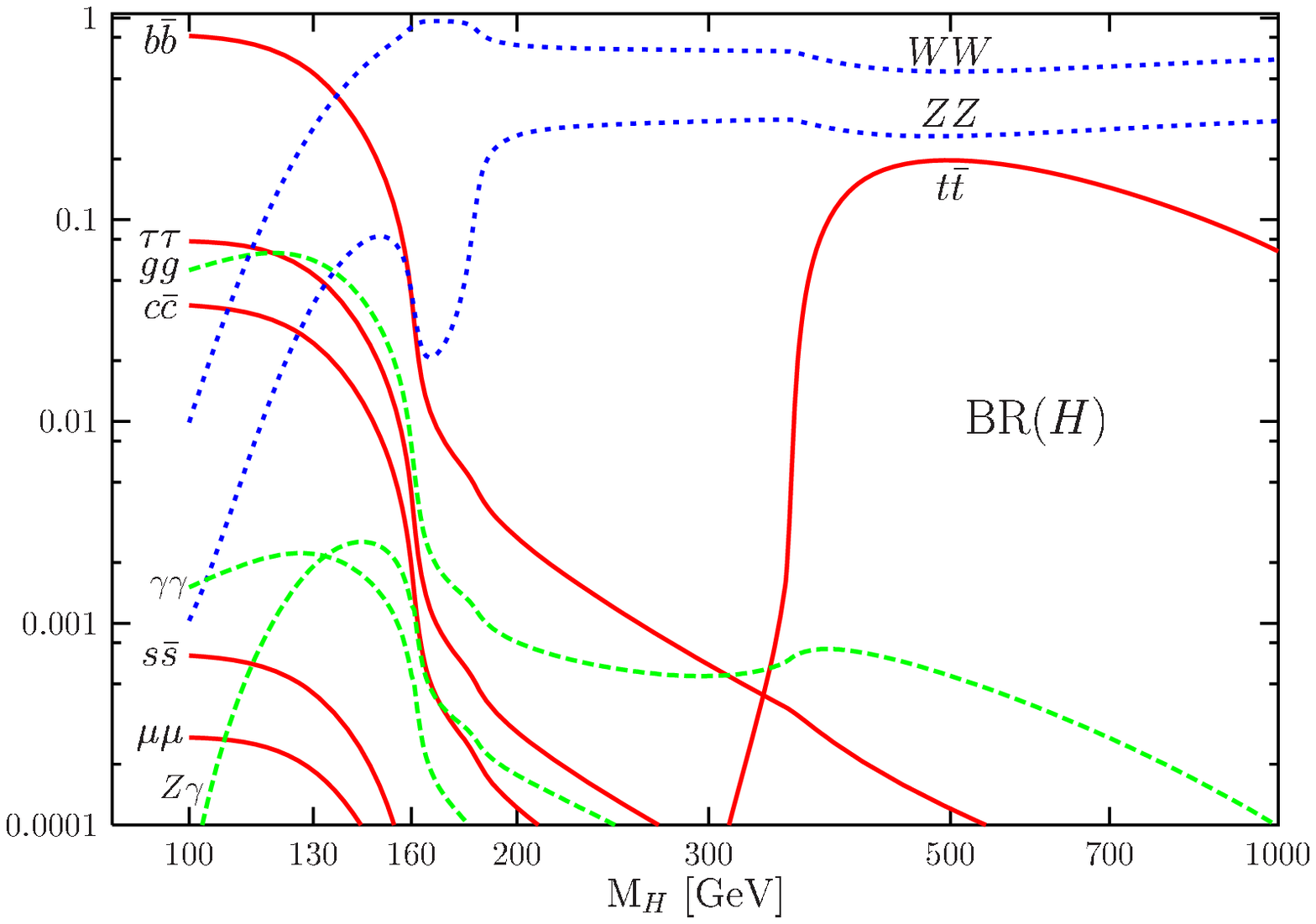, scale=0.47}
\end{center}
  \vspace{-0.5cm}
\caption{Left: the dominant mechanisms for producing a SM Higgs
boson at the LHC at 14~TeV, and right: the most important branching ratios for a SM Higgs
boson, taken from~\protect\cite{Djouadi}
\label{fig:Djouadi}}
\end{figure}

As already mentioned, the fact that Higgs
couplings to other particles are proportional to their masses implies that the Higgs
prefers to decay into the heaviest particles that are kinematically accessible. As
seen in Fig.~\ref{fig:Djouadi}, this
means that a Higgs lighter than $\sim 130$~GeV prefers to decay into ${\bar b}b$,
whereas a heavier Higgs prefers to decay into $W^+ W^-$ and $Z^0 Z^0$. However,
couplings to lighter particles can become important under certain circumstances.
For example, whilst there is no tree-level coupling to gluons because they are
massless, one is induced by loops of heavy particles such as the top quark. For the
same reason, there is no tree-level Higgs coupling to photons, but the Higgs boson
may decay into $\gamma \gamma$ {\it via} top and $W^\pm$ loops. Although this
decay has a very small branching ratio, it is very distinctive experimentally, and
may be detectable at the LHC if the SM Higgs weighs $< 130$~GeV.

\begin{figure}[t]
\begin{center}
  \epsfig{file=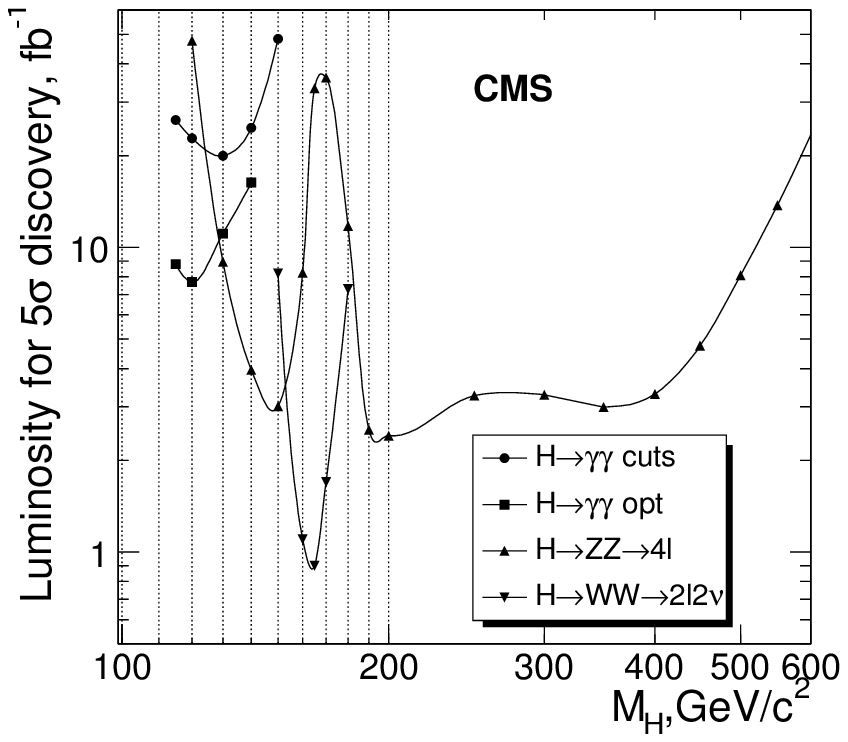, scale=0.9}
  \epsfig{file=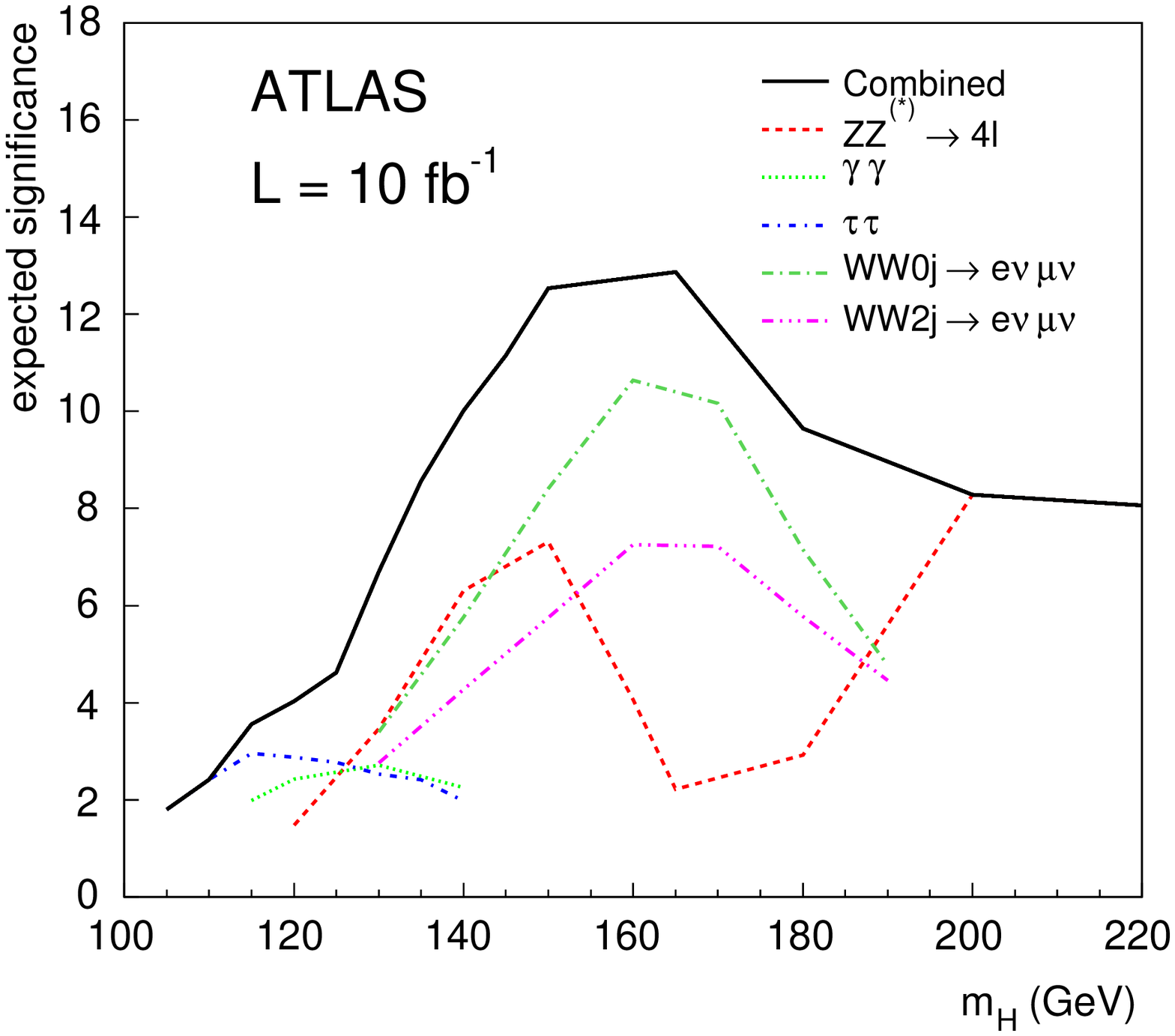, scale=0.4}
\end{center}
  \vspace{-0.5cm}
\caption{Left: the amount of integrated luminosity that would be
required by CMS~\protect\cite{CMS} to discover a SM Higgs boson as a function of $m_H$.
Right: the significance expected by ATLAS~\protect\cite{ATLAS} for a SM Higgs boson,
assuming 10/fb of data at 14~TeV.
\label{fig:LHCH}}
\end{figure}

Figure~\ref{fig:LHCH} displays estimates of the sensitivities of CMS (left)~\cite{CMS}
and ATLAS (right)~\cite{ATLAS} to a SM Higgs boson. A
fraction of an inverse femtobarn per experiment may suffice to exclude a Higgs boson over a
large range of masses from $\sim 150$~GeV to $\sim 400$~GeV. An integrated
luminosity $\sim 1$/fb per experiment would be needed to discover a Higgs boson with a mass
in a similar range, but more luminosity would be required if $m_H < 150$~GeV.
Indeed, a luminosity $\sim 5$/fb per experiment would be needed for discovery over
all the displayed range of $m_H$, down to the LEP limit. One way or another,
the LHC will be able determine whether or not there is a SM Higgs boson.

\subsection{Issues beyond the Standard Model}

The Standard Model, however, is not expected to be the final description of the fundamental 
interactions, but rather an effective low-energy (up to a few TeV) manifestation of a more 
complete theory.

Some of the outstanding questions in the Standard Model are:
\begin{itemize}
 \item \textbf{How is electroweak symmetry broken?} In other words, how do gauge bosons acquire mass? We have seen that the Standard Model incorporates the Higgs mechanism in the form of a single weak-isospin doublet with a non-zero v.e.v. in order to generate the gauge boson masses, but this is not the only possible way in which the electroweak symmetry can be broken. For instance, there could be more than one Higgs doublet, the Higgs could be a pseudo-Goldstone boson (with a low mass
relative to the mass scale of some new interaction) or electroweak symmetry could be broken by a condensate of new particles bound by a new strong interaction. We cover a few of the possibilities in Lecture~\ref{Section_EWSBBSM}.

 \item \textbf{How do fermions acquire mass?} Electroweak symmetry breaking is a necessary, but not a sufficient, condition to generate the fermion masses. There also needs to be a mechanism that generates the required Yukawa couplings [see Eq.~(\ref{EqLagYukawa})] between the fermions and the (effective) Higgs field. The separation between electroweak symmetry breaking and the generation of fermion masses is made evident in models of dynamical symmetry breaking, such as technicolour (see Section~\ref{Section_EWSBBSM}), where the breaking is carried out by the formation of a condensate of particles associated to a new interaction, a process which, while breaking electroweak symmetry and giving masses to the gauge bosons, does not necessarily give masses to the fermions. This situation is resolved by adding new interactions which are responsible for generating the fermion masses. Within the Standard Model, there are no predictions for the values of the Yukawa couplings. Moreover, the values required to generate the correct masses for the three charged leptons and the six quarks span six orders of magnitude, which presumably makes the mechanism for the generation of the couplings highly non-trivial. 

 \item \textbf{The hierarchy problem.} Why should the Higgs mass remain low, $m_H \lesssim 1$ TeV, in the face of divergent quantum loop corrections? Following~\cite{Quigg:2009vq}, the Higgs mass can be expanded in perturbation theory as
 \begin{equation}
  m_H^2 \left(p^2\right) = m_{0,H}^2 + \mathcal{C} g^2 \int_{p^2}^{\Lambda^2} dk^2 + \ldots ~,
 \end{equation}
where $m_{0,H}^2$ is the tree-level (classical) contribution to the Higgs mass squared, $g$ is the coupling constant of the the theory, $\mathcal{C}$ is a model-dependent constant, and $\Lambda$ is the reference scale up to which the Standard Model is assumed to remain valid. The integrals represent contributions at loop level and are apparently quadratically divergent. If there is no new physics, the reference scale is high, like the Planck scale, $\Lambda \sim M_{\text{Pl}} \approx 10^{19}$ GeV or, in Grand Unified Theories (GUTs), $\Lambda \sim M_{GUT} \approx 10^{15} \-- 10^{16}$ GeV
(see Lecture 4). Clearly, both choices result in large corrections to the Higgs mass. In order for these to be small, there are two alternatives: either the relative magnitudes of the tree-level and loop contributions are finely tuned to yield a net contribution that is small (a feature that is disliked by physicists, but which Nature might have implemented), or there is a new symmetry, like supersymmetry, that protects the Higgs mass, as discussed in Lecture~3. 

 \item \textbf{The vacuum energy problem}. The value of the scalar potential, Eq.~(\ref{EqScalarPot}), at the v.e.v. $\langle \phi \rangle_0$ of the Higgs boson is
 \begin{equation}
  V\left(\langle \phi^\dagger \phi\rangle_0\right) = \frac{\mu^2 v^2}{4} < 0 ~.
 \end{equation}
 Hence, because the Higgs mass is $m_H^2 = -2 \mu^2$, this corresponds to a uniform vacuum energy density
 \begin{equation}
  \rho_H = - \frac{m_H^2 v^2}{8} ~.
 \end{equation}
Taking $v = \left(G_F \sqrt{2}\right)^{-1/2} \approx 246$ GeV for the Higgs v.e.v. and using the current experimental lower bound on the Higgs mass \cite{Amsler:2008zzb}, $m_H \gtrsim 114.4$ GeV, we have
 \begin{equation}
 - \rho_H \gtrsim 10^8 ~\text{GeV}^4 ~.
 \end{equation}
 On the other hand, if the apparent accelerated expansion of the Universe --- originally inferred from observations of type 1A supernovae~\cite{SNae} --- is attributed to a non-zero cosmological 
 constant corresponding to $\sim 70\%$ of the total energy density of the Universe~\cite{Amsler:2008zzb}, the required energy density should be
 \begin{equation}
  \rho_{\text{vac}} \sim 10^{-46} ~\text{GeV}^4 ~,
 \end{equation} 
which is at least 54 orders of magnitude lower than the corresponding density from the Higgs field, and of the opposite sign! The character of this dark energy remains 
unexplained~\cite{Dobado:2008xn,Harvey:2009zz}, and will probably remain so until we have a full
quantum theory of gravity.
 \item \textbf{How is flavour symmetry broken?} Part of the flavour problem in the Standard Model is, of course, related to the widely different mass assignments of the fermions ascribed to the Yukawa couplings, which also set the mixing angles between flavour and mass eigenstates. Mixing occurs both in the quark and the lepton sectors, the former being parametrized by the Cabibbo--Kobayashi--Maskawa (CKM) matrix and the latter, by the 
 Maki--Nakagawa--Sakata (MNS) matrix. These are complex rotation matrices, and can each be written in terms of three mixing angles and one CP-violating phase ($\delta$)~\cite{Amsler:2008zzb}:
 \begin{equation}
  V =
  \left(\begin{array}{ccc}
   c_{12} c_{13}                                     & s_{12} c_{13}                                     & s_{13} e^{-i\delta} \\
   -s_{12} c_{23} - c_{12} s_{23} s_{13} e^{i\delta} & c_{12} c_{23} - s_{12} s_{23} s_{13} e^{i\delta}  & s_{23} c_{13} \\
   s_{12} s_{23} - c_{12} c_{23} s_{13} e^{i\delta}  & -c_{12} s_{23} - s_{12} c_{23} s_{13} e^{i\delta} & c_{23} c_{13}
  \end{array}\right) ~,
  \label{fmix}
 \end{equation}
 where $c_{ij} \equiv \cos\left(\theta_{ij}\right)$, $s_{ij} \equiv \sin\left(\theta_{ij}\right)$. While the off-diagonal elements in the quark sector are rather small (of order $10^{-1}$ to $10^{-3}$), so that there is little mixing between quark families, in the lepton sector the off-diagonal elements (except for $\left[V_{\text{MNS}}\right]_{e3}$, which is close to zero) are of order 1, so that the mixing between neutrino families is large. The Standard Model does not provide an explanation for this difference.


 \item \textbf{What is dark matter?} The observation that galaxy rotation curves do not fall off with radial distance from the galactic centre can be explained by postulating the existence of a new type of weakly-interacting matter, \textit{dark matter}, in the halos of galaxies.
Supporting evidence from the cosmic microwave background (CMB) indicates that the dark matter makes up $\sim 25\%$ of the energy density of the Universe~\cite{Komatsu:2008hk}. Dark matter is usually thought to be composed of neutral relic particles from the early Universe. Within the Standard Model, neutrinos are the only candidate massive neutral relics. However, they contribute only with a normalized density of $\Omega_\nu \gtrsim 1.2\left(2.2\right) \times 10^{-3}$ if the mass hierarchy is normal (inverted), or no more than $10\%$ if the lightest mass eigenstate lies around 1 eV, that is, if the hierarchy is degenerate~\cite{Quigg:2009vq}. On top of that, structure formation indicates that dark matter should be cold, i.e., non-relativistic at the time of structure formation, whereas neutrinos would have been relativistic particles. Within the Minimal Supersymmetric extension of the Standard Model (MSSM), the lightest supersymmetric partner, called a \textit{neutralino}, is a popular dark matter candidate~\cite{EHNOS}. 

 \item \textbf{How did the baryon asymmetry of the Universe arise?} The antibaryon density of the Universe is negligible, whilst the baryon-to-photon ratio has been determined, using WMAP data~\footnote{We use here values from the three-year WMAP analysis~\cite{Spergel:2006hy}, rather than the five-year analysis~\cite{Komatsu:2008hk}, in order to be consistent with the values quoted by the Particle Data Group~\cite{Amsler:2008zzb} summary tables.} of the CMB \cite{Spergel:2006hy} to be 
 \begin{equation}
  \eta = \frac{n_b-\overline{n}_b}{n_\gamma} \simeq \frac{n_b}{n_\gamma} = 6.12\left(19\right) \times 10^{-10} ~,
 \end{equation}
 where $n_b$, $\overline{n}_b$, and $n_\gamma$ are the number densities of baryons, antibaryons, and photons, respectively. The fact that the ratio is not zero is intriguing considering that, in a cosmology with an inflationary epoch, conventional thermal equilibrium processes would have yielded an equal number of particles and antiparticles. In 1967, Sakharov~\cite{Sakharov:1967dj} established three necessary conditions (more fully explained in~\cite{Cline:2006ts}) for the particle--antiparticle asymmetry of the Universe to be generated: 
 \begin{enumerate}
  \item violation of the baryon number, $B$;
  \item microscopic C and CP violation;
  \item loss of thermal equilibrium.
 \end{enumerate}
 Otherwise, the rate of creation of baryons equals the rate of destruction, and no net asymmetry results.
 In the perturbative regime, the Standard Model conserves $B$; however, at the non-perturbative level, $B$ violation is possible through the triangle anomaly~\cite{Theta}. The loss of thermal equilibrium may occur naturally through the expansion of the Universe, and CP violation enters the Standard Model through the complex phase in the CKM matrix~\cite{Amsler:2008zzb}. However, the CP violation observed so far, which is described by the Kobayashi--Maskawa mechanism of the Standard Model, is known to be insufficient to explain the observed value of the ratio $\eta$, and new physics is needed. One possible solution lies in leptogenesis scenarios, where the baryon asymmetry is a result of a previously existing lepton asymmetry generated by the decays of heavy sterile neutrinos~\cite{Pilaftsis:2009pk}. 

 \item \textbf{Quantization of the electric charge}. It is an experimental fact that the charges of all observed particles are simple multiples of a fundamental charge, which we can take to be the electron charge, $e$. Dirac \cite{Dirac:1931kp,Dirac:1948um,Dirac:1976bf} proved that the existence of even a single magnetic monopole (a magnet with only one pole) is sufficient to explain the quantization of the electric charge, but the particle content of the Standard Model (see Table~\ref{TabSMPartCont}) does not include magnetic monopoles. Hence, in the absence of any indication for a magnetic monopole, the explanation of charge quantization must lie beyond the Standard Model. 
Indeed, so far there has only been one candidate monopole detection event in a single superconducting loop~\cite{Cabrera:1982gz}, in 1982, and the monopole interpretation of the event has now been largely discounted. 
One expects monopoles to be very massive and non-relativistic at present, in which case time-of-flight measurements in the low-velocity regime ($\beta \equiv v/c \ll 1$) become important. The best current direct upper limit on the supermassive monopole flux comes from cosmic-ray 
observations~\cite{Amsler:2008zzb}:
 \begin{equation}
  \Phi_{\text{1pole}} < 1.0 \times 10^{-15} ~\text{cm}^{-2} \text{sr}^{-1} \text{s}^{-1} ~,
 \end{equation}
 for $1.1 \times 10^{-4} < \beta < 0.1$.
An alternative route towards charge quantization is {\it via} a Grand Unified Theory (GUT) (see Lecture~4).
Such a theory implies the existence of magnetic monopoles that would be so massive
that their cosmological density would be suppressed to an unobservably small value
by cosmological inflation.

\item \textbf{How to incorporate gravitation?} One of the most obvious shortcomings of the 
Standard Model is that it does 
not incorporate gravitation, which is described on a classical level by general relativity. 
However, the consistency of our physical theories requires a quantum theory of gravity. The 
main difficulty in building a quantum field theory of gravity is its non-renormalizability. String 
theory~\cite{Schwarz:1998ny} and loop quantum gravity~\cite{Ashtekar:2007tv} constitute 
attempts at building a quantized theory of gravity. If one could answer this question, one would 
surely also be able to solve the dark energy problem. Conversely, solving the dark energy
problem presumably requires a complete quantum theory of gravity.

\end{itemize}


\section{Electroweak symmetry breaking beyond the Standard Model}\label{Section_EWSBBSM}

\subsection{Theorists are getting cold feet}

After so many years, it seems that we will soon know whether a Higgs boson
exists in the way predicted by the Standard Model, or not. Closure at last!

Like the prospect of an imminent hanging, the prospect of imminent Higgs
discovery concentrates wonderfully the minds of theorists, and many theorists
with cold feet are generating alternative models, as prolifically as monkeys on 
their laptops. These serve the invaluable purpose of providing benchmarks
that can be compared and contrasted with the SM Higgs. Experimentalists
should be ready to search for reasonable alternatives, already at the Tevatron
and also at the LHC once it is up
and running, and they should be on the look-out for tell-tale deviations from
the SM predictions if a Higgs boson should appear.

Even within the SM with a single elementary Higgs boson, questions are
being asked. As discussed in the previous section, within this framework
the experimental data seem to favour a light Higgs boson. However, the
interpretation of the precision electroweak data has been challenged. Even
if one accepts the data at face value, the SM fit may need to take into
account non-renormalizable, higher-dimensional interactions that could
conspire to permit a heavier SM Higgs boson? In this section, in addition
to these possibilities, we explore several mechanisms of electroweak symmetry 
breaking beyond the 
minimal Higgs, i.e., a single elementary $SU(2)$ Higgs doublet whose potential is arranged 
to have a non-zero v.e.v.

Any successful model of electroweak symmetry breaking must give masses to
the matter fermions as well as the weak gauge bosons. This could be achieved
using either a single boson, as in the SM, or two of them, as in the 
Minimal Supersymmetric extension of the Standard Model (MSSM)~\footnote{We 
leave the treatment of the Higgs sector within the MSSM for a later section.},
or by some composite of new fermions with new strong interactions that
generate a non-zero v.e.v. as in (extended) technicolour models, or by some Higgsless mechanism. 

We do know, however, that the energy scale at which EWSB must occur is 
${\cal O}(1)$~TeV~\cite{Lane:1993wz}. This scale is set by the decay constant of the three 
Goldstone bosons that, through the Higgs mechanism, are transformed into the longitudinal 
components of the weak gauge bosons:
\begin{equation}
 F_\pi = \left(G_F \sqrt{2} \right)^{-1/2} \approx 246 ~\text{GeV} ~.
\end{equation}
If there is any new physics associated to the breaking of electroweak symmetry, it must occur 
near this energy scale. Another way to see how this energy scale emerges is to consider
$s$-wave $WW$ scattering. In the absence of a direct-channel Higgs pole, this
amplitude would violate the unitarity limit at an energy scale $\sim 1$~TeV (\ref{LQT1}).

It is the scale of 1 TeV, and the typical values of QCD and electroweak cross sections at this 
energy, $\sigma \simeq 1~\text{nb--}1~\text{fb}$, that set the energy and luminosity 
requirements of the LHC: $\sqrt{s} = 14$ TeV and $\mathcal{L} = 10^{34}$~cm$^{-2}$~s$^{-1}$
for $pp$ collisions~\cite{Amsler:2008zzb}. This energy scale is to be contrasted with the 
energy scale of the other unexplained broken symmetry in the SM, namely flavour symmetry, 
which is completely unknown: it may lie anywhere from 1 TeV up to the Planck scale, 
$M_{P} = 1.22 \times 10^{19}$~GeV.

There are some general constraints that any proposed model of electroweak symmetry 
breaking must satisfy~ \cite{Gunion:1990}. First, the model must predict a value of the 
$\rho$ parameter, Eq.~(\ref{EqRhoDef}), that agrees with the value $\rho \approx 1$ 
found experimentally. The desired value $\rho = 1$ is found automatically in models that 
contain only Higgs doublets and singlets, but would be violated in models with scalar
fields in larger $SU(2)$ representations. A second constraint comes from the strict 
upper limits on flavour-changing neutral currents (FCNCs). These are absent at tree 
level in the minimal Higgs model, a fact that is in general not true in non-minimal models. 

\subsection{Interpretation of the precision electroweak data}

\begin{figure}[t]
\begin{center}
  \epsfig{file=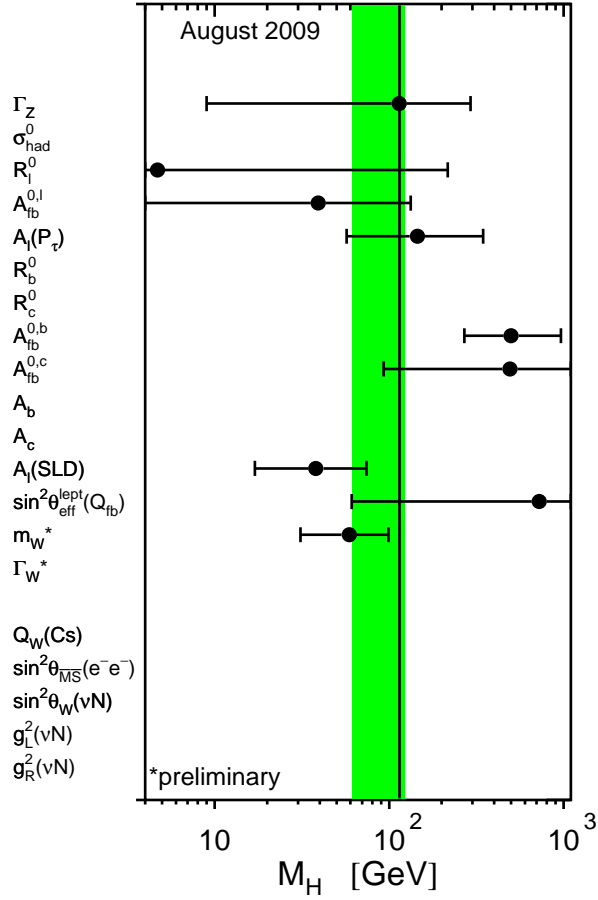, scale=0.4}
\end{center}
  \vspace{-0.5cm}
\caption{The 68\% confidence level ranges for $m_H$ that are indicated
by various individual electroweak measurements~\protect\cite{EWWG09}
\label{FigshowH}}
\end{figure}

It is notorious that the two most precise measurements at the $Z^0$ peak, namely 
the asymmetries measured with leptons (particularly $A_\ell(SLD)$) and hadrons
(particularly $A_{FB}^{0,b}$), do not agree very well~\cite{Mike},
as seen in Table~\ref{TabFitExpComp} and Fig.~\ref{FigFitExpComp}~\footnote{Another 
anomaly is exhibited by the NuTeV data on 
deep-inelastic $\nu - N$ scattering~\cite{NuTeV}, but this is easier to explain away as due 
to our lack of understanding of hadronic effects.}.
Within the SM, they favour different values of $m_H$, around 40 and 500~GeV, respectively,
as seen in Fig.~\ref{FigshowH}. 
Most people think that this discrepancy is just a statistical fluctuation, since the total
$\chi^2$ of the global electroweak fit is acceptable ($\chi^2 = 17.3$ for 13 d.o.f., 
corresponding to a probability of 18\%~\cite{EWWG09}), but it may also
reflect the existence of an underestimated systematic error. However, if there were a big
error in $A_{FB}^{0,b}$, the preferred value of $m_H$ would be pulled uncomfortably
low by the other data, whereas if there was a big error in the interpretation of the leptonic
data $m_H$ would be pulled towards much higher values. On the other hand, if we
take both pieces of data at face value, perhaps the discrepancy is evidence for new
physics at the electroweak scale. In this case there would be no firm basis 
for the prediction of a light Higgs boson, which is based on a Standard Model 
fit, and no fit value of $m_H$ could be trusted?

\subsection{Higher-dimensional operators within the SM}

The Standard Model should be regarded simply as an effective low-energy theory, 
to be embeded within some more complete and satisfactory theory. Therefore,
one should anticipate that the renormalizable dimension-four interactions of the SM 
could be supplemented by higher-dimensional operators of the general form:
\begin{equation} 
{\cal L}_{eff} \;  = \; {\cal L}_{SM} + \Sigma_i \frac{c_i}{\Lambda_i^p} {\cal O}^{4+p}_i ,
\label{higherD}
\end{equation}
where $\Lambda_i$ is a scale at which the supplementary
interaction ${\cal O}^{4+p}_i$ of dimension $4 + p$ appears to be generated.
A global fit to the precision electroweak data suggests that, if the Higgs is 
indeed light, the coefficients of these additional interactions are small: 
\begin{equation}
\Lambda_i \; > \; {\cal O}(10)~{\rm TeV}
\label{newLambda}
\end{equation}
for $c_i = \pm 1$. It is then a problem to understand the `little hierarchy' between
the electroweak scale and $\Lambda_i$.

\begin{figure}[t]
\begin{center}
  \epsfig{file=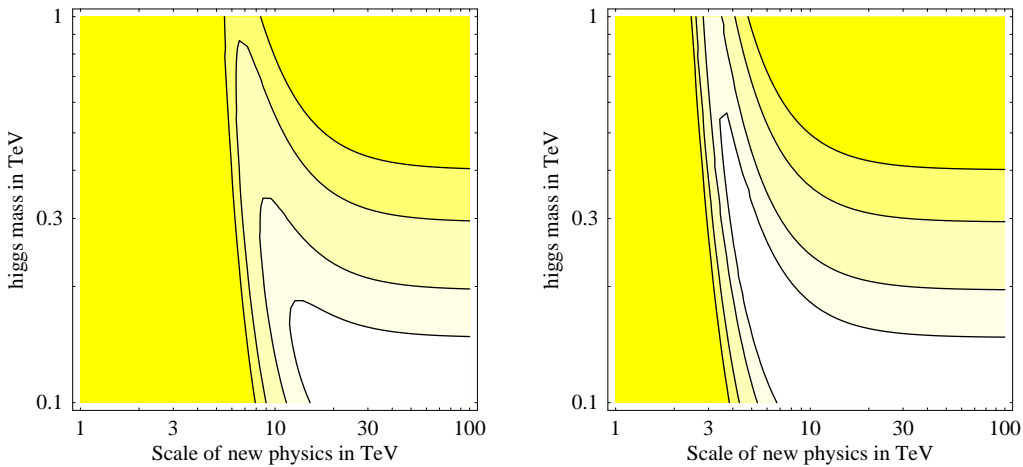, scale=0.8}
\end{center}
  \vspace{-0.5cm}
\caption{The 68\%, 90\%, 99\% and 99.9\% confidence levels fit for global
electroweak fits including two different types of higher-dimensional operators,
demosntrating that they might conspire with a relatively heavy Higgs boson to
yield and acceptable fit~\protect\cite{BS}
\label{fiGBS}}
\end{figure}

However, conspiracies are in principle possible, which could allow $m_H$ to be large, 
even if one takes the precision electroweak data at face value~\cite{BS}. Examples are 
shown in Fig.~\ref{fiGBS}, where one sees corridors of allowed parameter space 
extending up to a heavy Higgs mass, if $\Lambda_i \ll 10$~TeV. 
A theory that predicts a heavy Higgs boson but 
remains consistent with the precision electroweak data should predict a correlation 
of the type seen in Fig.~\ref{fiGBS}. At the moment, this may seem unnatural to 
us, but Nature may know better. In any case, any theory 
beyond the SM must link the value of $m_H$ and the scales of these 
higher-dimensional effective operators in some way. 

\subsection{Little Higgs}

One way to address the `little hierarchy problem' and
explain the lightness of the Higgs boson (if it is light) is by treating it as a 
pseudo-Goldstone boson corresponding to a spontaneously broken approximate 
global symmetry of a new strongly-interacting sector at some higher mass scale,
the `little Higgs' scenario~\cite{LH}.
Such a theory would work by analogy with the pions in QCD, which have masses
far below the generic mass scale of the strong interactions $\sim 1$~GeV. 

\begin{figure}[htb]
\begin{center}
\includegraphics[width=.4\textwidth]{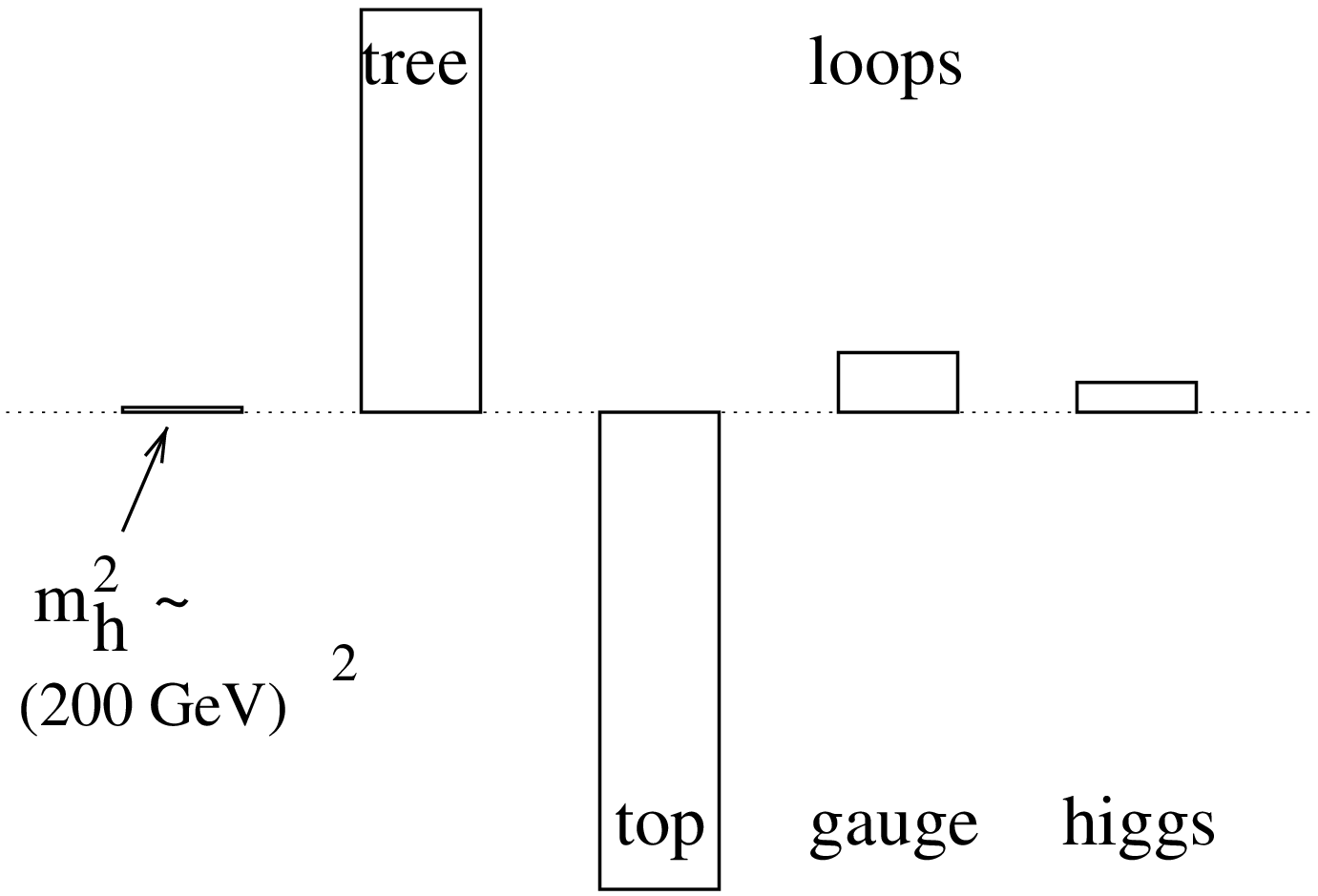}
\includegraphics[width=.4\textwidth]{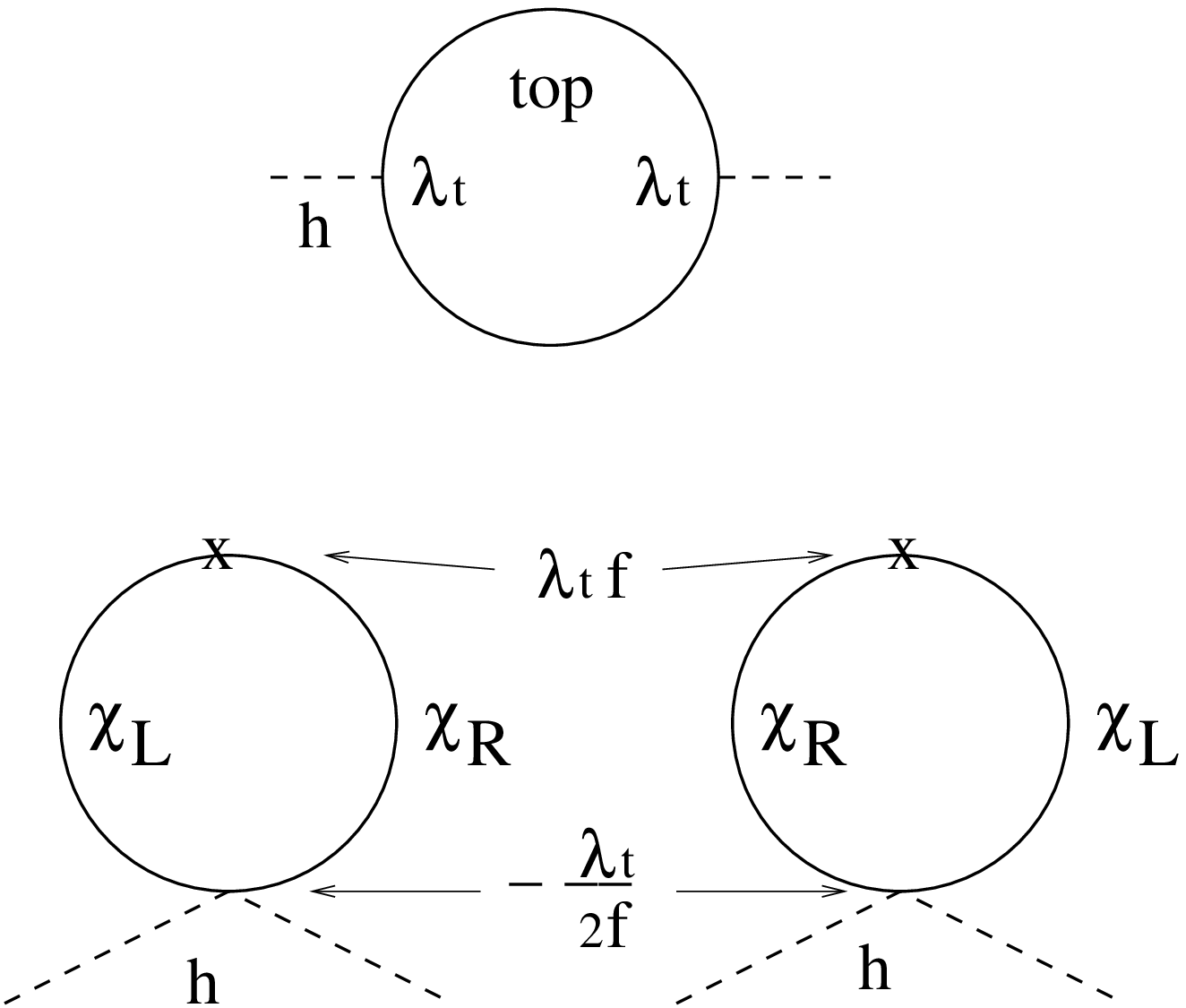}
\end{center} 
\caption{Left: If the Standard Model Higgs boson weighs around 200~GeV, the 
top-quark loop contribution to its physical mass (calculated here with a loop 
momentum cutoff of 10~TeV) must cancel delicately against 
the tree-level contribution. Right: In `little 
Higgs' models, the top-quark loop is cancelled by 
loops containing a heavier charge-2/3 quark~\protect\cite{LH}.} 
\label{FigLH}

\end{figure}

If the Higgs is a pseudo-Goldstone boson, its mass is protected from
acquiring quadratically-divergent loop corrections~\cite{ArkaniHamed:2001nc}. This occurs 
as a result of the particular manner in which the gauge and Yukawa couplings break the 
global symmetries: more than one couplng must be turned on at a time in order for the 
symmetry to be broken, a feature known as `collective symmetry 
breaking'~\cite{Cheng:2004yc,Perelstein:2005ka}. As a consequence, the quadratic
divergences that would normally appear in the SM are cancelled by new particles,
sometimes in unexpected ways. For example,  the top-quark loop contribution to the 
Higgs mass-squared has the general form
\begin{equation}
\delta m^2_{H,top} (SM) \; \sim \; (115~\text{GeV})^2 \left( \frac{\Lambda}{400~{\rm GeV}} \right)^2 . 
\label{topdivergence}
\end{equation}
As illustrated in Fig.~\ref{FigLH}, in little Higgs models this is cancelled by the loop 
contribution due to a new heavy top-like quark $T$ with charge +2/3 that is a singlet
of $SU(2)_L$, leaving a residual logarithmic
divergence:
\begin{equation}
\delta m^2_{H,top} (LH) \; \sim \; \frac{6 G_F m^2_t}{\sqrt{2 \pi^2}} m_T^2 
{\rm log} \frac{\Lambda}{m_T} . 
\label{Tcancel}
\end{equation}
Analogously, the quadratic loop divergences associated with the gauge bosons 
and the Higgs boson of the Standard Model are cancelled by loops of new 
gauge bosons and Higgs bosons in little Higgs models.

The net result is a spectrum containing a relatively light Higgs boson and 
other new particles that may be somewhat heavier: 
\begin{equation}
M_T < 2~{\rm TeV} \left( \frac{m_H}{200~{\rm GeV}} \right)^2,  
M_{W'} < 6~{\rm TeV} \left( \frac{m_H}{200~{\rm GeV}} \right)^2, 
M_{H^{++}} < 10~{\rm TeV}.
\label{littleHspectrum}
\end{equation} 
The extra $T$ quark, in particular, should be accessible to the LHC.
In addition, there should be more new strongly-interacting
physics at some energy scale at or above 10~TeV, 
to provide the ultra-violet completion of the theory.

\subsection{Technicolour}

Little Higgs models are particular examples of composite Higgs models,
of which the prototypes were technicolour models~\cite{FS,HS}.
In these models, electroweak symmetry is broken dynamically, by the introduction of a 
new non-Abelian gauge interaction~\cite{Weinberg:1979bn,Susskind:1978ms,Martin:2008cd}
that becomes strong at the TeV scale. The building blocks are massless fermions called 
technifermions and new force-carrying fields called technigluons. As in the SM,
the left-handed components of the technifermions are assigned to electroweak doublets, 
while the right-handed components form electroweak singlets, and both components carry 
hypercharge. At $\Lambda_{\text{EW}} \sim 1$ TeV the technicolour coupling becomes 
strong, which leads to the formation of condensates of technifermions with v.e.v.'s
\begin{equation}
 \langle \phi \rangle = \langle \overline{f}_L f_R \rangle \equiv v ~.
\end{equation}
Because the left-handed technifermions carry electroweak quantum numbers, 
but the right-handed ones do not, the formation of this technicondensate breaks 
electroweak symmetry.

The massless technifermions have the chiral symmetry group
\begin{equation}
 G_\chi = SU(2N_D)_L \otimes SU(2N_D)_R \supset SU(2)_L \otimes SU(2)_R ~,
\end{equation}
where $N_D$ is the number of technifermion doublets. When the condensate forms, 
this large global symmetry is broken down to 
\begin{equation}
 S_\chi = SU(2N_D) \supset SU(2)_V ~,
\end{equation}
where $V$ refers to the vector combination of left and right currents, 
and $4N_D^2 - 1$ massless Goldstone bosons appear, 
with decay constant $F_\pi^{\text{TC}}$. Similarly to the Higgs mechanism in the SM, 
three of these bosons are `eaten' and become the longitudinal components of the 
$W^\pm$ and $Z^0$ weak bosons, which acquire masses \cite{Lane:1993wz}
\begin{equation}
 m_W = \frac{g}{2} \sqrt{N_D} F_\pi^{\text{TC}} ~~~~, ~~~~~
 m_Z = \frac{1}{2} \sqrt{g^2+g^{\prime~2}} \sqrt{N_D} F_\pi^{\text{TC}} = \frac{m_W}{\cos\left(\theta_W\right)} ~.
\end{equation}
The scale $\Lambda_{\text{TC}}$ at which technicolour interactions become strong is related
to the magnitude of electroweak symmetry breaking, namely to the weak scale, by:
\begin{equation}
 \Lambda_{\text{TC}} = \text{few} \times F_\pi^{\text{TC}} ,
 F_\pi^{\text{TC}} = F_\pi / \sqrt{N_D} ~,
\end{equation}
where $F_\pi = v \approx 246$ GeV. The breaking of the chiral symmetry in technicolour 
is reminiscent of chiral symmetry in QCD, which provides a working precedent for the 
model~\footnote{The condensation phenomenon also occurs in solid-state physics: 
dynamical symmetry breaking in superconductors is achieved by the formation of Cooper 
pairs~\cite{Bardeen:1957mv}, which are condensates of electron pairs with charge $-2e$.}. 
Technicolour guarantees $\rho = m_W^2 / \left( m_Z^2 \cos\left(\theta_W\right) \right) = 1 + \mathcal{O}\left(\alpha\right)$ through a custodial $SU(2)_R$ flavour symmetry in $G_\chi$~\cite{Lane:1993wz},
which is traceable to the quantum numbers assigned to the technifermions.

Dynamical symmetry breaking addresses the problem of quadratic divergences in the
Higgs mass-squared, such as (\ref{topdivergence}), by introducing a composite Higgs
boson that `dissolves' at the scale $\Lambda_{\text{TC}}$. In this way, it makes loop
corrections to the electroweak scale `naturally' small. Moreover, technicolour
has a plausible mechanism for stabilizing the weak scale far below the 
Planck scale. The idea is that technicolour, being an asymptotically-free theory, couples 
weakly at very high energies $\sim 10^{16}$ GeV, and then evolves to become strong at 
lower energies $\sim 1$~TeV~\cite{FS}. However, writing down an explicit GUT scenario based
on this scenario has proved elusive. 

As described above, the simplest technicolour models could provide masses for
the gauge bosons $W^\pm$ and $Z^0$, but not to the matter fermions. Additions to 
technicolour could allow for quark and lepton masses by introducing new interaction with
technifermions, as in `extended technicolour' models~\cite{HS,APS}. However, these had severe
problems with flavour-changing neutral interactions~\cite{DE} and a proliferation of relatively
light pseudo-Goldstone bosons that have not been seen by experiment~\cite{ENRS}.

Moreover, a generic problem with technicolour models is presented by the global
electroweak fit discussed in the first Lecture. 
The preference within the SM for a relatively light Higgs boson (\ref{Higgsmass})
may be translated into constraints on the possible vacuum
polarization effects due to generic new physics models. 
QCD-like technicolour models have many strongly-interacting
dynamical scalar resonances in the TeV range, e.g., a scalar 
analogous to the $\sigma$ meson of QCD that corresponds naively to a
relatively heavy Higgs boson, which is disfavoured by the data~\cite{EFLTC}.
Such a model can be reconciled 
with the electroweak data only if some other effect is postulated to cancel the
effects of its large mass. One strategy for evading this problem
is offered by `walking technicolour' theories~\cite{Frandsen}, where the coupling strength
evolves slowly, i.e., walks. However, the loss of the close analogy with QCD makes it
more difficult to calculate so reliably in such models: lattice techniques may come
to the rescue here.

\subsection{Interpolating models}

So far, we have examined two extreme scenarios: the orthodox interpretation
of the SM in which the Higgs is elementary and relatively light, and hence
interacts only weakly, and strongly-coupled models exemplified by technicolour.
The weakly-coupled scenario would require additional TeV-scale particles to
stabilize the Higgs mass by cancelling out the quadratic divergences such as 
(\ref{topdivergence}). A prototype for such models is provided by supersymmetry,
as discussed in the next Lecture. On the other hand, strongly-coupled models
such as technicolour introduce many resonances that are required by unitarity and
generate important contributions to the oblique radiative corrections, e.g., a vector
resonance $\rho$ in $W^+ W^-$ scattering would induce
\begin{equation}
\delta \rho \sim \frac{m_W^2}{m_\rho^2} 
\end{equation}
where $\rho$ was defined in (\ref{EqRhoDef}),
and the experimental upper limit $|\rho| < 10^{-3}$ at the 95\% confidence level
imposes $m_\rho > 2.5$~TeV.

\begin{figure}[htb]
\begin{center}
\includegraphics[width=.48\textwidth]{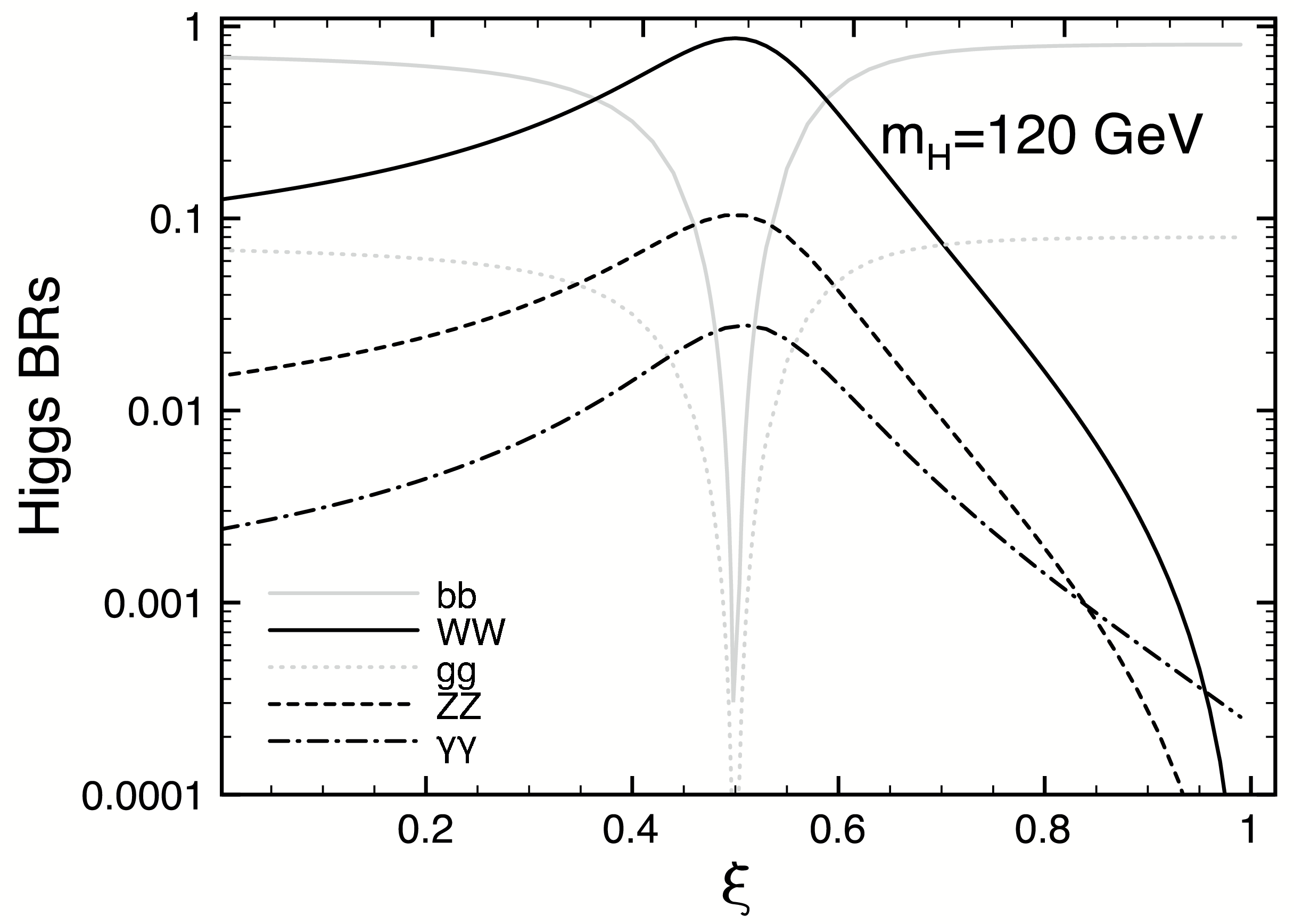}
\includegraphics[width=.48\textwidth]{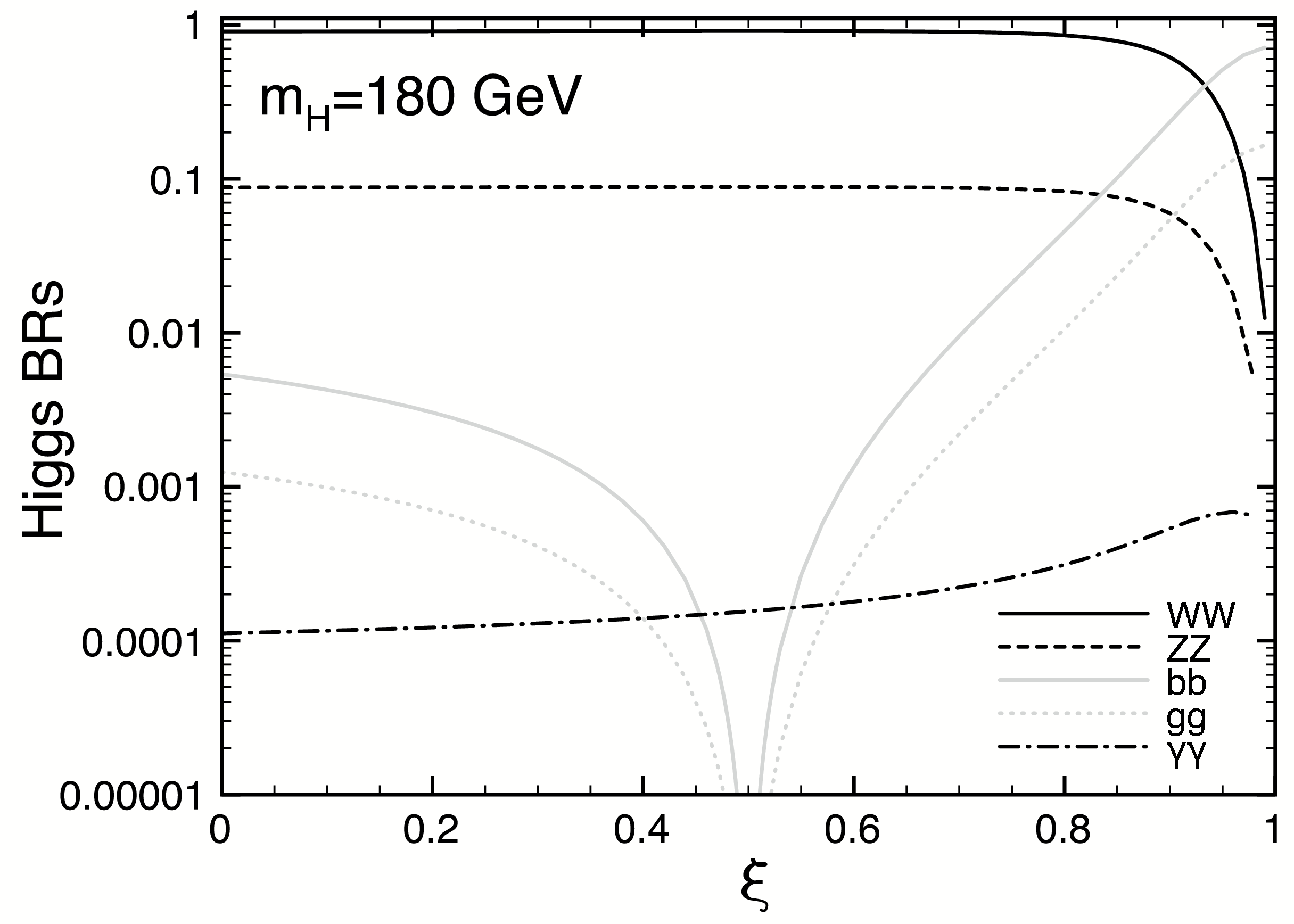}
\end{center} 
\caption{The dependences of Higgs branching ratios on the parameter $\xi$
(\protect\ref{xi}), for $m_H = 120$~GeV (left) and $180$~GeV (right)~\protect\cite{CGMPR}} 
\label{xiBR}
\end{figure}

One way to interpolate between these two extreme scenarios, and provide a
basis for determining how far from the light-SM-Higgs scenario the data permit
us to go, is to consider models in which the unitarization of the $W^+ W^-$
scattering amplitude is shared between a light Higgs boson with modified
couplings and a vector resonance with mass $m_\rho$ and coupling $g_\rho$,
whose relative importance is parametrized by the combination
\begin{equation}
\xi \; \equiv \; v \frac{g_\rho}{m_\rho}\ .
\label{xi}
\end{equation}
The SM is recovered in the limit $\xi \to 0$,
but its decay branching ratios may differ considerably as $\xi$ increases
towards the strong-coupling limit $\xi = 1$, as seen in Fig.~\ref{xiBR}. Thus,
one signature for such models at the LHC may be the observation of a Higgs
boson with couplings that differ from those of the SM.

\begin{figure}[htb]
\begin{center}
\includegraphics[width=.48\textwidth]{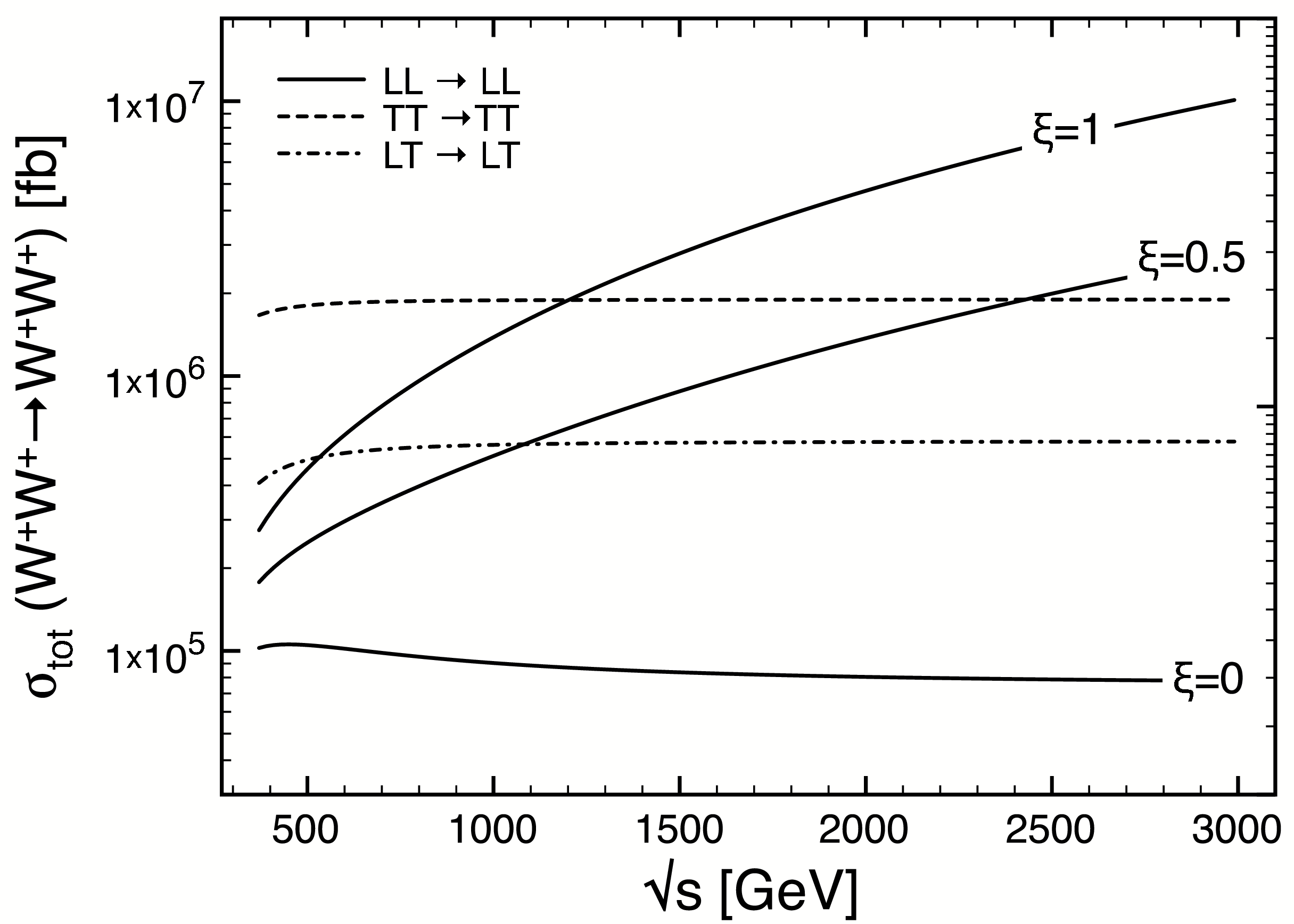}
\includegraphics[width=.48\textwidth]{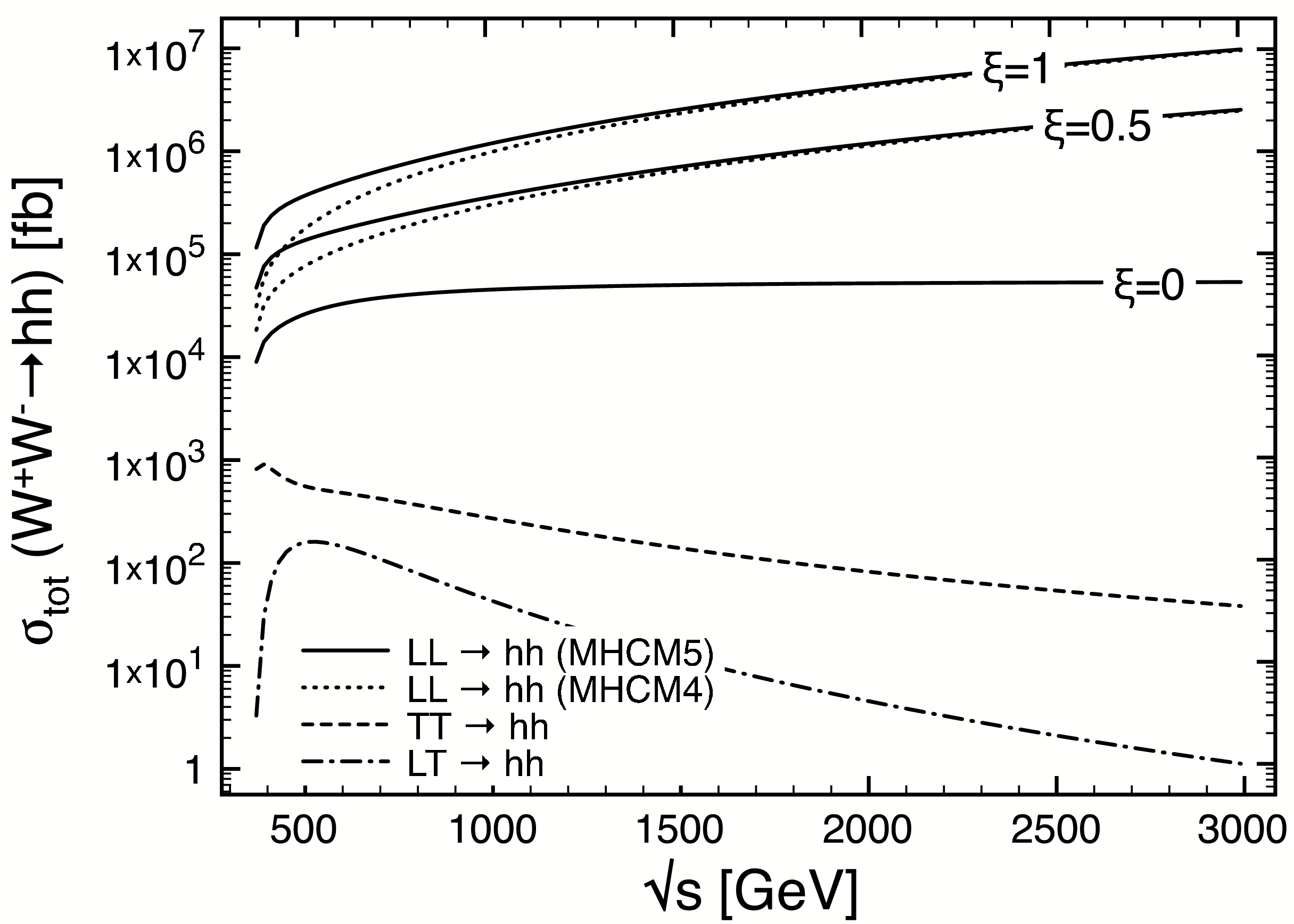}
\end{center} 
\caption{Left; the cross sections $\sigma ( W^+_T W^+_T \to W^+_T W^+_T)$,
$\sigma ( W^+_L W^+_T \to W^+_TLW^+_T)$, and $\sigma ( W^+_L W^+_L \to W^+_L W^+_L)$,
as functions of $\xi$ (\protect\ref{xi}). Right: cross sections for double Higgs 
production~\protect\cite{CGMPR}.} 
\label{fig:CG2}
\end{figure}

Another way to probe such models is to look for effects in $W^+_L W^+_L$
scattering. Unfortunately, at the LHC the $W^\pm$ bosons that are flashed
off from incoming energetic quarks: $q \to W q'$ have predominantly
transverse polarizations, so that $\sigma ( W^+_T W^+_T \to W^+_T W^+_T)
\gg \sigma ( W^+_L W^+_T \to W^+_TLW^+_T)$ and $\sigma ( W^+_L W^+_L \to W^+_L W^+_L)$
for  all $m_{W^+ W^+}$ in the SM, and there is an accidental very small factor~\cite{CGMPR}:
\begin{equation}
\frac{d\sigma^{LL}/dt}{d\sigma^{TT}/dt} \; = \; \frac{1}{2304} 
\left( \frac{m_{W^+ W^+}}{m_W} \right)^4 \xi^2 \ ,
\end{equation}
which implies that, even for $\xi = 1$, $\sigma ( W^+_L W^+_L \to W^+_L W^+_L) >
\sigma ( W^+_T W^+_T \to W^+_T W^+_T)$ only for $m_{W^+ W^+} > 1.2$~TeV,
which is unlikely to be accessible at the LHC, as seen in Fig.~\ref{fig:CG2}. 
An alternative possibility for
the LHC may be double-Higgs production {\it via} the reaction $W^+ W^- \to HH$,
which may be greatly enhanced as compared with its rate in the SM,
as also seen in Fig.~\ref{fig:CG2} --- though
its observability may be a different matter.

\subsection{Higgsless models and extra dimensions}

As has already been discussed, if there is nothing like a SM Higgs boson,
$s$-wave $W W$ scattering reaches the unitarity limit at $m_{W^+ W^-} \sim 1$~TeV
(\ref{LQT}).
An immediate reaction might be: Who cares? Some non-perturbative strong
dynamics will necessarily restore unitarity, even in the absence of a Higgs boson.
However, more detailed study in specific models has shown that this strong dynamics is apparently
incompatible with the precision data: one needs some perturbative mechanism
to break the electroweak symmetry.

How can one break a gauge symmetry? Breaking it explicitly would destroy the
renormalizability (calculability) of the gauge theory, whereas
breaking the symmetry spontaneously by the v.e.v. of
some field everywhere in space does retain the renormalizability (calculability) of the
gauge symmetry. But that is the Higgs approach that we are trying to escape: Is there
another way? The alternative is to break the electroweak symmetry {\it via} boundary
conditions. This is impossible in conventional $3 + 1$-dimensional space-time,
because it has no boundaries. However, it becomes an option if we postulate
finite-size (small) extra space 
dimensions~\cite{Rai:2005vy, Barbieri:2002uk,Gunion:2000gy}.

\begin{figure}[htb]
\begin{center}
\includegraphics[width=.55\textwidth]{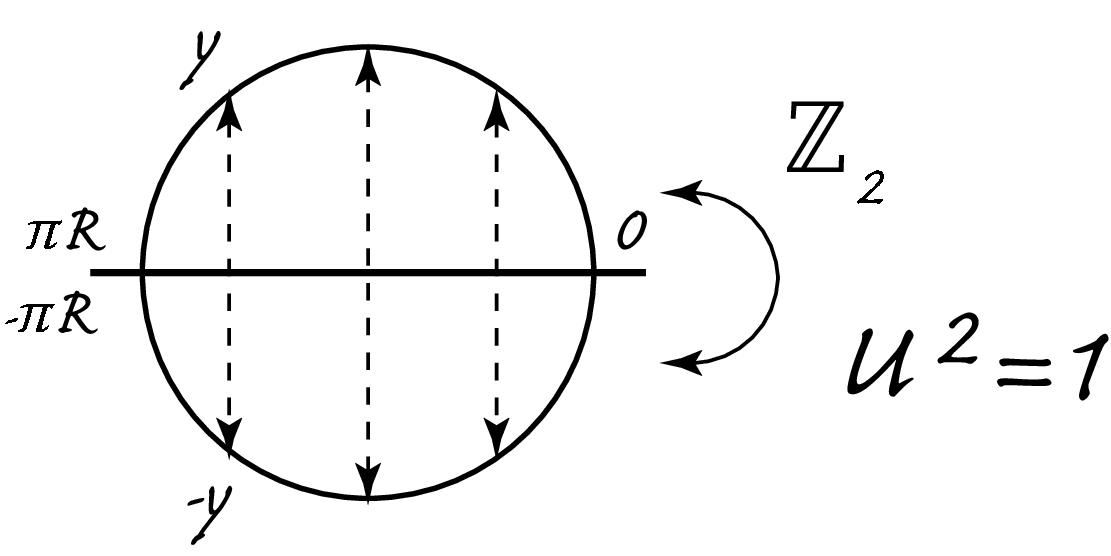}
\end{center} 
\caption{Compactification on a circle $S^1$ of radius $R$
with internal coordinate (fifth dimension) $y$, illustrating the possible orbifolding 
of this model {\it via} the identification $S^1/Z_2$} 
\label{fig:orbi}
\end{figure}

To see how this works, let us first consider the particle spectrum in the simplest
possible model with one extra dimension compactified on a circle $S^1$ of radius $R$
with internal coordinate (fifth dimension) $y$, as illustrated in Fig.~\ref{fig:orbi}.
In this case, the wave function of a boson $\phi$ at $y$ and $y + 2 \pi R$ must be identified:
\begin{equation}
\phi(y + 2 \pi R) \; = \; \phi(y) \ , 
\label{identify1}
\end{equation}
so that one can expand the five-dimensional field as follows:
\begin{equation}
\phi(x, y) \; = \; \sum_n \frac{1}{\sqrt{2^{\delta_{n0}} \pi R} }
\left( \cos \left(\frac{ny}{R}\right)\phi^+_n(x) + \sin \left(\frac{ny}{R}\right)\phi^-_n(x) \right) .
\label{expand}
\end{equation}
The $\phi^\pm_n$ are the four-dimensional Kaluza--Klein~\cite{Kaluza,Klein} modes of the field, 
which appear in four dimensions as particles with masses
\begin{equation}
m_n \; = \; p_y^n \; = \; \frac{n}{R} \ ,
\label{KKmasses}
\end{equation}
and the functions $ \cos, \sin (ny/R)$ describe the localizations of these modes
along the extra dimension. the lowest-lying mode has a flat wave function ($n = 0$),
and the excitations have $n > 0$.

We now consider what happens if we `fold' the circle by identifying $y \sim -y$.
Mathematically, this is the simplest {\it orbifold} $S^1/Z_2$, also illustrated
in Fig.~\ref{fig:orbi}. At the same time as identifying $y \sim -y$,
we can also identify the field $\phi$ up to a sign:
\begin{equation}
\phi ( -y) \; = \; U \phi(y) \; : \; U^2 \; = \; 1.
\label{identify2}
\end{equation}
This has the effect of projecting out half the Kaluza--Klein wave functions (\ref{expand}).
If we choose $U = +1$, we select the even wave functions $ \cos (ny/R)$ and hence the
Kaluza--Klein modes $\phi^+_n(x)$ whereas, if we choose $U = -1$, we select the odd
wave functions $ \sin (ny/R)$ and hence the Kaluza--Klein modes $\phi^-_n(x)$.
The `even' particles include the massless mode with $n=0$ whereas all the `odd'
particles are massive. The projection $U$ serves to give masses to all the states
that are asymmetric.

This mechanism can be extended to break gauge 
symmetry~\cite{Rai:2005vy, Barbieri:2002uk,
Gunion:2000gy}. Let us consider a 
five-dimensional theory with a gauge field $A_{\mu,5}$, and let us identify it on the
orbifold $y \sim -y$ up to a discrete gauge transformation $U: U^2 = 1$:
\begin{eqnarray}
A_\mu \; & = & \; + U A_\mu(y)U^\dagger , \\
A_5 \; & = & \; - U A_5(y)U^\dagger .
\label{5Dbreak}
\end{eqnarray}
The gauge symmetry group is broken at the end-points of the orbifold $y = 0, \pi R$: 
the surviving subgroup is the one that commutes with $U$, and asymmetric
particles acquire masses as described above. In this way, one could imagine breaking 
$SU(2) \otimes U(1) \to U(1)$ with a suitable orbifold construction.

\begin{figure}[htb]
\begin{center}
\includegraphics[width=.48\textwidth]{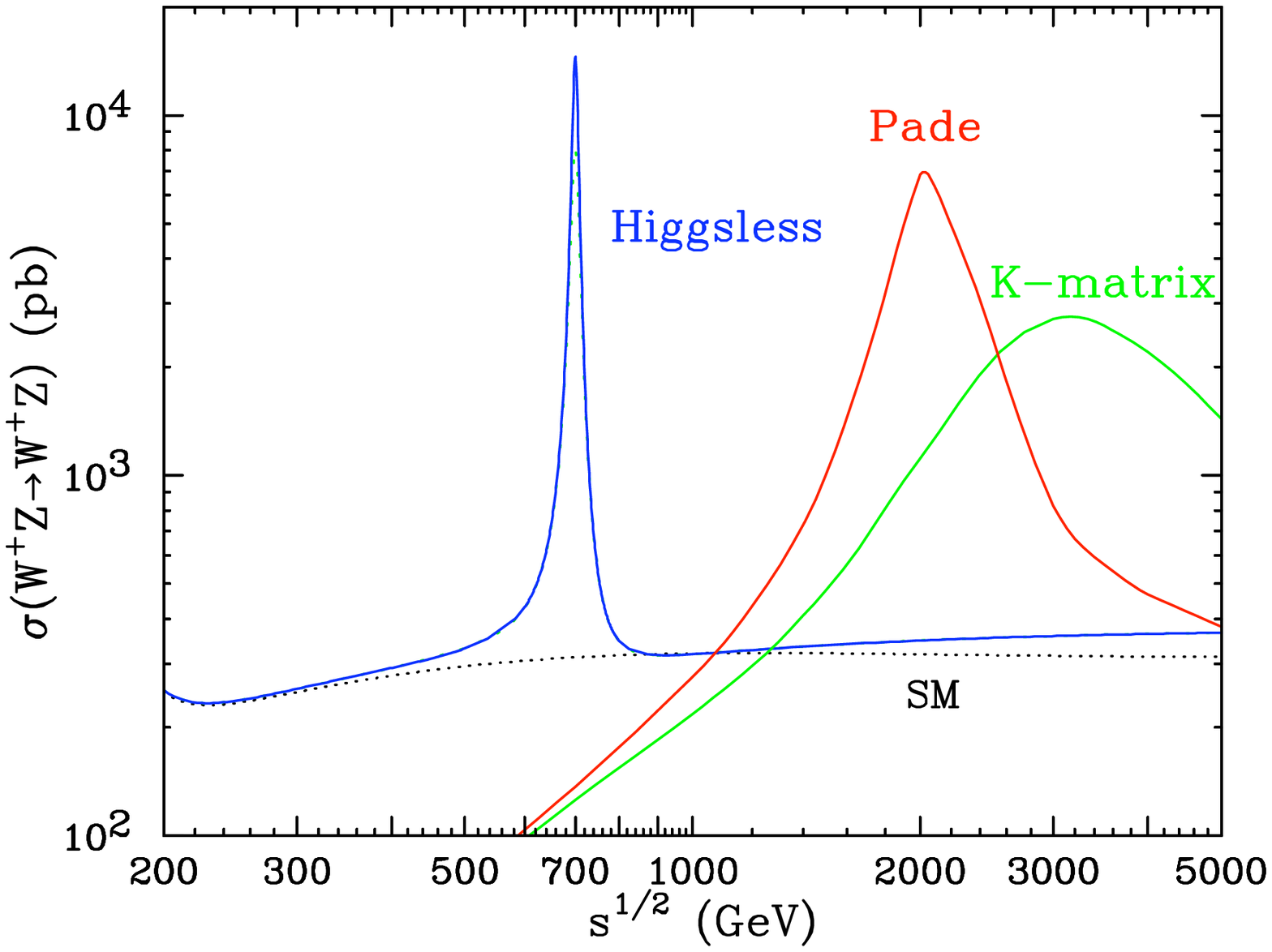}
\includegraphics[width=.48\textwidth]{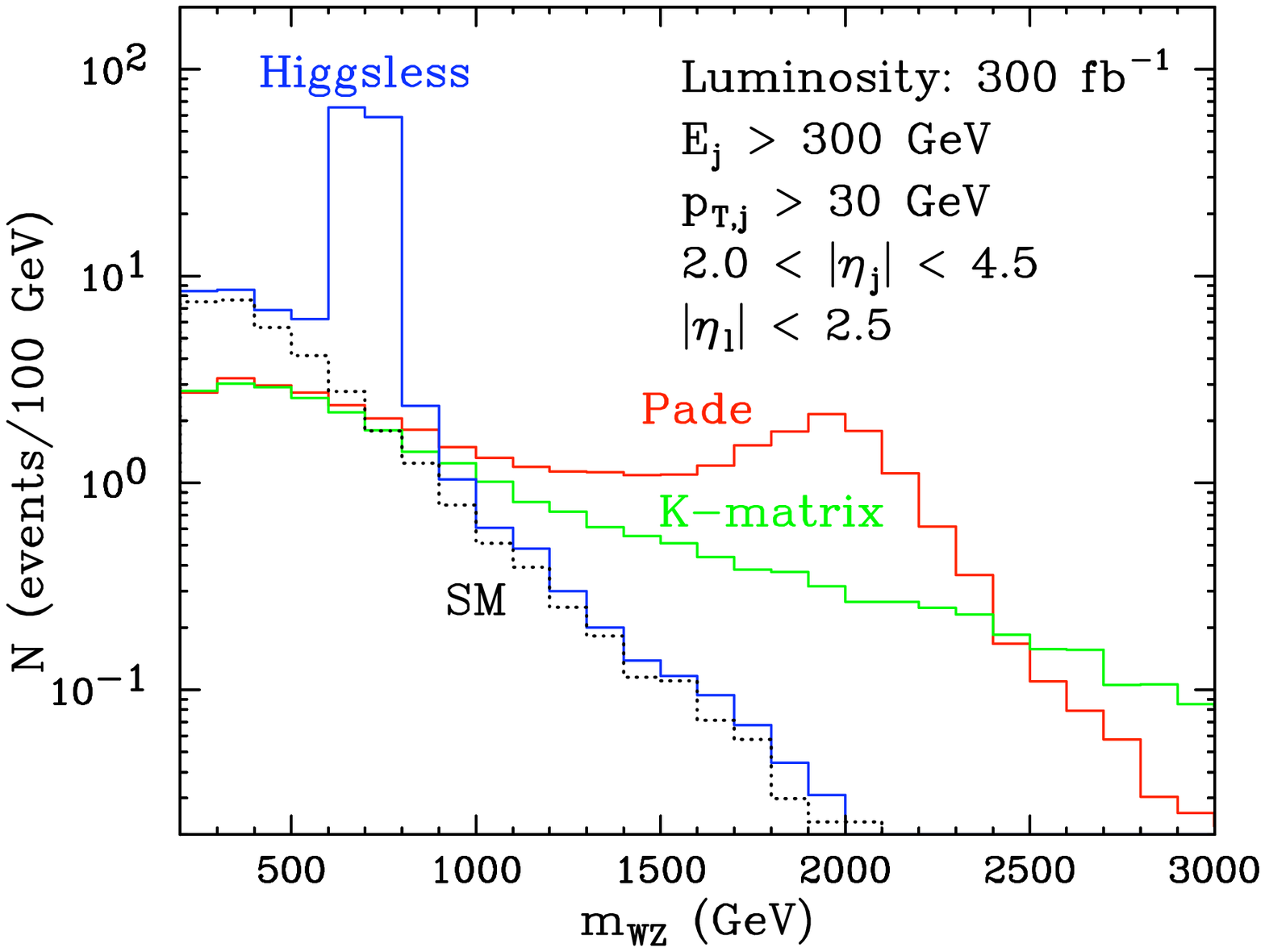}
\end{center} 
\caption{Left: calculations of the possible modifications of 
$\sigma ( W^+ Z^0 \to W^+ Z^0)$. Right: simulations of the possible numbers of events at the 
LHC~\protect\cite{CGMPR}.} 
\label{fig:CG1}
\end{figure}

It is a general feature of this construction that a vector resonance should appear 
in $W Z$ scattering, corresponding to the lowest-lying Kaluza--Klein excitation. The
production of such a particle at the LHC has been considered in the context of a
Higgsless model, and could well be observable, as seen in Fig.~\ref{fig:CG1}.

\begin{figure}[htb]
\begin{center}
\includegraphics[width=.7\textwidth]{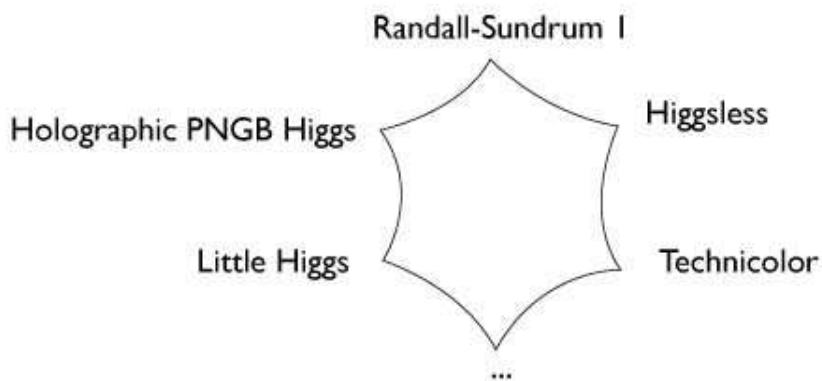}
\end{center} 
\caption{Relations between different models of electroweak symmetry breaking~\protect\cite{littleM}} 
\label{Fig:Mbreak}
\end{figure}

You might wonder whether this type of vector resonance bears any relation to
the vector resonances discussed previously in the context of new strong dynamics. The answer 
is yes: as was first emphasized in the context of string theory, a strong coupling is equivalent
to a new compactified dimension, and there is in general a `holographic' relation
between four- and five-dimensional theories, the former being considered as
boundaries of the five-dimensional `bulk' theory. These ideas enable the
strongly-interacting models of electroweak symmetry breaking discussed in this Lecture, and many
others, to be related through a unified description \`a la M-theory~\cite{Mtheory}, as seen in
Fig.~\ref{Fig:Mbreak}~\cite{littleM}. The alternative is a weakly-interacting model of electroweak
symmetry breaking, which is favoured, naively, by the indications from precision
electroweak data of a light Higgs boson. In the next Lecture we discuss
supersymmetry, which is the most developed such alternative.

\section{Supersymmetry}
\label{supersymm}

We have seen that the Standard Model is a valid description of 
physical phenomena at energies lower than a 
few hundreds of GeV. However, there are various reasons to think that supersymmetry might 
appear at the TeV scale, and hence play an 
important role in new discoveries at the LHC, which will explore energies of the order of a TeV.
In this Lecture we present and discuss supersymmetric models, with a focus on the 
phenomenological consequences of supersymmetry. 

We first give a brief historical introduction and summarize the motivations for supersymmetry
in particle physics. Subsequently we discuss the general formal structure of a physical 
supersymmetric theory. We then continue with some theoretical notions and applications to 
`low-energy' particle physics around the TeV scale. Among the possible models, we
focus on the Minimal Supersymmetric Standard Model (MSSM), which provides a basis 
for analysing supersymmetric phenomenology. Within the context of the MSSM,
we discuss the principal experimental constraints on supersymmetry, and then discuss 
possible aspects of the detection of supersymmetry.

\subsection{History and motivations}

\subsubsection{What is supersymmetry?}

Supersymmetry is a radically new type of symmetry that transforms a bosonic state into a fermionic 
state, or vice versa, with $\Delta S =\pm 1/2$, where $S$ is the spin. Denoting the supersymmetry generator by $Q$, we may write schematically:
\begin{eqnarray}
Q|Boson \rangle  & = &|Fermion \rangle\\
Q|Fermion \rangle  & = &|Boson \rangle.
\end{eqnarray}
Formally, supersymmetry is an extension of the space-time symmetry reflected in the 
Poincar\'e group, and this was a principal motivation leading to its discovery. Initially, it was also hoped
that one could use supersymmetry to combine the external space-time symmetries with internal
symmetries.  However, this prospect seems more distant, as discussed below.

\subsubsection{Milestones}

There were several attempts in the 1960s to combine internal and external symmetries,
but  Coleman and Mandula~\cite{CM}  showed in 1967 that it is impossible to combine these types of 
symmetry, {\it via} a famous no-go theorem that is discussed later in more detail. 
However, their proof assumed that the new
symmetry should be generated by bosonic charges of integer spin. In 1971, Golfand and 
Likhtman~\cite{GL} discovered an extension of the Poincar\'e group using fermionic charges of 
half-integer spin. In the same year, Ramond~\cite{Ramond}, 
Neveu and Schwarz~\cite{NS} proposed supersymmetric 
models in two dimensions, with the aim of obtaining strings with fermionic states that could accommodate baryons. A few years later, in 1973, Volkov and 
Akulov~\cite{VA} tried to apply a nonlinear realization of supersymmetry to neutrinos in four dimensions, but their 
theory did not describe correctly the low-energy interactions of neutrinos.

In the same year, Wess and Zumino~\cite{WZ1,WZ2} proposed the first 
four-dimensional supersymmetric field 
theories of interest from the phenomenological point of view. Specifically, they showed how
to construct supersymmetric field theories linking scalars with fermions of spin $1/2$~\cite{WZ1}, 
and also
fermions of spin $1/2$ with gauge particles of spin 1~\cite{WZ2}. Then, together with Iliopoulos and 
Ferrara, Zumino discovered that supersymmetry would eliminate many of the divergences present
in other field theories~\cite{IZ,FIZ}. At first, these ultraviolet properties were regarded as curiosities, in
particular because not all logarithmic divergences were eliminated, but attempts were made to
construct phenomenological supersymmetric models, for example theories unifying matter 
particles and Higgs fields in the same supermultiplet. Subsequently, in 1976, 
two groups~\cite{FvF,DZ} found a 
local version of supersymmetry in which the supersymmetry transformation depends on the 
space-time coordinates. This theory necessarily includes a description of gravitation, and hence
has been called supergravity.

\subsubsection{Why supersymmetry?}

Following these formal developments, the phenomenology of supersymmetry has been studied 
intensively, and models based on supersymmetry are considered to be among the most serious 
candidates for physics beyond the SM~\cite{FF,Nilles,HK}. 
Why introduce supersymmetry in particle  physics? 
What makes it so attractive for particle physicists? 

The reasons for its introduction in particle 
physics are principally physical, and quite diverse in nature, as we now discuss.

\begin{figure}
\begin{center}
\includegraphics[height=2.5cm]{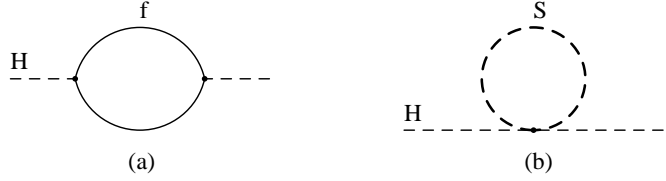}
\end{center}
\caption{One-loop quantum corrections to the mass-squared of the Higgs boson due to (a) 
a fermionic loop, (b) a scalar boson loop
\label{fig:higgscorr1}}
\end{figure}

$\bullet$
The very special properties of supersymmetric field theories are helpful in addressing the
naturalness of a (relatively) light Higgs boson. In the previous Lectures we have discussed the
existence of enormous radiative corrections to the Higgs mass-squared, $m_H^2$, which
feels the virtual effects of any particle that couples directly or indirectly to the Higgs field. For example,
the correction due to a fermionic loop such as that in Fig.~\ref{fig:higgscorr1}(a)
yields~\footnote{For this calculation, we define the Yukawa coupling of the Higgs boson to a fermion,
as usual, {\it via}: $y_f H\overline{\psi}\psi$.}:
\begin{equation}
\Delta m_H^2=-\frac{y_f^2}{8\pi^2}[2\Lambda^2 + 6m_f^2 \ln(\Lambda/m_f)+...],
\label{quadf}
\end{equation}
where $\Lambda$ is an ultraviolet cutoff used to represent the scale up to which
the SM remains valid, at which new physics appears. We see that the mass of the Higgs diverges
quadratically with $\Lambda$ and, if we suppose that the SM remains valid up to the Planck scale, 
$M_P\simeq 10^{19}$ GeV, then $\Lambda=M_P$ and this correction is $10^{30}$ times bigger 
than the reasonable value of the mass-squared of the Higgs, namely ${\cal}(10^2)$~GeV)$^2$!
Moreover, there is a similar correction coming from a loop of a scalar field $S$, such as that in 
Fig.~\ref{fig:higgscorr1}(b):
\begin{equation}
\Delta m_H^2=\frac{\lambda_S}{16\pi^2}[\Lambda^2 - 2m_S^2 \ln(\Lambda/m_S)+...],
\label{quadS}
\end{equation}
where $\Lambda_S$ is the quartic coupling to the Higgs boson.

Comparing (\ref{quadf}) and (\ref{quadS}), we see that the divergent contributions terms
$\propto \Lambda^2$ are cancelled if, for every fermionic loop of the theory there is also a 
scalar loop with $\lambda_S= 2 y_f^2$. {\it We will see later that supersymmetry imposes
exactly this relationship!} Thus supersymmetric field theories have no quadratic
divergences, at both the one- and multi-loop levels, which enables a large hierarchy
between different physical mass scales to be maintained in a natural way. In addition, other
logarithmic corrections to couplings also vanish in a supersymmetric theory~\cite{Martin:1997ns}.

$\bullet$ 
A second circumstantial hint in favour of supersymmetry is the fact,
discussed in the previous Lecture, that precision electroweak 
data prefer a relatively light Higgs boson weighing less than about
150~GeV~\cite{EWWG09}. This is perfectly consistent with calculations in
the minimal supersymmetric extension of the Standard Model (MSSM), in
which the lightest Higgs boson weighs less than about
130~GeV~\cite{susyHiggs}.

$\bullet$ 
A third motivation for supersymmetry is provided by the astrophysical necessity of cold dark
matter, which has a density of $\Omega_{CDM}h^2 = 0.1099\pm 0.0062$
according to the recent measurements of WMAP~\cite{Komatsu:2008hk}. 
This dark matter could be provided by a neutral, weakly-interacting particle
weighing less than about 1~TeV, such as the lightest supersymmetric
particle (LSP) $\chi$~\cite{EHNOS}. In many supersymmetric models,
a conserved quantum number called $R$ parity guarantees that
the LSP is stable. As the Universe expanded and cooled, 
all the particles present at high energies and densities would have 
annihilated, disintegrated, or combined to form baryons, atoms, etc., 
except for stable weakly-interacting particles such as the neutrinos and the LSP.
The latter would be present in
the Universe as a relic from the Big Bang, and could have the right density
to constitute the majority of the cold dark 
matter favoured by cosmologists. 

\begin{figure}[htbp!]
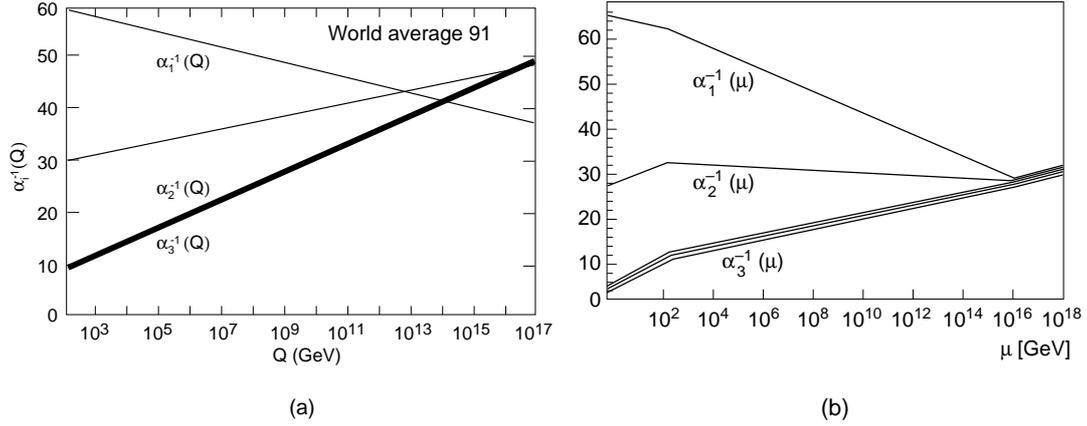

\begin{center}
{\includegraphics[height=2.25in]{EllisScotfig28a.eps}}
{\includegraphics[height=2.25in]{EllisScotfig28b.eps}}
\end{center}
\caption[]{The measurements of the gauge coupling strengths at LEP 
(a) do not evolve to a unified value if there is no supersymmetry but do (b) if
supersymmetry is included~\cite{ADF}} 
\label{28}
\end{figure}

$\bullet$ 
Fourthly, let us consider the couplings that characterize each of the fundamental forces. As
seen in the left panel of Fig.~\ref{28}, it has been known for a long time now that if we 
evolve them with energy according to the renormalization-group equations of the Standard Model, 
we find that they never quite become equal at the same scale. However, as seen in the
right panel of Fig.~\ref{28}, when we include 
supersymmetric particles in the evolution of the couplings, they appear to intersect at exactly the same
energy scale (about $2\times 10^{16}$ GeV)~\cite{ADF}. Nobody is forced to believe in such a `Grand 
Unification' on the basis of this possible unification of the couplings, but it is very intriguing
that supersymmetry favours unification with high precision. 

$\bullet$ 
Fifthly, supersymmetry seems to be essential for the consistency of string
theory~\cite{GSW}, although this argument does not really restrict the
mass scale at which supersymmetric particles should appear.

\begin{figure}[htbp!]
\begin{center}
{\includegraphics[height=3in]{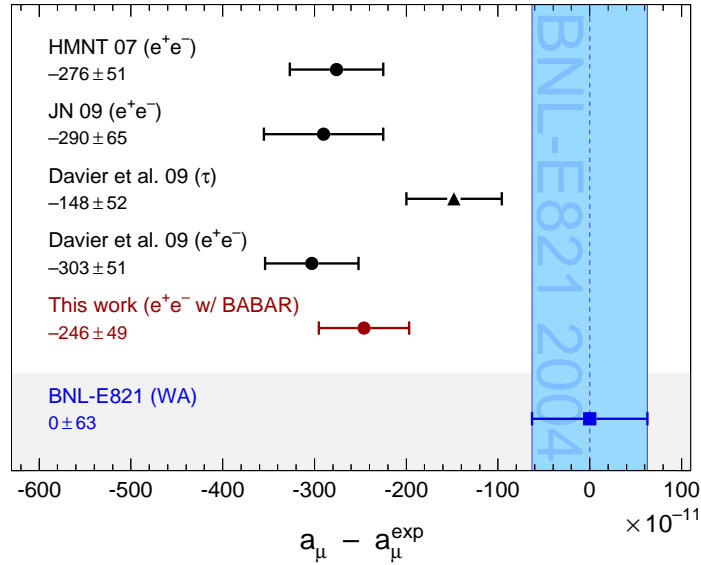}}
\end{center}
\caption[]{SM calculations of $a_\mu \equiv (g_\mu - 2)/2$ 
disagree with the experimental measurement~\protect\cite{BNL},
particularly if they are based on low-energy $e^+ e^-$ data~\protect\cite{CM}.} 
\label{FigDavier}
\end{figure}

$\bullet$
A final hint for supersymmetry may be provided by the anomalous magnetic moment
of the muon, $g_\mu - 2$, whose experimental value~\cite{BNL} seems to differ from that calculated
in the SM, in a manner that could be explained by contributions from supersymmetric
particles. The amount of this discrepancy depends on how one calculates the SM contributions to
$g_\mu - 2$, in particular that due to low-energy hadronic vacuum polarization, and to
a lesser extent that due to light-by-light scattering. The most direct way to calculate the
hadronic vacuum polarization contribution is to use low-energy data on $e^+ e^- \to$
hadrons: these do not agree perfectly, but may be combined to yield a discrepancy~\cite{Davier}
\begin{equation}
\delta a_\mu \; \equiv \; \delta \left(\frac{g_\mu - 2}{2} \right) \; = \; (24.6 \pm 8.0) \times 10^{-10} ,
\label{g-2disc}
\end{equation}
a discrepancy of 3.1 $\sigma$, as illustrated in Fig.~\ref{FigDavier}.
Alternatively, and less directly, one may use $\tau$
decay data, in which case the discrepancy is reduced to about 2 $\sigma$.

As we have seen, there are several arguments that motivate the study of 
supersymmetry~\footnote{Other extensions of the SM also address
some of these issues, though perhaps none do so as naturally as supersymmetry.}.
Although there are no experimental proofs of its existence, supersymmetry combines 
so many attractive and useful characteristics that it deserves to be studied in detail.

\subsection{The structure of a supersymmetric theory}
\subsubsection{Interlude on `spinorology'}

In order to lay the basis for the theoretical description of supersymmetry~\cite{FF}, 
we first present the notations and conventions that we use in the rest of the 
section~\cite{Welzel:2005cb,Martin:1997ns}.

\noindent $\bullet$ We choose the {\it Weyl representation} for the $\gamma$ matrices:
\begin{equation}
 \gamma^{\mu}=\left( \begin{array}{cc} 0 & \sigma^{\mu} \\
\overline{\sigma}^{\mu} & 0 \end{array}\right) ,
\end{equation}
with $\sigma^{\mu}=(\mathbf{1}_2,\sigma^{i}),\ \overline{\sigma}^{\mu}
=(\mathbf{1}_2,-\sigma^{i})$ where $\sigma_{i}$ are the Pauli matrices, and $\gamma_5=i\gamma^0\gamma^1\gamma^2\gamma^3=\mathrm{diag}(-\mathbf{1}_2,\mathbf{1}_2)$.
We also use $\{\gamma^{\mu},\gamma^{\nu}\}=2\eta_{\mu\nu}$, where
$\eta_{\mu\nu}=\mathrm{diag}(+1,-1,-1,-1)$ is the Minkowski metric, that may be used to lower or to 
raise Lorentz indexes.

\noindent $\bullet$ A {\it Weyl spinor} describes a particle of spin $1/2$ and given chirality. It has two components, which we label with Greek letters, $\psi_{\alpha}$, $\xi_{\beta}$, \ldots where $\alpha,\ \beta,...= 1,2$.
A spinor $\psi_{\alpha}$ or $\psi_L$ will denote a particle with left chirality, whereas we denote by 
$\overline{\psi}^{\dot{\alpha}}$ or  $\psi_R$ a spinor with right chirality. These are related by 
complex conjugation:
\begin{eqnarray}
&(\psi_{\alpha})^*=\overline{\psi}_{\dot{\alpha}}&,
\\
&(\overline{\psi}^{\dot{\alpha}})^*=\psi^{\alpha}&.
\end{eqnarray}
We also use the matrix
$\varepsilon_{\alpha\beta}=\varepsilon_{\dot{\alpha}\dot{\beta}} \equiv i\sigma_2$ and 
$\varepsilon^{\alpha\beta}=\varepsilon^{\dot{\alpha}\dot{\beta}} \equiv -i\sigma_2$, which allows us
to raise and lower the spinorial indices $\alpha$ and 
$\beta$.
 
\noindent $\bullet$ A {\it Dirac spinor} is constructed out of two Weyl spinors, and describes a particle with
both chiralities. It is a spinor of four components, which we denote here using capital Greek 
letters: $\Psi$, $\chi$, $\Phi$, ... In terms of Weyl spinors, we have 
\begin{equation}
 \Psi=\left(\begin{array}{c} \psi_L \\ \psi_R \end{array}\right)=
\left(\begin{array}{c} \psi_{\alpha} \\ \overline{\eta}^{\dot{\alpha}} 
\end{array}\right) .
\end{equation}
The projection operators  $P_{R,L}=\frac{1}{2}(1\pm\gamma_5)$ allow us to select the right or left 
chiralty, respectively: $\Psi_{R,L}=P_{R,L}\Psi$.

\noindent $\bullet$ A {\it charge conjugate spinor} is a spinor to which charge conjugation has been applied. 
It describes the antiparticle of a given particle, with opposite internal opposite charge.
\begin{equation}
\Psi^c=C \overline{\Psi}^T = 
\left(\begin{array}{c} \eta_{\alpha} \\ \overline{\psi}^{\dot{\alpha}} 
\end{array}\right) ,
\end{equation}
where the charge conjugation matrix  $C$ can be written:
\begin{equation}
 C=i\gamma^0\gamma^2 .
\end{equation}

\noindent $\bullet$ A  {\it Majorana spinor} is constructed out of a single Weyl spinor, but possesses 
four components that are interrelated by charge conjugation, so that $\Psi_M=\Psi^c_M$:
\begin{equation}
 \Psi_M=\left(\begin{array}{c} \psi_L \\ -i\sigma_2(\psi_L)^* \end{array}\right)=
\left(\begin{array}{c} \psi_{\alpha} \\ \overline{\psi}^{\dot{\alpha}} 
\end{array}\right) .
\end{equation}

\subsubsection{The supersymmetry algebra and supermultiplets}

As was described before, supersymmetry combines the space-time transformations of the Poincar\'e 
group with transformations of an internal symmetry. Prior to the advent of supersymmetry,
there had been many previous attempts to combine internal and external symmetries,
but they had always failed, for a reason demonstrated by Coleman and Mandula~\cite{CM}. All
the previous attempts used bosonic charges, scalar (or vector) such as the electromagnetic
charge (or momentum operator):
\begin{eqnarray}
\langle {\rm Spin} J | Q | {\rm Spin} J \rangle \; & = & \; q  , \\
\langle {\rm Spin} J | P_\mu | {\rm Spin} J \rangle \; & = & \; p_\mu  .
\label{CM1}
\end{eqnarray}
Conservation of momentum in any $2 \to 2$ collision implies
\begin{equation}
p_\mu^{(1)} + p_\mu^{(2)} \; = \; p_\mu^{(3)} + p_\mu^{(4)} .
\label{momcons}
\end{equation}
Consider now a tensor charge $\Sigma_{\mu\nu}$: by Lorentz invariance, its diagonal
matrix elements in any particle state $| a \rangle$ must be of the form
\begin{equation}
\langle a | \Sigma_{\mu \nu} | a \rangle \; = \; \alpha g_{\mu \nu} + \beta p_\mu p_\nu .
\label{CM3}
\end{equation}
Conservation of the tensor charge during a $2 \to 2$ collision would require
\begin{equation}
p_\mu^{(1)} p_\nu^{(1)} + p_\mu^{(2)} p_\nu^{(2)}\; = \; p_\mu^{(3)} p_\nu^{(3)}+ 
p_\mu^{(4)} p_\nu^{(4)} .
\label{mom2cons}
\end{equation}
This is compatible with the linear relation (\ref{momcons}) of conventional momentum 
conservation iff
\begin{equation}
p_\mu^{(1)} \; = \; p_\mu^{(3)} \; {\rm or} \; p_\mu^{(4)} ,
\label{CM5}
\end{equation}
implying that only exactly forward and backward scattering are allowed: no need to
place any detectors at large angles! This proof can easily be extended to bosonic
charges with any number of indices. However, it makes the crucial assumption
that the diagonal matrix element $\langle a | Q | a \rangle \ne 0$, which is not true
in supersymmetry, enabling it to evade the Coleman--Mandula no-go theorem.

Supersymmetry is generated by {\it spinorial} charges $Q_{\alpha}$ which have vanishing 
diagonal matrix elements: $\langle a | Q_\alpha | a \rangle = 0$.
Being spinors, the $Q_{\alpha}$ anti-commute in the same way as other fermionic fields. 
It is possible to introduce more generators, but in the simplest version of supersymmetry there is just a pair of generators, $Q_{\alpha}$ and 
$\bar{Q}^{\dot{\alpha}}$, that are  complex spinors transforming inequivalently under the Lorentz group. 
This is $\mathcal{N}=1$ supersymmetry, which is essentially the only case that 
we consider in these notes. The initial reason for this choice is pedagogical, but in the following section we give some physical reasons for such a choice.

The algebra of the supersymmetry (like that of any other symmetry) is summarized in the
commutation (and anticommutation) relations of its generators, i.e., its Lie (super)algebra. In addition
to the commutation relations of the Poincar\'e algebra, the supersymmetry algebra includes the 
following relations for the generators $Q_{\alpha}$ y $\bar{Q}^{\dot{\alpha}}$:
\begin{eqnarray}
&[P^{\mu},Q_{\alpha}]&=0=[P^{\mu},\bar{Q}^{\dot{\alpha}}] , \label{com1}
\\
&\lbrace Q_{\alpha},\bar{Q}_{\dot{\beta}}\rbrace &
=2(\sigma_{\mu})_{\alpha\dot{\beta}}P^{\mu}  , \label{com2}
\\
&\lbrace Q_{\alpha},Q_{\beta}\rbrace & =
\lbrace \bar{Q}^{\dot{\alpha}},\bar{Q}^{\dot{\beta}}\rbrace=0 , \label{com3}
\\
&\lbrace M_{\mu\nu}, Q_{\alpha}\rbrace & =
\frac{1}{2}(\sigma_{\mu\nu})_{\alpha}^{\beta}Q_{\beta} , \label{com4}
\\
&\lbrace M_{\mu\nu}, \bar{Q}_{\dot{\alpha}}\rbrace & =
\frac{1}{2}(\overline{\sigma}_{\mu\nu})^{\dot{\beta}}_{\dot{\alpha}}\bar{Q}_{\dot{\beta}} .
\label{com5}
\end{eqnarray}
What is the significance of $Q_{\alpha}$? First, $Q$ is a charge in the sense of Noether's theorem, i.e, it is
the charge conserved by the symmetry. As a conserved charge, it commutes with the Hamiltonian  of the system and is invariant under translations, see (\ref{com1}). Since it  possesses spin 1/2 and has two 
complex components, it can be written as a Weyl spinor, or alternatively
as a Majorana spinor with 4 components: as such, its commutation relations with the Lorentz
generators are completely determined, see (\ref{com4}) and (\ref{com5}). The non-trivial
anticommutation relation above is (\ref {com2}): schematically $\lbrace Q, \bar {Q} \rbrace \sim P$, 
which means that $Q$ is the `square root' of a space-time translation.

If we want to apply supersymmetry to particle physics, we must know how to arrange particles in irreducible representations (supermultiplets), and their transformation properties. Therefore, we now study the supermultiplets and detail their 
contents. We recall that the Poincar\'e group has two  Casimir invariant elements,
the spin invariant $W^2=W^{\mu} W_{\mu}$, where 
$W^{\mu} = \frac{1}{2} \epsilon^{\mu\nu\rho\sigma} P_{\nu} M_{\rho\sigma}$ is the Pauli-Lubanski vector, 
and the mass invariant $P^2=P^{\mu} P_{\mu}$, where $P^{\mu}$ is the four-momentum. In a multiplet of 
the Poincar\'e group, the particles have the same masses and the same spins. However, in the case of
supersymmetry,  $W^2$ is not an invariant of the algebra, so only mass is conserved, not spin:
\begin{eqnarray}
&[P^2,Q_{\alpha}]&=0 ,
\\
&[W^2,Q_{\alpha}]&\not=0 .
\end{eqnarray}
Thus, in a supermultiplet, the particles have the same mass but  different spins. We can nevertheless 
modify $W$ to obtain a new invariant whose eigenvalues are of the form $2 j (j+1) m^4$ with 
$j=0, \frac{1}{2}, 1,...$ the quantum number of this `superspin'. This modified $W$ is an invariant, so 
every irreductible representation can be characterized by a pair $[m,j]$, and the relation between the 
spin $S$ and $j$ is deduced from the relation: $M_S=M_j, M_j + \frac{1}{2},  M_j-\frac{1}{2}, M_j$. 
Within a given supermultiplet, there are particles of the same mass and the same superspin. In 
addition, an important property of any supermultiplet is that there are equal numbers of bosonic
and fermionic degrees of freedom:
$n_B=n_F$.

We can construct now two different supermultiplets:

$\noindent \triangleright$  The fundamental representation $[m, 0]$ is called a chiral supermultiplet. 
The value $j=0$ implies $M_S=0, + \frac{1}{2},- \frac{1}{2},0$, and this supermultiplet $\Psi$
contains two real scalar  fields described by a single complex scalar field (the sfermion), $\phi$,  
and a two-component Weyl fermionic field of spin 1/2, $\psi$ with the same mass:
\begin{equation} 
\Psi=(\phi, \psi_{\alpha}, F). 
\end{equation} 
What is $F$? In order that the 
supersymmetry be preserved in loops, where the particles are not on-shell, i.e., 
$P^2\not=M^2$, it is necessary  that the fermionic and bosonic degrees of freedom be balanced also 
\textit{off-shell}. This is an issue because an off-shell Weyl fermion  possesses 4 spin degrees of 
freedom, as opposed to 2 on-shell. It is necessary to add to the on-shell content of this representation 
another scalar complex field $F$ that does not propagate, and does not correspond to a physical particle. 
This is termed an auxiliary field, and does not have a kinetic term, and the equation of motion 
$F=F^* = 0$ may be used to eliminate it when on-shell.

$\noindent \triangleright$ The second representation we use later is the vector (or gauge)
supermultiplet $[m, 1/2]$, denoted by $\Phi$. Its field content is obtained in the same way: a Weyl fermion (or,
equivalently, a Majorana fermion), called the gaugino $\lambda^a _ {\alpha}$, a gauge boson 
(of zero mass) $A_a^{\mu}$, and in the presence of any chiral supermultiplet, an auxiliar real
scalar field,  $D^a$:
\begin{equation} 
\Phi = (\lambda^a_{\alpha}, A^a_{\mu},D^a), 
\end{equation}
where $a$  is an  index of the gauge group.

These two representations may be used to accommodate the particles of the SM and their superpartners. 
However, before doing so, we first construct with these two representations generic supersymmetric 
field theories.

\subsection{Supersymmetric field theories}

Before discussing supersymmetric models in general, and particularly the minimal supersymmetric 
extension of the SM (the MSSM), we first present, without detailed derivations, the general structure 
of a field theory with supersymmetry. We first introduce the model of Wess and Zumino~\cite{WZ1} without 
interactions to see how the fields transform.
Then we introduce the interactions, which will lead us to the new notion of the superpotential. 
Finally, we discuss gauge fields in a supersymmetric theory. At the end of this section, we will have
accumulated enough theoretical baggage to understand the structure of the MSSM, and be able
to study concretely its experimental predictions.

\subsubsection{The action for free bosons and fermions is globally supersymmetric}

The simplest supersymmetric action is the combination of actions for a 
non-interacting massless complex scalar $\phi$and a spin-$1/2$ fermion $\psi$:
\begin{eqnarray}
S & \; = \; & \int d^4x\ (\mathcal{L}_{scalar}+\mathcal{L}_{fermion}) : \label{twosixteen}
\\
\mathcal{L}_{scalar} & \; = \; & -\partial^{\mu}\phi\,\partial_{\mu}\phi^{*} ,
\\
\mathcal{L}_{fermion} & \; = \; & -i\psi^{\dag}\bar{\sigma}^{\mu}\,\partial_{\mu}\psi .
\end{eqnarray}
If we introduce an infinitesimal supersymmetric global transformation parameter $\epsilon_{\alpha}$, 
which is a Weyl fermion independent of the space-time coordinates ($\partial ^ {\mu} \epsilon _ {\alpha} =0$), and apply it to the scalar field $\phi$, the result must be proportional to the fermionic field $\psi$:
\begin{equation}
\delta\phi= \epsilon^{\alpha}\psi_{\alpha} \ \ \mathrm{and}\ \ 
\delta\phi^*= \bar{\epsilon}_{\dot{\alpha}}\,\bar{\psi}^{\dot{\alpha}} ,
\label{deltascalar}
\end{equation}
leading to
\begin{equation}
\delta \mathcal{L}_{scalar} = -\epsilon^{\alpha}\,
(\partial^{\mu}\psi_{\alpha})\,\partial_{\mu}\phi^*
-\partial^{\mu}\phi\,\bar{\epsilon}_{\dot{\alpha}}\,
(\partial_{\mu}\bar{\psi}^{\dot{\alpha}}) .
\label{twoseventeen0}
\end{equation}
Since the mass dimensions of free boson and fermion fields are
\begin{equation}
[\phi]=1,\ \ [\psi]=\frac{3}{2} ,
\label{dims}
\end{equation}
the infinitesimal fermion $\epsilon_{\alpha}$ must have the dimensionality 
$(mass)^{-1/2}$:
\begin{equation}
[\epsilon]=-\frac{1}{2} ,
\end{equation}
in contrast to an usual Weyl fermion that has dimension $(mass)^{3/2}$ (\ref{dims}).
By simple dimensional counting, the infinitesimal transformation of the fermion field
must therefore be proportional to the derivative of the boson field:
\begin{equation}
\delta\psi_{\alpha}=i(\sigma^{\mu}\epsilon^{\dag})_{\alpha}\,
\partial_{\mu}\phi \ \ \mathrm{and}\ \ \delta\bar{\psi}^{\dot{\alpha}}=
- i(\epsilon\,\sigma^{\mu})^{\dot{\alpha}}\,\partial_{\mu}\phi^{*} .
\label{twoseventeen1}
\end{equation}
Combining (\ref{deltascalar}) and (\ref{twoseventeen1}) and using the equations of motion, we see
that the sum $\delta L_{scalar} + \delta L_{fermion}$ is a total divergence. This implies that the 
combined action, which is the
space-time integral of the two free Lagrangians $L_{scalar} + L_{fermion}$,
is invariant under this pair of transformations.

Does this transformation correspond to a supersymmetry transformation? To convince ourselves
that this is the case, it is enough to start from a fermion $\psi$ or from a boson  $\phi$, and to apply 
these transformations twice. We find the following chain:
\begin{equation}
\phi\to\psi\to\partial\phi,\ \ \psi\to\partial\phi\to\partial\psi,
\label{twoseventeen}
\end{equation}
which means that in both cases the combined effects of two successive supersymmetry 
transformations are equivalent to a space-time derivative $\partial^{\mu}$, and hence
to the momentum operator $P^{\mu} \sim i \partial^{\mu}$. Thus we recover the result of the 
previous section, namely $Q^2\sim P$, and our transformations satisfy the supersymmetric algebra.
This free Lagrangian model is actually the simplest Wess--Zumino model with a single
chiral supermultiplet, without mass and without interactions.

If we wish to preserve supersymmetry  \textit{off-shell}, which will be essential once we
include interactions, we cannot use the equations of motion to demonstrate supersymmetry.
To overcome this problem, as discussed earlier,
the action $S$ must be modified by the addition of a term that contains an
auxiliary field $F$:   
\begin{eqnarray}
&S=\int d^4x\ (\mathcal{L}_{scalar}+\mathcal{L}_{fermion}+\mathcal{L}_{aux}),&
\\
&\mathcal{L}_{aux}=F^*\,F,&
\label{Laux}
\end{eqnarray}
In the \textit{on-shell} case, the equation of motion for $F$ would yield
$F=F^* = 0$. However, its introduction modifies the supersymmetry transformations of the 
fields $\psi$ and $\phi$ \textit{off-shell}.
Specifically, the transformation of the field $\psi$ is affected by the scalar field  $F$. To see this,
we first observe that the dimension of the field $F$ is of $(mass)^2$, so that its only possible
transformation law is
\begin{equation}
\delta F= i\,\bar{\epsilon}^{\dot{\alpha}}
\,(\overline{\sigma}^{\mu})_{\dot{\alpha}}^{\beta}
\,\partial_{\mu}\psi_{\beta} \ \ \mathrm{and}\ \ 
\delta F^*=-i\,\partial_{\mu}\bar{\psi}^{\dot{\beta}}
\,(\bar{\sigma}^{\mu})_{\dot{\beta}}^{\alpha}
\,\epsilon_{\alpha}\ .
\label{varF}
\end{equation}
The variation of the term $\mathcal{L}_{aux}$ in $S$ therefore gives
\begin{equation}
\delta\mathcal{L}_{aux}=i\,\bar{\epsilon}
\,(\overline{\sigma}^{\mu})
\,\partial_{\mu}\psi\,F^* - i\,\partial_{\mu}\bar{\psi}
\,(\bar{\sigma}^{\mu})\,\epsilon\,F.
\label{varL}
\end{equation}
In the  \textit{on-shell} case, as we have already seen,
the equation of motion for $F$ would yield
$F=F^* = 0$, and the variation (\ref{varF}) would also vanish, thanks to the equation of
motion for $\psi$. To compensate the variation (\ref{varL}) in the \textit{off-shell} case, 
we see that we require a supplementary term in the transformation law for $\psi$:
\begin{equation}
\delta\psi_{\alpha}=i(\sigma^{\mu}\bar{\epsilon})_{\alpha}\,
\partial_{\mu}\phi+\epsilon_{\alpha}F \ \ \mathrm{et}\ \ 
\delta\bar{\psi}^{\dot{\alpha}}=-
i(\epsilon\,\sigma^{\mu})^{\dot{\alpha}}\,\partial_{\mu}\phi^{*}+
\bar{\epsilon}^{\dot{\alpha}} F^*.
\end{equation}
Once again, the supplementary term vanishes when the on-shell condition $F = 0$ is
applied. For simple dimensional reasons, the transformations of $\phi$ are not affected. 
It is easy to check that $\delta S=0$ without using the equations of motion, and hence
supersymmetry continues to be satisfied off-shell, thanks to the appearance of the
auxiliary field $F$.

In fact, the auxiliary field plays an additional role. We must not forget that we have not observed supersymmetry  in the range of energies explored so far. Hence, if supersymmetry exists at all
in Nature, it must be broken in some way. The auxiliary field $F$ (and the other auxiliary field $D$
that we meet later) serve to break supersymmetry if their v.e.v.s are non-zero, as we will see
in the last part of this section.

\subsubsection{Interactions of the chiral multiplets}

We now add to the theory interactions between the scalar
and fermion fields that comprise chiral supermultiplets. The most general form
of interaction that is at most quadratic in the fermion fields is
\begin{equation}
\mathcal{L}_{int}=-\frac{1}{2}W^{ij}(\phi, \phi^*)\psi_i\psi_j + V(\phi,\ \phi^*) + c.c. 
\label{LW} 
\end{equation}
We do not demonstrate it in detail, but the quantity $W^{ij}$ must be an analytic 
function of the fields $\phi_i$, i.e., it does not depend on the $\phi_i^*$,
in order to ensure that the variation due to a supersymmetry transformation of the first term of 
$\mathcal{L}_{int}$  can be compensated by the variation of another term
(basically because supersymmetry transforms $\psi_i$ into $\phi_i$ and {\it vice versa}). For the same reason, $W^{ij}$ must be completely symmetric. Hence $W^{ij}$ must be of the form:
\begin{equation}
W^{ij}=\frac{\partial^2 W(\phi)}{\partial\phi_i\,\partial\phi_j},
\end{equation}
where the object $W$ is called the {\it superpotential}. In order for the model to be
renormalizable, the term in (\ref{LW}) that is bilinear in the fermion fields $\psi_i$
can have at most a linear dependence on the scalar fields $\phi_i$,
implying that $W$ can be at most cubic:
\begin{equation}
W=\frac{1}{2}m^{ij}\phi_i\phi_j+\frac{1}{6}y^{ijk}\phi_i\phi_j\phi_k
\label{superpotentiel}
\end{equation}
in the context of a renormalizable theory. Remarkably, apart from wave-function 
renormalization of the fields, there is no intrinsic renormalization of the superpotential
parameters. 

In general, the superpotential has dimension $(mass)^3$. 
The quadratic term in $W$ (\ref{superpotentiel}) provides the (symmetric) mass matrix 
$m^{ij}$ of the fermions, which is equal to the mass matrix of the scalar bosons, by virtue of
supersymmetry. The trilinear term in $W$ provides the matrix of Yukawa couplings $y^{ijk}$
betweeen a scalar and two fermions, and summarizes all the interactions that are not gauge 
interactions. As already noted, $W$ is an analytical function of the complex fields $\phi_i$,
which has an importance that we discuss later.

The requirement that $\mathcal{L}_{int}$ be invariant under supersymmetry transformations
also determines the form of the potential $V$. In presence of interactions, i.e., if the superpotential is non-zero, the auxiliary fields $F^i$ introduced earlier (\ref{Laux}) can be written in the form:
\begin{equation}
F_i=-\frac{\partial W(\phi)}{\partial\phi^i}=-W^*_i,\ \ \ \ \ 
F^{*i}=-\frac{\partial W(\phi)}{\partial\phi_i}=-W^i.
\label{FW}
\end{equation}
We may therefore write the Lagrangian without introducing explicitly the $F$ fields, in which
case the potential $V$ of the theory is:
\begin{equation}
 V=W^*_i W^i=F_iF^{*i} .
 \label{VFF}
\end{equation}                                                                                                                           
That is automatically non-negative, since it is a sum of modulus-squared terms. 
If we use the general form~(\ref{superpotentiel}) of the superpotential, we have the general
Lagrangian:
\begin{equation}
\mathcal{L}=
-\partial^{\mu}\phi\,\partial_{\mu}\phi^{*}
-i\psi^{\dag}\bar{\sigma}^{\mu}\,\partial_{\mu}\psi
-\frac{1}{2}m^{ij}\psi_i\psi_j-\frac{1}{2}m^{*}_{ij}\psi^{\dag i}\psi^{\dag j}
-V-\frac{1}{2}y^{ijk}\phi_i\psi_j\psi_k
-\frac{1}{2}y^{*}_{ijk}\phi^{*i}\psi^{\dag j}\psi^{\dag k} ,
\end{equation}
where $V$ is given by (\ref{VFF}), (\ref{FW}) and (\ref{superpotentiel}). It is easy to see
from (\ref{superpotentiel}) that the boson and fermion masses are equal, as one would
expect from supersymmetry.

\subsubsection{Supersymmetric gauge theories}

In addition to chiral fermions (quarks, leptons), the SM contains gauge fields of spin 1 ($W$ and $Z$ bosons, photons and gluons). In the section dedicated to the supersymmetry algebra, we saw that vector supermultiplets would provide the appropriate frameworks for such gauge fields. We now study the properties of such a supermultiplet, both with and without interactions~\cite{WZ2}. We recall that a vector 
supermultiplet  contains a massless gauge boson $A^{\mu}_a$ and a massless Weyl fermion, the gaugino $\lambda_a$, both in the adjoint representation of the gauge group. In order to go off-shell, one must introduce an auxiliary real scalar field $D_a$ analogous to the auxiliary field $F$ introduced for the chiral supermultiplet.  

The form of the Lagrangian is completely determined by the condition of gauge invariance and of renormalizability:
\begin{equation}
\mathcal{L}_{gauge}=
-\frac{1}{4}F_{\mu\nu}^a F^{a \mu\nu}
-i\lambda^{a\dag}\bar{\sigma}^{\mu}D_{\mu}\lambda^{a}
+\frac{1}{2}D^aD^a , \label{Ljauge}
\end{equation}
where the gauge covariant derivative  $D_{\mu}$ and $F_{\mu\nu}^a$ take the forms:
\begin{eqnarray}
&F_{\mu\nu}^a&=\partial_{\mu}A^a_{\nu}-\partial_{\nu}A^a_{\mu}
-gf^{abc}A^b_{\mu}A^c_{\nu}, \\
&D_{\mu}\lambda^{a}&=\partial_{\mu}\lambda^{a}-gf^{abc}A^b_{\mu},
\end{eqnarray}
as usual for a gauge theory. Remarkably, this Lagrangian is already supersymmetric,
as can be checked using the following supersymmetry transformations for the fields of the
vector supermultiplet:
\begin{eqnarray}
&\delta A^a_{\mu}&=
\frac{1}{\sqrt{2}}\left(\epsilon^{\dag}\bar{\sigma}^{\mu}\lambda^{a}
+\lambda^{a\dag}\bar{\sigma}^{\mu}\epsilon\right), \\
&\delta \lambda^{a}_{\alpha}&=
-\frac{i}{2\sqrt{2}}(\sigma^{\mu}\bar{\sigma}^{\nu}\epsilon)_{\alpha}F_{\mu\nu}^a
+\frac{1}{\sqrt{2}}\epsilon_{\alpha}D^a, \\
&\delta D^a&=
\frac{i}{\sqrt{2}}\left(\epsilon^{\dag}\bar{\sigma}^{\mu}D_{\mu}\lambda^{a}
-D_{\mu}\lambda^{a\dag}\bar{\sigma}^{\mu}\epsilon\right).
\end{eqnarray}
In the absence of any interactions with chiral supermultiplets, the equation of motion for the 
auxiliary field $D^a$ is simply $D^a=0$, as seen directly from the Lagrangian~(\ref{Ljauge}),
since it does not have a kinetic term  and therefore does not propagate. 

However,
in the SM the gauge fields do interact with the chiral fermions. Hence, in our supersymmetric 
version we have to consider interactions between chiral supermultiplets and vector supermultiplets.
As in the SM, the usual derivatives $\partial^{\mu}$ of the fermions must be replaced by gauge-covariant derivatives $D^{\mu}$, and the same applies to their scalar supersymmetric partners.
The supersymmetric transformation laws of the chiral  supermultiplets must be changed to take into account the variations of these new terms. As a result, the equation of motion for $D^a$  becomes:
\begin{equation}
D^a=-g(\phi^{*}T^a\phi),
\end{equation}
where the $T^a$ are the generators of the gauge group and $g$ is its coupling constant,
and the full scalar potential is
\begin{equation}
V=F_iF^{*i}+\frac{1}{2}\sum_aD^aD^a=
W^*_iW^i+\frac{1}{2}\sum_ag^2(\phi^{*}T^a\phi)^2 . \label{pot}
\end{equation}
This potential is completely determined by the Yukawa couplings ({\it via} the  $F$ term) and by the gauge interactions ({\it via} the $D$ term). The full scalar potential is automatically non-negative, 
which is important for the spontaneous breaking of the symmetry.

In a globally supersymmetric theory, spontaneous breaking may occur {\it via} a v.e.v. for
the $D$ term or the $F$ term, either of which would give a positive contribution
to the vacuum energy. However, it is difficult to construct models that are interesting
for phenomenology, and most model-builders pursue the spontaneous breaking
of local supersymmetry in the context of a supergravity theory, in which this
positive contribution may be cancelled.

\subsection{Low-energy supersymmetric models}

In this section we apply the results obtained in the previous section, with the objective of supersymmetrizing the Standard Model while preserving its successful characteristics.
The minimal supersymmetric extension of the SM is called the MSSM~\cite{Nilles,HK}. 
We will present its particle
content (including the nomenclature of the new particles), we will discuss how the electroweak symmetry may broken, and we will outline an effective framework for describing the breaking of supersymmetry. Later we will present typical predictions of the MSSM. Along the way, we will also mention possible variants of the MSSM, because Nature might very well have chosen a path more complex than this minimal model.

\subsubsection{How many supersymmetries?}

As well as mentioned already, the number of supersymmetric generators $Q_{\alpha}$ may be 
$\mathcal{N}  \geq 1$. Supersymmetric theories with $\mathcal{N}\geq 2$ have some characteristic
advantages, e.g., they have fewer divergences, which make them very interesting theoretically.
Specifically, in the $\mathcal{N}= 2$  case there is only a finite number of divergent Feynman diagrams, and in the  $\mathcal{N} =4$ case there are none, i.e., any theory with $\mathcal{N} =4$ 
supersymmetries is intrinsically finite, and it is easy to construct finite $\mathcal{N}= 2$. 

Unfortunately, it is not possible to construct realistic models with $\mathcal{N}\geq 2$, because
they do not allow the violation of parity that is observed in the weak interactions. This is
because a supermultiplet of a theory with $\mathcal {N} \geq 2$ supersymmetries necessarily incorporates both left- and right-handed fermions in the same supermultiplet: applying a
supersymmetry charge $Q$ changes the helicity by 1/2, so applying two charges relates states
with helicity $\pm 1/2$, implying that they are in the same representation of the gauge group, and hence have the same interactions.
This contradicts experimental observations, which tell us, for example, that the left-handed electron (which forms part of a doublet in the SM) does not have the same interaction with $W$ bosons as the right-handed electron (which is a singlet with zero electroweak isospin that does not feel the 
$SU(2)$ weak interaction). Models with $\mathcal{N}\geq 2$ cannot describe the physics of the SM
particles observed at low energy.

\subsubsection{The particle content in the MSSM}

The supermultiplets in the minimal $\mathcal {N} =1$ case are

$\bullet$ the chiral supermultiplet  that includes a fermion of spin 1/2 and a boson of spin 0, 

$\bullet$ the vector supermultiplet that includes a boson of spin 1 and one fermion of spin 1/2.

Could we link the particles of the SM in such multiplets, i.e., could we associate quarks 
and leptons with the bosons $W$, $Z$, the photon, and so on? The answer is no, 
because this 
would raise problems for the conservation of their quantum numbers. Specifically, the  gauge bosons 
and the fermions do not have the same transformation properties under the SM gauge group, since 
they possess  different quantum numbers, e.g., quarks are triplets of the colour group whereas
gauge bosons are either octets (the gluons) or singlets (the other gauge bosons), and leptons carry
lepton numbers whereas gauge bosons do not. Simple 
$\mathcal{N} = 1$ supersymmetry does not modify these quantum numbers, so we cannot associate any
gauge boson with a known fermion or {\it vice versa}.
Therefore, we have to postulate unseen supersymmetric partners for all the known particles. 
Table~\ref{sparticules} lists, for every SM particle, the name, spin and notation for its spartner.

\begin{table}[htbp!]
\caption[Particle content of the MSSM]{Particle content of the MSSM}
\label{sparticules}
\vspace{-.8cm}
\begin{center}
\begin{tabular}[t]{lcc}
\hline
\hline
{Particle}& {Spartner}& {Spin}	\\
\hline
\hline
quarks q & squarks $\tilde{q}$ & 0 \\
$\to$ top t & stop $\tilde{t}$ & \\
$\to$ bottom b & sbottom $\tilde{b}$ & \\
...& & \\
leptons l & sleptons $\tilde{l}$ & 0 \\
$\to$ electron $e$ & selectron $\tilde{e}$ & \\
$\to$ muon $\mu$ & smuon $\tilde{\mu}$ & \\
$\to$ tau $\tau$ & stau $\tilde{\tau}$ & \\
$\to$ neutrinos $\nu_\ell$ & sneutrinos $\tilde{\nu_\ell}$ & \\
\hline
gauge bosons  & gauginos & 1/2\\
$\to$ photon $\gamma$ & photino $\tilde{\gamma}$ & \\
$\to$ boson $Z$ & Zino $\tilde{Z}$ & \\
$\to$ boson $B$ & Bino $\tilde{B}$ & \\
$\to$ boson $W$ & Wino $\tilde{W}$ & \\
$\to$ gluon $g$ & gluino $\tilde{g}$ & \\
\hline\\[-.3cm]
Higgs bosons  $H_i^{\pm,0}$ & higgsinos $\tilde{H}_i^{\pm,0}$& 1/2 \\[.1cm]
\hline
\hline
\end{tabular}
\end{center}
\end{table}

Before going on to the following sections, we make a few observations. First, 
we note that the spartners of SM fermions and gauge bosons are
of lower spin. \textit{ A priori}, one could have considered associating the 
fermions of the SM with spartners of spin 1, and the gauge bosons with spartners of spin 3/2. 
However, to introduce a particle of spin 1 would require introducing a new gauge interaction, and
hence a non-minimal model. Also, introducing particles of spin $> 1$ would make the theory 
non-renormalizable, i.e., it would no longer be possible to absorb the divergences in perturbation 
theory in a finite number of physical quantities~\footnote{Supergravity does allow a 
restricted number ${\cal N} \leq 8$ of spin-3/2 gravitino partners of the spin-2 graviton to be 
introduced, but they do not carry conventional gauge interactions.}.

Secondly, we recall that in the SM the right-handed fermions have different interactions
from the left-handed fermions, e.g., being singlets of $SU(2)$ instead of doublets.
In supersymmetry, the left- and right-handed must belong to different supermultiplets, and 
have distinct spartners, e.g., $q_L \to \tilde{q}_L$ and $q_R\to \tilde{q}_R$. These two squarks are 
quite different, and we use the chirality index $L$ or $R$ to identify them, even though the
concept of handedness does not make physical sense for a scalar particle, whose only 
helicity is $\lambda=0$. In general, the $\tilde{f}_L$ and $\tilde{f}_R$ mix, and the physical
mass eigenstates are combinations of them. In constructing the Yukawa interactions of the
MSSM, it is often convenient to work with superfields that comprise conjugates of the 
$\tilde{f}_R$ and their scalar spartners: these are left-handed chiral supermultiplets
denoted by $F^c$.

Thirdly, we note that, besides the new spartners, at least two doublets of Higgs bosons
are required. To understand why, we recall that, in the study of supersymmetric theories, 
we introduced the notion 
of the superpotential. This governs all the possible Yukawa interactions of the matter particles 
with the Higgs fields. In the SM, if we use a Higgs field
$h$ to give masses to the quarks of type `down', {\it via} Yukawa couplings $q {\bar d} h$, we 
could use the complex conjugate field $h^*$ to give masses to quarks of type `up', {\it via} 
couplings $q {\bar u} h^*$. However, we recall that in a supersymmetric theory the superpotential is 
an analytic function of the superfields that cannot depend on their complex conjugates.
Therefore, we must use separate Higgs supermultiplets (denoted by capital letters)
with opposite hypercharge quantum numbers,
and interactions of the forms $Q D^c H_d$ and $Q U^c H_u$. Charged leptons may
acquire masses through interactions of the form $L E^c H_d$.
We also note that pairs of Higgs superfields are needed in order to cancel the triangle 
anomalies that would be generated by higgsino fermion loops.

Fourthly, we note that in general the $\tilde {\gamma},  \tilde{Z},  \tilde{W}$ and $\tilde{H}$ mix,
and the experimentally observable mass eigenstates are combinations of these gauginos and 
higgsinos that are generally named neutralinos $\tilde{N}^0_{1,2,3,4}$, which have zero electrical charge, and charginos $\tilde{C}_{1,2}^{\pm}$~\footnote{These are often denoted by 
$\tilde{\chi}^0_{1,2,3,4}$ and  $\tilde{\chi}_{1,2}^{\pm}$, respectively.}, which are electrically charged
and mix the $\tilde{W}^{\pm}$ and the $\tilde{H}^{\pm}$.

\subsubsection{Interactions in the MSSM}

The MSSM is the minimal supersymmetric extension of the Standard Model~\cite{Nilles,HK}. 
The quarks and the leptons are put together in chiral superfields with their superpartners that have the same charges under 
$SU(3)_C$, $SU(2)_L$ y $U(1)_Y$. The gauge bosons are placed with their fermionic 
superpartners in vector superfields. The superpotential of the MSSM is
\begin{equation}
 \mathcal{W}=\mathcal{Y}_u Q U^c H_u + \mathcal{Y}_d Q D^c H_d +\mathcal{Y}_e
L E^c H_d + \mu H_u H_d,
\label{superpot}
\end{equation}
where we recall that the $Q$ and $L$ are the superfields containing the left-handed quarks and leptons,
respectively, and the $U^c, D^c$ and $E^c$ are the superfields containing the left-handed
antiquarks and antileptons, which are the charge conjugates of the right-handed quarks and leptons.
Note that, for clarity, we have suppressed the $SU(2)$ indexes. The $\mathcal{Y}$ are $3\times 3$ Yukawa matrices  in flavour space, and do not have dimensions. After electroweak symmetry
breaking, they give the masses to the quarks and leptons as well as the CKM angles and phases.
As already mentioned, two Higgs doublets, $H_u$ y $H_d$, are needed because of the analytical form of the superpotential.

The $\mu H_u H_d$ term is permitted by the symmetries of the MSSM and is required in order to 
have a suitable vacuum after electroweak symmetry breaking. The quantity $\mu$ has the
dimension of a mass, and phenomenology requires it to be of the order of a TeV. The origin of
$\mu$ is a puzzle: it might be associated to the scale of supersymmetry breaking.

The superpotential (\ref{superpot}) determines all the non-gauge interactions of the MSSM, 
thanks to the formula~(\ref{LW}), and the form of the effective potential of the theory is
given by formula~(\ref{pot}).

The next-to-minimal supersymmetric extension of the Standard Model (NMSSM)~\cite{Gunion} 
is the simplest 
extension of the MSSM. In this model, the particle content is modified by the addition of a new 
singlet chiral supermultiplet $S$, with some additional superpotential terms:
\beq \mathcal{W}_{NMSSM}=\frac{1}{6}k S^3 +\frac{1}{2}\mu_S S^2 + \lambda S H_u
H_d +\mathcal{W}_{MSSM} \label{NMSSM}.\eeq 
The principal interest of the NMSSM is to propose a solution to the $\mu$ problem. Specifically,
if the scalar part of $S$ has a non-zero vacuum espectation value $\langle S \rangle$, the 
last term in (\ref{NMSSM}) gives an effective $\mu$ term:
$\mu_{eff} = \lambda \langle S \rangle$. Assuming that a soft supersymmetry-breaking
scalar mass for $S$ also appears in $\mathcal{L}_{soft}$, its v.e.v. is naturally  of the order of 
$m_{soft} \sim \mathcal{O}(1)$~TeV, the typical mass scale of the other
scalars and gauginos. Thus the effective value of $\mu$ is of the order of 1~TeV, rather than 
being a parameter whose magnitude is independent of the scale of supersymmetry breaking.

Phenomenologically the NMSSM differs from the MSSM because it allows the lightest Higgs 
boson to become heavier. In addition, the fermionic partner of $S$ can mix with the four
neutralinos of the MSSM. Thus the experimental signatures of the NMSSM may differ
significantly from those of the MSSM.

\subsubsection {Soft supersymmetry breaking} 

We have discussed so far the supersymmetric aspects of the MSSM. However, we
know that supersymmetry must be broken: the selectron weighs more than the 
electron, squarks weigh more than quarks, etc. Therefore, we must introduce into the 
model the breaking of supersymmetry. However, the mechanism and the effective scale of its 
breaking are still unknown. Hence we adopt the {\it ad hoc} strategy of parametrizing the
breaking of supersymmetry in terms of effective soft~\footnote{Here, the adjective
`soft' means that they do not introduce quadratic divergences.}
low-energy supersymmetry-breaking terms
that are added to the Lagrangian~\cite{DG}. For a general supersymmetric theory,
the form of these soft supersymmetry-breaking 
terms $\mathcal{L}_{soft}$ in the Lagrangian is
\begin{equation}
\mathcal{L} \supset
\mathcal{L}_{soft}=-\frac{1}{2}(M_{\lambda}\lambda^a\lambda^a +\ c.c) - m_{ij}^2
\phi_j^{*}\phi_i +
(\frac{1}{2}b_{ij}\phi_i\phi_j+\frac{1}{6}a_{ijk}\phi_i\phi_j\phi_k+\ c.c) .
\end{equation}
This breaks supersymmetry explicitly, since only the the gauginos $\lambda^a$
and the scalars $\phi_i$  have mass terms, and the trilinear terms with coefficients
$a_{ijk}$ are also not of supersymmetric form. In the case of the MSSM, $\mathcal{L}_{soft}$ 
takes the following general form in terms of the spartner fields of the MSSM:
\begin{eqnarray}
-\mathcal{L}_{soft} &=&\frac{1}{2}(M_3 \tilde{g}\tilde{g} + M_2\tilde{W}\tilde{W}+M_1\tilde{B}\tilde{B}+\ c.c) \nonumber \\ 
&+&  \tilde{Q}^{\dag}m_Q^2\tilde{Q} + \bar{\tilde{U}}^{\dag}m_{\bar{U}}^2\bar{\tilde{U}} +\bar{\tilde{D}}^{\dag}m_{D}^2\bar{\tilde{D}} +\bar{\tilde{L}}^{\dag}m_{L}^2\bar{\tilde{L}} +\bar{\tilde{E}}^{\dag}m_{\bar{E}}^2\bar{\tilde{E}} \nonumber \\ 
&+& (\bar{\tilde{U}}^{\dag}a_U \tilde{Q}H_u - \bar{\tilde{D}}^{\dag}a_D\tilde{Q}H_d -\bar{\tilde{E}}^{\dag}a_E\tilde{L}H_d +\ c.c) \nonumber \\ 
&+& m_{H_u}^2 H_u^{*}H_u +m_{H_d}^2 H_d^{*}H_d + (b H_uH_d +\ c.c) .
\label{superpotentielMSSM}
\end{eqnarray}
The masses $M_3$, $M_2$ and $M_1$ of the gauginos are complex in general, which introduces 6 parameters. The quantities 
$m_Q$, $m_L$ and $m_{\bar{u}}$, are the mass matrices of the squarks and sleptons,
which are hermitian $3\times 3$ matrices in family space, adding 45 more
unknown parameters. The couplings  $a_U$, $a_D$, ..., are also complex $3\times 3$ 
matrices, characterized by 54 parameters. In addition, the quadratic couplings of the Higgs
bosons introduce 4 more parameters, so that the whole $\mathcal{L}_{soft}$ contains 
a total of 109 unknown parameters, including many that violate CP!
 
Supersymmetry itself is a very powerful principle whose implementation introduces only one new parameter ($\mu$) in the MSSM. However, in our present state of ignorance, the breaking of
supersymmetry introduces many new parameters. On the other hand, the number of soft 
parameters can be reduced by postulating symmetries or making supplementary hypotheses.
Measuring the parameters of soft supersymmetry breaking would allow us to go beyond the
phenomenological parametrization (\ref{superpotentielMSSM}), and open the way to testing
models of the high-energy dynamics that breaks supersymmetry.

\subsubsection{Electroweak symmetry breaking and supersymmetric Higgs bosons}

As we have already seen, the Higgs sector of the MSSM contains two complex doublets:
\begin{equation}
H_u=\left(\begin{array}{c} H_u^0 \\ H_u^- \end{array}\right),
\ H_d=\left(\begin{array}{c} H_d^+ \\ H_d^0 \end{array}\right).
\end{equation}
Electroweak symmetry breaking is a little bit more complicated than its analogue in the 
Standard Model. At tree level, we can write the effective scalar potential (after 
simplifications whose details we do not reproduce):
\begin{eqnarray}
V&=&(|\mu|^2+m_{H_u}^2)|H_u^0|^2 + (|\mu|^2+m_{H_d}^2)|H_d^0|^2 - b(H_u^0 H_d^0
+c.c) \nonumber \\
&&+\frac{1}{8}(g_2^2+g_1^2)(|H_u^0|^2-|H_d^0|^2)^2. 
\end{eqnarray}
The terms proportional to $|\mu|^2$ originate from the $F$ terms  in the supersymmetric
effective potential, and the terms proportional to the gauge couplings $(g_1,\ g_2)$  originate 
from the  $D$ terms. The other terms originate from
$\mathcal{L}_{soft}$ (without mentioning the other scalars that do not play any role here).
Spontaneous electroweak symmetry breaking can arise with this form of potential 
if the $b$ parameter satisfies:
\begin{equation}
b^2>(|\mu|^2+m_{H_u}^2)(|\mu|^2+m_{H_d}^2), \label{EWSBb1}
\end{equation}
In addition, we want the potential to be bounded from below. Thus 
\begin{equation}
2b<2|\mu|^2+m_{H_u}^2+m_{H_d}^2 \label{EWSBb2}
\end{equation}
at tree level~\footnote{As we shall see shortly, radiative corrections to the
effective potential play important roles.}. After electroweak symmetry breaking, both 
the fields $H_u^0$ and $H_d^0$ must develop v.e.v.'s, in order to give masses to all the
quarks and leptons:
\begin{equation}
<H_u^0>=v_u,\ <H_d^0>=v_d.
\end{equation}
Comparing with the Standard Model, we have
\begin{equation}
v^2=v_u^2+v_d^2 =\frac{2m_Z^2}{(g_2^2+g_1^2)}.
\end{equation}
Conventionally, one defines also the $\tan\beta$ parameter:
\begin{equation}
\tan\beta=\frac{v_u}{v_d}:\ 0<\beta<\frac{\pi}{2}.
\label{tanbeta}
\end{equation}
At the minimun of the potential
\begin{equation}
\frac{\partial V}{\partial H_u^0}=\frac{\partial V}{\partial H_d^0}=0,
\end{equation}
giving the two relations
\begin{eqnarray}
|\mu|^2+m_{H_u}^2 & = & b\tan\beta -\frac{m_Z}{2}\cos^2\beta, \nonumber \\
|\mu|^2+m_{H_d}^2 & = & b\cot\beta +\frac{m_Z}{2}\cos^2\beta.
\label{fixmu}
\end{eqnarray}
These expressions are important because they relate a measurable quantity, $m_Z$, to 
the soft parameters. We note that some amount of fine-tuning would be
required if the soft parameters were much larger than $m_Z$. We note also
that the vacuum conditions (\ref{fixmu}) do not depend on the phase of $\mu$.

The two complex Higgs doublets of the MSSM have a total of $8$ degrees of freedom. 
However, the Higgs mechanism for electroweak breaking uses 3 degrees of freedom to give 
longitudinal polarization states, and hence masses, to the two $W$ bosons and to the $Z$ 
boson. Therefore, five physical Higgs bosons remain in the spectrum. Of these, two are
neutral Higgs bosons that are even under the CP transformation, called $h^0$and $H^0$.
In addition, there is one neutral Higgs boson that is odd under CP, called $A^0$. 
The final two Higgs bosons are charged, the $H^{\pm}$.

At tree level, the masses of the supersymetric Higgs bosons are:
\begin{eqnarray}
m^2_{h^0, H^0} &=& \frac{1}{2}\left(m_{A^0}^2+m_Z^2 \mp 
\sqrt{(m_{A^0}^2+m_Z^2)^2 -4m_{A^0}^2m_Z^2\cos^22\beta}\,\right), \\
m_{A^0}^2 &=& \frac{2b}{\sin2\beta}, \\
m_{H^{\pm}}^2 &=& m_{A^0}^2 + m_W^2, 
\end{eqnarray}
and the mass of the $h^0$ is bounded from above by:
\begin{equation}
m_{h^0}<|\cos2\beta|m_Z.
\end{equation}
This upper limit on $m_{h^0}$ may be traced to the fact that the quartic Higgs
coupling $\lambda$ is fixed in the MSSM, being equal to the square of the electroweak gauge
coupling (up to numerical factors). This means that $\lambda$ and hence $m_{h^0}$
cannot be very large.

\begin{figure}[htbp!]
\begin{center}
\includegraphics[height=7cm]{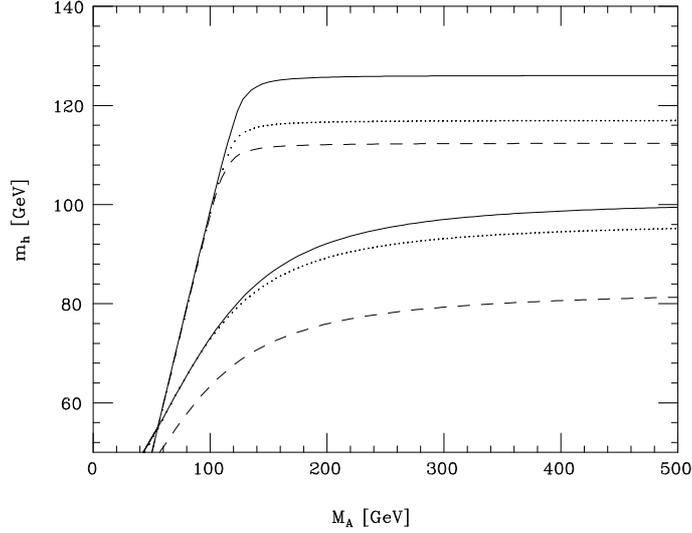}
\end{center}
\caption{The mass of the lightest supersymmetric Higgs boson as a function of $m_{A^0}$ for 
different values of $\tan\beta$
\label{mHSUSY}}
\end{figure}

However, the above relations are valid only at tree level, and the masses of Higgs scalars have
one-loop radiative corrections that are not negligible~\cite{susyHiggs}. 
The most important corrections for $m_h$
are those due to the top quark and squark:
\begin{equation}
\Delta m_h^2=\frac{3m_t^4}{4\pi^2 v^2}\ln\left(\frac{m_{\tilde{t}_1}m_{\tilde{t}_2}}{m_t^2}\right) 
+ \frac{3m_t^4}{8\pi^2
v^2}\mathrm{f}(m_{\tilde{t}_1}^2,m_{\tilde{t}_2}^2,\mu,\tan\beta),  
\label{deltamh}
\end{equation}
where $m_{\tilde{t}_{1,2}}$ are the physical masses of the stops (that are mixtures of 
$\tilde{t}_R$ and $\tilde{t}_L$), and $\mathrm{f}(m_{\tilde{t}_1}^2, m_{\tilde{t}_2}^2, \mu, \tan\beta)$ is 
a non-logarithmic function that can be found in~\cite{JE-SUSY}. The correction $\Delta m_h^2$ 
depends quartically on the mass of the top, making it more important than the one-loop
corrections due to other quarks, leptons, and gauge multiplets. After including this correction, 
the mass of the lightest Higgs boson may  be as large as
\begin{equation}
m_h \lesssim 130\ \mathrm{GeV}\ ,
\label{hmass}
\end{equation}
for masses of sparticles about a TeV.
This is seen in Fig.~\ref{mHSUSY}, which shows $m_h$ as a function of $m_{A^0}$ for 
different values of $\tan\beta$. 
As noted,
the range (\ref{hmass}) for the mass of the lightest supersymetric Higgs boson is in perfect
agreement with the indications provided by the electroweak data, as discussed in Lecture 1! This is
just one of many attractive features of supersymmetry that we review here.

\subsubsection{$R$ parity and dark matter}

We introduced above the superpotential (\ref{superpotentielMSSM}) of the MSSM,
which includes only the Yukawa interactions of the SM. However, gauge invariance, Lorentz invariance,
and analyticity in the SM  fields would allow us to introduce in the superpotential other terms that do 
not have any correspondence with the SM, and do not preserve either baryon number and/or 
lepton number~\footnote{The conservation of $B$ and $L$ in the SM is an accidental symmetry
of its renormalizable interactions that is {\it a priori} not obligatory. As we see later in the
context of Grand Unified Theories, the SM, non-renormalizable terms that violate $L$ or $B$ may
be added to the SM Lagrangian. In the MSSM, such $L$- and $B$-violating may appear at the
renormalizable level.}. These terms are
\begin{equation}
 \mathcal{W}_{RPV}=\lambda_{ijk}L_iL_jE_k + \lambda'_{ijk} L_iQ_j D^c_k +
\lambda''_{ijk} U^c_i D^c_j D^c_k + \mu'_i L_i H_u, \label{WRPV}
\end{equation}
where $\lambda,\lambda'$ and $\lambda''$ are arbitrary dimensionless coupling constants,
and the $\mu'_i$ are parameters with the dimension of a mass.

These parameters are subject to strong phenomenological restrictions.
For example, a combination of the second and third terms would induce rapid disintegration of the 
proton {\it via} squark exchange, whereas the proton is very stable, with a lifetime exceeding 
$\sim 10^{33}$ years. This implies that the product of such terms must be strongly
suppressed~\cite{RPVSUSY}: 
\begin{equation}
|\lambda' \lambda''| \; < \; {\cal O}(10^{-9}). 
\end{equation}
One way to avoid all such terms is to add to the MSSM a new symmetry called $R$-parity,
given by the following combination baryon number, lepton number, and spin $S$:
\begin{equation}
 R=(-1)^{3(B-L)+2S}.
\end{equation}
This is a multiplicatively-conserved quantum number in the SM, 
since all the SM particles and Higgs bosons have even $R$ parity: $R = + 1$.
On the other hand, all the sparticles have odd $R$ parity ($R = - 1$). 

Conservation of $R$ parity would have important phenomenological consequences:

$\bullet$  The sparticles are produced in even numbers (usually two at time), for example:
$\bar{p}\,p\to\tilde{q}\,\tilde{g}\,X$, $e^+\,e^-\to \tilde{\mu}^+\,
\tilde{\mu}^-$.

$\bullet$  Each sparticle decays into another sparticle (or into an odd number of them), for example: $\tilde{q}\to q\, \tilde{g}$, $\tilde{\mu}\to\mu\,
\tilde{\gamma}$.

$\bullet$ The lightest sparticle (LSP) must be stable, since it has $R =-1$. If it is electrically 
neutral, it can interact only weakly with ordinary matter, and may be a good candidate for the 
non-baryonic dark matter that is required by cosmology~\cite{EHNOS}.

The dark matter particles should have neither electric charge nor strong interactions,
otherwise they would be visible or detectable, e.g., through their binding to ordinary
matter to form what would look like anomalous heavy nuclei, which have never been
seen. We therefore expect any dark matter particle to have only weak interactions, in
which case, if it was produced at a collider such as the LHC, it would carry
energy--momentum away invisibly. Accordingly, most LHC searches for supersymmetry
focus on events with missing transverse momentum, though searches for signatures
of $R$-violating models are also considered.

The existence of a stable, weakly-interacting LSP is a very important prediction of the MSSM,
but its nature and its total contribution to the density of dark matter depend on the 
parameters of the MSSM. One weakly-interacting candidate was the lightest sneutrino, 
but this has already been excluded by direct searches at LEP and by experiments searching 
directly for dark matter. The remaining candidate particles are the lightest neutralino $\chi$
of spin 1/2, and the gravitino of spin 3/2.  As we discuss later,
there are chances to detect a neutralino LSP
at the LHC in events with missing energy, or directly as astrophysical dark matter.
On the other hand, the interactions of the gravitino are so weak that it could not be 
detected as astrophysical dark matter, and could only be detected indirectly in collider experiments.

\subsection{Phenomenology of supersymmetry}

As we have seen, the soft supersymmetry-breaking sector of the MSSM has over
a hundred parameters. This renders very difficult the interpretation of experimental constraints and
(hopefully) the extraction of the experimental values of these parameters. A simplifying hypothesis
is to assume \textit{universality} at a certain scale before renormalization, 
leading us to the constrained MSSM (CMSSM): 

$\bullet$ The gaugino masses are assumed to be equal at some input GUT or
supergravity scale: $M_3=M_2=M_1=m_{1/2}$;

$\bullet$ The scalar masses of squarks and sleptons are assumed to be universal at
the same scale: 
$m_Q^2=m_{U^c}^2=...=m_0^2$, as are the soft supersymmetry-breaking contributions
to the Higgs masses $m_{H_u}^2=m_{H_d}^2=m_0^2$;

$\bullet$ The trilinear couplings  are related by a universal coefficient $A_0$ to the
corresponding Yukawa couplings: $a_u=A_0y_u$, $a_d=A_0y_d$, $a_e=A_0y_e$.

Simplifying the MSSM to the CMSSM reduces the number of parameters from over
one hundred to only 4: $m_{1/2}, m_0, A_0, \tan \beta$ and the sign of $\mu$ [the
magnitude of $\mu$ is fixed by the electroweak vacuum conditions: see (\ref{fixmu}].
The CMSSM hypothesis is very practical from a phenomenological point of view, though 
questionable from a purely theoretical point of view. The CMSSM and the simplification of 
$\mathcal{L}_{soft} $ are inspired by simple supergravity models where the breaking 
of supersymmetry is mediated by gravity, though minimal supergravity
models actually impose two additional constraints. On the other hand, generic string
models often lead to different patterns of soft supersymmetry breaking.

Dropping universality for squarks or sleptons with the same quantum numbers but
in different generations would lead to problems with flavour-changing neutral
interactions, and Grand Unified Theories relate the soft supersymmetry-breaking
masses of squarks and sleptons with different quantum numbers. However, there
is no strong theoretical or phenomenological reason to postulate universality
for the soft supersymmetry-breaking contributions to the Higgs masses.
One may relax this assumption for the Higgs scalar masses-squared $m_H^2$ by assuming
the same single-parameter non-universal Higgs mass parameter (the NUHM1), or by
allowing the non-universal Higgs mass parameters to be different (the NUHM2).

\subsection{Renormalization of the soft supersymmetry-breaking parameters}

In our ignorance of the underlying mechanism of supersymmetry breaking,
it is usually assumed that this occurs at some large mass scale far above a TeV,
perhaps around the grand unification or Planck scale. The soft supersymmetry-breaking 
parameters therefore undergo
significant renormalization between this input scale and the electroweak scale.
Although quadratic divergences are absent from a softly-broken supersymmetric 
theory, it still has logarithmic divergences that may be treated using the
renormalization group (RG).

At leading order in the RG, which resums the leading one-loop logarithms,
the renormalizations of the soft gaugino masses $M_a$ are the same as for the
corresponding gauge couplings:
\begin{equation}
Q \frac{d M_a}{d Q} \; = \; \beta_a M_a ,
\label{rengaugino}
\end{equation}
where $\beta_a$ is the standard one-loop renormalization coefficient including
supersymmetric particles that is discussed in more detail in the next Lecture.
As a result of (\ref{rengaugino}), to leading order
\begin{equation}
M_a(Q) \; = \; \frac{\alpha_a(Q)}{\alpha_{GUT}} m_{1/2}
\label{gauginomasses}
\end{equation}
if the gauge couplings $\alpha_a$ and the gaugino masses are assumed to
unify at the same large mass scale $M_{GUT}$.
As a consequence of (\ref{gauginomasses}), one expects the gluino to be
heavier than the wino: $m_{\tilde g}/m_{\tilde W} = \alpha_3/\alpha_2$ at
leading order.

The soft supersymmetry-breaking scalar masses-squared $m_0^2$ acquire renormalizations
related to the gaugino masses {\it via} the gauge couplings, and to the scalar masses
and trilinear parameters $A_\lambda$ {\it via} the Yukawa couplings:
\begin{equation}
\frac{Q d m^2_{0}}{d Q} \; = \; \frac{1}{16 \pi^2} \left[ - g_a^2 M_a^2 + 
\lambda^2 (m_0^2 + A_\lambda^2) \right] .
\label{renm0}
\end{equation}
The latter effect is significant for the stop squark, one of the Higgs
multiplets, and possibly the other third-generation
sfermions if $\tan \beta$ is large. For the other sfermions, at leading order one has
\begin{equation}
m^2_0(Q)  \; = \; m_0^2 + C m_{1/2}^2 ,
\label{scalarmasses}
\end{equation}
where the coefficient $C$ depends on the gauge quantum numbers of the
corresponding sfermion. Consequently, one expects the squarks to be
heavier than the sleptons. Specifically, in the CMSSM one finds at the
electroweak scale that
\begin{eqnarray}
{\rm squarks:} \; m^2_{\tilde q} \; & \sim & \; m_0^2 + 6 m^2_{1/2} , \\
\textrm{left-handed sleptons:} \; m^2_{\tilde \ell_L} \; & \sim & \; m_0^2 + 0.5 m^2_{1/2} , \\
\textrm{right-handed sleptons:} \; m^2_{\tilde \ell_R} \; & \sim & \; m_0^2 + 0.15 m^2_{1/2} .
\label{numm0}
\end{eqnarray}
The difference between the left and right slepton masses may have
implications for cosmology, as we discuss later.
A small difference is also expected between the masses of the left and right
squarks, but this is relatively less significant numerically.

The CKM mixing between quarks is related in the SM to off-diagonal
entries in the Yukawa coupling matrix, and shows up in leading-order
charged-current interactions and flavour-changing neutral current (FCNC) interactions
induced at the loop level. One would expect additional FCNCs to be induced by
similar loop diagrams involving squarks, which would propagate through
the RGEs (\ref{renm0}) and induce flavour-violating terms in the sfermion
mass matrices. However, experiment imposes important
upper limits on such additional supersymmetric flavour effects. As already discussed, these 
would be suppressed (though non-zero) if the soft supersymmetry-breaking scalar
masses of all sfermions with the same quantum numbers were the same before
renormalization. The hypothesis of Minimal Flavour Violation (MFV) is that flavour
mixing of squarks and sleptons is induced only by the CKM mixing in the quark
sector and the corresponding MNS mixing in the lepton sector: see the next Lecture.
The MFV hypothesis requires also that the soft supersymmetry-breaking trilinear
parameters $A$ be universal for sfermions with the same quantum numbers:
$A_\lambda = A_0 \lambda$. However, the MFV hypothesis does permit the appearance of
6 additional phases beyond those in the CKM model for quarks: 3 phases for the
different gaugino mass parameters, and 3 phases for the 
different $A_0$ coefficients~\cite{MCPMFV}.

\begin{figure}[htbp!]
\begin{center}
{\includegraphics[height=3in]{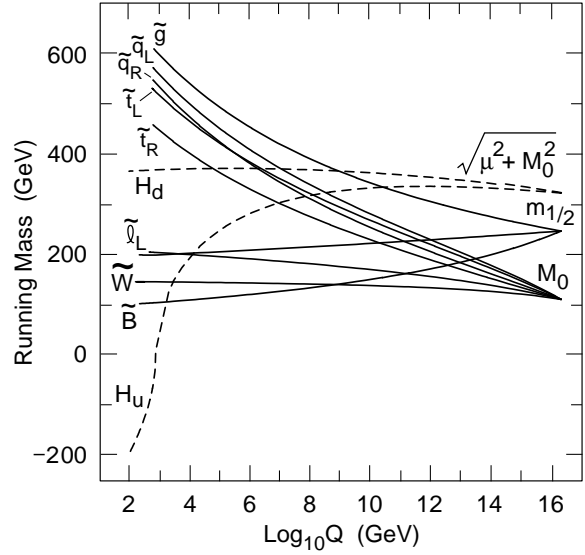}}
\end{center}
\caption[]{Calculations of the renormalization of soft supersymmetry-breaking
sparticle masses, assuming universal scalar and gaugino masses $m_0, m_{1/2}$
at the GUT scale. Note that strongly-interacting sparticles have larger physical
masses at low scales, and the $m^2_{H_u}$ is driven negative, triggering
electroweak symmetry breaking.} 
\label{FiGPeskin}
\end{figure}

Results of typical numerical calculations of these renormalization effects
in the CMSSM are shown in Fig.~\ref{FiGPeskin}. An important effect illustrated
there is that the RGEs may drive $m_{H_u}^2$ negative at some low
renormalization scale $Q_N$, thanks to the top quark
Yukawa coupling appearing in (\ref{renm0})~\footnote{The effect of the Yukawa
coupling is to {\it increase} $m_0^2$ as $Q$ increases, i.e., to {\it decrease} $m_0^2$
as $Q$ decreases.}. A negative value of $m_{H_u}^2$ would trigger electroweak
symmetry breaking at a scale $\sim Q_N$. Since the negative value of
$m^2_{H_u}$ is due to the logarithmic renormalization by the top quark Yukawa
coupling, electroweak symmetry breaking appears at a scale exponentially
smaller than the input GUT or Planck scale:
\begin{equation}
\frac{m_W}{M_{GUT, P}} \; = \; \exp \left( - \frac{{\cal O}(1)}{\alpha_t} \right): 
\; \alpha_t \; \equiv \; \frac{\lambda_t^2}{4 \pi} .
\label{smallmW}
\end{equation}
In this way, it is possible for the electroweak scale to be generated naturally
at a scale $\sim 100$~GeV if the top quark is heavy:
$m_t \sim 60$ to 100~GeV, a realization that
long predated the discovery of just such a heavy top quark.

\subsubsection{Sparticle masses and mixing}

There are aspects of sparticle masses and
mixing that are important for phenomenology, as we now discuss. \\
~ \\
\noindent
{\bf Sfermions}: As we have seen, each flavour of charged lepton or quark has both
left- and right-handed components $f_{L,R}$, and these have separate
spin-0 boson superpartners $\tilde f_{L,R}$. These have different
isospins $I = {1\over 2},~0$, but may mix as soon as the electroweak
gauge symmetry is broken. Thus, for each flavour we should consider a
$2\times 2$ mixing matrix for the 
$\tilde f_{L,R}$, which takes the following general form:
\beq
M^2_{\tilde f} \equiv \left( \begin{matrix}
m^2_{\tilde f_{LL}} & m^2_{\tilde
f_{LR}} \cr \cr m^2_{\tilde f_{LR}} & m^2_{\tilde f_{RR}}
\end{matrix}
\right) .
\label{threeten}
\eeq
The diagonal terms may be written in the form
\beq
m^2_{\tilde f_{LL,RR}} = m^2_{\tilde f_{L,R}} + m^{D^2}_{\tilde
f_{L,R}} + m^2_f ,
\label{threeeleven}
\eeq
where $m_f$ is the mass of the corresponding fermion, $\tilde
m^2_{\tilde f_{L,R}}$ is the soft supersymmetry-breaking mass discussed
in the previous section, and $m^{D^2}_{\tilde f_{L,R}}$ is a
contribution due to the quartic $D$ terms in the effective potential:
\beq
m^{D^2}_{\tilde f_{L,R}} = m^2_Z~\cos 2\beta~~(I_3 + \sin^2\theta_WQ_{em}) ,
\label{threetwelve}
\eeq
where the term $\propto I_3$ is non-zero only for the $\tilde f_L$.
Finally, the off-diagonal mixing term takes the general form
\beq
m^{2}_{\tilde f_{L,R}} = m_f \left(A_f +
\mu^{\tan\beta}_{\cot\beta}\right)~~{\rm for}~~f =
^{e,\mu,\tau,d,s,b}_{u,c,t} .
\label{threethirteen}
\eeq
It is clear that $\tilde f_{L,R}$ mixing is likely to be important for
the $\tilde t$, and it may also be important for the $\tilde b_{L,R}$ and
$\tilde\tau_{L,R}$ if $\tan\beta$ is large. 

We also see from (\ref{threeeleven}) that the diagonal entries for the
$\tilde t_{L,R}$ would be different from those of the $\tilde u_{L,R}$ and
$\tilde c_{L,R}$, even if their soft supersymmetry-breaking masses were
universal, because of the $m^2_f$ contribution. In fact, we also expect
non-universal renormalization of $m^2_{\tilde t_{LL,RR}}$ (and also
$m^2_{\tilde b_{LL,RR}}$ and $m^2_{\tilde \tau_{LL,RR}}$ if $\tan\beta$ is
large), because of Yukawa effects analogous to those discussed previously
for the renormalization of the soft Higgs masses. For
these reasons, the $\tilde t_{L,R}$ are not usually assumed to be
degenerate with the other squark flavours.  \\

\noindent
{\bf Charginos}: These are the supersymmetric partners of the
$W^\pm$ and $H^\pm$, which mix through a $2\times 2$ matrix
\beq
-{1\over 2} ~(\tilde W^-, \tilde H^-)~~M_C ~~\left(
\begin{matrix}
\tilde
W^+\cr\tilde H^+
\end{matrix}
\right)~~+~~{\rm herm.conj.}
\label{threeforteen}
\eeq
where
\beq
M_C \equiv \left(
\begin{matrix}
M_2 & \sqrt{2} m_W\sin\beta \cr \sqrt{2}
m_W\cos\beta & \mu
\end{matrix}
\right) .
\label{threefifteen}
\eeq
Here $M_2$ is the unmixed $SU(2)$ gaugino mass and $\mu$ is the Higgs
mixing parameter introduced previously.  \\

\noindent
{\bf Neutralinos}: These are characterized by a $4\times 4$ mass
mixing matrix~\cite{EHNOS}, which takes the following form in the $(\tilde
W^3, \tilde
B, \tilde H^0_2, \tilde H^0_1)$ basis :
\beq
m_N = \left( 
\begin{matrix}
M_2 & 0 & {-g_2v_2\over\sqrt{2}} & {g_2v_1\over\sqrt{2}}\cr\cr
0 & M_1 & {g^\prime v_2\over\sqrt{2}} & {-g^\prime 
v_1\over\sqrt{2}}\cr\cr
{-g_2 v_2\over\sqrt{2}} & {g^\prime v_2\over\sqrt{2}} & 0 & \mu \cr\cr
{g_2v_1\over\sqrt{2}} & {-g^\prime v_1\over\sqrt{2}} & \mu & 0
\end{matrix}
\right)
\label{threesixteen}
\eeq
Note that this has a structure similar to $M_C$ (\ref{threefifteen}), but 
with its entries replaced by $2\times 2$ submatrices. As has already been
mentioned, one often assumes that the $SU(2)$ and $U(1)$ gaugino
masses $M_{1,2}$ are universal at the GUT or supergravity scale, so that
\beq
M_1 \simeq M_2~~{\alpha_1\over\alpha_2} ,
\label{threeseventeen}
\eeq
so the relevant parameters of (\ref{threesixteen}) are generally taken to
be $M_2 = (\alpha_2/ \alpha_{GUT}) m_{1/2}$, $\mu$ and
$\tan\beta$.

In the
limit $M_2\rightarrow 0$, the lightest neutralino 
$\chi$ would be approximately a photino, and it
would be approximately a higgsino in the limit $\mu\rightarrow 0$.
However, these idealized limits are excluded by unsuccessful LEP
and other searches for neutralinos and charginos. Possibilities that
persist are that $\chi$ be approximately a Bino, ${\tilde B}$, or that it has
a substantial higgsino component.

\subsection{Constraints on the MSSM}

\begin{figure}[ht]
\resizebox{0.48\textwidth}{!}{
\includegraphics{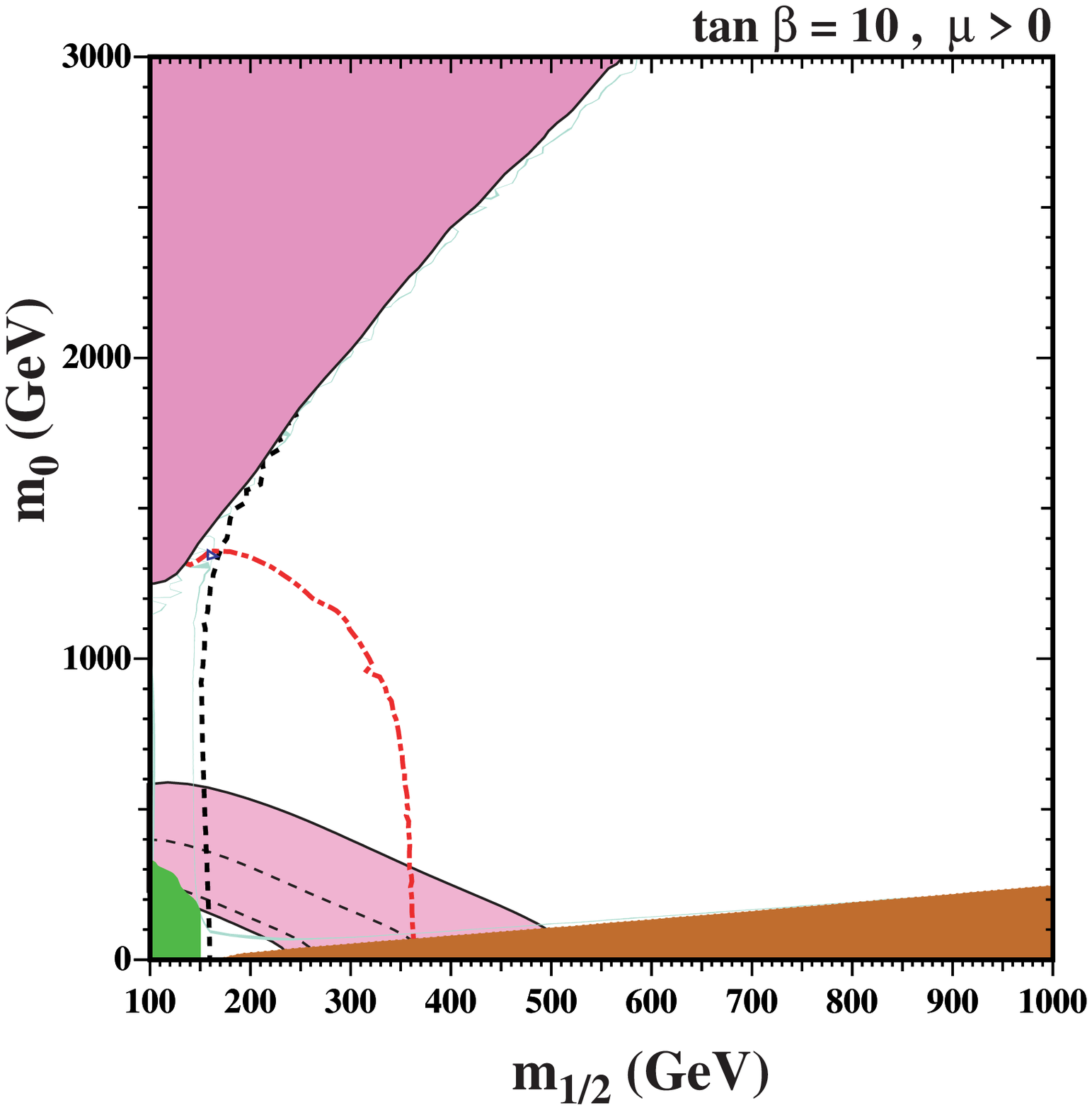}}
\resizebox{0.48\textwidth}{!}{
\includegraphics{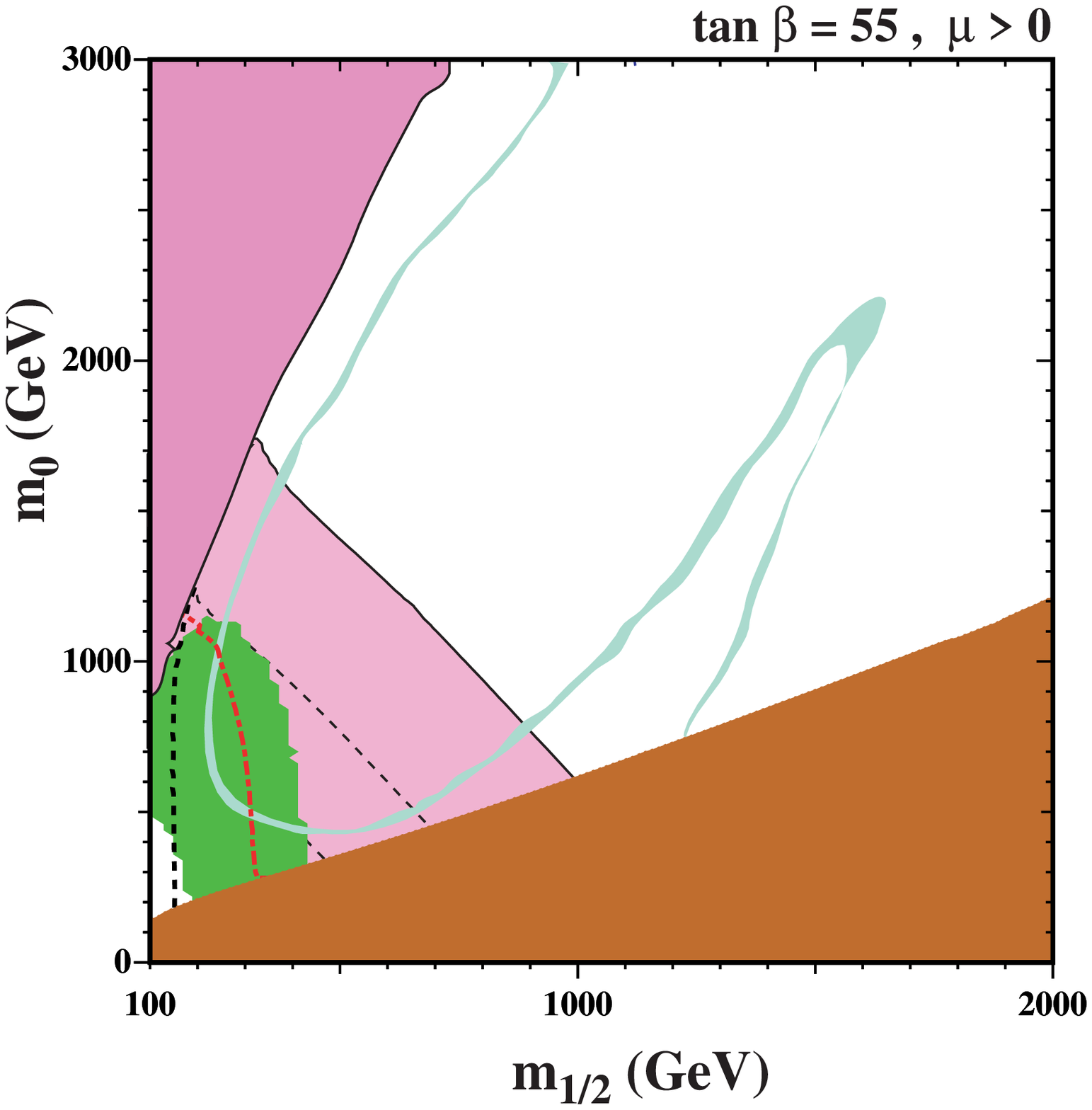}}
\caption{\label{fig:CMSSM1}
{The CMSSM $(m_{1/2}, m_0)$ planes for  (a) $\tan \beta = 10$ and (b) $\tan \beta = 55$,
assuming $\mu > 0$, $A_0 = 0$, $m_t = 173.1$~GeV and
$m_b(m_b)^{\overline {MS}}_{SM} = 4.25$~GeV. The near-vertical (red)
dot-dashed lines are the contours for $m_h = 114$~GeV, and the near-vertical (black) dashed
line is the contour $m_{\chi^\pm} = 104$~GeV. Also
shown by the dot-dashed curve in the lower left is the region
excluded by the LEP bound $m_{\tilde e} > 99$ GeV. The medium (dark
green) shaded region is excluded by $b \to s
\gamma$, and the light (turquoise) shaded area is the cosmologically
preferred region. In the dark
(brick red) shaded region, the LSP is the charged ${\tilde \tau}_1$. The
region allowed by the measurement of $g_\mu - 2$ at the 2-$\sigma$
level, assuming the $e^+ e^-$ calculation of the Standard Model contribution,
is shaded (pink) and bounded by solid black lines, with dashed
lines indicating the 1-$\sigma$ ranges (updated from~\cite{Ellis:2003cw}).}}
\end{figure}

Most of the current constraints on possible physics beyond the SM are negative and, specifically,
no sparticle has ever been detected. The concordance with the SM predictions means that,
in general, one can only set lower limits on the possible masses of supersymmetric particles.
However, there are two observational indications of physics beyond the SM that may, 
in the supersymmetric context, be used for setting {\it upper} limits of the masses of the 
supersymmetric particles. As discussed earlier,
these two hints for new physics  are the anomalous magnetic moment 
of the muon, $g_\mu - 2$, which seems to disagree with the prediction of the SM (at least if
this is calculated using low-energy $e^+e^-$ data as an input), and the density of cold dark matter 
$\Omega_{CDM}$. However, these discrepancies may be explained 
either with supersymmetry or with other 
possible extensions of the SM, so their interpretations require special care. Nevertheless,
these may be regarded as additional phenomenological motivations for supersymmetry, in addition 
to the more theoretical motivations described in the beginning of this section, such as
the naturalness of the hierarchy of mass scales in physics, grand unification, string theory, etc.
Therefore, in addition to considering the more direct searches for supersymmetry, 
it is also natural to ask what $g_\mu - 2$ and $\Omega_{CDM}$ may imply for the parameters of supersymmetric models.
Figure~\ref{fig:CMSSM1} compiles the impacts of various 
constraints on supersymmetry,
assuming that the soft supersymmetry-breaking contributions $m_{1/2}, m_0$ 
to the different scalars and gauginos are each universal at the GUT scale
(the scenario called the CMSSM), and that the
lightest sparticle is the lightest neutralino $\chi$. 

Experiments at LEP and the Tevatron collider, in particular, have made direct searches 
for supersymmetry using the missing-energy-momentum signature. LEP established
lower limits $\sim 100$~GeV on the masses of many charged sparticles without
strong interactions, such as sleptons and charginos. The Tevatron collider has
established the best lower limits on the masses of squarks and gluinos, $\sim 400$~GeV.
In view of the greater renormalization of the squark and gluino masses than for charginos
and sleptons, see (\ref{gauginomasses}) and (\ref{numm0}), these two sets of limits
are quite complementary.

Another important constraint is provided by the LEP lower limit on the
Higgs mass: $m_H > 114.4 \textrm{ GeV}$~\cite{LEPsearch}. This holds in the
Standard Model, for the lightest Higgs boson $h$ in the general MSSM for
$\tan\beta
\lappeq 8$, and almost always in the CMSSM for all $\tan\beta$, at least
as long as CP is conserved~\footnote{The lower bound on the lightest MSSM
Higgs boson may be relaxed significantly if CP violation feeds into the
MSSM Higgs sector~\cite{CEPW}.}. Since $m_h$ is sensitive to sparticle
masses, particularly $m_{\tilde t}$ {\it via} the loop corrections (\ref{deltamh}),
the Higgs limit also imposes important constraints on the soft 
supersymmetry-breaking CMSSM parameters,
principally $m_{1/2}$~\cite{Ellis:2003cw},
as seen in Fig.~\ref{fig:CMSSM1}.

Important constraints are imposed on the CMSSM parameter space
by flavour physics, specifically the agreement with data of the SM prediction
for the decay $b \to s \gamma$, as well as the upper limit on the decay
$B_s \to \mu^+ \mu^-$, which is important at large $\tan \beta$ in particular.

We see in Fig.~\ref{fig:CMSSM1} that narrow strips of
the $(m_{1/2}, m_0)$ planes are compatible~\cite{Ellis:2003cw} with the range of the 
astrophysical cold dark matter density favoured by WMAP and other
experiments. However, these strips vary with $\tan \beta$ and $A_0$. 
In fact, foliation by these WMAP strips covers large
fractions of the $(m_{1/2}, m_0)$ plane as $\tan \beta$ and $A_0$ are varied. Away from
these narrow strips, the relic neutralino density exceeds the WMAP range over most of 
the $(m_{1/2}, m_0)$ planes
shown in Fig.~\ref{fig:CMSSM1}. In its left panel, the relic density is reduced into the
WMAP range only in the shaded strip at $m_0 \sim 100$~GeV that extends to
$m_{1/2} \sim 900$~GeV. This reduction is brought about by co-annihilations between
the LSP $\chi$ (which is mainly a Bino)
and sleptons that are only slightly heavier, most notably the lighter stau
and the right selectron and smuon, which are significantly lighter than the left
sleptons, as discussed earlier.
In the right panel of Fig.~\ref{fig:CMSSM1} for $\tan \beta = 50$, 
this co-annihilation strip moves to larger $m_0$. Also, it is extended to larger $m_{1/2}$, as
a result of a reduction in the relic density due to
rapid $\chi - \chi$ annihilations though direct-channel heavy Higgs ($H, A$) states.
In addition to these visible WMAP regions, there is in principle another allowed strip
at very large values of $m_0$, called the focus-point region, where the LSP becomes
relatively light and acquires a substantial higgsino component, favouring annihilation {\it via}
$W^+ W^-$ final states.

Finally, also shown in the two panels of Fig.~\ref{fig:CMSSM1} are the regions
favoured by the supersymmetric interpretation of the discrepancy (\ref{g-2disc}) between the
experimental measurement of $g_\mu - 2$ and the value calculated in the SM
using low-energy $e^+ e^-$ data~\cite{Ellis:2003cw}. The favoured regions are displayed as bands
corresponding to $\pm 2 \sigma$. We see that they can be used to set {\it upper}
limits on the sparticle masses! In particular, $g_\mu - 2$ disfavours the focus-point
region, where $m_0$ is so large that the supersymmetric contribution to $g_\mu - 2$
is negligible, and also the region at large $\tan \beta$ and large $m_{1/2}$ where
the neutralinos may annihilate rapidly though direct-channel heavy-Higgs states.

\subsection{Frequentist analysis of the supersymmetric parameter space}

In a recent paper~\cite{Buchmueller:2009fn} the likely range of parameters of the CMSSM and NUHM1
has been estimated using a frequentist approach, by building a $\chi^2$ likelihood function
with contributions from the various relevant observables, including precision electroweak
physics, $g_\mu - 2$, the lower limit on the lightest Higgs boson mass (taking into
taking into account the theoretical uncertainty in the {\tt FeynHiggs}
calculation of $\Mh$~\cite{FeynHiggs}), the experimental measurement of \bsg (which agrees with the SM), 
the experimental upper limit on \bmm, and $\Omega_{CDM}$.
This frequentist analysis used a Markov chain Monte Carlo technique to sample
thoroughly the $(m_0, m_{1/2})$ plane up to masses of several TeV, including the
focus-point and rapid-annihilation regions, for a wide
range of values of $A_0$ and $\tan \beta$.

\begin{figure*}[htb!]
{\resizebox{8cm}{!}{\includegraphics{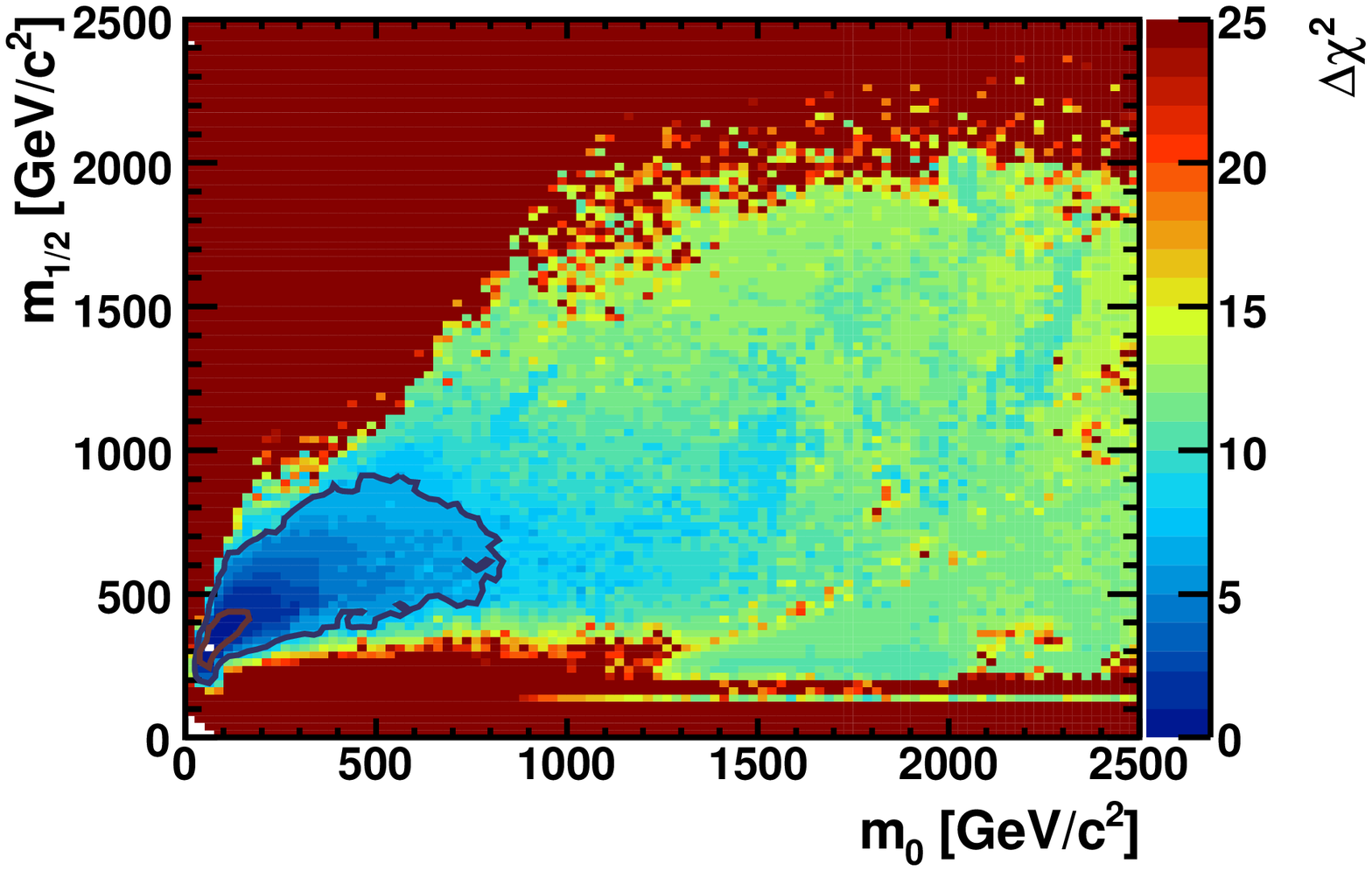}}}  
{\resizebox{8cm}{!}{\includegraphics{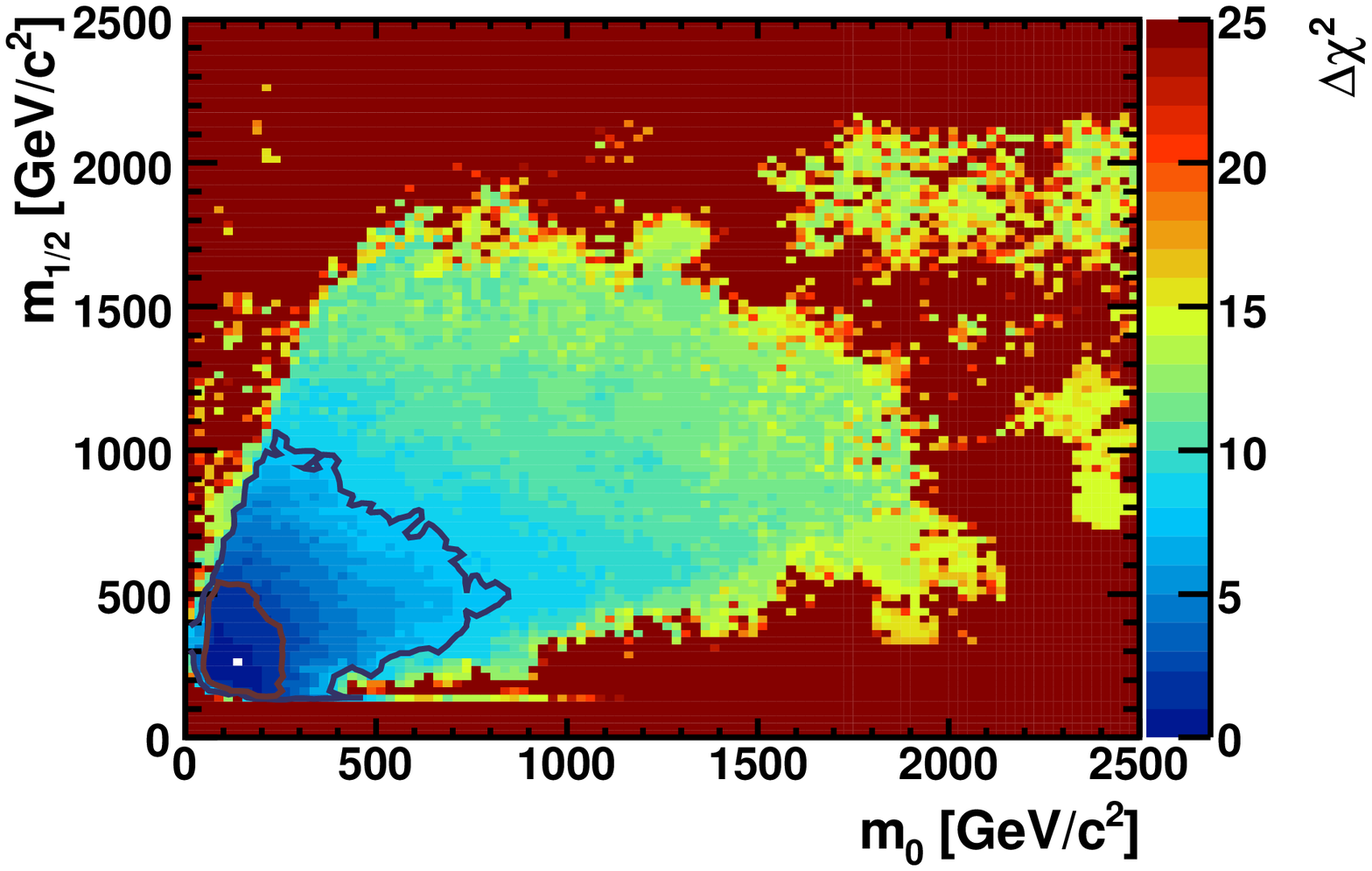}}}
\caption{The $\Delta\chi^2$ functions in the $(m_0, m_{1/2})$ planes for
  the CMSSM (left plot) and 
  for the NUHM1 (right plot), as found in frequentist analyses of the
  parameter spaces. We see that the co-annihilation regions at
  low $m_0$ and $m_{1/2}$ are favoured in both cases~\protect\cite{Buchmueller:2008qe}.
}
\label{fig:m0m12like}
\end{figure*}

We display in Fig.~\ref{fig:m0m12like} the $\Delta\chi^2$ functions in the 
$(m_0, m_{1/2})$ planes for the CMSSM (left plot) and for the NUHM1 (right plot).
The parameters of the best-fit CMSSM point are
$m_0 = 60 \gev$,  $m_{1/2} = 310 \gev$,  $A_0 = 130 \gev$, $\tb = 11$,
and $\mu = 400 \gev$ (corresponding nominally to $\Mh = 114.2$~GeV
and an overall $\chi^2 = 20.6$ for 19 d.o.f. with a probability of 36\%), 
which are very close to the ones previously reported in Ref.~\cite{Buchmueller:2008qe}.
The corresponding parameters of
the best-fit NUHM1 point are $m_0 = 150 \gev$, $m_{1/2} = 270 \gev$,
$A_0 = -1300 \gev$, $\tb = 11$, and
$m_{h_1}^2  = m_{h_2}^2 = - 1.2 \times 10^6 \gev^2$ or, equivalently,
$\mu = 1140 \gev$, yielding
$\chi^2 = 18.4$ (corresponding to a similar fit probability to the CMSSM)
and $\Mh = 120.7 \gev$. 
The similarities between the
best-fit values of $m_0$, $m_{1/2}$ and $\tb$ in the CMSSM and the NUHM1
suggest that the model frameworks used are reasonably stable: if they
had been very different, one might well have wondered what would be
the effect of introducing additional parameters, as in the
NUHM2 with two non-universality parameters in the
Higgs sector.

These best-fit points are both in the co-annihilation region of the
$(m_0, m_{1/2})$ plane, as can be seen in Fig.~\ref{fig:m0m12like}. 
The C.L.\ contours extend to slightly larger
values of $m_0$ in the CMSSM, while they extend to slightly larger
values of $m_{1/2}$ in the NUHM1, as was
already shown in Ref.~\cite{Buchmueller:2008qe} for
the 68\% and 95\% C.L.\ contours.
However, the qualitative features of the $\Delta\chi^2$ contours are quite
similar in the two models, indicating that the preference for small
$m_0$ and $m_{1/2}$ are quite stable and do not depend on details
of the Higgs sector. We recall that it was found in Ref.~\cite{Buchmueller:2008qe} that
the focus-point region was disfavoured at beyond the 95\% C.L. in both
the CMSSM  and the NUHM1. We see in Fig.~\ref{fig:m0m12like} that this
region is disfavoured at the level $\Delta\chi^2 \sim 8$ in the CMSSM
and $> 9$ in the NUHM1.

\begin{figure*}[htb!]
\begin{center}
\includegraphics[width=0.65\textwidth]{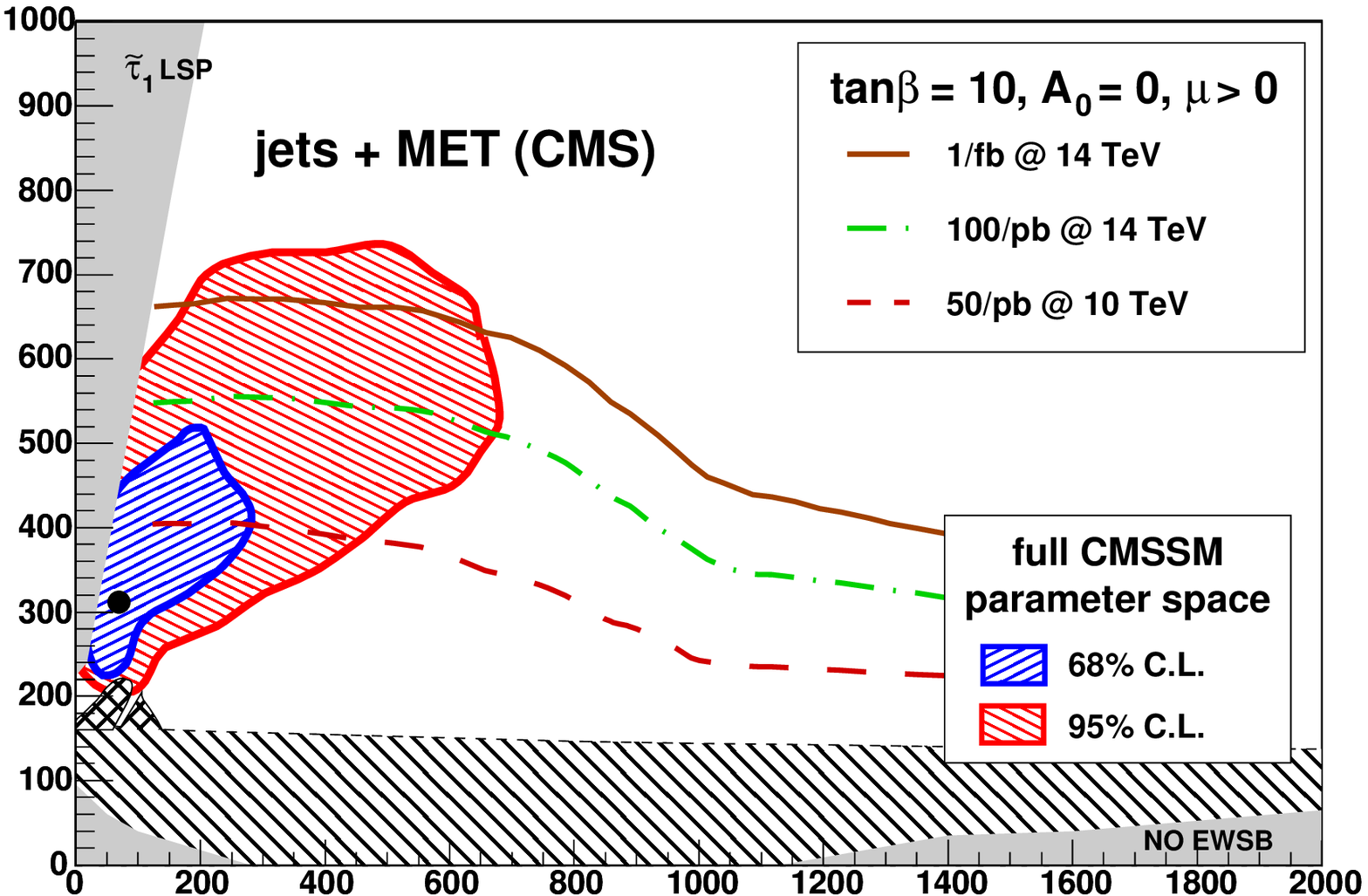}
\includegraphics[width=0.65\textwidth]{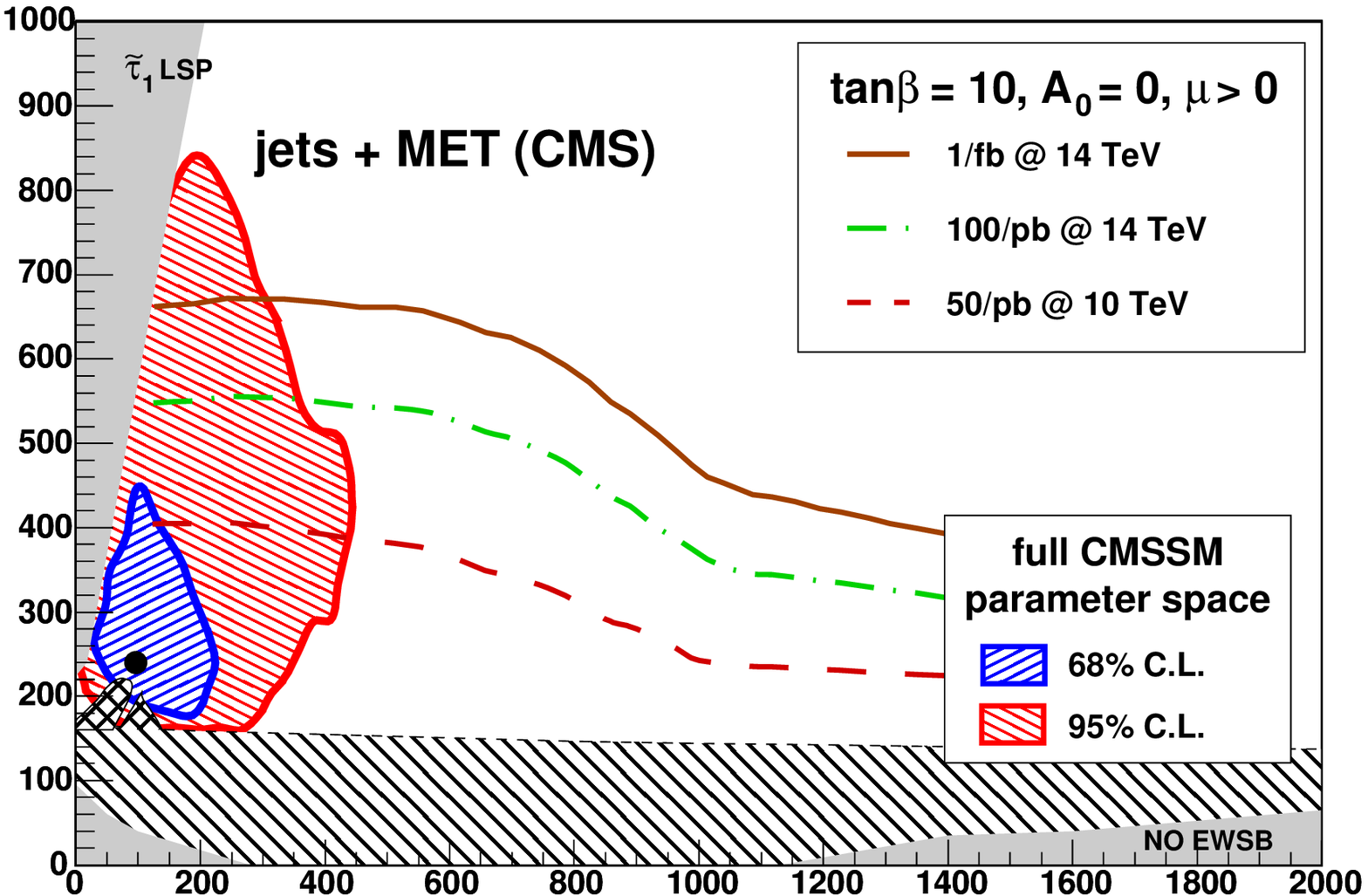}
\end{center}
\vspace{-0.5em}
\caption {The $(m_0, m_{1/2})$ planes in the CMSSM (upper) and
the NUHM1 (lower) for $\tan\beta = 10$
  and $A_0 = 0$. The dark shaded areas at low $m_0$ and high $m_{1/2}$ are
  excluded due 
  to a scalar tau LSP, the light shaded areas at low $m_{1/2}$ do not
  exhibit electroweak symmetry breaking. The nearly horizontal line at
  $m_{1/2} \approx 160$~GeV in the lower panel 
  has $m_{\tilde \chi_1^\pm} = 103$~GeV, and the area
  below is excluded by LEP searches. Just above this contour at low $m_0$
  in the lower panel is the region that is
  excluded by trilepton searches at the Tevatron.
  Shown in each plot is the best-fit point~\protect\cite{Buchmueller:2008qe}, 
  indicated by a star, and the
  68 (95)\%~C.L.\ contours from the fit as dark grey/blue (light
  grey/red) overlays, scanned over all $\tan\beta$ and $A_0$ values.
 The plots also show some $5\,\sigma$ discovery contours for 
 CMS~\cite{CMS} with
  1~fb$^{-1}$ at 14~TeV, 100~pb$^{-1}$ at 14~TeV and
  50~pb$^{-1}$ at 10~TeV centre-of-mass energy~\protect\cite{Buchmueller:2008qe}.
} 
\label{fig:MC2LHC2}

\end{figure*}

The favoured values of the particle masses in both models are such
that there are good prospects for detecting supersymmetric particles
in CMS~\cite{CMS} and ATLAS~\cite{ATLAS}
even in the early phase of the LHC running with reduced centre-of-mass
energy and limited luminosity, as seen in Fig.~\ref{fig:MC2LHC2}.
The best-fit points and most of the 68\% confidence level regions are
within the region of the $(m_0, m_{1/2})$ plane that could be explored
with 100/pb of data at 14~TeV in the centre of mass, and hence
perhaps with 200/fb of data at 10~TeV~\footnote{The comparisons are
made with experimental simulations for $\tan \beta = 10$ and $A_0 = 0$,
whereas the frequentist analysis sampled all values of $\tan \beta$ and
$A_0$. As it happens, the preferred values of $\tan \beta$ in both the
CMSSM and the NUHM1 are quite close to 10: the value of $A_0$ is
relatively unimportant for the experimental analysis.}. Almost all the
95\% confidence level regions would be accessible to the LHC with
1/fb of data at 14~TeV. As seen in Fig.~\ref{fig:MC2LHC2},
in substantial parts of these regions there are
good prospects for detecting ${\tilde q} \to q \ell^+ \ell^- \chi$ decays,
which are potentially useful for measuring sparticle mass parameters,
and the lightest supersymmetric Higgs boson may also be detectable
in ${\tilde q}$ decays.

\begin{figure*}[htb!]
\begin{center}
~~~CMSSM~~~~~~~~~~~~~~~~~~~~~~~~~~~~~~~~~~~~~~~~~~~~~~~~~~~~~~~~~~~~~~~~~~NUHM1
\includegraphics[width=0.48\textwidth]{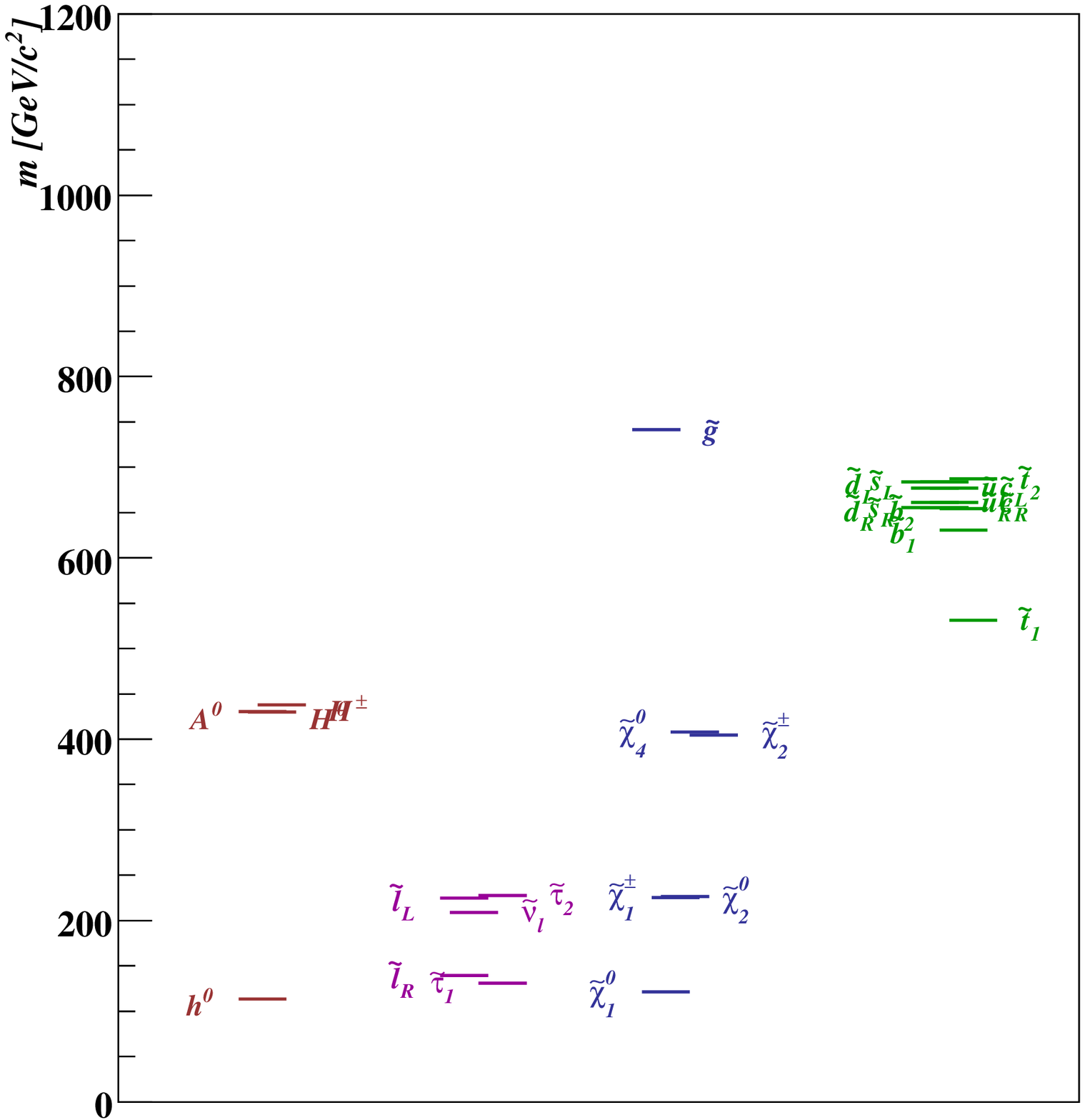}
\includegraphics[width=0.48\textwidth]{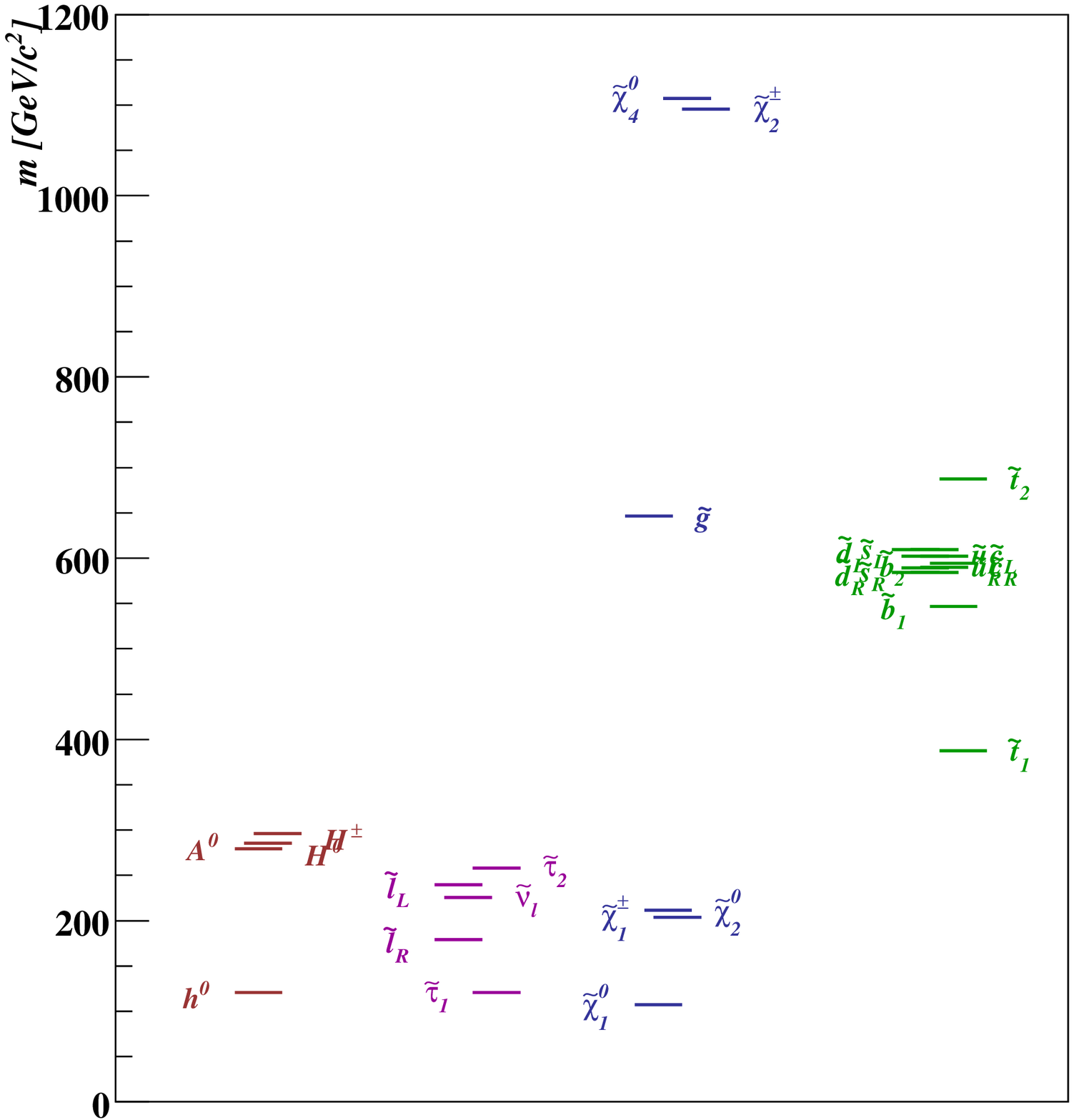}
\end{center}
\caption {The spectra at the best-fit points: left --- in the CMSSM with 
$m_{1/2} = 311$~GeV, $m_0 = 63$~GeV, $A_0 = 243$~GeV, $\tan\beta = 11.0$, and
right --- in the NUHM1 with $m_{1/2} = 265$~GeV, $m_0 = 143$~GeV, 
$A_0 = -1235$~GeV, $\tan \beta = 10.4$, and $\mu = 1110$~GeV~\protect\cite{Buchmueller:2008qe}.}
\label{fig:spectra}
\end{figure*}

\begin{figure*}[htb!]
\resizebox{8cm}{!}{\includegraphics{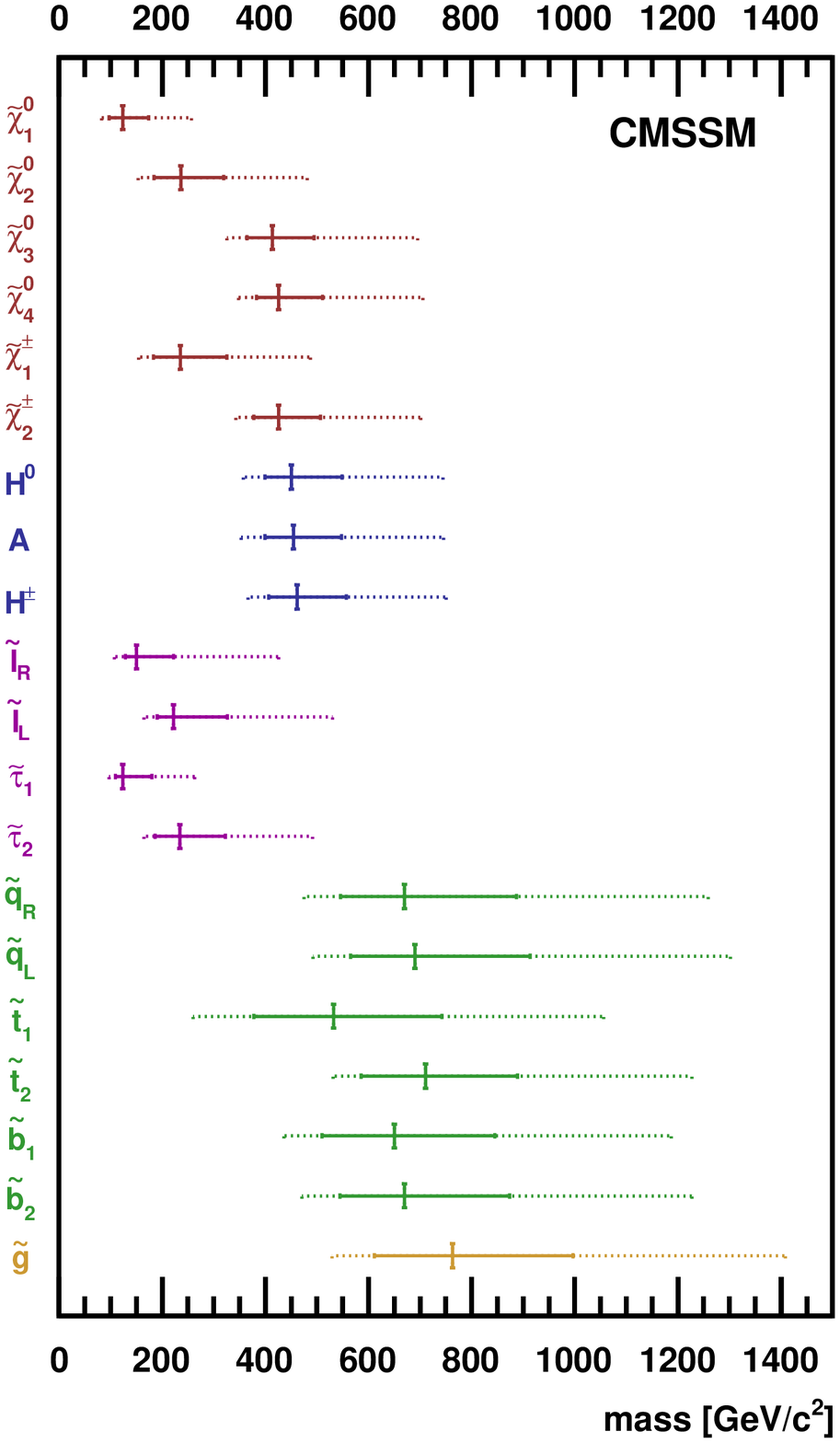}}
\resizebox{8cm}{!}{\includegraphics{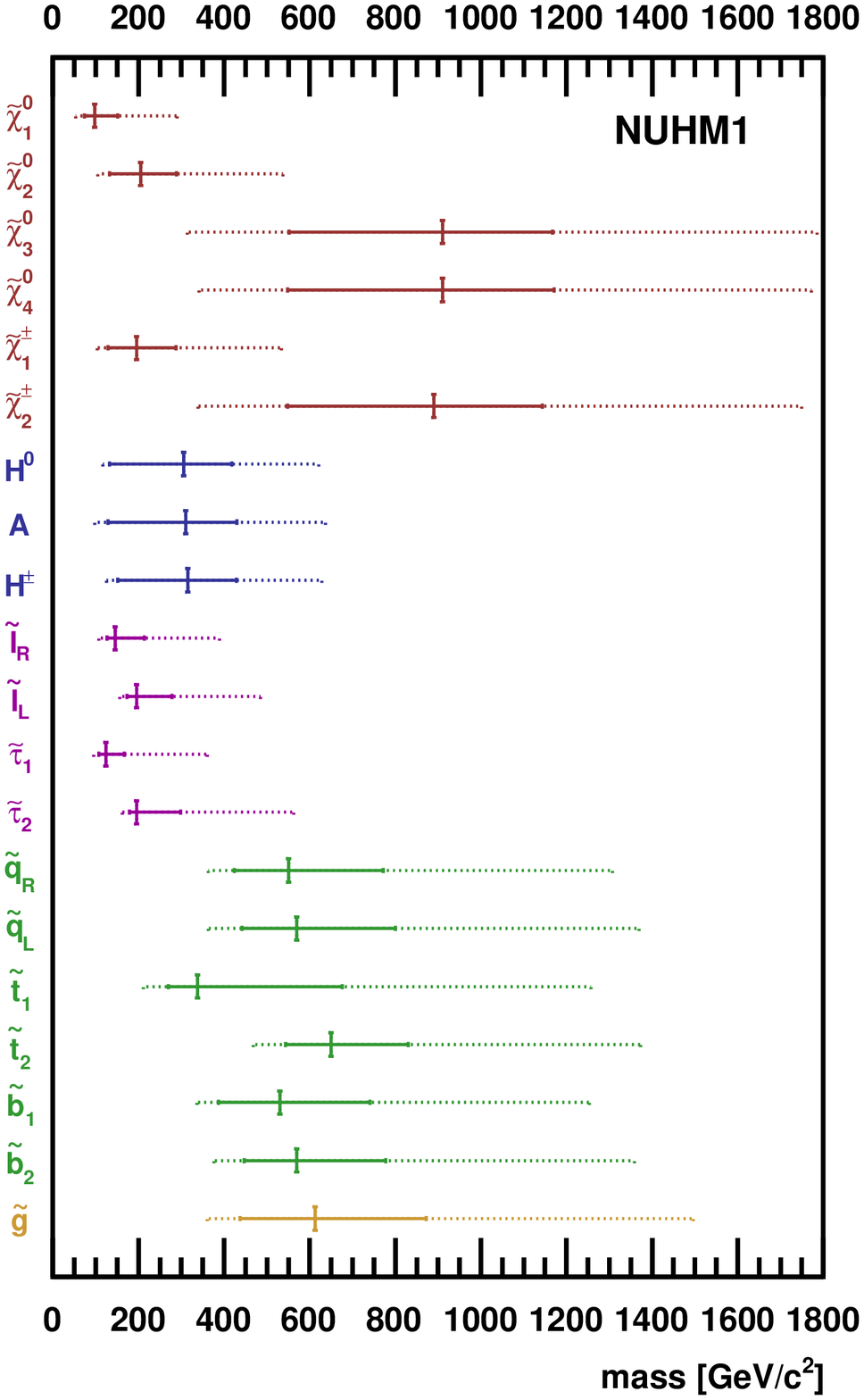}}
\caption{Spectra in the CMSSM (left) and the NUHM1 (right). The vertical
solid lines indicate the best-fit values, the horizontal solid lines
are the 68\% C.L.\
ranges, and the horizontal dashed lines are the 95\% C.L.\ ranges for the
indicated mass parameters~\protect\cite{Buchmueller:2009fn}. 
}
\label{fig:spectra2}
\end{figure*}

\begin{figure*}[htb!]
\resizebox{8cm}{!}{\includegraphics{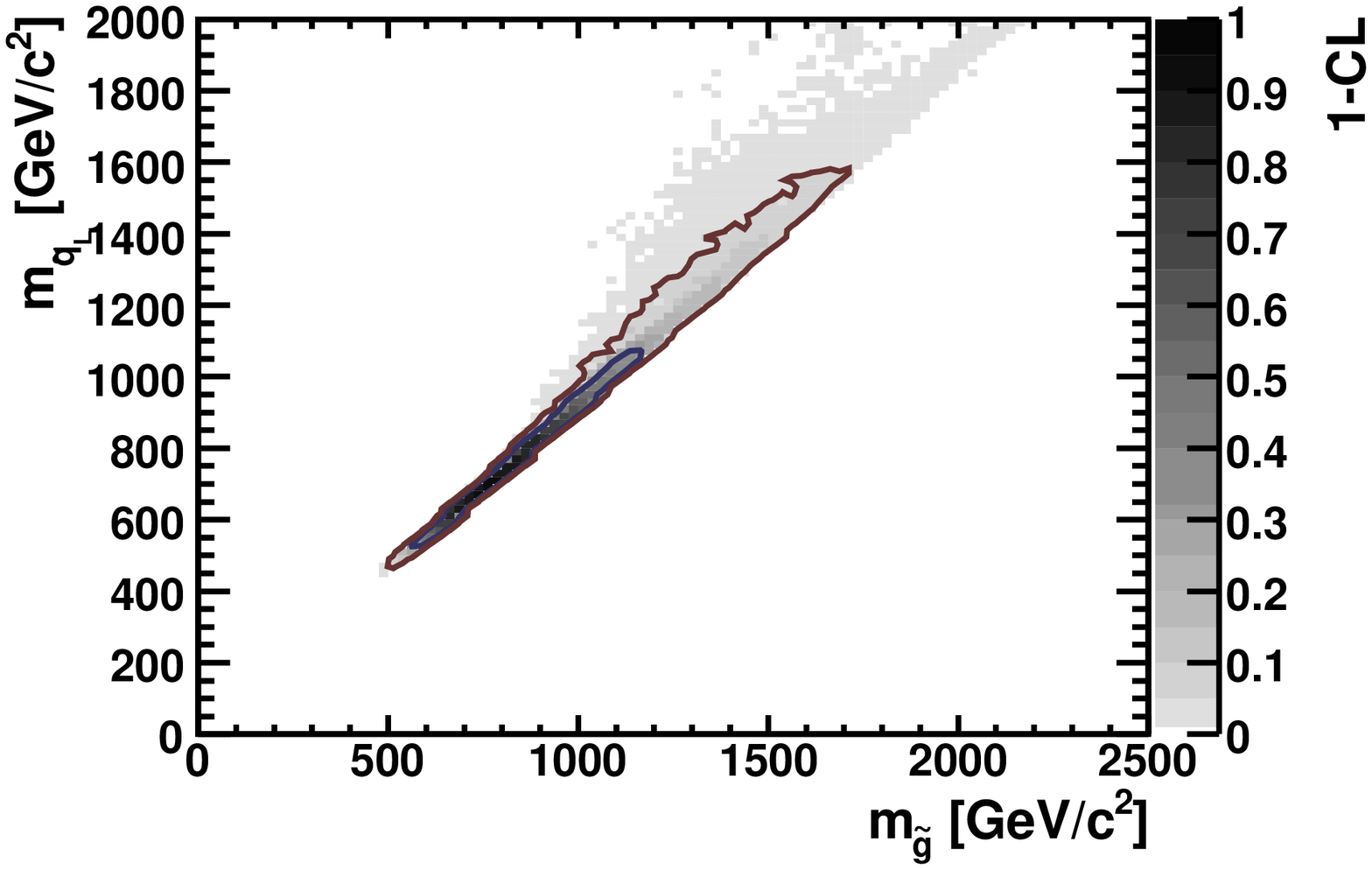}}
\resizebox{8cm}{!}{\includegraphics{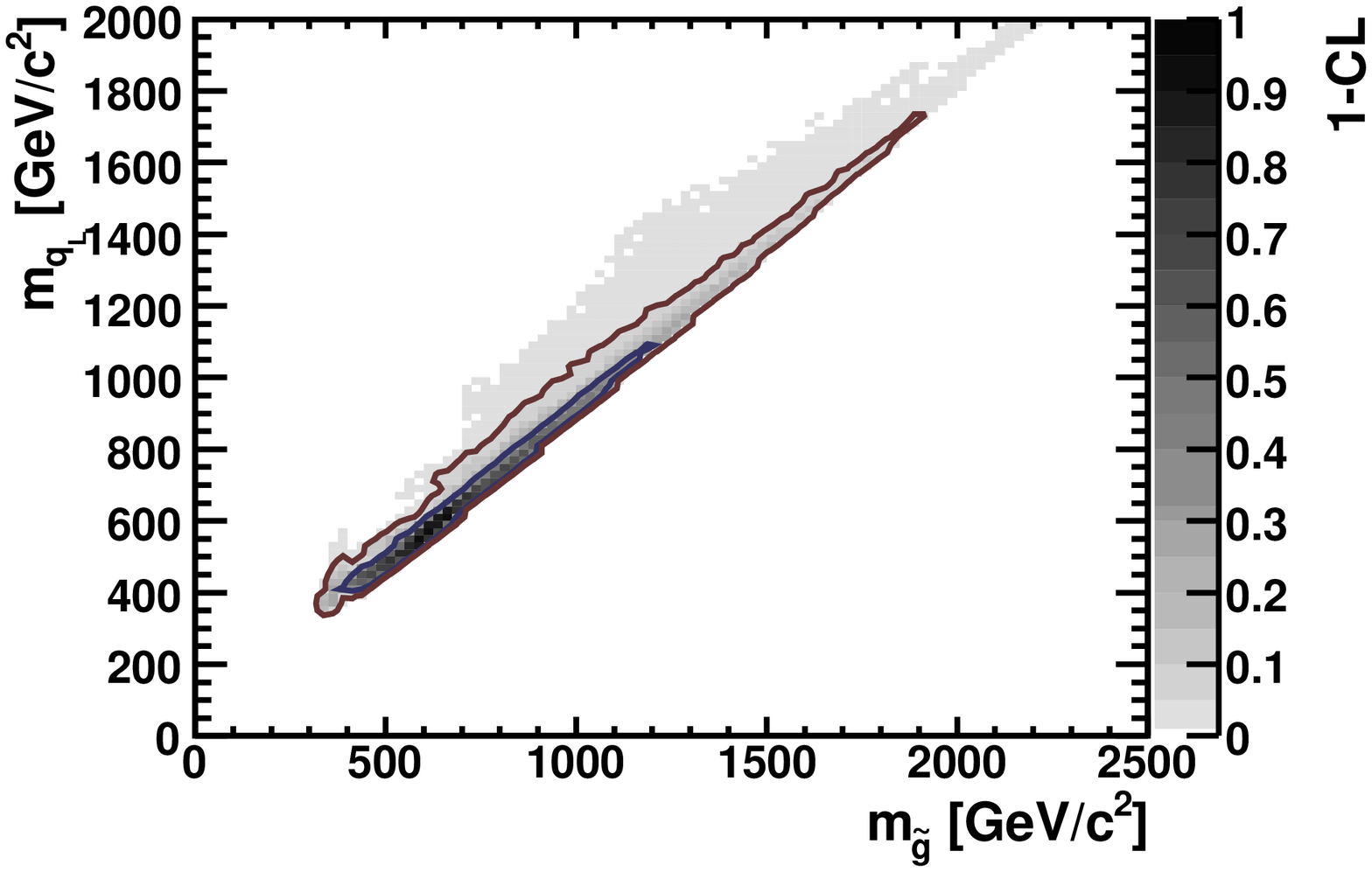}}
\caption{The correlations between the gluino mass, $m_{\tilde g}$,
and the masses of the the left-handed partners
of the five light squark flavours, $m_{\tilde q_{L}}$, are shown  
in the CMSSM (left panel) and in the NUHM1
(right panel)~\protect\cite{Buchmueller:2009fn}. 
}
\label{fig:correlation}

\end{figure*}

The best-fit spectra in the CMSSM and NUHM1 are shown in 
Fig.~\ref{fig:spectra}: they are relatively similar, though the heavier Higgs
bosons, the gluinos, and the squarks may be somewhat heavier in the CMSSM,
whereas the heavier charginos and neutralinos may be heavier in the 
NUHM1~\cite{Buchmueller:2008qe}.
There are considerable uncertainties in these spectra, as seen in 
Fig.~\ref{fig:spectra2}~\cite{Buchmueller:2009fn}.
However, in general there are strong correlations between the different sparticle
masses, as exemplified in Fig.~\ref{fig:correlation}, though the correlation is
weaker, e.g., for the lighter stau and the LSP in the NUHM1~\footnote{This
reflects the possible appearance of rapid direct-channel annihilations also at
low $m_{1/2}$ and low $\tan \beta$, allowing an escape from the co-annihilation
region where $m_\chi \sim m_{\tilde \tau_1}$.}.

\begin{figure*}[htb!]
\resizebox{8cm}{!}{\includegraphics{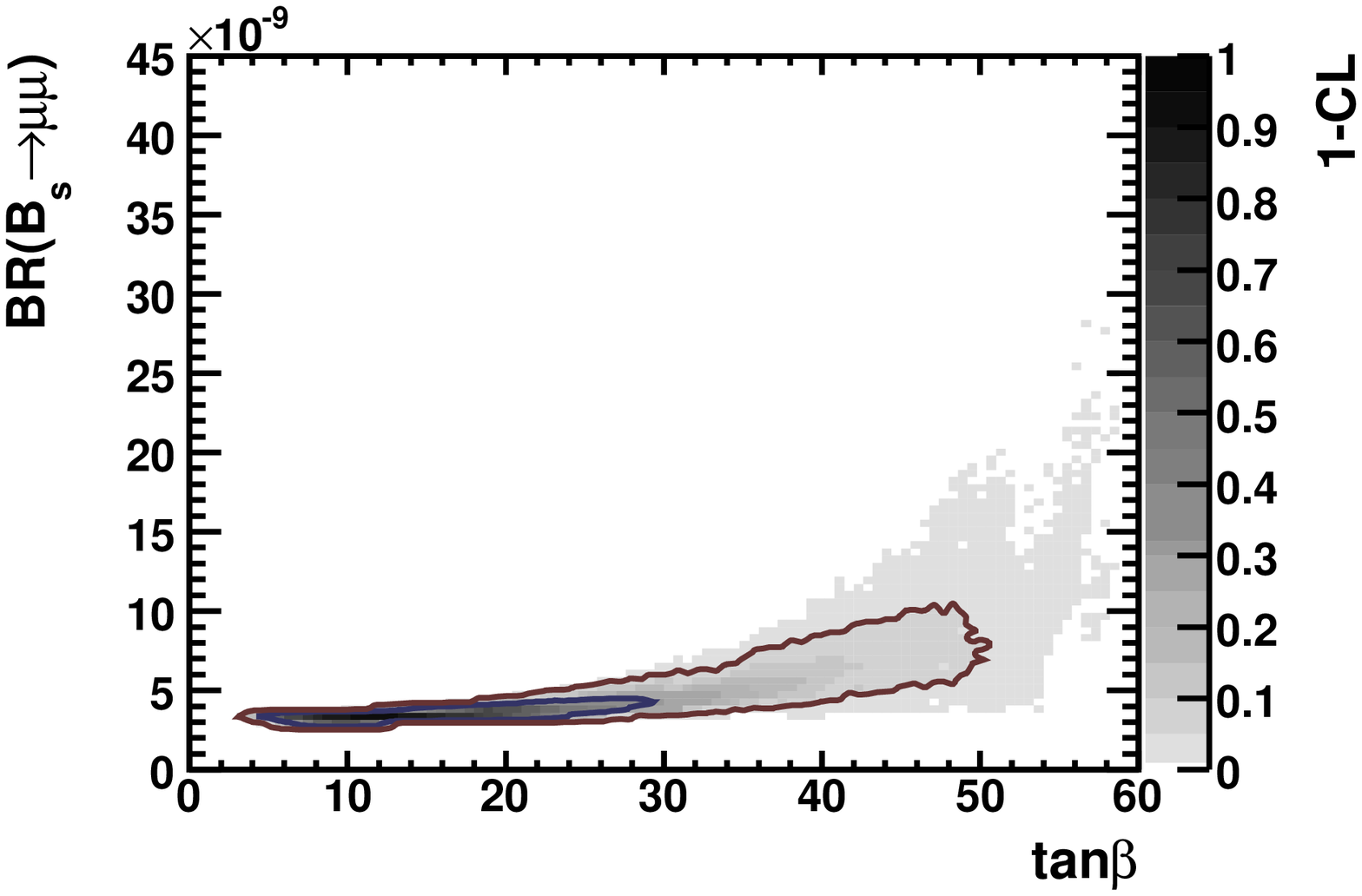}}
\resizebox{8cm}{!}{\includegraphics{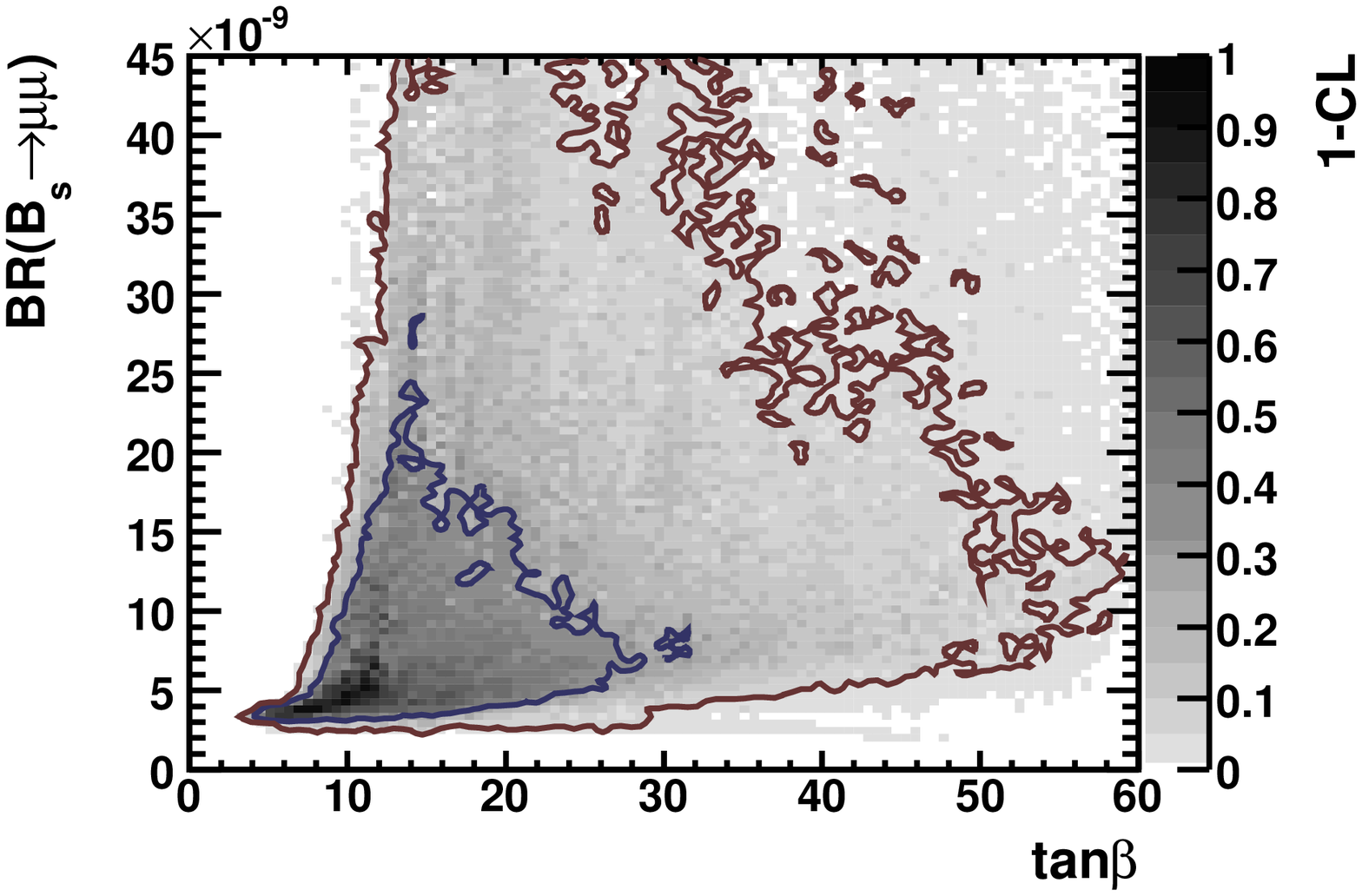}}
\caption{The correlation between the branching ratio for $B_s \to \mu^+ \mu^-$ and $\tan \beta$
in the CMSSM (left panel) and in the NUHM1 (right panel)~\protect\cite{Buchmueller:2009fn}.
}
\label{fig:bsmumu}
\end{figure*}

Finally, a result from this frequentist analysis that also concerns LHC physics,
but away from the high-energy frontier. We see in Fig.~\ref{fig:bsmumu} that the
branching ratio for $B_s \to \mu^+ \mu^-$ may well exceed
considerably its value in the SM, particularly at large $\tan \beta$. This is true to some extent
in the CMSSM, and even more so in the NUHM1. Particularly in the latter case,
this decay might perhaps be accessible to the LHCb experiment during
initial LHC running. Therefore, there may be
important competition for ATLAS and CMS in their quest to discover supersymmetry!

\section{Further beyond: GUTs, string theory and extra dimensions}

\subsection{Grand unification}

Gauge theories, particularly non-Abelian Yang--Mills theories, are the only 
suitable framework for describing interactions in particle physics. In the SM, there are three different 
gauge groups  $SU(3)_C$, $SU(2)_L$, and $U(1)_Y$, and correspondingly there are
three different couplings. It is logical to look for a single, more powerful non-Abelian
grand unified gauge group
with a single coupling $g_{GUT}$ that would enable us to unify the three couplings, and might
provide interesting relations between the other different SM parameters such as Yukawa
couplings and hence fermion masses~\footnote{In this section, we denote the 
couplings by $g_1$ for the $U(1)$ subgroup, $g_2$ for $SU(2)$, and $g_3$ for $SU(3)$,
which have the appropriate normalizations for grand unification [see later].}.
As a first approximation, we assume that the effects of the gravitational interaction are negligible, 
which is generally true if the grand unification scale $M_{GUT}$ is significantly smaller that the 
Planck mass.
As we see later, it turns out that typical estimations, based on extrapolation to very high energies
of the known physics of the SM~\cite{GQW}, give a grand unification scale of the order of $10^{16}$~GeV, 
which is about a thousand times smaller than the Planck scale $M_{Pl}=\mathcal{O}(10^{19})$ GeV. 

Postulating a single group to describe all the interactions of particle physics also implies new 
relations between the matter particles themselves, as well as new gauge bosons. Specifically, 
if the symmetry changes then the representations, and hence the organization of the particles 
into multiplets, also change. There are some hints for this in low-energy physics, such as
charge quantization and the correlation of fractional electrical charges with colour charges, 
and the cancellation of anomalies between the leptons and the quarks that also lead us to
anticipate an organization simpler than the SM.

Clearly, one must recover the Standard Model at low energy, implying that in these
Grand Unified Theories (GUTs) one must also study the breaking of the GUT group $G
\to SU(3)_C \otimes SU(2)_L\otimes U(1)_Y$. 

This section begins  with a presentation of the renormalization-group evolution
equations of the three SM gauge couplings and studies their possible unification
at some GUT scale.
Subsequently, some specific examples of GUTs are discussed, notably the prototype
based on the group $SU(5)$, which makes possible a simple discussion of many properties 
of GUTs. This is followed by a short discussion of typical predictions of these models,
such as the decay of the proton and the relations between the masses of the quarks and leptons. 
We finish by discussing some of the advantages, problems,  and perspectives of GUT models.

\subsubsection{The evolution equations for gauge couplings}

The first apparent obstacle to the philosophy of grand unification is the
fact that the strong coupling strength $\alpha_3 = g^2_3/ 4\pi$ is
much stronger than the electroweak couplings at present-day energies:
$\alpha_3 \gg \alpha_2, \alpha_1$. However, the strong coupling
is asymptotically free~\cite{Ellis:1998eh}:
\begin{equation}
\alpha_3(Q) \simeq {12\pi\over (33-2N_q) \ln (Q^2/\Lambda^2_3)} + \ldots ,
\label{fourthree}
\end{equation}
where $N_q$ is the number of quarks, $\Lambda_3 \simeq$ few hundred MeV
is an intrinsic scale of the strong interactions, and the dots in
(\ref{fourthree}) represent higher-loop corrections to the leading
one-loop behaviour shown. The other SM gauge couplings also exhibit
logarithmic violations analogous to (\ref{fourthree}). For example, the
fine-structure constant $\alpha_{em} = 1/137.035999084(51)$ is renormalized to
effective value of $\alpha_{em}(m_Z) \sim 1/ 128$ at the $Z$ mass scale. 
The renormalization-group
evolution for the $SU(2)$ gauge coupling corresponding to (\ref{fourthree}) is
\begin{equation}
\alpha_2(Q) \simeq {12\pi\over (22-2N_q - N_{H/2}) \ln
(Q^2/\Lambda^2_2)} + \ldots ,
\label{fourfour}
\end{equation}
where we have assumed equal numbers of quarks and leptons, and $N_H$ is
the number of Higgs doublets. Taking the inverses of (\ref{fourthree})
and (\ref{fourfour}), and then taking their difference, we find
\begin{equation}
{1\over\alpha_3(Q)} - {1\over \alpha_2(Q)} = \left({11+N_{H/2}\over
12\pi}\right) \ln \left({Q^2\over m^2_X}\right) + \ldots .
\label{fourfive}
\end{equation}
Note that we have absorbed the scales $\Lambda_3$ and $\Lambda_2$ into a single
grand unification scale $M_X$ where $\alpha_3 = \alpha_2$.

Evaluating (\ref{fourfive}) when $Q = {\cal O}(M_W)$, where $\alpha_3 \gg
\alpha_2 = 0(\alpha_{em})$, we derive the characteristic feature~\cite{GQW}
\beq
{m_{GUT}\over m_W} = \exp \left( {\cal
O}\left({1\over\alpha_{em}}\right)\right) ,
\label{fourone}
\eeq
i.e., the grand unification scale is exponentially large.
As we see in more detail later, in most GUTs there are new interactions 
mediated by bosons weighing ${\cal O}(m_X)$ that cause protons to decay with
a lifetime $\alpha m^4_X$. In order for the proton lifetime to exceed the
experimental limit, we need $m_X \gappeq 10^{14}$ GeV and hence
$\alpha_{em} \lappeq 1/120$ in (\ref{fourone})~\cite{ENnature}. On the
other hand,
if the neglect of gravity is to be consistent, we need $m_X \lappeq
10^{19}$ GeV and hence $\alpha_{em} \gappeq 1/ 170$ in
(\ref{fourone})~\cite{ENnature}. The fact that the measured value of the
fine-structure constant $\alpha_{em}$ lies in this allowed range
may be another hint favouring the GUT philosophy.

Further empirical evidence for grand unification is provided by the
prediction it makes for the neutral electroweak mixing angle~\cite{GQW}.
Calculating
the renormalization of the electroweak couplings, one finds
\begin{equation}
\sin^2\theta_W = {\alpha_{em}(m_W)\over\alpha_2(m_W)} \simeq {3\over
8}~~\left[ 1 - {\alpha_{em}\over 4\pi}~~{110\over 9} \ln {m^2_X\over
m^2_W}\right] ,
\label{foursix}
\end{equation}
which can be evaluated to yield $\sin^2\theta_W \sim$ 0.210 to 0.220,
if there are only SM particles with masses $\lappeq m_X$~\cite{GQW}.
This is to be
compared with the experimental value $\sin^2\theta_W = 0.23120 \pm 0.00015$
in the $\overline{\rm MS}$ renormalization scheme. Considering that $\sin^2\theta_W$ could
{\it a~priori}  have had any value between 0 and 1, this is an
impressive qualitative success. The small discrepancy can be removed by
adding some extra particles, such as the supersymmetric particles in the
MSSM. 

To see this explicitly, we may write
\begin{equation}
\sin^2 \theta (m_Z) \; = \; \frac{{g'}^2}{g_2^2 + {g'}^2} \; = \; \frac{3}{5}
\frac{g_1^2(m_Z)}{g_2^2(m_Z) + \frac{3}{5} g_1^2(m_Z)} ,
\label{b1}
\end{equation}
where $g_1$ is defined in such a way that its quadratic Casimir coefficient,
summed over all the particles in a single generation, is the same as for $g_2$
and $g_3$, which is the appropriate normalization within a GUT. Using the
one-loop RGEs, we can then write
\begin{equation}
\sin^2 \theta (m_Z) \; = \; \frac{1}{1 + 8 x} \left[ 3 x + \frac{\alpha_{em}(m_Z)}{\alpha_3(m_Z)} \right]
\; = \; \frac{1}{5} \left(\frac{b_2 - b_3}{b_1 - b_2} \right)\ ,
\label{b2}
\end{equation}
where the $b_i$ are the one-loop coefficients in the RGEs for the different SM couplings.
Their values in the SM (on the left) and the MSSM (on the right) are:
\begin{eqnarray}
\frac{4}{3} N_G - 11 \; \leftarrow & b_3 & \rightarrow \; 2 N_G - 9 \; = \; -3 \\
\frac{1}{6} N_H + \frac{4}{3} N_G - \frac{22}{3} \; \leftarrow & b_2 & \rightarrow \; \frac{1}{2} N_H + 2 N_G - 6 \; = \; +1 \\
\frac{1}{10} N_H + \frac{4}{3} N_G \; \leftarrow & b_1 & \rightarrow \; \frac{3}{10} N_H + 2 N_G \; = \; \frac{33}{5} \\
\frac{23}{218} \; = \; 0.1055 \; \leftarrow & x & \rightarrow \; \frac{1}{7}.
\label{b3}
\end{eqnarray}
Experimentally, using $\alpha_{em}(m_Z) = 1/128, \alpha_3 = 0.119 \pm 0.003, \sin^2 \theta_W (m_Z) = 0.2315$, we find
\begin{equation}
x \; = \; \frac{1}{6.92 \pm 0.07} ,
\label{b4}
\end{equation}
in striking agreement with the MSSM prediction in (\ref{b3})!

Another qualitative success is the prediction of the $b$ quark
mass~\cite{CEG,BEGN}. In many
GUTs, such as the minimal $SU(5)$ model, discussed shortly, the $b$ quark
and the $\tau$ lepton have equal Yukawa couplings when renormalized at
the GUT sale. The renormalization group then tells us that
\begin{equation}
{m_b\over m_\tau} \simeq \left[\ln \left({m^2_b\over m^2_X}
\right)\right]^{12\over 33-2N_q} .
\label{fourseven}
\end{equation}
Using $m_\tau =$ 1.78 GeV, we predict that $m_b\simeq$ 5 GeV, in
agreement with experiment.
Happily, this prediction remains
successful if the effects of supersymmetric particles are included in the
renormalization-group calculations~\cite{susymb}.


To examine the GUT predictions for $\sin^2\theta_W$ etc.  in more detail,
one needs to study the renor\-ma\-li\-za\-tion-group equations beyond the leading
one-loop order. Through two loops, one finds that
\begin{equation}
Q~~{\partial\alpha_i(Q)\over\partial Q} = -{1\over 2\pi}~~\left( b_i +
{b_{ij}\over 4\pi}~~\alpha_j(Q)\right)~~\left[\alpha_i(Q)\right]^2 ,
\label{fourseven1}
\end{equation}
where the $b_i$ receive the one-loop contributions
\begin{eqnarray}
b_i = \left(
\begin{matrix} 
0 \cr -\frac{22} {3} \cr -11
\end{matrix}
\right) 
+ N_g
\left(
\begin{matrix}
\frac{4}{ 3} \cr\cr \frac{4}{ 3} \cr\cr \frac{4}{ 3}
\end{matrix}
\right) 
+ N_H
\left(
\begin{matrix}
\frac{1} {10} \cr\cr \frac{1}{6} \cr \cr 0
\end{matrix}
\right)
\label{foureight}
\end{eqnarray}
from gauge bosons, $N_g$ matter generations and $N_H$ Higgs doublets,
respectively, and at two loops
\begin{eqnarray}
b_{ij} = \left(
\begin{matrix}
0&0&0\cr\cr 0&-\frac{136}{ 3} & 0 \cr\cr
0&0&-102
\end{matrix}
\right) 
+ N_g
\left(
\begin{matrix}
\frac{19}{15} & \frac{3}{ 5} &\frac {44}{15} \cr\cr \frac{1}{5} & \frac{49}{3} & 4 \cr\cr 
\frac{4}{30} & \frac{3}{ 2} & \frac{76}{ 3}
\end{matrix}
\right) +  N_H \left(
\begin{matrix}
\frac{9}{50} &\frac{9}{10} & 0 \cr\cr \frac{3}{10} & \frac{13}{ 6} & 0 \cr\cr 0 & 0 & 0
\end{matrix}
\right) .
\label{fournine}
\end{eqnarray}
It is important to note that these coefficients are all independent of any specific GUT model,
depending only on the light particles contributing to the
renormalization. 

Including supersymmetric particles as in the MSSM, one
finds~\cite{DRW}
\begin{eqnarray}
b_i = \left(
\begin{matrix}0 \cr\cr -6 \cr\cr -9
\end{matrix}
\right) + N_g
\left(
\begin{matrix}
2\cr\cr 2 \cr\cr 2
\end{matrix}\right) + N_H
 \left(
 \begin{matrix}
 \frac{3}{10}
\cr\cr frac{1}{2}\cr\cr 0
\end{matrix}\right) ,
\label{fourten}
\end{eqnarray}
and
\begin{eqnarray}
b_{ij} = \left(
\begin{matrix}
0&0&0\cr\cr 0&-24 & 0 \cr\cr 0&0&-54
\end{matrix}\right) + N_g
\left(
\begin{matrix}
\frac{38}{15} & \frac{6}{5} & \frac{88}{15} \cr\cr \frac{2}{5} & 14 & 8
\cr\cr \frac{11}{5} & 3 & \frac{68}{3}
\end{matrix}
\right) +  N_H \left( 
\begin{matrix}
\frac{9}{50} &
\frac{9}{10} & 0 \cr\cr \frac{3}{10} & \frac{7}{2} & 0 \cr\cr 0 & 0 & 0
\end{matrix} \right) ,
\label{foureleven}
\end{eqnarray}
again independent of any specific supersymmetric GUT.

One can use these two-loop equations to make detailed calculations of 
$\sin^2\theta_W$ in different GUTs. These
confirm that non-supersymmetric models are not consistent with the 
determinations of the
gauge couplings from LEP and elsewhere~\cite{EKN}. 
Previously, we argued that these models predicted a
wrong value for $\sin^2\theta_W$, given the experimental 
value of $\alpha_3$. In Fig.~\ref{28}(a) we
see the converse, namely that extrapolating the experimental 
determinations of the
$\alpha_i$ using the non-supersymmetric 
renormalization-group equations (\ref{foureight}),
(\ref{fournine}) does not lead to a common value of the gauge couplings at 
any renormalization scale. In contrast,
we see in Fig.~\ref{28}(b) that extrapolation using the 
supersymmetric renormalization-group
equations (\ref{fourten}), (\ref{foureleven}) {\bf does} lead to possible
unification at $M_{GUT} \sim 10^{16}$ GeV~\cite{ADF},
if the spartners of the SM particles weigh $\sim 1$~TeV.

Turning this success around, and assuming 
$\alpha_3 = \alpha_2 = \alpha_1$ at $M_{GUT}$ with
no threshold corrections at this scale, one may estimate that~\cite{EKN2}
\bea
\sin^2\theta_W(M_Z)\bigg\vert_{\overline{\rm MS}} &=& 
0.2029 + {7 \alpha_{em}\over 15
\alpha_3}+ {\alpha_{em}\over 20\pi}
\left[ -3 \ln \left({m_t\over m_Z}\right) + {28\over 3}
\ln \left({m_{\tilde g}\over m_Z}\right) \right.\nonumber \\ \nonumber \\
&&\left. - {32\over 3} \ln \left({m_{\tilde
W}\over m_Z}\right) - \ln \left({m_A\over m_Z}\right) - 
4\ln \left({\mu\over m_Z}\right) +
\ldots
\right] .
\label{fourtwelve}
\eea
Setting all the sparticle masses to 1 TeV reproduces approximately the value of
$\sin^2\theta_W$ observed experimentally. 
Can one invert this successful argument to
estimate the supersymmetric particle mass scale? 
One can show~\cite{lump} that the sparticle mass
thresholds in (\ref{fourtwelve}) can be lumped into the parameter
\beq
T_{susy} \equiv \vert\mu\vert \left({m^2_W\over m_{\tilde
g}}\right)^{14/19}~~\left({m^2_A\over \mu^2}\right)^{3/38} ~~
\left({m^2_{\tilde W}\over \mu^2}\right)^{2/19}~~\prod^3_{i=1}~~
\left( {m^3_{\tilde \ell_{Li}} m^7_{\tilde q_i}\over m^2_{\tilde\ell_{R_i}} m^5_{\tilde u_i}
m^3_{\tilde d_i}}\right)^{1/19} .
\label{fourthirteen}
\eeq
If one assumes sparticle mass universality at the GUT scale,
then~\cite{lump}
\beq
T_{susy} \simeq \vert\mu\vert \left({\alpha_2\over 
\alpha_3}\right)^{3/2} \simeq {\mu\over 7} ,
\label{fourforteen}
\eeq
approximately. The measured value of 
$\sin^2\theta_W$ is consistent with $T_{susy} \sim$ 100
GeV to 1 TeV, roughly as expected from the 
hierarchy argument. However, the uncertainties are
such that one cannot use this consistency to 
constrain $T_{susy}$ very tightly~\cite{Zich}. In
particular, even if one accepts the universality hypothesis, 
there could be important
model-dependent threshold corrections around the GUT
scale~\cite{EKN2,GUTthresh}.

\subsubsection{Specific GUTs}

What groups may be used to construct a GUT~\cite{GG}?

First, suitable groups must be sufficiently large to include the SM. The latter is of rank four, 
i.e., there are four simultaneously-diagonalizable symmetry generators~\footnote{Each 
one is associated with a quantum number, a `charge', that may be used to label particle states.}: 
 $SU(3)_C$ have two,  $SU(2)_L$ one, and $U(1)_Y$ one also. It is striking that
all of the diagonal generators are traceless: this is trivial for the non-Abelian groups
$SU(3)_C$ and $SU(2)_L$, but non-trival for $U(1)_Y$, and a possible hint that it
should be embedded in a non-Abelian GUT group. Therefore, we must first find in the 
Cartan classification of Lie groups a group of rank higher than or equal to four. Secondly, a GUT group 
must possess complex representations, in order that the matter particles and their
antiparticles (described by complex conjugate spinors) could be in inequivalent 
representations. Thirdly, we should also keep track of the hypercharges $Y = Q - T_3$.
One of the major puzzles of the SM is why
\beq
\sum_{q,\ell} Q_i = 3Q_u + 3Q_d + Q_e = 0 .
\label{fourtwo}
\eeq
In the SM, the hypercharge assignments are {\it a priori}
independent of the $SU(3)\times SU(2)_L$ assignments, although 
constrained by the fact that quantum
consistency requires the resulting triangle anomalies to cancel. In a
simple GUT group, the relation (\ref{fourtwo}) is automatic: whenever $Q$ is a
generator of a simple gauge group, $\sum_RQ = 0$ for particles in any
representation $R$, cf., the values of $I_3$ in any
representation of $SU(2)$.

There are only two groups of rank 4 that have complex representations
and hence are suitable {\it a priori} for GUTs, namely $SU(5)$ and 
$SU(3)\otimes SU(3)$. However, $SU(3)\otimes SU(3)$  does not allow 
simultaneously the leptons to have an integer electric charge and the quarks 
to have a fractional electric charge.  Moreover, if
one tried to use $SU(3)\times SU(3)$, one would need 
to embed the electroweak gauge group in
the second $SU(3)$ factor. This would be possible 
only if $\sum_q Q_q = 0 = \sum_\ell
Q_\ell$, which is not the case for the known 
quarks and leptons. Therefore, attention has
focused on
$SU(5)$~\cite{GG} as the only possible rank-4 GUT group.

The group $SU(5)$ is the simplest  GUT group capable of including the SM. 
Other possible GUT groups have higher rank, and groups that are commonly used
are $SO(10)$, the only suitable simple group of rank 5 with complex representations, 
and the exceptional group $E_6$ of rank 6. As examples that  may help understand 
the new physics that appears when the symmetry of the SM is enhanced, we are first going 
to study key aspects of the group $SU(5)$ and then, more briefly, some aspects of the 
group $SO(10)$.

\textbf{The $SU(5)$ group}

As in the SM, particles must be arranged in suitable representations of $SU(5)$.
This group has a fundamental spinorial representation of dimension 5 and a
2-index antisymmetric spinorial representation of dimension 10. Together they
are suitable for accommodating the fermions of a given generation, which
consist of $3\times2\times2=12$ quarks + 2 charged leptons + 1 neutrino. To
see how this may be done, we first decompose the smallest
representations of $SU(5)$ in terms of representations of $SU(3)\otimes SU(2)$:
\beqn
\bf{\bar{5}} &=& (\bf{\bar{3}},\bf{1}) + (\bf{1},\bf{2}) , \label{rep5barofSU(5)}\\
\bf{10} &=& (\bf{\bar{3}},\bf{1}) +(\bf{3},\bf{2})+ (\bf{1},\bf{1}) .
\eeqn
For example, in (\ref{rep5barofSU(5)}) the representation ${\bf \bar 5}$ of $SU(5)$
can accommodate a colour antitriplet that is also an $SU(2)$ singlet, and a colour singlet
that is also an $SU(2)$ doublet. In addition, it is necessary that the sum of the charges in 
each of these two multiplets be zero. The only possible combination of first-generation
fermions in the SM is:
\beq {\bf \bar{5}} : (\psi_i)_L=\left(\begin{array}{c} \bar{d}_1 \\ \bar{d}_2 \\
\bar{d}_3 \\ e^- \\ -\nu_e \end{array}\right)_L, \eeq
and the rest of the first-generation fermions may be accommodated uniquely, as follows:
\beq {\bf 10} : (\chi^{ij})_L=\frac{1}{\sqrt{2}}\left(
\begin{array}{ccccc} 
0 & \bar{u}_3 & -\bar{u}_2 & u_1 & d_1 \\ 
-\bar{u}_3 & 0 & \bar{u}_1 & u_2 & d_2 \\ 
u_2 & -\bar{u}_1 & 0 & u_3 & d_3 \\ 
-u_1 & -u_2 & -u_3 & 0 & e^+ \\ 
-d_1 & -d_2 & -d_3 & -e^+ & 0 
\end{array}\right)_L , 
\label{foureighteen}
\eeq
where we neglect the eventual mixings between the fermions in different generations.
We must repeat the previous classification of fermions in $\bf{10} + \bf{\bar{5}}$
representations for the other two generations: there is no explanation  in $SU(5)$
for the presence of three generations~\footnote{The pairing of ${\bf \bar{5}}$ and
${\bf 10}$ representations is free of triangle anomalies.}.

After discussing the matter fermions,  we now discuss the GUT gauge bosons.
Groups of type $SU(N)$ have $N^2-1$ symmetry generators in an adjoint
representation (e.g., $SU(3)_C$ 
has 8 gluons, $SU(2)$ has 2 $W$ bosons, etc.), so that $SU(5)$ has 24 
gauge bosons. Of these 24 gauge bosons, 12 correspond to the SM gluons,
$W^\pm$, $Z^0$ and $\gamma$, and 12 are new. Decomposing this 
24-dimensional adjoint representation into representations of 
$SU(3)\otimes SU(2)\otimes U(1)$, we find
\beq
{\bf 24} =\underbrace{({\bf 3},{\bf 2},\frac{5}{3}) \oplus ({\bf \bar{3}},{\bf 2},-\frac{5}{3})
}_{new\ bosons}  \oplus \underbrace{({\bf 8},{\bf 1},0)}_{gluons\ G_a} \oplus 
\underbrace{({\bf 1},{\bf 3},0)}_{W_i} \oplus \underbrace{({\bf 1},{\bf
1},0)}_{B}\ ,
\eeq
where the third numbers in the parentheses are the hypercharges of the multiplets.
The new bosons, called $X$ and $Y$, have electric charges 4/3 and 2/3, respectively,
carry leptoquark quantum numbers, are coloured and have isospin 1/2~\footnote{They have 
direct interactions with quarks and leptons, which we discuss in the next section.}.
In matrix notation,
\beq A=\sum_{a=1}^{24} T_a A^a=\left(
\begin{array}{ccccc}
G_i&G_i&G_i& \bar{X} & \bar{Y} \\
G_i& G_i &G_i& \bar{X} & \bar{Y} \\
G_i&G_i&G_i& \bar{X} & \bar{Y} \\
X & X & X &W_i&W_i\\
Y & Y & Y &W_i& W_i\\
\end{array}\right),
\label{fourseventeen}
\eeq
where the $T_a$  are the generators of $SU(5)$ represented by $5\times 5$ matrices 
(the equivalents for $SU(5)$ of the Pauli matrices of $SU(2)$). The basis is chosen
so that $SU(3)_C$ corresponds to the first three lines and columns, and $SU(2)_L$ to the last
two lines. The top-left and bottom-right blocks therefore contain the gluons and $W$ bosons,
respectively, and the $U(1)$ boson $B$ (not shown) corresponds to a traceless diagonal
generator. 

The remaining steps in constructing an $SU(5)$ GUT 
are the choices of representations for
Higgs bosons, first to break $SU(5)\rightarrow 
SU(3)\times SU(2)\times U(1)$ and
subsequently to break the electroweak $SU(2)\times 
U(1)_Y\rightarrow U(1)_{em}$. The
simplest choice for the first stage is an 
adjoint $\bf{24}$ of Higgs bosons $\Phi$
with a v.e.v.
\beq
<0\vert\Phi\vert 0 > = \left(
\begin{matrix}
1 & 0 & 0 &\vdots & 0 & 0 \cr
0 & 1 & 0 & \vdots & 0 & 0 \cr
0 & 0 & 1 & \vdots & 0 & 0 \cr
\multispan6 \dotfill \cr
0 & 0 & 0 & \vdots & -{3\over 2} & 0 \cr
0 & 0 & 0 & \vdots & 0 & -{3\over 2}
\end{matrix}
\right) \times {\cal O} (m_{GUT}) .
\label{fourtwenty}
\eeq
It is easy to see that this v.e.v. preserves 
colour $SU(3)$, which reshuffles the first three rows
and columns, weak $SU(2)$, which reshuffles the last two 
rows and columns, and the hypercharge
$U(1)$,  which is a diagonal generator. The subsequent breaking 
of  $SU(2)\times U(1)_Y\rightarrow
U(1)_{em}$ is most economically accomplished by a 
$\bf{5}$ representation of Higgs
bosons $H$:
\beq
< 0 \vert\phi\vert 0 > = (0,0,0,0,1) \times 0 (m_W) .
\label{fourtwentyone}
\eeq
It is clear that this v.e.v. has an $SU(4)$ symmetry which yields~\cite{CEG} the
relation $m_b = m_\tau$ before renormalization
that leads, after renormalization (\ref{fourseven}), to a 
successful prediction for $m_b$
in terms of $m_\tau$. However, the same trick does not 
work for the first two generations,
indicating a need for epicycles in this simplest GUT model~\cite{EG}.

Making the minimal $SU(5)$ GUT supersymmetric, as motivated 
by the naturalness of the
gauge hierarchy, is not difficult~\cite{DG}. One must replace the above
GUT multiplets by supermultiplets:
$\bf{\bar 5}$ $\bar F$ and $\bf{10}$ $T$ for the matter particles,
$\bf{24}$ $\Phi$ for the GUT Higgs fields 
that break $SU(5) \rightarrow SU(3)\times
SU(2) \times U(1)$. The only complication is that one needs both $\bf{5}$ and
$\bf{\bar 5}$ Higgs representations $H$ 
and $\bar H$ to break $SU(2)\times U(1)_Y
\rightarrow U(1)_{em}$, just as two doublets were 
needed in the MSSM to cancel anomalies and give masses to all the
matter fermions. The simplest possible form of the Higgs potential
is specified by the superpotential~\cite{DG}:
\beq
W = (\mu + {3\lambda\over 2} M) + \lambda \bar H \Phi H + f(\Phi )
\label{fourtwentytwo}
\eeq
where $\mu = {\cal O}(1)$~TeV and $M = {\cal O}(M_{GUT})$,
and $f(\Phi )$ is chosen so that $\partial f/ \partial\Phi = 0$ when
\beq
<0\vert\Phi\vert 0 > = M \left( 
\begin{matrix}
1 & 0 & 0 &\vdots & 0 & 0 \cr
0 & 1 & 0 & \vdots & 0 & 0 \cr
0 & 0 & 1 & \vdots & 0 & 0 \cr
\multispan6 \dotfill \cr
0 & 0 & 0 & \vdots & -\frac{3}{2} & 0 \cr
0 & 0 & 0 & \vdots & 0 & -\frac{3}{2} 
\end{matrix}
\right) .
\label{fourtwentythree}
\eeq
Inserting this into the second term of (\ref{fourtwentytwo}), 
one finds terms $\lambda M
\bar H_3 H_3,~~-3/ 2 \lambda M \bar H_2 H_2$ for the 
colour-triplet and weak-doublet
components of $\bar H$ and $H$, respectively. Combined 
with the bizarre coefficient of the
first term, these lead to terms
\beq
W \ni (\mu + \frac{5\lambda}{2} M) \bar H_3 H_3 + \mu \bar H_2 H_2 .
\label{fourtwentyfour}
\eeq
Thus we have heavy Higgs triplets with masses ${\cal O}(M_{GUT})$
and light Higgs doublets with masses ${\cal O}(\mu)$. However,
this requires fine tuning the 
coefficient of the first term in
$W$ (\ref{fourtwentytwo}) to about 1 part in $10^{13}$! 
In the absence of supersymmetry, such fine tuning would be
destroyed by quantum loop corrections~\cite{BEGN}.

A primary advantage of supersymmetry
is that its no-renormalization theorems~\cite{IZ,FIZ} guarantee that this
fine tuning is {\it natural}, in the sense 
that quantum corrections do not
destroy it, unlike the situation without supersymmetry. 
On the other hand, supersymmetry
alone does not explain the {\it origin} of the hierarchy.
A second advantage of supersymmetry, as we saw earlier in this section, is that
it would make possible
a much more precise unification of the gauge couplings.
However, a potential snag is that the exchanges of the supersymmetric
partners of the heavy Higgs triplets $\bar H_3, H_3$ may cause rapid
proton decay, as discussed later.

Another possible GUT group that is frequently studied is $SO(10)$~\cite{GG,FM}.
It is a group of rank 5, that contains $SU(5) \otimes U(1)$. 
The principal advantage of $SO(10)$ over $SU(5)$ is that it possesses a fundamental 
spinorial representation of dimension 16 that can accommodate all the fermions of one 
generation, as well as a singlet right-handed neutrino, thanks to its decomposition
in terms of  $SU(5)$ representations~\footnote{The $SO(10)$ group is anomaly-free, so
this decomposition explains finally the freedom from anomalies of $SU(5)$ and the SM.}
\beq 
{\bf 16} = {\bf 10} \oplus {\bf \bar{5}} \oplus {\bf 1}. 
\eeq
The appearance of an $SU(5)$ singlet provides a natural framework
for the physics of the neutrinos and the seesaw mechanism~\footnote{In $SU(5)$, singlet right-handed
neutrinos could be added `by hand', in which case they would have no gauge interactions. 
In the case of $SO(10)$, the gauge interactions of $SO(10)$ do not have any direct
influence on accessible neutrino phenomenology, but may provide interesting
restrictions on their Yukawa interactions.}. 
In $SO(10)$ the number of gauge bosons rises to 45, which includes 33 additional
gauge bosons beyond the SM, and therefore many possible interactions, including
additional options for proton decay. In addition, the breaking of $SO(10)$ is more complicated
than that of $SU(5)$, because it is done in two steps. One may 
pass from $SO(10)$ to $SU(5)\otimes U(1)$ or $SU(4)\otimes SU(2)_L \otimes SU(2)_R$, 
and then to $SU(2)\otimes U(1)$. The Higgs sector is potentially
quite extensive, and may include large multiplets of dimensions 10, 16, 45, 54, 120 and 126,
depending on the model.

\subsubsection{Baryon decay}

\begin{figure}
\centerline{\includegraphics[height=1.5in]{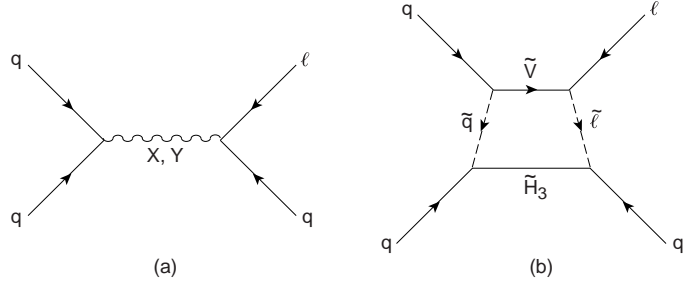}}
\caption[]{Diagrams contributing to baryon decay (a) in minimal $SU(5)$
and (b) in minimal supersymmetric $SU(5)$}
\label{29}

\end{figure}

Baryon instability is to be expected on general grounds, 
since there is no exact gauge
symmetry to guarantee that baryon number $B$ is conserved. 
Indeed, baryon decay is a
generic prediction of GUTs, which we illustrate with the 
simplest $SU(5)$ model, that is
anyway embedded in larger and more complicated GUTs. 
We see in (\ref{fourseventeen}) that
there are two species of gauge bosons in $SU(5)$, called $X$ and $Y$, 
that couple the colour $SU(3)$ indices
(1,2,3) to the electroweak $SU(2)$ indices (4,5). 
As we can see from the
matter representations (\ref{foureighteen}), 
these may enable two quarks or a quark and
lepton to annihilate, as seen in Fig.~\ref{29}(a). 
Combining these possibilities leads to an
interaction with $\Delta B  =
\Delta L = 1$. The forms of effective 
four-fermion interactions mediated by the exchanges of
massive $Z$ and $Y$ bosons, respectively, are~\cite{BEGN}
\bea
&&\left(\epsilon_{ijk} u_{R_k} \gamma_\mu u_{L_j}\right)~~\frac{g^2_X}{8 m^2_X} ~~ \left(2 e_R
~\gamma^\mu ~d_{L_i} + e_L~\gamma^\mu~d_{R_i} \right)~, \nonumber \\
&&\left(\epsilon_{ijk} u_{R_k} \gamma_\mu d_{L_j}\right)~~\frac{g^2_Y}{8 m^2_X} ~~ \left(\nu_L
~\gamma^\mu ~d_{R_i}\right)~,
\label{fourtwentyfive}
\eea
up to generation mixing factors.

\noindent
Since the gauge couplings $g_X = g_Y = g_{3,2,1}$ 
in an $SU(5)$ GUT, and $m_X \simeq m_Y$,
we expect that
\beq
G_X \equiv \frac{g^2_X}{8m^2_X}\simeq G_Y \equiv \frac{g^2_Y}{8m^2_Y} .
\label{fourtwentysix}
\eeq
It is clear from (\ref{fourtwentyfive}) that the baryon 
decay amplitude $A\propto G_X$, and
hence the baryon $B\rightarrow \ell +$ meson decay rate
\beq
\Gamma_B = c G^2_X m^5_p ,
\label{fourtwentyseven}
\eeq
where the factor of $m^5_p$ comes from dimensional 
analysis, and $c$ is a coefficient that
depends on the GUT model and the non-perturbative 
properties of the baryon and meson.

The decay rate (\ref{fourtwentyseven}) corresponds to a proton lifetime
\beq
\tau_p = \frac{1}{c} ~\frac{m^4_X}{m^5_p} .
\label{fourtwentyeight}
\eeq
It is clear from (\ref{fourtwentyeight}) that 
the proton lifetime is very sensitive to
$m_X$, which must therefore be calculated very 
precisely. In minimal $SU(5)$, the best
estimate was
\beq
m_X \simeq (1~{\rm to}~2) \times 10^{15} \times \Lambda_{QCD}
\label{fourtwentynine}
\eeq
where $\Lambda_{QCD}$ is the characteristic 
QCD scale in the $\overline{\rm MS}$
prescription with four active flavours. 
Making an analysis of the generation mixing factors~\cite{SU5mix}, one
finds that the preferred proton (and bound neutron) decay
modes in minimal
$SU(5)$ are
\bea
&&p \rightarrow e^+\pi^0~,~~e^+\omega~,~~\bar\nu \pi^+~,
~~\mu^+K^0~,~~\ldots \nonumber \\
&& n \rightarrow e^+\pi^-~,~~ e^+\rho^-~,~~\bar\nu \pi^0~,~~\ldots ,
\label{fourthirty}
\eea
and the best numerical estimate of the lifetime is
\beq
\tau (p\rightarrow e^+\pi^0) \simeq 2\times 10^{31\pm 1} \times
\left(\frac{\Lambda_{QCD}}{400~{\rm MeV}}\right)^4~~y \ .
\label{fourthirtyone}
\eeq
This is in {\it prima facie} conflict with the latest experimental lower limit
\beq
\tau (p \rightarrow e^+\pi^0) > 8.2 \times 10^{33}~y
\label{fourthirtytwo}
\eeq
from super-Kamiokande~\cite{SK:2009gd}. However, this failure of minimal
$SU(5)$ is not as conclusive as
the failure of its prediction for $\sin^2\theta_W$.

We saw earlier that supersymmetric GUTs, including $SU(5)$, fare better with
$\sin^2\theta_W$. They also predict a larger GUT scale~\cite{DRW}:
\beq
m_X \simeq 2\times 10^{16}~{\rm GeV} ,
\label{fourthirtythree}
\eeq
so that $\tau (p\rightarrow e^+\pi^0)$ is considerably 
longer than the experimental lower
limit. However, this is not the dominant proton 
decay mode in supersymmetric $SU(5)$~\cite{susySU5pdk}. In
this model, there are important $\Delta B = \Delta L = 1$ 
interactions mediated by the
exchange of colour-triplet higgsinos $\tilde H_3$, 
dressed by gaugino exchange as seen in
Fig.~\ref{29}(b)~\cite{WSY}, these give
\beq
G_X\rightarrow {\cal O}~\left(\frac{\lambda^2g^2}{16\pi^2}\right)~\frac{1}{m_{\tilde
H_3}\tilde m} ,
\label{fourthirtyfour}
\eeq
where $\lambda$ is a generic Yukawa coupling. 
Taking into account colour factors and the values
of $\lambda$ for more massive particles, 
it was found~\cite{susySU5pdk} that decays into neutrinos and
strange particles should dominate:
\beq
p\rightarrow \bar\nu K^+~,~~n\rightarrow\bar\nu K^0~,~~\ldots
\label{fourthirtyfive}
\eeq
Because there is only one factor of a heavy mass 
$m_{\tilde H_3}$ in the denominator of
(\ref{fourthirtyfour}), these decay modes are 
expected to dominate over $p\rightarrow
e^+\pi^0$ etc.  in minimal supersymmetric $SU(5)$. 
The current experimental 
limit is $\tau(p\rightarrow \bar\nu
K^+) > 10^{33} y$~\cite{Kobayashi:2005pe}.
Calculating carefully the other factors
in (\ref{fourthirtyfour})~\cite{EGR}, it seems that the modes
(\ref{fourthirtyfive}) may be close to
detectability in this model, possibly even too close for comfort,
in which case a more complicated supersymmetric GUT might
be needed.

There are non-minimal supersymmetric GUT models 
such as flipped $SU(5)$~\cite{AEHN} in which the $\tilde
H_3$- exchange mechanism 
(\ref{fourthirtyfour}) is suppressed. 
In such models, $p\rightarrow e^+\pi^0$ may again be
the preferred decay mode~\cite{aspects}. However, this is not necessarily
the case, as colour-triplet
Higgs boson exchange may also be important, in 
which case $p\rightarrow \mu^+K^0$ could be
dominant~\cite{CER}, or there may be non-intuitive generation mixing in
the couplings of the $X$ and
$Y$ bosons, offering the possibility 
$p\rightarrow \mu^+\pi^0$ etc.  Therefore, the
continuing search for proton decay should be open-minded about the
possible decay modes. The current experimental 
limits for these process are $\tau(p\rightarrow e^+
\pi^0) > 10^{33} y$~\cite{SK:2009gd}, $\tau(p\rightarrow \mu^+
K^0) > 10^{33} y$~\cite{Kobayashi:2005pe}, and $\tau(p\rightarrow \mu^+
\pi^0) > 10^{33}y$~\cite{SK:2009gd}.

\subsubsection{Neutrino masses and oscillations}

The experimental upper limits on neutrino 
masses are far below the corresponding lepton
masses~\cite{Amsler:2008zzb}. From studies of the end-point of tritium $\beta$
decay,
we have
\beq
m_{\nu_e} \lappeq 2~{\rm eV} ,
\label{fourtyirthsix}
\eeq
to be compared with $m_e = 0.511$ MeV. 
Neglecting mixing effects, from studies of $\pi\rightarrow \mu\nu_\mu$ decays,
we have
\beq
m_{\nu_\mu} < 190~{\rm keV} ,
\label{fourthirtyseven}
\eeq
to be compared with $m_\mu$ = 105 MeV, and 
from studies of $\tau\rightarrow$ pions +
$\nu_\tau$, again neglecting mixing effects, we have
\beq
m_{\nu_\tau} < 18.2~{\rm MeV} ,
\label{fourthirtyeight}
\eeq
to be compared with $m_\tau$ = 1.78 GeV. 

On the other
hand, there is no good symmetry reason to expect 
the neutrino masses to vanish. We expect
masses to vanish only if there is a corresponding 
exact gauge symmetry, cf., $m_\gamma$ =
0 in QED with an unbroken $U(1)$ gauge symmetry.

However, although there is no candidate gauge symmetry to 
ensure $m_\nu = 0$, this is a prediction
of the SM. We recall that the neutrino couplings 
to charged leptons take the form
\beq
J_\mu = \bar e\gamma_\mu (1-\gamma_5) \nu_e + \bar\mu \gamma_\mu (1-\gamma_5)\nu_\mu +
\bar\tau\gamma_\mu (1-\gamma_5)\nu_\tau ,
\label{fourthirtynine}
\eeq
and that only left-handed neutrinos have 
ever been detected. In the cases of charged
leptons and quarks, their masses arise in 
the SM from couplings between left- and
right-handed components {\it via} a Higgs field:
\beq
g_{H\bar ff}~H_{\Delta I = \frac{1}{2},\Delta L = 0}~~
\bar f_R f_L + h.c. \rightarrow
m_f = g_{H\bar ff} \langle 0\vert H_{\Delta I = \frac{1}{2},\Delta L = 0} \vert 0 \rangle .
\label{fourforty}
\eeq
Such a left--right coupling is conventionally 
called a Dirac mass. The following questions
arise for neutrinos: if there is no $\nu_R$, 
can one have $m_\nu \not= 0$? On the other hand, if there is
a $\nu_R$, why are the neutrino masses so small?

The answer to the first question is positive, 
because it is possible to generate neutrino
masses {\it via} the Majorana mechanism that involves the $\nu_L$ alone. 
This is possible
because an $(\overline{f_R})$ field is in fact left-handed:
$(\overline{f_R}) = (f^c)_L = f^T_L C$, where the 
superscript $T$ denotes a transpose, and
$C$ is a $2\times 2$ conjugation matrix. We can therefore imagine replacing
\beq
(\overline{f_R}) f_L \rightarrow f^T_L~C~ f_L ,
\label{fourfortyone}
\eeq
which we denote by $f_L \cdot f_L$. In the 
cases of quarks and charged leptons, one
cannot generate masses in this way, because 
$q_L \cdot q_L$ has $\Delta Q_{em}$, 
$\Delta \textrm{(colour)} \not= 0$ and $\ell_L\cdot \ell_L$ 
has $\Delta Q_{em}\not= 0$. However, the
coupling
$\nu_L\cdot\nu_L$ is not forbidden by such 
exact gauge symmetries, and would lead to a
neutrino mass:
\beq
m^M~\nu_L^T~C~\nu_L = m^M (\overline{\nu^c})_L\nu_L \equiv m^M~\nu_L\cdot\nu_L .
\label{fourfortytwo}
\eeq
Such a combination has non-zero net lepton 
number $\Delta L = 2$ and weak isospin $\Delta I
= 1$. There is no corresponding Higgs field in 
the SM or in the minimal $SU(5)$ GUT, but
there is no obvious reason to forbid one. 
If one were present, one could generate a
Majorana neutrino mass {\it via} the renormalizable coupling
\beq
\tilde g_{H\bar\nu\nu} ~~H_{\Delta I=1, 
\Delta L = L}~~\nu_L\cdot\nu_L \Rightarrow m^M =
\tilde g_{H\bar\nu\nu} \langle 0 \vert H_{\Delta I = 1,\Delta L = 2}\vert 0 \rangle .
\label{fourfortythree}
\eeq
However, one could also generate a Majorana mass 
without such an additional Higgs field,
{\it via} a non-renormalizable coupling to the 
conventional $\Delta I = \frac{1}{2}$ SM Higgs
field:
\beq
\frac{1}{M}~~\left(H_{\Delta I = \frac{1}{2}} 
\nu_L\right) \cdot \left( H_{\Delta I =
\frac{1}{2}} \nu_L\right) \Rightarrow m^M = 
\frac{1}{M} \langle 0\vert H_{\Delta I = \frac{1}{2}}
\vert 0 \rangle ^2 ,
\label{fourfortyfour}
\eeq
where $M$ is some (presumably heavy 
mass scale: $M \gg m_W)$. 

The simplest possibility for
generating a non-renormalizable interaction of 
the form (\ref{fourfortyfour}) would be {\it via}
the exchange of a heavy field $N$ that is a 
singlet of $SU(3)\times SU(2)\times U(1)$ or
$SU(5)$:
\beq
\frac{1}{M} \rightarrow \frac{\lambda^2}{M_N} ,
\label{fourfortyfive}
\eeq
where one postulates a renormalizable coupling $\lambda H_{\Delta I=1/2}
\nu_L\cdot N$. As already mentioned, such a heavy singlet field 
appears automatically in extensions of the $SU(5)$
GUT, such as $SO(10)$, though it does not actually {\it require}
the existence of any new GUT gauge bosons.

We now have all the elements we need for 
the see-saw mass matrix~\cite{GRY} favoured by GUT
model-builders:
\beq
(\nu_L , N) \cdot \left(
\begin{matrix}
m^M & m^D\cr 
m^D & M^M
\end{matrix}
\right)~~\left(
\begin{matrix}
\nu_L \cr
N
\end{matrix}
\right) ,
\label{fourfortysix}
\eeq
where the $\nu_L\cdot\nu_L$ Majorana mass 
$m^M$ might arise from a $\Delta I = 1$ Higgs
with coupling $\tilde g_{H\bar\nu\nu}$, 
(\ref{fourfortythree}), the $\nu_L\cdot N$ Dirac
mass $m^D$ could arise from a conventional 
Yukawa coupling $\lambda$ (\ref{fourfortyfive})
and should be of the same order as a conventional 
quark or lepton mass, and $M^M$ could {\it
a priori} be ${\cal O}(M_{GUT})$~\footnote{It is often assumed that
there are three singlet neutrinos $N$, but this need not be the case. If there were only two,
one of the light neutrinos would be massless. On the other hand, 
there could be many more than three~\cite{EL}.}. Diagonalizing 
(\ref{fourfortysix}) and assuming that $m^M =
0$ or that $\langle 0\vert H_{\Delta I = 1}\vert 0 \rangle = 
{\cal O}(m^2_W/ m_{GUT})$, as generically
expected in GUTs, one obtains the mass
eigenstates
\bea
\nu_L + 0\left(\frac{m_W}{m_X}\right) N &:& 
m = {\cal O}\left(\frac{m^2_W}{M_{GUT}}\right) ,
\label{fourfortysevenone} \\
N + 0\left(\frac{m_W}{m_X}\right) \nu_L &:& M = {\cal O} (M_{GUT}) .
\label{fourfortyseven}
\eea
We see that one mass eigenstate (\ref{fourfortysevenone}) is
naturally much lighter than the electroweak scale, whereas the
other (\ref{fourfortyseven}) is naturally much heavier.

There is evidence for
atmospheric neutrino oscillations~\cite{SK}
between $\nu_\mu$ and $\nu_\tau$ with 
$\Delta m^2_A \sim (10^{-2}$ to $10^{-3}$) eV$^2$
and a large mixing angle: $\sin^2\theta_{23} 
\gappeq 0.9$. In addition, there is
evidence~\cite{solar} for solar neutrino oscillations with
$\Delta m^2_S \simeq 10^{-5}$ eV$^2$ 
and $\sin^2\theta_{12} \sim 0.6$.
We also know that the third neutrino mixing angle $\theta_{13}$
must be small, but it is an open experimental question just how small it may be.
The pattern of MNS neutrino mixing seems very different from that of CKM quark
mixing, perhaps reflecting special ingredients related to the see-saw mechanism.
Other open questions include the magnitude of the CP-violating phase
in the neutrino mixing matrix (analogous to the Kobayashi--Maskawa
phase in quark mixing), and also the sequence of neutrino mass
eigenstates. 

CP-violating decays of heavy singlet neutrinos provide a simple mechanism
for generating the baryon number of the Universe~\cite{FY}, by first providing a
lepton asymmetry that is subsequently converted partially into a baryon
asymmetry by non-perturbative electroweak interactions~\cite{Theta}. Essential ingredients
in this scenario are the violation of lepton number {\it via} Majorana neutrino masses
and CP violation~\cite{Pilaftsis:2009pk}. 
The CP-violating phase observable in neutrino oscillations does
not play a direct role in this scenario for baryogenesis~\cite{ERaidal}, but its observation would
nevertheless be of great conceptual importance.

\subsection{Local supersymmetry and supergravity}

Why study a local theory of supersymmetry~\cite{FvF,DZ}?  
One motivation is the analogy with gauge
theories, in which bosonic symmetries are 
made local.  Another is that local supersymmetry
necessarily involves the introduction of gravity.  
Since both gravity and (surely!)
supersymmetry exist, this seems an inevitable step.  
It also leads to the possibility of
unifying all the particle interactions including 
gravity, which was one of our original
motivations for supersymmetry.  Moreover, it is interesting
that local supersymmetry (supergravity) admits 
an elegant mechanism for supersymmetry
breaking~\cite{superHiggs}, analogous to the Higgs mechanism 
in gauge theories, which allows us to address
more seriously the possible existence of a cosmological constant.

The basic building block in a supergravity theory~\cite{FvF,DZ} is the
graviton supermultiplet, which contains particles with 
helicities $(2, 3/2)$, the latter being the
gravitino of spin $3/2$.  Why is this 
required when one makes supersymmetry local?

We recall the basic global supersymmetry 
transformation laws (\ref{twoseventeen1}, \ref{twoseventeen}) for bosons
and fermions. Consider
now the combination of two such global supersymmetry transformations
\beq
[\delta_1 , \delta_2 ]~(\phi \,~ \mbox{or} ~\, \psi ) = 
- (\bar{\xi}_2 \gamma_{\mu} \xi_1 )~(i \, 
\partial_{\mu})~(\phi \, ~\mbox{or} ~\,
\psi ) + \ldots
\label{fivethirteen}
\eeq
The operator $(i \, \partial_{\mu})$ 
corresponds to the momentum $P_{\mu}$, and we see again
that the combination of two global supersymmetry 
transformations is a translation. Consider 
now what happens when we consider local 
supersymmetry transformations characterized
by a varying spinor $\xi(x)$.  It is evident 
that the infinitesimal translation 
$\bar{\xi}_2 \gamma^{\mu} \xi_1$ in 
(\ref{fivethirteen}) is now $x$-dependent,
and the previous global translation becomes 
a local coordinate transformation, as occurs in
General Relativity.

How do we make the theory invariant under 
such local supersymmetry transformations? 
Consider again the simplest globally 
supersymmetric model containing a free spin-1/2 fermion
and a free spin-0 boson (\ref{twosixteen}),
and make the local versions of the transformations 
(\ref{twoseventeen}), we can obtain
\beq
\delta {\cal L} = \partial_{\mu} (\cdots) + 2 \bar{\psi} \gamma_{\mu} \, 
\partial\llap{$/$} S(\partial^{\mu} \xi(x)) + \mbox{herm. conj.} 
\label{fivefifteen}
\eeq
In contrast to the global case, the action 
$A = \int d^4 x {\cal L}$ is not invariant,
because of the second term in (\ref{fivefifteen}).  
To cancel it out and restore invariance,
we need more fields.

We proceed by analogy with gauge theories.  In order to make the kinetic term 
$(i \bar{\psi} \partial\llap{$/$} \psi )$ 
invariant under gauge transformations $\psi \to
e^{i \epsilon (x)} \psi$, we need to cancel a variation
\beq
- \bar{\psi} \partial_{\mu} \psi \partial^{\mu} \epsilon (x) ,
\label{fivesixteen}
\eeq
which is done by introducing a coupling to a gauge boson
\beq
g \bar{\psi} \gamma_{\mu} \psi A^{\mu} (x)\ ,
\label{fiveseventeen}
\eeq
and the corresponding transformation
\beq
\delta A_{\mu} (x) = \frac{1}{g} \partial_{\mu} \epsilon (x) .
\label{fiveeighteen}
\eeq
In the supersymmetric case, we cancel the 
second term in (\ref{fivefifteen}) by a coupling
\beq
\kappa \bar{\psi} \gamma_{\mu} \partial\llap{$/$} S \psi^{\mu} (x)
\label{fivenineteen}
\eeq
to a spin-3/2 spinor $\psi^{\mu} (x)$, 
representing a gauge fermion or gravitino, with the
corresponding transformation
\beq
\delta \psi^{\mu} = - \frac{2}{\kappa} \, \partial^{\mu} \xi (x) ,
\label{fivetwenty}
\eeq
where $\kappa \equiv 8 \pi / m^2_P$.

For completeness, let us at least write 
down the Lagrangian for the graviton--gravitino
supermultiplet
\beq
L = - \frac{1}{2\kappa^2} \, \sqrt{-g} R   - 
\frac{1}{2} \, \epsilon^{\mu \nu \rho \sigma}
\bar{\psi}_{\mu} \gamma_{5} \gamma_{\nu} {\cal D}_{\rho} \psi_{\sigma} ,
\label{fivetwentyone}
\eeq
where $g$ denotes the determinant of the metric tensor
\beq
g_{\mu \nu} = \epsilon^m_{\mu} \eta_{mn} \epsilon^{\mu}_{\nu} ,
\label{fivetwentytwo}
\eeq
$\epsilon^m_{\mu}$ is the vierbein and 
$\eta_{mn}$ the Minkowski metric tensor, and
${\cal D_{\rho}}$ is a covariant derivative
\beq
{\cal D_{\rho}} \equiv \partial_{\rho} + \frac{1}{4} \, \omega^{mn}_{\rho}
[\gamma_m , \gamma_n ] ,
\label{fivetwentythree}
\eeq
where $\omega^{mn}_{\rho}$ is the spin 
connection.  This is the simplest possible
generally-covariant model of a spin-3/2 field.  
It is remarkable that it is invariant under
the local supersymmetry transformations
\bea
\delta \epsilon^m_{\mu} & = & \frac{x}{2} \, \bar{\xi} (x) \gamma^m
\psi_{\mu} (x), \nonumber \\
\delta \omega_{\mu}^{mn} & = & 0, 
\delta \psi_{\mu} = \frac{1}{x} \, {\cal D}_{\mu} \xi (x) ,
\label{fivetwentyfour}
\eea
just as the simplest possible $(1/2, 0)$ theory (\ref{twosixteen}) was globally
supersymmetric, and also the action of an 
adjoint spin-1/2 field in a gauge theory.

As already remarked, supergravity 
admits an elegant analogue of the Higgs mechanism of
spontaneous symmetry breaking~\cite{superHiggs}.  Just as one combines the
two polarization states of a
massless gauge field with the single state 
of a massless Goldstone boson to obtain the three
polarization states of a massive gauge boson, 
one may combine the two polarization states of
a massless gravitino $\psi_{\mu}$ with the 
two polarization states of a massless Goldstone
fermion $\lambda$ to obtain the four polarization 
states of a massive spin-3/2 particle
$\tilde{G}$.  This super-Higgs mechanism 
corresponds to a spontaneous breakdown of local
supersymmetry, since the massless graviton $G$ 
has a different mass from the gravitino
$\tilde{G}$:
\beq
m_G = 0 \not= m_{\tilde{G}}.
\label{fivetwentyfive}
\eeq
This is the only known consistent way of 
breaking local supersymmetry, just as the Higgs
mechanism is the only way to generate $m_W \not= 0$.

Moreover, this can be achieved while keeping zero 
vacuum energy (cosmological constant), at
least at the tree level.  The reason for 
this is the appearance in local supersymmetry
(supergravity) of a third term in the effective 
potential (\ref{pot}), which has a
{\it negative} sign~\cite{superHiggs}.  There is no time 
in these lectures to discuss this exciting feature
in detail:  the interested reader is referred 
to the original literature and the simplest
example~\cite{CFKN}.  In this particular case, $\Lambda = V = 0$ for {\it any}
value of the gravitino mass,
for which reason it was named no-scale supergravity~\cite{ELNT}.

Again, there is no time to discuss here details of 
the coupling of supergravity to matter~\cite{superHiggs}.
However, it is useful to have in mind the general 
features of the theory in the limit where
$\kappa \to 0$,  but the gravitino mass 
$m_{\tilde{G}} \equiv m_{3/2}$ remains fixed.  One
generally has non-zero gaugino masses $m_{1/2} 
\propto m_{3/2}$, and their universality is
quite generic.  One also has non-zero 
scalar masses $m_0 \propto m_{3/2}$, but their
universality is much more problematic, and 
even violated in generic string models.  It was
this failing that partly refuelled interest in gauge-mediated models.
A generic supergravity theory also yields non-universal
trilinear soft supersymmetry-breaking couplings 
$A_{\lambda} \lambda \phi^3 : A_{\lambda}
\propto m_{3/2}$ and bilinear scalar 
couplings $B_{\mu} \mu \phi^2 : B_{\mu} \propto
m_{3/2}$.  Therefore, supergravity may generate the full menagerie of soft
supersymmetry-breaking terms:
\beq
- \frac{1}{2} \, \sum_a \, m_{1/2_a}\, \tilde{V}_a \tilde{V}_a - 
\sum_i \, m^2_{0_i} |\phi_i|^2 -
\left(\sum_{\lambda} A_{\lambda} \lambda \phi^3 + \mbox{h.c.}\right)
- \left(\sum_{\mu} B_{\mu} \mu \phi^2 + \mbox{h.c.}\right) .
\label{fivetwentysix}
\eeq
In a minimal supergravity (mSUGRA) framework, the gaugino masses $m_{1/2}$,
scalar masses $m_0$, and trilinear couplngs $A$
are universal, as assumed in the CMSSM, but there are specific conditions:
$B = A - 1$, and the gravitino mass is fixed: $m_{3/2} = m_0$. The
former condition is more restrictive than in the CMSSM, and the latter
condition implies that the gravitino is the LSP in significant regions of
parameter space. Hence, the CMSSM and mSUGRA are distinct scenarios~\cite{mSUGRA}.

Since these soft supersymmetry-breaking parameters are generated at the supergravity 
scale near $m_P \sim 10^{19}$ GeV, the soft
supersymmetry-breaking parameters are 
renormalized as discussed earlier.  The analogous
parameters in gauge-mediated models would 
also be renormalized, but to a different extent,
because the mediation scale $\ll m_P$. This 
difference may provide a signature of such
models, as discussed elsewhere~\cite{massdiff,EOSandick}.

Also renormalized is the vacuum energy 
(cosmological constant), which is a potential
embarassment.  Loop corrections in a 
non-supersymmetric theory are quartically divergent,
whereas those in a generic supergravity theory 
are only quadratically divergent, suggesting
a contribution to the cosmological constant 
of order $m^2_{3/2} m^2_P$, perhaps
$O(10^{-32})m^4_P$!  Particular models 
may have a one-loop quantum correction of order
$m^4_{3/2} = O(10^{-64})m^2_P$, but more 
magic (a new symmetry?) is needed to suppress the
cosmological constant to the required level 
\beq
\Lambda \lappeq 10^{-123} m^4_P.
\eeq
This is one of the
motivations for seeking a fundamental
Theory of Everything including gravity.

Once upon a time, supergravity was considered a possible candidate
for such a Theory of Everything, particularly the maximal ${\cal N} = 8$
supergravity in 4 dimensions. However, this candidature would need
two elements that are still lacking: a proof that the theory is finite, or at
least renormalizable, and a demonstration of how it could lead to a
low-energy theory resembling the SM, e.g., {\it via} the formation of
bound states: see Ref.~\cite{AEN} for a review of these issues. In the
meantime, string theory~\cite{GSW} is the most plausible
candidate for a Theory of Everything.

\subsection{Towards a Theory of Everything}

\subsubsection{Problems in quantum gravity}

One of the most important unfinished tasks for 
understanding the Universe and the fundamental interactions is the 
unification of the two great theories of the 20th century: general relativity and quantum 
mechanics. To write such a unified Theory of Everything is one of the major 
challenges for physicists in our century. The solution of the problem of the 
cosmological constant, for example, will have to find a place in the frame of such
a Theory of Everything. 

Gravity is a puzzle for conventional quantum theory, in particular because incontrollable,
non-renormalizable infinities appear when one tries to calculate Feynman diagrams that 
contain loops with gravitons. These correction terms diverge increasingly rapidly as the 
order of the perturbative calculation increases, essentially because the coupling of gravity
has negative mass dimensionality, being $\propto 1/M_P^2$, where $M_P \simeq
1.2 \times 10^{19}$~GeV.

\begin{figure}
\centerline{\includegraphics[height=1.5in]{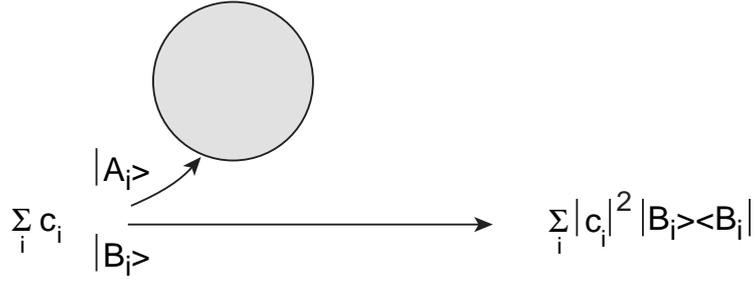}}
\caption[]{If a pair of particles $\vert A\rangle~\vert B\rangle$ is produced near
the horizon of a black hole, and one of them ($\vert A\rangle$, say) falls in,
the remaining particle $\vert B\rangle$ will appear to be in a mixed state,
since the state of $\vert A\rangle$ is unobservable}
\label{34} 
\end{figure}

There are also non-perturbative problems in the quantization of gravity, 
which first appeared in connection with black holes. We recall that a black hole is a 
non-perturbative solution of the equations of General Relativity, in which the curvature 
of space-time induced by gravitational forces becomes so strong that no particle can 
escape the event horizon. The existence of this horizon is linked to the existence
of entropy $S$ and a non-zero temperature $T$ of the black hole.
From the pioneering work of Bekenstein and Hawking~\cite{BekHawk} on
black-hole thermodynamics, we know that the mass of a black 
hole is proportional to the surface area $A$ of its horizon, which is related in turn to its entropy:
\begin{equation}
S\;=\frac{1}{4}\;A \ .
\end{equation}
The appearance of non-zero entropy means that the quantum
description of a black hole must involve 
mixed states.  The intuition underlying this
feature is that information can be lost 
through the event horizon.  To see how this may happen, consider, for example, a
pure quantum-mechanical pair state $|A,B\rangle \equiv 
\sum_i c_i |A_i \rangle |B_i \rangle$ prepared near the
horizon, and what happens if one of the 
particles, say $A$, falls through the horizon while
$B$ escapes, as seen in Fig. \ref{34}.  In this case,
all the information about the component $|A_i \rangle$ of the wave function is lost, so that
\beq
\sum_i \, c_i |A_i B_i \rangle \to \sum_i |c_i|^2 |B_i \rangle \langle B_i|
\label{fivetwentyeight}
\eeq
and $B$ emerges in a mixed state, as in Hawking's 
original treatment of the black-hole
radiation that bears his name~\cite{BekHawk}.
The problem is that conventional quantum 
mechanics does not permit the evolution of a pure
initial state into a mixed final state.

For a discussion of these and other open problems in quantum black hole
physics, see Ref.~\cite{Strominger:2009aj}.
Many theorists consider that these problems point to a fundamental conflict 
between the proudest achievements of 
early-twentieth-century physics, namely quantum mechanics 
and General Relativity.  One or the other
should be modified, and perhaps both.  
Since quantum mechanics is sacred to field theorists,
most particle physicists prefer to modify General 
Relativity by elevating it to string
theory, as we now discuss.

\subsubsection{Introduction to string theory}

As was just mentioned, one of the major issues of quantum gravity is that it has an
infinite number of infinities. These divergences can be traced to the absence of a 
short-distance cut-off in conventional field theories, where the particles are points.
The problem is that one can in principle approach infinitely near a point particle, 
giving rise to interactions of infinite strength:
\begin{equation}
\int^{\Lambda \to \infty} \, d^4k \left (\frac{1}{k^2} \right )
\leftrightarrow \int_{1/ \Lambda \to 0} \, d^4x
\left ( \frac{1}{x^6} \right ) \sim \Lambda^2 \to \infty.
\end{equation}
Such divergences can be avoided or removed 
if one replaces point particles by extended
objects.  The simplest possibility is to 
extend in just one dimension, leading to a
theory of strings.  In such a theory, instead 
of point particles moving along one-dimensional
world lines, one has strings moving over 
two-dimensional world sheets.  Historically, closed loops of string
have been the most popular, and the corresponding world sheet
would be tubes.  The `wiring diagrams'
generated by the Feynman rules of conventional 
point-like particle theories become
`plumbing circuits' generated by the 
junctions and connections of these tubes of closed
string.  One could imagine generalizing this 
idea to higher-dimensional extended objects
such as membranes describing world volumes, 
etc., and we return later to this option.

Back in the early 1960s, there existed a quantum theory of the 
electromagnetic force (QED), but successful descriptions of the weak and strong forces 
were not yet known. At that time, theoretical efforts were concentrated on developing a 
theory that would determine the scattering ($S$) matrix, which describes on-mass-shell 
scattering amplitudes, which should possess certain properties abstracted from 
quantum field theory, such as unitarity and maximal analytic properties. These 
characteristics would ensure the requirements of causality and non-negative probabilities. 
A key idea in those years was maximal analyticity in the 
angular momentum plane, i.e., that the conventional partial-wave amplitudes $a_l(s)$ defined in the 
first instance for discrete angular momenta $l=0,1,...$, can be extended uniquely to
analytic functions of $l$, $a(l,s)$. These have isolated `Regge' poles
that move along Regge trajectories $l=\alpha(s)$ in the complex
angular-momentum plane. The values of $s$ for which $l$ take suitable 
discrete values correspond to a physical hadron states. Experimental results indicated
that the Regge trajectories are approximately linear, with a common slope $\alpha'$:
\beq 
\alpha(s) = \alpha(0) +  \alpha ' s,
\eeq 
where $\alpha ' \sim 1.0(\textrm{GeV})^{-2}$.
These ideas were insufficient to determine the $S$ matrix, and additional
principles were invoked, such as the \textit{bootstrap} idea, according to
which the exchanges of hadrons in crossed channels provide forces that are 
responsible for forming hadronic bound states. In the narrow-resonance 
approximation, i.e., if resonance decay widths are negligible compared to their masses,
the scattering amplitude can be expanded in an infinite series of $s$-channel poles, 
and this should give the same result as its expansion in an infinite series of $t$-channel 
poles due to exchanged particles. The narrow-resonance version of the
bootstrap idea, which was called duality, had a precise formulation with a definite
solution.

The decisive contribution to the solution was made by Veneziano in 1968~\cite{Ven}:
he gave an analytic formula that exhibited duality with linear Regge trajectories. 
Its structure was the sum of three Euler beta functions~\cite{Schwarz:2007yc}:
\begin{equation}
T \; = \;  A(s,t) + A(s, u) + A(t,u): \; \;
A(s,t) \; = \; \frac{\Gamma(-\alpha(s)) \Gamma(- \alpha(t))}{\Gamma(-\alpha(s) -\alpha(t)) },
\end{equation}
where $\alpha$ is a linear Regge trajectory, with $\alpha(s) = \alpha(0) + \alpha' s$
as described above. In the course of the next few years, several further
breakthroughs were achieved. Virasoro~\cite{Virasoro} showed how to generalize the Veneziano 
formula to one with full symmetry in the three Mandelstam invariants $s, t, u$. 
Multi-particle generalizations of the Veneziano and Virasoro formulas were constructed
and shown to factorize consistently on a finite spectrum of single-particle states at
each energy level, which could be described by an infinite number of simple
harmonic oscillators. This surprising result led to the first ideas of strings~\cite{Goddard}:
they could be interpreted as the scattering modes of a relativistic string: open strings in the 
Veneziano case and closed strings in the Virasoro case~\footnote{It still seems
amazing that the mathematical formulae preceded the string interpretation~\cite{Schwarz:2007yc}.}.

While looking for a way to incorporate baryons into
the string framework, in 1971 Ramond~\cite{Ramond} constructed a dual-resonance model 
generalization of the Dirac equation. The solutions of this equation gave the 
spectrum of a noninteracting fermionic string. In combination with work by 
Neveu and Schwarz~\cite{NS}, 
this led to a unified interacting theory of bosons and fermions, which was
essentially a prototype for what later came to be known as superstring theory. 
The action of this theory has two-dimensional global supersymmetry on the
world-sheet, described by infinitesimal fermionic transformations of the type
discussed in the previous Lecture.

Initially, it was regarded as a disadvantage that this first incarnation of
string theory was not able to accommodate the point-like partons seen 
inside hadrons at this time. In retrospect, this was the converse of the
quantum-gravity motivation for string theory 
mentioned at the beginning of this section, which disfavours point-like structures. 
Then  in 1973 along  came QCD which incorporated 
these point-like scaling properties and provided a
qualitative understanding of confinement
that has now become quantitative with the advent
of modern lattice calculations.  Thus string 
theory languished as a candidate model of the
strong interactions, though there is still hope that
some as yet undiscovered variant of string theory might provide a useful
alternative description of the strong interactions. In the mean time, interest
was sparked in 1973 by the realization that string theory predicted the
existence of a massless spin-2 state~\cite{TOE}. Could this be the graviton? It was
known that in any consistent theory of a massless spin-2 particle its
low-energy interactions would be identical with those of general relativity.
Might string theory be a consistent high-energy completion of this theory,
in which case it might be the longsought Theory of Everything?

As already mentioned, one of the primary reasons 
for studying extended objects in connection
with quantum gravity is the softening of 
divergences associated with short-distance
behaviour.  Since the string propagates 
on a world sheet, the basic formalism is
two-dimensional.  Accordingly, string 
vibrations may be described in terms of left- and
right-moving waves:
\beq
\phi (r,t) \to \phi_L (r-t), \, \phi_R (r+t) .
\label{fivethirty}
\eeq
If the string has no boundary, as for a 
closed string, the left- and right-movers are
independent.  When quantized, they may be 
described by a two-dimensional field theory.
Compared to a four-dimensional theory, it 
is relatively easy to make a two-dimensional field
theroy finite.  In this case, it has conformal 
symmetry, which has an infinite-dimensional
symmetry group in two dimensions.  However, 
as you already know from gauge theories, one must be careful to ensure
that this classical symmetry is not broken at 
the quantum level by anomalies.  If the quantum
string theory is to be consistent in a flat 
background space-time, the conformal anomaly
fixes the number of left- and right-movers 
each to be equivalent to 26 free bosons if the
theory has no supersymmetry, or 10 boson/fermion 
supermultiplets if the theory has $N = 1$
supersymmetry on the world sheet. There 
are other important quantum consistency conditions,
and it was the demonstration by Green and Schwarz~\cite{GrSch} that
certain string theories are
completely anomaly-free that opened the 
floodgates of theoretical interest in string theory
as a potential Theory of Everything.

Among consistent string theories, one 
may enumerate the following.  The {\it bosonic string}
exists in 26 dimensions, but this is not even its worst 
problem! It contains no fermionic matter
degrees of freedom, and the flat-space 
vacuum is intrinsically unstable. {\it Superstrings}
exist in 10 dimensions, have fermionic matter 
and also a stable flat-space vacuum.  On the
other hand, the ten-dimensional theory is 
left-right symmetric, and the incorporation of
parity violation in four dimensions is not 
trivial. The {\it heterotic string} was originally
formulated in 10 dimensions, with parity 
violation already incorporated, since the left- and
right movers were treated differently.  
This theory also has a stable vacuum, but still suffers
from the disadvantage of having too many 
dimensions. {\it Four-dimensional heterotic
strings} may be obtained either by compactifying the six
surplus dimensions: $10 = 4 + 6$
compact dimensions with size $R \sim 1/m_P$, or by direct
construction in four dimensions,
replacing the missing dimensions by other 
internal degrees of freedom such as fermions or
group manifolds or ...?  In this way it was 
possible to incorporate a GUT-like gauge group~\cite{AEHN}
or even something resembling the Standard Model.

What are the general features of such string models? 
First, they predict there are no more
than 10 dimensions, which agrees with 
the observed number of 4. Secondly, they suggest that
the rank of the four-dimensional gauge group 
should not be very large, in agreement with the
rank 4 of the Standard Model~\footnote{However,
the number of gauge symmetries may be enhanced by non-perturbative
effects.}.  Thirdly, the 
simplest four-dimensional string models do not
accommodate large matter representations~\cite{nobig}, 
such as an \textbf{8} of SU(3) or a
\textbf{3} of SU(2), again in agreement 
with the known representation structure of the
Standard Model. Fourthly, simple string models 
predict fairly successfully the mass of the
top quark, from the requirement that the theory make
sense at all energies up to the Planck mass. Fifthly, string theory makes a 
fairly successful prediction for the gauge
unification scale in terms of $m_P$.  If 
the intrinsic string coupling $g_s$ is weak, one
predicts
\beq
M_{GUT} = O(g) \times \frac{m_P}{\sqrt{8 \pi}} \simeq
\mbox{few} \times 10^{17} \mbox{GeV} ,
\label{fivethirtytwo}
\eeq
where $g$ is the gauge coupling, which is 
${\cal O}(20)$ higher than the value calculated on the basis 
of LEP measurement of the gauge couplings.
Nevertheless, it would be nice to obtain 
closer agreement, and this provides the major
motivation for considering strongly-coupled 
string theory, which corresponds to a large
internal dimension $l > m^{-1}_{GUT}$, as we discuss next.

\subsubsection{M theory}

As was already said, the bosonic string model has many more disadvantages than
other models. It has 26 dimensions, does not contain fermions, and has an unstable vacuum.
Consequently, physicists focused on superstring models, of which five types exist:
\begin{itemize}
\item Type IIA, that reduces at low energy to a non-chiral $N=2$ supergravity in $d=10$ dimensions;

\item Type IIB, that reduces at low energy to a chiral $N=2$ supergravity in $d=10$ dimensions;

\item The heterotic $E(8)\times E(8)$ theory, that reduces at low energy to an $N=1$ 
supergravity  in $d=10$, connected to a Yang--Mills gauge theory with an $E(8)\times E(8)$
gauge group;

\item The heterotic theory $SO(32)$, that  reduces at low energy to an $N=1$ supergravity in 
$d=10$, connected to a Yang--Mills gauge theory with an $SO(32)$ gauge group;

\item Type I, that contains simultaneously opened and closed strings, and that reduces
at low energy to an $N=1$ supergravity in $d=10$ connected to a Yang--Mills gauge
theory with an $SO(32)$ gauge group.

\end{itemize}

These theories all look different. For example, the Type I theory is the only one that 
contains simultaneously open and closed strings, whereas the others contain only 
closed strings. In addition, the  low-energy gauge structures of the five theories are different.
It seems then, that we have five distinct theories that may describe gravity at the quantum
level. How may we understand this? Is it possible that there is a link between the different theories?

Current developments involve going beyond 
strings to consider higher-dimensional extended
objects, such as generalized membranes with 
various numbers of internal dimensions.  These
can be regarded as solitons (non-perturbative 
classical solutions) of string theory~\cite{Polch}, with
masses
\beq
m \propto \frac{1}{g_s} ,
\label{fivethirtythree}
\eeq
somewhat analogously to monopoles in gauge theory.  
It is evident from (\ref{fivethirtythree}) that
such membrane-solitons become light in the 
limit of strong string coupling: $g_s \to \infty$.

\begin{figure}
\centerline{\includegraphics[height=3in]{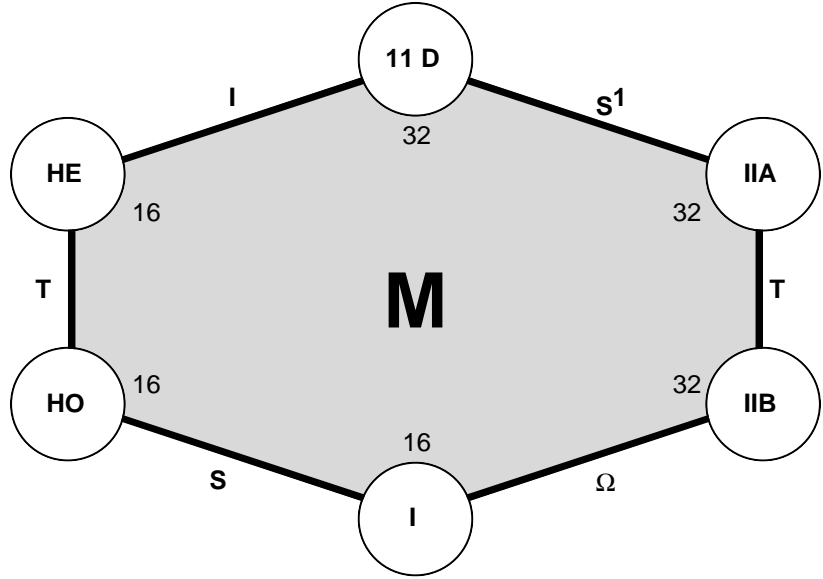}}
\caption[]{The different limits of the $M$ theory are joined by different duality relations. 
The numbers $16$ and $32$ are the numbers of spinor components in the theory.}
\label{M}
\end{figure}

It was observed some time ago that there should 
be a strong-coupling/weak-coupling duality between
elementary excitations and monopoles in supersymmetric 
gauge theories.  These ideas were
confirmed in a spectacular 
solution of $\mathcal{N} = 2$ supersymmetric gauge theory in
four dimensions~\cite{SW}. Similarly, it was shown that
there are analogous dualities in
string theory~\cite{HT}, whereby solitons in some 
strongly-coupled string theory are equivalent to
light string states in some other weakly-coupled 
string theory. Indeed, it appears that all
string theories are related by such dualities.  
A peculiarity of this discovery is that
the string coupling strength $g_s$ is 
related to an extra dimension in such a way that its
size $R \to \infty$ as $g_s \to \infty$.  
This then leads to the idea of an underlying
11-dimensional framework called $M$ theory~\cite{Mtheory} 
that reduces to the different string theories in
different strong/weak-coupling linits, and reduces to eleven-dimensional
supergravity in the low-energy limit (see Fig.~\ref{M}).

A particular class of string solitons called $D$-branes 
offers a promising approach to the
black hole information paradox mentioned previously.  
According to this picture, black holes
are viewed as solitonic balls of string, 
and their entropy simply counts the number of
internal string states. These are in principle 
countable, so string theory may provide an
accounting system for the information contained 
in black holes.  Within this framework, the
previously paradoxical process (\ref{fivetwentyeight}) becomes
\beq
|A,B \rangle + |BH \rangle \to |B^\prime \rangle + |BH^\prime \rangle
\label{fivethirtyfour}
\eeq
and the final state is pure if the initial 
state was.  The apparent entropy of the final
state in (\ref{fivetwentyeight}) is now 
interpreted as entanglement with the state of the
black hole.  The `lost'
information is encoded in 
the black-hole state, and this information could in principle be
extracted if we measured all properties of this ball of string~\cite{Giddings}.

In practice, we do not know how to recover this 
information from macroscopic black holes, so
they appear to us as mixed states. What 
about microscopic black holes, namely fluctuations
in the space-time background with $\Delta 
E = O(m_P)$, that last for a period $\Delta  t =
O(1/m_P)$ and have a size $\Delta x = O(1/m_P)$?  
Do these steal information from us, or do
they give it back to us when they decay?  
Most people think there is no microscopic leakage
of information in this way, but not all of 
us~\cite{EMN} are convinced.  The neutral kaon system is
among the most sensitive experimental areas for
testing this speculative possibility.

\begin{figure}
\centerline{\includegraphics[height=3in]{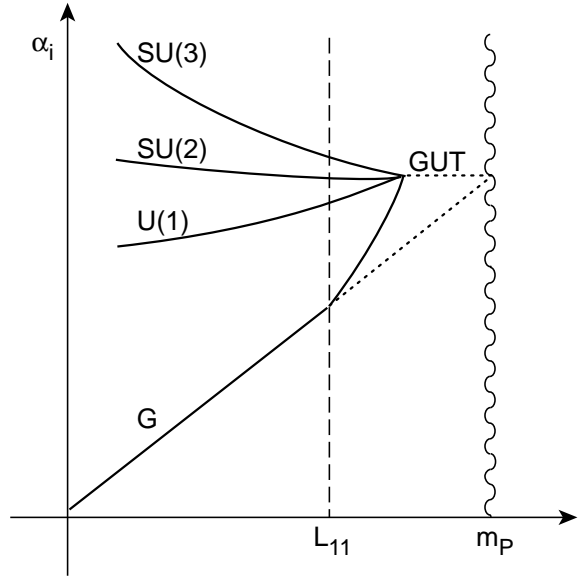}}
\caption[]{Sketch of the possible evolution of the gauge couplings and
the gravitational coupling $G$: if there is a large fifth dimension with
size $\gg m^{-1}_{GUT}$, $G$ may be unified with the gauge couplings at
the GUT scale~\protect\cite{Horava}}
\label{36} 
\end{figure}

\begin{figure}
\centerline{\includegraphics[height=2in]{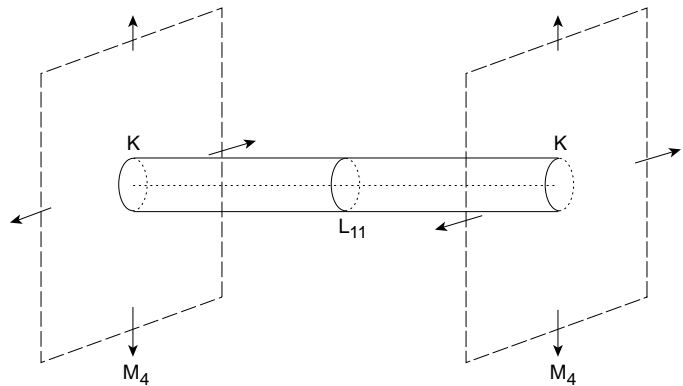}}
\caption[]{The capacitor-plate scenario favoured in
eleven-dimensional $M$ theory. The eleventh dimension has a size
$L_{11} \gg M_{GUT}^{-1}$, whereas dimensions $5, ... , 10$ are
compactified on a small manifold $K$ with characteristic size
$\sim M_{GUT}^{-1}$. The remaining four dimensions form
(approximately) a flat Minkowski space $M_4$~\cite{Horava}.}
 \label{37} 

\end{figure}

How large might the extra dimension be in $M$ theory? 
Remember that the na\"\i ve string
unification scale~(\ref{fivethirtytwo}) is about 20 times 
larger than $m_{GUT}$ as inferred from LEP data. If one wants to maintain
consistency of LEP data with supersymmetric GUTs, it seems that the extra
dimension may be relatively large, 
with size $L_{11} \gg 1/m_{GUT} \simeq 1/10^{16}~
\mbox{GeV} \gg 1/m_P$~\cite{Horava}.   This may be traced to the
fact that the gravitational interaction strength, 
although growing rapidly as a power of
energy 
\beq
\sigma_G \sim E^2/m^4_P,
\label{fivetwentyseven}
\eeq
is still much smaller than the gauge coupling strength at
$E = m_{GUT}$.  However, if an extra space-time 
dimension appears at an energy $E <
m_{GUT}$, the gravitational interaction 
strength grows faster, as indicated in Fig. \ref{36}. 
Unification with gravity around $10^{16}~\mbox{GeV}$ then becomes
possible, {\it if} the gauge couplings do not also acquire a
similar higher-dimensional kick.  Thus we are led to the 
startling capacitor-plate framework for fundamental physics shown in
Fig.~\ref{37}.

Each capacitor plate is {\it a priori} ten-dimensional, 
and the bulk space between them is {\it a
priori} eleven-dimensional.  Six dimensions 
are compactified on a scale $L_6 \sim
1/m_{GUT}$, leaving a theory which is 
effectively five-dimensional in the bulk and
four-dimensional on the walls.  Conventional 
gauge interactions and observable matter
particles are hypothesized to live on one 
capacitor plate, and there are other hidden gauge
interactions and matter particles living on 
the other plate.  The fifth dimension has a
characteristic size which is estimated to 
be ${\cal O}(10^{12}~ \mbox{to}~ 10^{13}~ \mbox{GeV})^{-1}$. 
Physics at smaller energies (large distances) looks effectively
four-dimensional, whereas gravitational physics at larger energies (smaller distances)
looks five-dimensional, and the strength of the gravitational coupling rises rapidly
to unify with the gauge couplings.  Supersymmetry breaking 
is expected to originate on the hidden capacitor
plate in this scenario, and to be transmitted 
to the observable wall by gravitational-strength interactions in the bulk.

The phenomenological richness of this speculative $M$-theory 
approach is only beginning to be explored,
and it remains to be seen whether it offers a 
realistic phenomenological description. 
However, it does embody all the available 
theoretical wisdom as well as offering the
prospect of unifying all the observable gauge 
interactions with gravity at a single effective scale
$\sim m_{GUT}$, including the interactions of the Standard Model.  As such, it constitutes our 
best contemporary guess about the Theory of
Everything within and beyond the Standard Model.

\subsection{Extra dimensions}

We have seen that string theories suggest that there may be extra unseen
dimensions of space, but this speculation did not originate with string theorists.
The idea of extra dimensions was first developed by Kaluza~\cite{Kaluza}
and Klein~\cite{Klein}. They noticed that gravitational and electromagnetic interactions, 
being so alike in many ways, could be descendants of a common ancestor. Indeed,
if we formulate a theory with extra spatial dimensions, it is possible to unify gravity and 
electromagnetism. In the same way, non-Abelian gauge fields can be unified with 
Einstein's gravity in more complicated models with extra dimensions. Thus, the 
first reason why extra dimensions were studied was to unify the gravitational 
and gauge interactions. These initial discussions concerned gravitation at the
classical level. If you want to quantize gravity, you would be well advised to 
look at the best available candidate, namely string or M-theory, which, as we have seen, 
can be formulated consistently in a space with six or seven extra dimensions. From this
point of view, the quantization of gravitational interactions becomes a second reason for 
extra dimensions. 

In all the scenarios considered above, the extra dimensions were very small, close to the Planck size or
perhaps somewhat larger, but undetectable in conceivable experiments.

However, it was suggested by Antoniadis~\cite{IA} that an extra dimension might be a good
way to break supersymmetry, in which case its size would be $\sim 1/$~TeV, in which
case it might have some observable manifestations at the LHC.

Another suggestion, discussed in Lecture 2, was the possibility that boundary conditions
in an extra dimension might be used to break the electroweak gauge symmetry. In this
case also, the size of the extra dimension should be $\sim 1/$~TeV, and potentially
detectable at the LHC~\cite{Rai:2005vy,Barbieri:2002uk,Gunion:2000gy}.

Arkani-Hamed, Dimopoulos and Dvali (ADD)~\cite{ADD} went even further, observing that the 
Higgs mass hierarchy problem might be addressed in models with large extra dimensions,
if they were of a millimetre or micron in size. Because the extra dimensions are so
large in the ADD framework, their effects might be measurable even in low-energy
table-top experiments. These models can be embedded in string theory framework,
as discussed in Ref.~\cite{AADD}. The main ingredients of the simplest ADD scenario 
are~\cite{Gabadadze:2003ii}:

\begin{itemize}
\item The particles of the SM live on a 3-brane, while gravity spreads to all 4+N dimensions;

\item There is a new fundamental scale of gravity in extra dimensions, $M_{*}$,
which together with the ultraviolet completion scale of the SM is around a few TeV or so, 
thus eliminating the Higgs mass hierarchy problem;

\item  $N$ extra dimensions are compactified.
\end{itemize}
If we define in this context the 4-dimensional Planck mass
\beq
M^2_{Pl}=M^{2+N}_{*}(2\pi L)^N,
\label{ADDformula}
\eeq 
and postulate that the quantum gravity scale $M_{*} \sim$~TeV, we can estimate the size of 
the extra dimensions to be
\beq 
L\sim 10^{-17+30/N} \textrm{cm}\ .
\eeq 
For one extra dimension, $N=1$, we obtain $L\sim 10^{13}$~cm, which is excluded within the 
ADD framework, because gravity would have become higher-dimensional at distances
$\sim 10^{13}$~cm. On the other hand, for $N=2$ we get $L\sim 10^{-2}$~cm. 
This case is very interesting, because it predicts a modification of the 4-dimensional 
laws of gravity at submillimeter distances --- which has become the subject of active experimental 
studies~\cite{Gabadadze:2003ii}. For larger $N$, the value of $L$ should decrease
but, even for $N=6$, $L$ is very large compared to $1/M_{P}$.

Randall and Sundrum (RS) went much further still~\cite{RS2}, showing that a model with an
{\it infinite} warped extra dimension could provide an attractive way to reformulate the 
hierarchy problem. In this scenario, 4-dimensional gravity on a brane is obtained
through the phenomenon of localization of gravity. The brane is embedded in a
5-dimension bulk space with negative cosmological constant. In this case we find a relation 
between the 4-dimensional Planck mass and $M_{*}$
\beq
M^2_{Pl}=M^{3}_{*}(2 L).
\eeq 
This is similar to the relation between the fundamental scale $M_{*}$, the size $L$
of the extra dimension, and the Planck mass $M_{P}$ in the ADD model with one extra 
dimension (\ref{ADDformula}). This similarity is based on the fact that in both theories the 
effective size of the extra dimension that is felt by the zero-mode graviton is finite and $\sim L$.

So, are extra dimensions very small, small, large or infinite, and how do we tell?
There are several ways to search for extra dimensions in experiments
at the TeV scale at the LHC.

Typical examples in theories with TeV-scale extra
dimensions are the appearance of Kaluza--Klein
excitations, corresponding to particle wave functions that wrap themselves around
the extra dimension. These show up as resonances that can appear in cross sections
at specific energies related to the compactification scale. These Kaluza--Klein
excitations occur in `towers' that can be understood by analogy with a 
quantum-mechanical particle in a potential well. Its energy is quantized due to the 
boundary conditions at the walls of the well. In our case, the supplementary 
dimension plays the role of the wall of the well.

In models with very large extra dimensions, there are many Kaluza--Klein
excitations of the graviton, which may be detectable {\it via} missing-energy events.

\begin{figure}[htb!]
\resizebox{8.1cm}{!}{\includegraphics{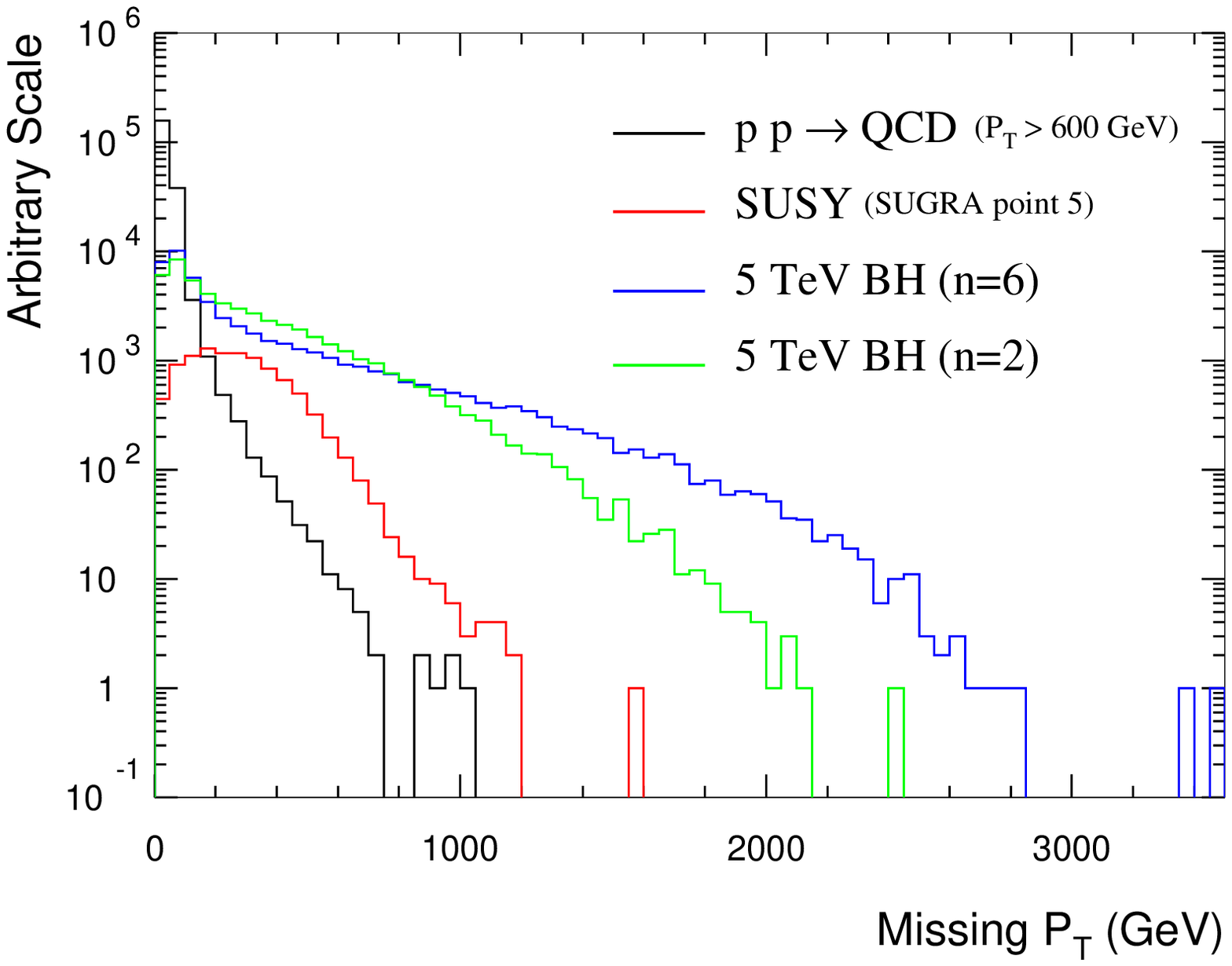}}
\resizebox{7.9cm}{!}{\includegraphics{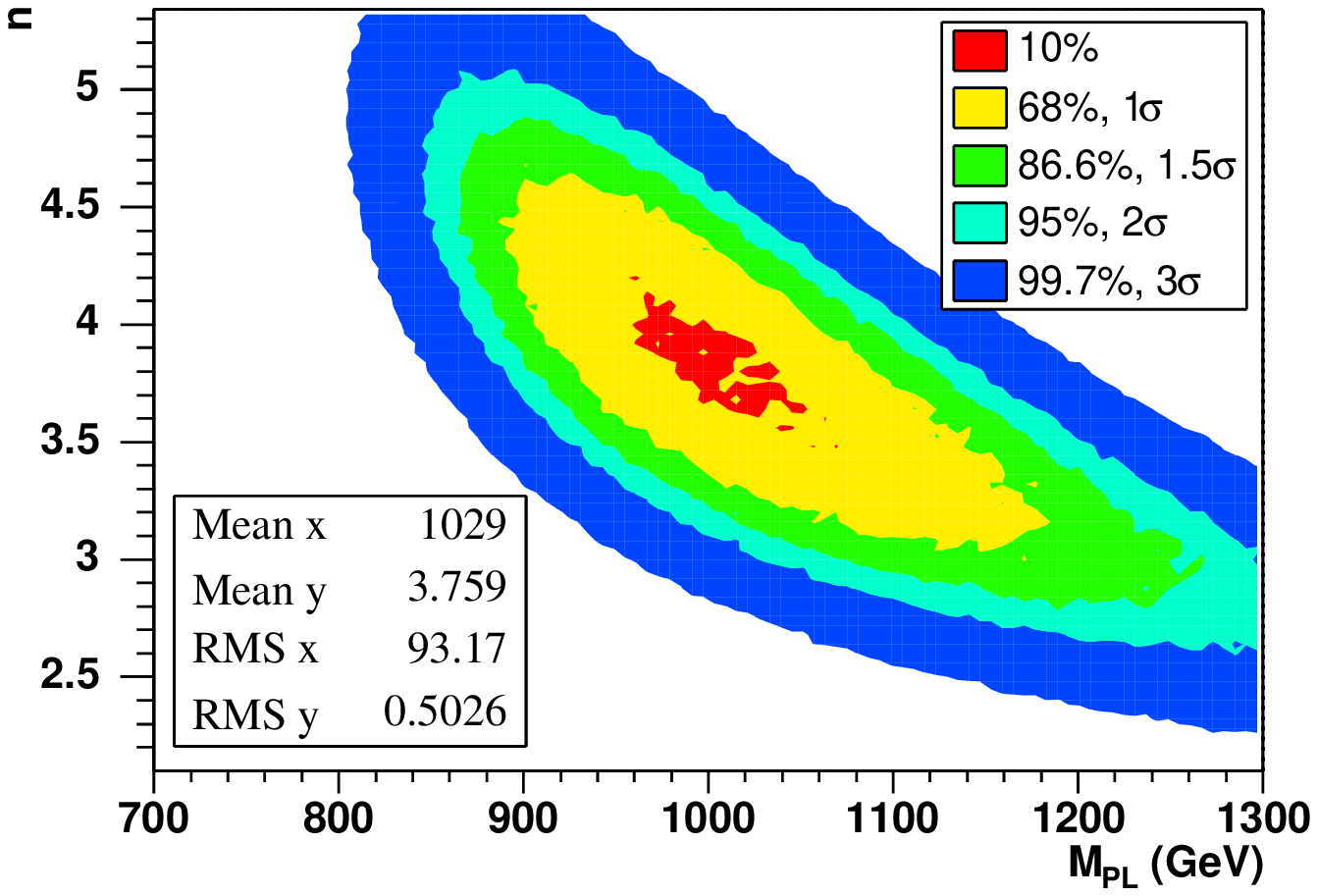}}
\caption{Left: a comparison of the missing transverse momentum spectra in the SM, in a 
typical supersymmetric model, and in two black hole scenarios, and right: the results of a fit to
the number of extra dimensions $n$ and the higher-dimensional Planck mass $M_{PL}$
on the basis of simulated black hole production at the LHC, taken from Ref.~\cite{Cambridge}.
}
\label{fig:CamBH}
\end{figure}

Another speculative possibility is the creation of a microscopic black hole~\cite{LHCBH}. Any 
concentration of energy or mass $m$ will be transformed into a black hole if it is squeezed 
below its Schwarzschild radius: $G/m$. The larger the mass, the easier it can
be squeezed below its Schwarzschild radius. Moreover, as we have seen, 
extra dimensions can increase the value of $G$. Hence, if there are a few extra dimensions 
of sufficient size, it is conceivable that collisions in the LHC might squeeze a
pair of partons below their combined Schwarzschild radius, and hence create a
microscopic black hole. These should evaporate rapidly, since Hawking radiation 
implies that the black hole loses energy at a rate inversely proportional to its mass. 
Studies performed by the CMS~\cite{CMS} and ATLAS~\cite{ATLAS} 
collaborations have demonstrated that
such Hawking radiation would be visible in the LHC {\it via} energetic jets, leptons and photons,
as well as missing energy carried away by neutrinos. See Fig.~\ref{fig:CamBH} for some
results for simulated black hole production at the LHC~\cite{Cambridge}.

\subsection{And now for something completely different?}

In 1982, Prime Minister Thatcher of the United Kingdom visited CERN: I was
placed in the receiving line, and introduced as a theoretical physicist. ``So what
do theoretical physicists {\it do}?'' she boomed. I replied that ``We think of things for the experimentalists
to look for, and we hope they find something different''. Mrs Thatcher was not sure
about this, and asked ``Wouldn't it be better if they found what you had predicted?''
My response was that ``In that case, we would not be learning anything new.''
In the same spirit, let us hope that new experiments, particularly at the LHC, will
soon reveal new physics beyond the Standard Model. Perhaps it will look something
like the possibilities discussed in these Lectures, but let us hope that it will take us
beyond the beyonds imagined by theorists.


\end{document}